# Error-Tolerant Big Data Processing

Dissertation Submitted to

**Tsinghua University**

in partial fulfillment of the requirement

for the degree of

**Doctor of Philosophy**

in

**Computer Science and Technology**

by

**Deng Dong**

Dissertation Supervisor : Professor Li Guoliang

**June, 2016**

# 大数据处理中的容错技术研究

（申请清华大学工学博士学位论文）

培 养 单 位：计 算 机 科 学 与 技 术 系

学　　　科：计 算 机 科 学 与 技 术

研　究　生：邓　栋

指 导 教 师：李　国　良　副　教　授

二〇一六年六月

# Abstract

Real-world data contains various kinds of errors, e.g., typos, different formats and inconsistencies. Before analyzing data, one usually needs to process the raw data to get useful data. However, traditional data processing based on exactly match often misses lots of valid information or introduces data with errors. To get high-quality analysis results and fit in the big data era now, this thesis studies the error-tolerant big data processing. As most of the data in real world can be represented as a sequence or set, this thesis utilizes the widely-used sequence-based similarity functions and set-based similar functions to tolerate errors in data processing and studies the approximate entity extraction, similarity join and similarity search problems. The main contributions of this thesis include:

**1. Approximate Entity Extraction**: This thesis proposes a unified framework to support approximate entity extract on both sequence-based and set-based similarity functions simultaneously. Based on this unified framework, this thesis designs efficient filtering algorithms to avoid unnecessary computation and share the other computation. This thesis also proposes efficient and effective pruning techniques to further improve the performance. The experiments show that the method proposed in this thesis can improve the state-of-the-art methods by 1 to 2 orders of magnitude.

**2. Similarity Join**: This thesis designs two partition-based methods respectively for the sequence similarity join and the set similarity join. For the sequence similarity join, this thesis proposes to evenly partition each sequence to multiple segments and guarantees only if a sequence contains a subsequence that matches a segment of another sequence, they can be similar. This thesis proposes effective subsequence selection methods to generate candidates. This thesis proves that the proposed subsequence selection method can select the minimum number of subsequences among all the methods satisfy completeness. This thesis develops an extension-based method and early-termination techniques for efficient candidate verification. For the set similarity join, this thesis proposes to partition all the sets into segments based on the universe and guarantees two sets are similar only if they share a common segment. This thesis proposes to use the mixture of segments and their 1-deletion neighborhoods to improve the pruning power. This thesis evaluates the allocation strategy of the mixture and designs a dynamical programming method to choose the optimal allocation. This thesis develops a greedy algorithm with approximate ratio of 2 and adaptive grouping mechanism to speedup the allocation selection. These two techniques together reduce the time complexity of selecting the allocation for a set with size $s$ from $O(s^3)$ to $O(s \log s)$. This thesis further extends the two partition-based methods to support the large-scale data processing framework, MapReduce and Spark. The partition-based method won the string similarity join competition held by EDBT and beat the second place by 10 times.

**3. Similarity Search**: This thesis proposes a pivotal prefix filter technique to solve the sequence similarity search problem. This thesis shows that the pivotal prefix filter has stronger pruning power and less filtering cost compared to the state-of-the-art filters. This thesis designs a dynamical programming method to efficiently select high-quality pivotal prefix, which is effective for the non-consecutive errors. This thesis also proposes an alignment filter, which is effective for the consecutive errors. The experiments show that these two techniques can prune most of the dissimilar sequence and improve the performance of the state-of-the-art methods.

# 摘 要

现实世界中的数据存在着各种各样的错误，例如拼写错误、格式错误、数据不一致等。在分析数据之前，往往需要先对原始数据进行处理和转化，从而得到可用的数据。传统数据处理方式可能丢失很多有效信息甚至引入错误信息，为了得到最佳的分析结果并适应当今大数据时代的需求，论文研究大数据处理中的容错技术。鉴于现实世界中的数据很多都可以用序列或者集合表示，论文利用广泛应用的序列相似函数和集合相似函数来容忍数据的错误。针对数据处理的最典型三个操作：抽取、连接、检索，论文研究了近似抽取、近似连接和近似检索技术来实现错误容忍的数据处理，并设计了高效的索引和算法。论文的主要贡献包括：

1. **近似抽取**：论文提出了一个统一的框架来同时支持序列相似函数和集合相似函数下的近似抽取。基于该统一框架，论文设计了高效的过滤算法来避免不必要的计算并设计了堆算法来共享计算。论文提出了快速有效的剪枝策略来进一步提高抽取性能。实验表明，论文提出的方法比现有最好的方法快 1-2 个数量级。

2. **近似连接**：论文设计了一个基于划分的框架来支持近似连接，并针对序列相似函数和集合相似函数设计了高效的连接算法。对于序列近似连接，论文把序列平均划分为不相交的片段并保证仅当一个序列的子序列与另一个序列的片段匹配时它们才可能相似。论文提出了有效的子序列选取技术并证明了该技术选取的子序列数量是最少的。论文提出了扩展验证技术来快速验证候选结果。对于集合近似连接，论文根据全集把集合划分为不相交的片段（子集），提出混合使用片段和 1-删集（移除片段中 1 个元素后的子集）来提高过滤能力，设计了动态规划算法来选取最优的混合分配，提出了近似比为 2 的贪心算法和多长度分组机制来把分配选取时间复杂度从 $O(s^3)$ 降低到 $O(s \log s)$，其中 $s$ 是集合大小。论文扩展了这两个算法以运行在 MapReduce 和 Spark 上来支持大数据的近似连接。基于划分的算法在 EDBT 大数据融合竞赛中以绝对优势取得冠军，并且效率比亚军高 10 倍。

3. **近似检索**：论文提出了一个关键前缀过滤技术来解决基于序列相似性的近似检索问题。相比现有最好的前缀过滤技术，关键前缀过滤技术不但剪枝能力更强而且过滤代价更小。论文设计了动态规划算法来快速选取高质量的关键前缀，能够更好的检测序列中离散的错误。论文还提出了一个对齐过滤技术，能够检测序列中连续的错误。实验表明，关键前缀过滤技术能够过滤掉绝大部分不相似的序列并大幅提高了现有过滤技术的性能。

# 目 录

# 第 1 章　绪论

## 1.1　选题背景和意义

现在正处于数据洪流之中，在各个领域里，数据正在以空前的规模被采集。之前基于精细的模型或者猜测作出的决定现在可以通过大数据分析产生。大数据分析正在驱动着现代社会的方方面面，例如移动服务、零售业、制作业、金融业、生命科学以及医疗科学等。根据数据库研究者的共识[1–3]，大数据至少具有数量大（Volumn）、来源多（Variety）以及更新快（Velocity）三个特性。

然而，现实世界中的数据存在着各种各样的错误。例如，在通过光学字符识别（Optical Character Recognition, OCR）数字化的文本集合中包含 7% 到 16% 的错误，在打字（1% 到 3.2%）与拼写（1.5% 到 2.5%）的过程中也存在着大量的错误[4]。在一个由荷兰人输入荷兰人姓氏的实验中，实验者的拼写错误率高达 38%[4]。此外，现在随着手持设备、迷你键盘和虚拟键盘的普及，拼写错误在现实世界中也更加普遍[5,6]。另一方面，大数据来源多的特性加剧了错误蔓延。有研究表明，即使是人们通常认为比较可靠的股票数据和航班数据中，也存在着大量的不一致信息。例如，在一个月内，通过 55 个数据源采集的股票信息和 38 个数据源采集的航班信息中，70% 的信息拥有两种或两种以上不同的值[7]。此外，不同的数据格式也会导致数据的不一致错误，例如日期和姓名的不一致表示格式等。根据一个对工业界的调研[8]，世界顶级公司中超过 25% 的关键数据包含显著的数据错误。

在进行数据分析之前，人们往往需要对原始数据进行处理转化，从而得到可用的数据。然而，由于错误的普遍存在，人们在进行数据处理的过程中可能丢失很多有效的信息或者引入很多错误的信息，从而导致分析结果质量下降，甚至得到错误的结论，从而造成严重后果。例如，根据一项来自 Experian QAS Inc. 的研究[9]，在 2011 年，因为低下的客户数据质量，英国商业损失了超过 80 亿英镑的收入。因此，本文希望在数据处理过程中能够容忍错误。为了达到这一目的，首先需要衡量数据的相似性。

现实世界中很多数据都可以用序列或者集合的形式来表示。例如姓名、DNA 和蛋白质等都可以用*序列* (sequence) 来表示，其中元素有顺序。字符串也是一种序列，本文不加区分的使用字符串和序列；文档、向量和图像等都可以用*集合* (set) 来表示，其中的元素没有顺序，可以加权。近百年来，研究者们提出了多种不同的相似函数[10]，其中最常用的包括衡量两个序列近似度的编辑距离 (Edit Distance), Edit Similarity 和汉明距离 (Hamming Distance) 等，以及衡量两个集合近似度的 Jac-

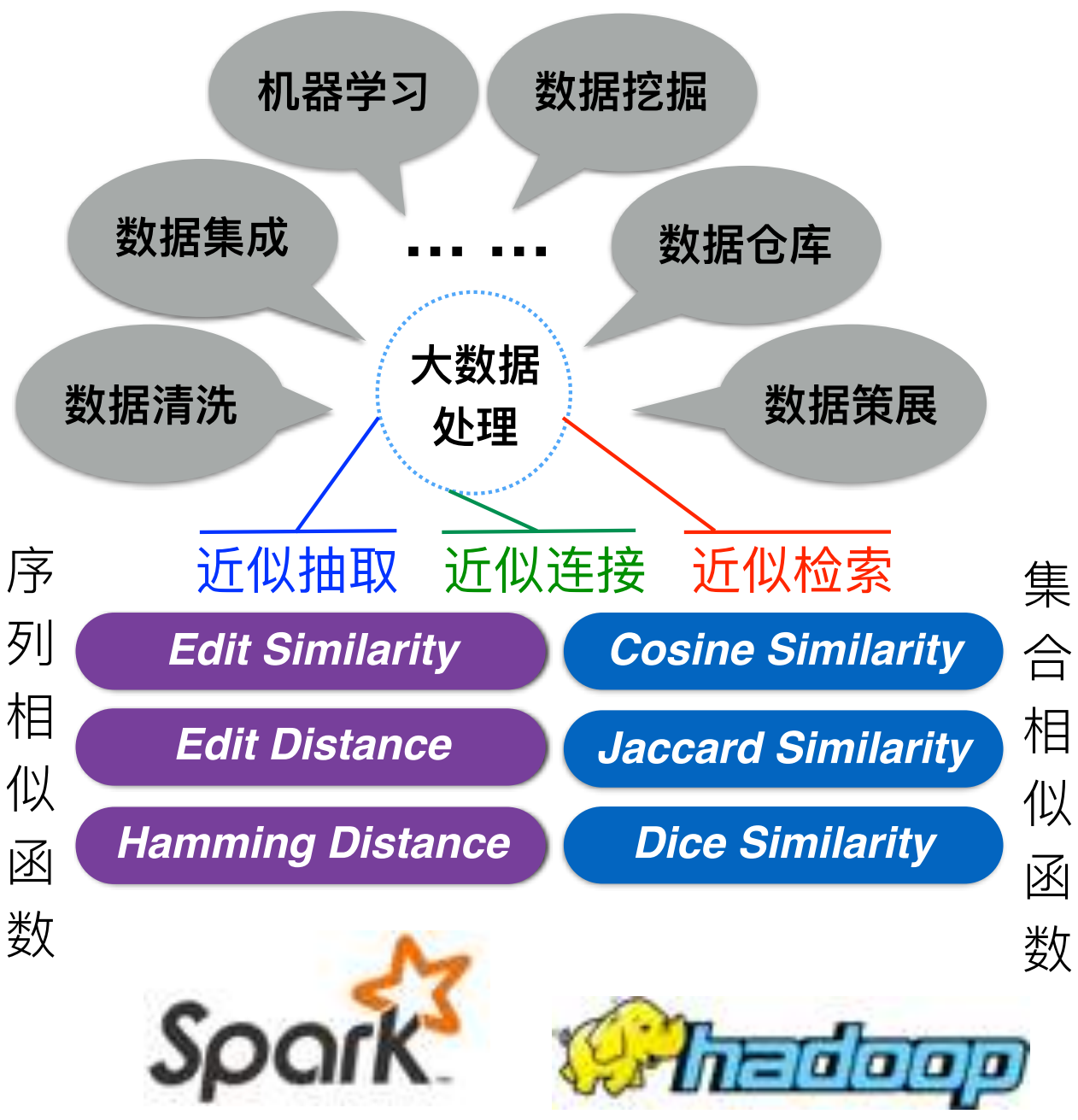

图 1.1　容错数据处理系统架构图

card Similarity, Cosine Similarity, Dice Similarity 和 Overlap Similarity。这些相似函数被证明是十分有效的。例如，很多广泛应用的系统都支持编辑距离，比如数据库系统 `Oracle`[11] 和 `PostgreSQL`[12]，信息检索系统 `Lucene`[13] 和 `Elasticsearch`[14]，机器翻译系统 `OmegaT`[15]，数据清洗系统 `OpenRefine`[16]（前身为 `GoogleRefine`[17]）和 `DataWrangle`[18,19]（`Trifacta`[20] 的前身）以及 Unix 文件对比工具 `diff`[21]。又比如，Spertus 等[22] 表明 Cosine Similarity，又叫 L2-norm，在为 Orkut[23] 社交网络用户推荐在线社区任务中击败了其它五种近似函数，取得了最好的实验结果。Word2Vec[24] 使用 Cosine Similarity 来衡量两个特征向量的距离。

总体来说，数据处理是数据收集和控制从而产生有意义的信息的过程，其中抽取、连接以及检索是数据处理中的三个重要的基本操作，在数据策展、数据仓库、数据清洗、抽取转换加载 (ETL) 中都有极为重要的应用。因此本文主要研究这三个数据处理操作中的容错问题，即基于序列相似性和基于集合相似性的近似抽取、近似连接和近似检索问题。

另一方面，随着现在数据量的增大，单机算法已经很难处理全部的数据了，为此，并行数据处理显得十分迫切和必要。为了简化并行计算编程并提高计算的可扩展性，研究者们设计并实现了多种大规模数据处理系统，例如 MapReduce 和 Spark 等。本文还讨论了如何在流行的分布式数据处理框架 MapReduce 和 Spark 上解决近似连接问题。图1.1展示了本文提出的容错数据处理系统架构图。

### 1.1.1 近似抽取

基于字典的实体抽取可以从文档中识别出字典中预先给定的实体（比如人名和地名等）。例如，考虑一个推文（Tweet）文档 "*I'm in the mood for something Coka-Cola flavored but I don't want an actual Coke*'' 和一个包含两个实体 "`Coca Cola`'' 和 "`Coke`'' 的字典。基于字典的实体抽取可以从该文档中找出预先定义的实体 "`Coke`''。这个问题在现实世界中有很多的重要应用场景，例如命名实体识别、数据清洗、分子生物学、生物信息学（DNA 以及蛋白质的序列检测）和自然语言处理等[25]。

然而，很多文档中存在排版错误或者拼写错误，而且同一个实体也可能有不同的表示形式[26]。例如，以上文档中的子字符串 "`Coka-Cola`'' 就存在拼写错误。传统的基于完全匹配的实体抽取从这个文档中很难找到这个子字符串，因为这个子字符串与预定义的实体 "`Coca Cola`'' 并不完全匹配。为了解决这个问题，现在的一个趋势是支持近似实体抽取，即找到文档中所有和给定字典中的实体近似的子字符串

为了提高实体抽取的召回率，论文研究基于字典的近似实体抽取问题。给定一个包含多个实体的字典和一个文档，基于字典的近似实体抽取问题找到该文档中所有与字典中某个实体近似的子字符串。为了衡量两个字符串的相似度，人们提出了很多相似函数，例如 Jaccard Similarity 和编辑距离。在以上例子中，假如使用编辑距离函数来衡量相似度，近似实体抽取可以找到子字符串 "*Coka-Cola*''，它与实体 "`Coca Cola`'' 相似。

尽管近似实体抽取存在很多已有工作[26,27]，但是它们要么只支持基于标记集合的近似函数（例如 Jaccard Similarity ），要么只支持基于字符序列的近似函数（例如编辑距离）。设计一种统一的解决方案来支持不同的近似函数是非常有必要的，这是因为一个统一的解决方案不但能够减少编程工作量，而且可以降低硬件要求以及节省维护多份代码的人力资源。此外，很多现实世界中的应用对于近似实体抽取都有高性能的要求。例如，在数据流应用中 (例如，微博、推文和在线广告)，显然高性能的近似实体抽取是非常重要的。比如，产品分析师（如：*Coca Cola* 员工) 希望在推特上看到所有与他们产品相关的推文和评论。在这种情况下，产品分析师可能会注册一些与他们产品相关的实体，近似实体抽取系统必须能够快速的把相关推文传送给产品分析师。

此外，文档中很多子字符串都有交叠，例如，考虑以上文档，很多子字符串 (例如 "*Coka-Cola*", "*oka-Cola*'' 和 "*Cola flavored*") 都有交叠（例如 "*Cola*''）。论文可以利用这个特性来避免不同子字符串重叠部分的冗余计算。例如，通过在不同

的子字符串之间共享 “*Cola*” 部分的计算可以提高性能。

为了达到以上目的，论文在第 2 章提出了一个统一的框架来同时解决不同类型的相似函数下的基于字典的近似实体抽取问题。为了避免冗余计算，论文研究了高效的过滤算法。论文还提出了有效的剪枝技术来进一步提高性能。实验结果表明，论文提出的方法能够大幅提高现有最好算法的性能。

### 1.1.2　近似连接

论文设计了一个基于划分的框架来支持近似连接，并针对序列相似函数和集合相似函数设计了高效的连接算法。

序列相似函数下的近似连接（字符串近似连接）找到两个字符串集合中所有的相似字符串元组。例如，考虑两个字符串集合 {vldb, sigmod, . . . } 和 {pvldb, icde, . . . }，字符串近似连接找到其中所有的近似元组，比如 ⟨vldb, pvldb⟩。人们提出很多相似函数来衡量两个字符串之间的相似度，论文主要关注编辑距离和 Edit Similarity，即给定两个字符串集合，字符串近似连接找到其中所有编辑距离或者 Edit Similarity 在给定的阈值内的字符串元组。字符串近似连接在很多应用中都是一个关键操作，例如数据集成，数据清洗，抄袭检测以及协同过滤等。

解决字符串近似连接问题的现有方法大体上被分为两类。第一类方法使用过滤加验证的框架，例如 PartEnum[28]，All-Pairs-Ed[29] 和 ED-Join[30]。在过滤步骤，它们为每个字符串生成一个特征（signature）集合并使用这些特征来产生候选元组。在验证步骤，它们验证候选元组以产生最终结果。然而，这些方法对短字符串数据集（例如人名和地名等）效果较差[31]，主要原因是它们很难从短字符串中选取出高质量的特征，因此会产生大量的候选元组，导致验证时间过长。第二类方法（TrieJoin[31]）采用基于 trie 的框架，它们使用 trie 结构来索引字符串并共享相同前缀上的计算，它们利用前缀剪枝来提高性能。然而 TrieJoin 对于长字符串数据集（例如文章名字或者摘要等）十分低效，这是因为长字符串的共同前缀的长度相对来说比较短。

如果一个系统希望同时支持短字符串数据集和长字符串数据集，它可能需要实现和维护两份代码并调整许多参数来选择最好的方法。为了解决这个问题，一个能够同时高效的支持短字符串和长字符串的算法是非常必要的。论文设计了一个基于划分的方法，该方法把字符串划分为多个片段并保证如果一个字符串 $s$ 与另一个字符串 $r$ 相似，那么 $s$ 一定有一个子字符串与 $r$ 的一个片段完全匹配。根据这个特性，论文提出了一个基于划分的框架来解决字符串近似连接问题，并称之为 PassJoin。PassJoin 首先为所有片段构建一个倒排索引，然后对每个字符串 $s$，

PassJoin 选择它的一些子字符串并利用这些子字符串去倒排索引中检索相对应的倒排列表。这些子字符串对应的倒排列表中的每个字符串 $r$ 都可能与 $s$ 相似（因为 $r$ 也包含这个子字符串），因此 PassJoin 把 $r$ 和 $s$ 作为一个候选元组。接下来 PassJoin 验证这些元组并产生最终结果。论文提出了有效的技术来选择子字符串并证明在保证不丢失结果的前提下，论文提出的子字符串选取方法选取的子字符串数量是最少的。论文还设计了新颖的剪枝技术来快速验证候选元组，包括一个基于扩展的验证方法、提早终止验证技术以及共享计算技术。

集合相似函数下的近似连接（集合串近似连接）从两组集合中找到所有的近似集合元组。集合近似连接在很多应用中扮演至关重要的角色，例如个性化推荐和协同过滤[32,33]、实体解析[34]、重复检测[35]、数据清洗[36]、数据集成[37] 以及机器学习[38] 等。集合相似函数被应用于衡量两个集合的相似性，当两个集合的相似度超过了给定的阈值时，它们才被认为是相似的。论文主要致力于三种广泛应用的相似函数：Jaccard Similarity 、Cosine Similarity 和 Dice Similarity。

现行的方法当中大部分都采用了基于前缀过滤的框架[29,35,39,40]。前缀过滤首先将每个集合中所有的元素按照全局顺序排序。然后将每个集合中最开始的几个元素作为前缀。这种方法可以保证如果两个集合的前缀不相交，那么它们一定不相似。然而，基于前缀过滤的框架的剪枝能力很有限。因为，一旦两个不相似的集合在前缀中共享一个相同的元素，这些方法就不起作用了。

为了解决这个问题，论文提出了一个基于划分的框架来解决集合近似连接问题。论文提出平均划分全集并据此把所有集合划分为不相交的片段，论文证明仅当两个集合共享一个相同的片段时，它们才可能相似。论文将相同大小的集合分组从而共享计算，并基于这些片段建立倒排索引以快速进行集合近似连接。为了减少所访问的倒排列表的总长度，论文提出混合使用片段与片段的 1-删集（移除片段中一个元素而得到的子集)。因为存在多种策略来分配这个混合，论文评估了不同的分配策略并设计了动态规划算法来选择最优的分配策略。然而为大小为 $s$ 的集合产生最优分配的时间复杂度是 $O(s^3)$，为了加快分配策略的选择，论文提出了一种近似比为 2 的贪心算法来选取分配策略，并设计多长度分组机制。这两种方法共同将分配选择的时间复杂度降低到 $O(s \log s)$。

### 1.1.3　近似检索

序列相似函数下的近似检索（字符串近似检索）是从一个字符串集合中找到与查询字符串近似的所有数据字符串。在数据清洗和集成中，它是一个很重要的操作。由于它在现实世界中广泛的应用，例如拼写检查，抄袭检测，实体连接以及

macromolecules 序列对齐[41–43] 等，它引起了众多研究者的关注。

解决该问题的现有工作通常采用过滤加验证的框架[41–46]。在过滤阶段，它们设计有效的过滤条件，利用过滤条件来剪枝掉大量与查询串不相似的数据字符串从而得到剩下的候选字符串。在验证阶段，它们通过计算候选字符串与查询串的编辑距离来进行验证，进而得到最终的结果。现有方法主要致力于设计有效的过滤条件以取得强效的剪枝能力。前缀过滤[29,36] 是一个重要的过滤技术。它首先为每个字符串产生一个前缀（特征集合）并保证如果两个字符串是相似的，它们的前缀一定共享一个相同的特征。然后，它利用这个特性来过滤大量与查询串没有共同特性的不相似的数据串。最近，研究者们提出了一些技术来改进前缀过滤，比如基于位置的前缀过滤[30,35]，自适应的前缀过滤[39] 以及非对称过滤[47] 等。

值得注意的是特征的数量对剪枝能力和过滤代价有重要的影响。一方面，减少特征的数量将会减少两个字符串之间特征相同的概率，因此剪枝能力将变强。另一方面，减少特征的数量将减少检查两个字符串是否共享相同的特征的比对代价。过滤代价也会减少。因此，减少特征的数量不但可以增加剪枝能力还能降低过滤代价。然而，简单的移除字符串的一些特征将导致结果的丢失，所以需要设计一个有效的方法来减少特征的数量并保证不丢失任何结果。

为了解决这个问题，论文提出了一个关键前缀过滤技术，它可以显著的减少特征的数量并仍然能够找到所有的结果。论文论证了关键前缀过滤比现有的过滤技术在剪枝能力和过滤代价上都要更好。因为存在多种策略来产生关键前缀，论文提出了一个动态规划算法来选择高质量的关键前缀，从而剪枝掉大量因为离散的编辑错误而与查询串不相似的数据串。当数据串中存在大量的连续的编辑错误时，关键前缀过滤以及现有的过滤技术可能产生很多候选字符串，导致验证时间过长。为了解决这个问题，论文提出了一个对齐过滤技术，它通过考虑特征与子字符串之间的对齐情况来进一步剪枝掉大量因为连续的编辑错误而与查询串不相似的数据串。

### 1.1.4 大规模数据处理框架

大数据的另一个特性是体量大。例如，2013 年，全世界每天收发的电子邮件超过 1 千亿封[48]；2012 年，Facebook 每天产生的数据超过 15TB[49]；2014 年，用户每分钟向 Youtube 上传的视频长度长达 100 小时[50]。显然单机的处理能力已经适应不了大数据时代的需求。为了应对大数据的挑战，人们提出了多种大规模数据处理系统，它们通常部署在集群上，集群中每台计算机都有自己的内存和磁盘并负责一部分计算任务，所有的计算机并行的计算并通过网络进行通信。现在主

流的大规模数据处理框架有 MapReduce[51] 和 Apache Spark[52] 两种[53,54]。

MapReduce 是由谷歌（Google[55]）提出并实现的一个在分布式集群环境下处理大规模数据集的框架，它被广泛的应用到很多任务中，例如搜索引擎索引的构建、文档聚类、访问日志分析以及各种其它形式的数据分析中。MapReduce 的数据文件存储在分布式文件系统中，并按照固定大小分块。通常数据块在多个不同节点上有冗余拷贝，因此 MapReduce 框架具有很好的容灾能力。一个 MapReduce 程序通常包括一个 map 函数以及一个 reduce 函数，每个函数都可以在数千个节点上并行的执行。两个阶段中间有个 shuffle 过程，它将各个 map 函数产生的输出分发至各个节点作为 reduce 函数的输入。每个 map 函数读取该节点上的数据块作为输入，执行该用户自定义的 map 函数，并产生一些中间键值对 (intermediate key-value pair)。shuffle 过程根据中间键值对的键来分发，它保证拥有相同键的中间键值对一定会被发送到同一个 reduce 函数中。每个 reduce 节点将收到一个相同键的中间键值对列表，reduce 函数以这个列表作为输入，执行用户自定义的逻辑，最后产生结果键值对并将其存储在分布式文件系统中。Hadoop[56] 是 MapReduce 的开源实现，HDFS 是其相应的文件系统。

然而，MapReduce 不适合执行一些特定类型的任务，例如需要迭代计算的任务等[57–59]。为了解决这个问题，Zaharia 等人[52] 设计并实现了 Spark，一个在集群上进行的基于内存计算的分布式计算框架。Spark 把数据存储抽象为分布式弹性数据集（Resilient Distributed Dataset, RDD）。RDD 是一个横跨多台计算机的记录集合。Spark 任务与 MapReduce 任务很类似：RDD 从 HDFS 中构建并作为 map 函数（或其他函数）的输入，map 函数（或其他函数）被用来处理 RDD 分块中的每条记录。reduce 函数被用来收集数据和做其他计算。Spark 能够缓存中间 RDD 并在之后的迭代中重用它们。此外，这些中间 RDD 能被缓存到各台计算机的内存中以进一步加速计算。

## 1.2　主要研究内容和贡献

1. **近似抽取**：目前支持容错的实体抽取方法只能支持基于集合的近似函数或者基于序列的近似函数，如果有一种方法能够同时支持多种不同近似函数，那么它不但能够减少编程代价和硬件需求还能节省人力资源。因此论文在第 2 章提出了一种高效的统一框架来同时支持基于集合和序列的近似函数，例如 Jaccard Similarity, Cosine Similarity, Dice Similarity, Edit Similarity 以及 Edit Distance 等。此外，因为文档中很多子字符串都有交叠，所以可以共享交叠部分的计算从而避免不必要的重复计算。论文在第 2 章中设计了高效的过滤

算法来共享计算。论文还提出了有效的剪枝技术来提高算法的性能。实验结果表明论文提出的方法取得了高性能并超过了现有的方法。已发表的与近似抽取问题相关的论文[60–63] 如下:

- **Dong Deng**, Guoliang Li, Jianhua Feng, Yi Duan, Zhiguo Gong. *A Unified Framework for Approximate Dictionary-based Entity Extraction.* VLDB Journal 2014:143-167.
- **Dong Deng**, Guoliang Li, Jianhua Feng. *An Efficient Trie-based Method for Approximate Entity Extraction with Edit-Distance Constraints.* ICDE 2012:762-773.
- Guoliang Li, **Dong Deng**, Jianhua Feng. *Faerie: Efficient Filtering Algorithms for Approximate Dictionary-based Entity Extraction.* SIGMOD 2011:529-540.

2. **序列近似连接**：现有的算法只对长的字符串或者短的字符串有效，但是没有一个算法能够同时高效的处理短的字符串和长的字符串。为了解决这个问题，论文提出了一个基于划分的方法（PassJoin）来处理序列近似连接问题。PassJoin 首先把字符串划分为多个片段，然后基于这些片段构建一个倒排索引，之后对于每个字符串，PassJoin 选择它的一些子字符串并使用这些被选取的子字符串来探测倒排索引从而找到近似元组。论文设计了高效的技术来选择子字符串并证明论文提出方法能够最小化选取的子字符串的数量。论文还研究了新颖的剪枝技术来高效的验证候选元组。实验结果表明，论文提出的算法对于短的字符串和长的字符串都非常高效，并比现有最好的方法效率更高。已发表的与序列近似连接相关的论文[64–66] 如下：

- **Dong Deng**, Guoliang Li, Shuang Hao, Jiannan Wang, Jianhua Feng. *MassJoin: A MapReduce-based Algorithm for String Similarity Joins.* ICDE 2014:340-351.
- Guoliang Li, **Dong Deng**, Jiannan Wang, Jianhua Feng. *Pass-Join: A Partition based Method for Similarity Joins.* VLDB 2012:353-364.
- Guoliang Li, **Dong Deng**, Jianhua Feng. *A Partition-based Method for String Similarity Joins with Edit-Distance Constraints.* ACM TODS, 2013,38(2)1-47.

3. **集合近似连接**：现有的集合近似连接算法通常采用前缀过滤框架：它们首先为每个集合产生一个前缀，然后剪枝掉所有前缀不相交的集合元组。然而，这种方法的剪枝能力十分有限，因为两个不相似的集合只要在前缀中共享一个相同的元素，那么它们就不能被剪枝掉。为了解决这个问题，论文提出了一个基于划分的框架，论文设计了一个划分方式来把集合划分为多个子集并保证两个集合只有在它们共享一个相同的子集时才可能相似。为了加强剪枝能力，论文提出在框架中混合使用子集以及 1-删集（删除集合中一个元素

后的子集)。因为存在多种分配策略来产生这个混合，论文评估了不同的分配策略并设计了一个动态规划算法来选择最优的一个。然而，为大小为 $s$ 的一个集合产生最优分配策略的时间复杂度是 $O(s^3)$，为了加速分配策略选择，论文提出了一种近似比为 2 的贪心算法。为了进一步减少复杂度，论文还设计了一个多长度分组机制。这两个技术一起可以把算法的时间复杂度降低到 $O(s\log s)$。已发表的与近似检索相关的论文[67,68] 如下：

- **Dong Deng**, Guoliang Li, He Wen, Jianhua Feng. *An Efficient Partition based Method for Set Similarity Join.* VLDB 2016:360-371
- **Dong Deng**, Yu Jiang, Guoliang Li, Jian Li, Cong Yu. *Scalable Column Concept Determination for Web Tables Using Large Knowledge Bases.* VLDB 2014:1606-1617.

4. **近似检索**：现有的算法都采用基于特征的框架，它们首先为每个字符串产生一些特征，然后剪枝掉所有与查询串没有共同特征的字符串。然而这些算法需要产生大量不必要的的特征。减少特征的数量不但能增加剪枝能力而且能减少过滤代价，为了达到这个目的，本文在第 5 章提出了一个新颖的关键前缀过滤技术，它能够显著的减少特征的数量。论文证明了关键前缀过滤比现有的过滤技术的过滤代价更小而剪枝能力更强。论文提出了一个动态规划算法来选择高质量的关键前缀特征来剪枝掉因为连续的编辑错误而与查询串不近似的字符串。论文还提出了一个对齐过滤技术来过滤大量因为离散的编辑错误而与查询串不近似的字符串。实验结果表明论文提出的方法能够过滤掉绝大部分不相似的字符串并且取得了比现有的方法更高的性能。已发表的与近似检索相关的论文[44,69] 如下：

   - **Dong Deng**, Guoliang Li, Jianhua Feng. *A Pivotal Prefix Based Filtering Algorithm for String Similarity Search.* SIGMOD 2014:673-684.
   - **Dong Deng**, Guoliang Li, Jianhua Feng, Wen-Syan Li. *Top-k String Similarity Search with Edit-Distance Constraints.* ICDE 2013:925-936.

## 1.3 章节安排

论文的章节安排如下。首先在第 2 章研究了基于字典的近似实体抽取问题，论文提出了一个可以同时支持集合相似函数和序列相似函数的统一近似抽取框架。然后在第 3 章和第 4 章提出了基于划分的框架来分别解决序列近似连接问题和集合近似连接问题。接下来在第 5 章介绍了关键前缀过滤技术以及对齐过滤技术来解决序列近似检索问题。最后在第 6 章总结全文并展望未来的工作。

# 第 2 章　支持多相似性函数的近似抽取统一框架

## 2.1　引言

本章研究基于字典的近似实体抽取问题，即给定一个实体字典以及一个文档，找到文档中所有与字典中实体近似的子字符串。这个问题有很多应用，例如命名实体识别、数据清洗、分子生物学、生物信息学和自然语言处理等[25]。比如，图 2.1展示了一个维基百科文档，当用户在维基百科中编辑提交这个文档后，基于字典的实体抽取可以抽取出这个文档中在维基百科里已存在的词条并自动添加超链接（图中蓝色颜色的文字）。由于用户在撰写文档的过程中可能引入一些拼写错误，又或者用户使用的实体名称与维基百科中对应的词条格式不同，传统的基于完全匹配的实体抽取仍然不能取得令人满意的结果，因此，为了提高实体抽取的召回率，论文提出研究近似实体抽取问题。

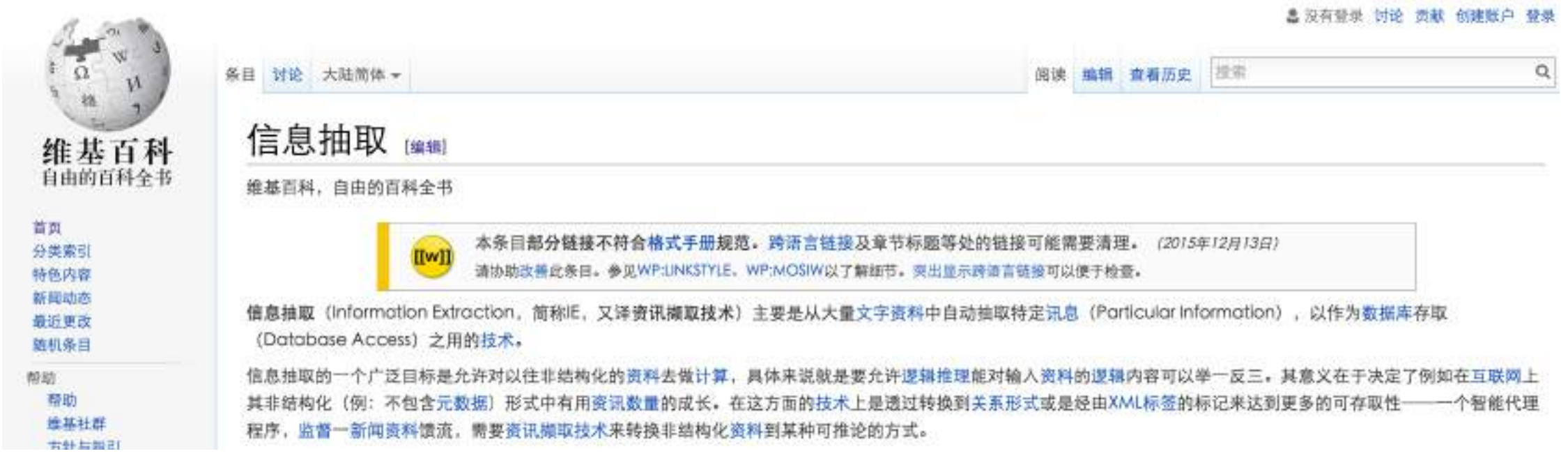

图 2.1　基于实体字典的近似抽取的应用：自动抽取词条

给定一个包含多个实体的字典和一个文档，基于字典的近似实体抽取问题找到该文档中所有与字典中某个实体近似的子字符串。为了衡量数据的相似性，研究者们提出了很多相似函数，例如 Jaccard Similarity 和编辑距离。其中两类最重要的相似函数是集合相似函数与序列相似函数，它们分别被用来衡量两个集合和两个序列的相似度。尽管近似实体抽取存在很多已有工作[26,27]，但是它们要么只支持基于集合相似性的近似抽取，要么只支持基于序列相似性的近似抽取。设计一种统一的解决方案来支持不同的近似函数是非常有必要的。

为了解决这个问题，本章提出一个支持多相似性函数的近似抽取统一框架，它可以同时解决不同类型的相似函数下的基于字典的近似实体抽取问题（第 2.2节）。

为了避免不必要的计算和冗余计算，论文研究了高效的过滤算法（第 2.3节）。论文还提出了基于二分查找的剪枝技术来进一步提高算法性能（第 2.4节）。实验结果表明，论文提出的方法能够大幅提高现有最好算法的性能（第 2.6节）。

## 2.2 统一的框架

### 2.2.1 问题定义

**定义** 2.1 (**基于字典的近似实体抽取**)： 给定一个实体字典 $E = \{e_1, e_2, \ldots, e_n\}$，一个文档 $D$，一个近似函数以及一个阈值，基于字典的近似实体抽取找出所有相似的元组 $\langle s, e_i\rangle$，其中 $s$ 和 $e_i$ 在给定的近似函数下的相似度大于给定的阈值。其中 $s$ 是文档 $D$ 的一个子字符串，$e_i$ 是字典 $E$ 中的一个实体。

两个字符串的相似性一般用相似函数（近似函数）来衡量，本文中主要关注基于字词集合的近似函数：Jaccard Similarity，Cosine Similarity 和 Dice Similarity ；以及基于字符序列的近似函数编辑距离和 Edit Similarity。

**基于集合的近似函数**：基于集合的近似函数可以衡量两个集合的相似度，因此首先将字符串分词为单词集合。代表性的基于集合的近似函数包括 Jaccard Similarity (JAC)，Cosine Similarity (COS) 和 Dice Similarity (DICE)。给定两个字符串 $r$ 和 $s$，用 $|r|$ 来表示 $r$ 中单词的数量，有 $\textsc{jac}(r,s) = \frac{|r\cap s|}{|r\cup s|}$, $\textsc{cos}(r,s) = \frac{|r\cap s|}{\sqrt{|r|\cdot|s|}}$ 和 $\textsc{dice}(r,s) = \frac{2|r\cap s|}{|r|+|s|}$。例如 JAC(“vldb journal 2013’’，“vldb journal”) = $\frac{2}{3}$，COS(“vldb journal 2013’’，“vldb journal”) = $\frac{2}{\sqrt{6}}$，DICE(“vldb journal 2013’’，“vldb journal”) = $\frac{4}{5}$。

**基于序列的近似函数**：编辑距离（编辑距离）是代表性的基于序列的近似函数。两个字符串 $r$ 和 $s$ 的编辑距离是把 $r$ 转化为 $s$ 所需要的最少的单字符编辑操作（包括替换、插入和删除一个字符）的次数，记作 $\textsc{ed}(r,s)$。例如，ED(“surajit”, “surauijt”) = 2。注意，两个字符串的编辑距离越大，它们就越不相似。Edit Similarity 是另一个代表性的基于序列的近似函数。给定两个字符串 $r$ 和 $s$，他们的 Edit Similarity 定义为 $\textsc{eds}(r,s) = 1 - \frac{\textsc{ed}(r,s)}{\max\left(\text{len}(r),\text{len}(s)\right)}$，其中 len($s$) 是字符串 $s$ 的长度。例如，EDS(“surajit”, “surauijt”) = $1 - \frac{2}{8} = \frac{3}{4}$。

在本文中如果两个字符串的 Jaccard Similarity，Cosine Similarity，Dice Similarity 或者 Edit Similarity 不小于给定的阈值 $\delta$；又或者它们的编辑距离不大于给定的阈值 $\tau$，那么就认为这两个字符串是相似的。例如，考虑表 2.1中的文档 $D$ 和字典 $E$，假设给定的编辑距离阈值为 $\tau = 2$, 那么 ⟨“*venkaee sh*", “venkatesh”⟩，⟨“*surauijt ch*", “surajit ch”⟩ 以及 ⟨“*chadhuri*’’，“chaudhuri”⟩ 就是其中三个结果。虽然文档 $D$ 的子字符串 “*chadhurisigmod*’’ 由于笔误缺少了一个空格，论文提出的方法仍然能够找

表 2.1　一个实体字典以及一个文档

(a) 字典 $E$

| ID | 实体字符串 $e$ | $e$ 的长度 | $\|e\|$ (2-gram 的数量) |
|---|---|---|---|
| 1 | kaushik ch | 10 | 9 |
| 2 | chakrabarti | 11 | 10 |
| 3 | chaudhuri | 9 | 8 |
| 4 | venkatesh | 9 | 8 |
| 5 | surajit ch | 10 | 9 |

(b) 文档 $D$

| |
|---|
| *an efficient filter for approximate membership checking. venkaee shga kamunshik kabarati, dong xin, surauijt chadhurisigmod.* |

到其中的实体 “*chadhuri*"（与字典实体 “chaudhuri’’ 相似）。

有研究表明近似实体抽取能够提高实体抽取质量[26]，例如，在蛋白质名称识别任务中，通过近似实体识别技术，召回率能够从 65.4% 提升到 71.4%。本文关注如何提高近似实体抽取的效率并且主要关注抽取文本实体，本章假设阈值（$\delta$ 和 $\tau$）是预先给定的。

### 2.2.2　统一框架

本节提出一个统一的框架来支持多种近似函数。本章把实体和文档看成标记的集合，对于 Jaccard Similarity，Cosine Similarity 以及 Dice Similarity，把文档和实体中的单词作为一个标记；特别地，对于 Edit Similarity 和编辑距离，把文档和实体中字符串的$q$-gram 作为标记。一个字符串 $s$ 的$q$-gram 就是它的长度为 $q$ 的子字符串。$s$ 的$q$-gram 集合是它所有的$q$-gram 组成的集合，记作 $G(s)$。例如 “surajit_ch” 的 2-gram 集合就是 {su, ur, ra, aj, ji, it, t_, _c, ch}。在下文中，如果没有歧义，不加区分的使用 $e$ 和 $G(e)$。因此，用 $e \cap s$ 表示 $G(r) \cap G(s)$，用 $|e|$ 表示 $|G(e)|$，也就是说 $|e| = len(e) - q + 1$。

给定一个实体 $e$ 和一个子字符串 $s$，首先把各种相似函数都转化为 *Overlap Similarity*($|e \cap s|$)，然后用 Overlap Similarity 作为一个统一的过滤条件：如果 $e$ 和 $s$ 相似，那么 $|e \cap s|$ 一定不小于阈值 $T > 0$。不同的近似函数的阈值 $T$ 如下：

- Jaccard Similarity: $T = \lceil (|e| + |s|) * \frac{\delta}{1+\delta} \rceil$。
- Cosine Similarity: $T = \lceil \sqrt{|e| \cdot |s|} * \delta \rceil$。
- Dice Similarity: $T = \lceil (|e| + |s|) * \frac{\delta}{2} \rceil$。
- 编辑距离: $T = \max(|e|, |s|) - \tau * q$。
- Edit Similarity: $T = \lceil \max(|e|, |s|) - (\max(|e|, |s|) + q - 1) * (1 - \delta) * q \rceil$。

引理 2.1给出了这些阈值的正确性证明。

**引理** 2.1： 给定一个实体 $e$ 和一个子字符串 $s$，有：

- *Jaccard Similarity:* 如果 $\textsc{jac}(e,s) \geq \delta$，那么 $|e \cap s| \geq \lceil (|e| + |s|) * \frac{\delta}{1+\delta} \rceil$。
- *Cosine Similarity:* 如果 $\cos(e,s) \geq \delta$，那么 $|e \cap s| \geq \lceil \sqrt{|e| \cdot |s|} * \delta \rceil$。
- *Dice Similarity:* 如果 $\textsc{dice}(e,s) \geq \delta$，那么 $|e \cap s| \geq \lceil (|e| + |s|) * \frac{\delta}{2} \rceil$。
- 编辑距离*:* 如果 $\textsc{ed}(e,s) \leq \tau$，那么 $|e \cap s| \geq \max(|e|,|s|) - \tau * q$。
- *Edit Similarity:* 如果 $\textsc{eds}(e,s) \geq \delta$，那么 $|e \cap s| \geq \lceil \max(|e|,|s|) - (\max(|e|,|s|) + q - 1) * (1-\delta) * q \rceil$。

**证明** (1) Jaccard Similarity：因为 $\textsc{jac}(e,s) = \frac{|e \cap s|}{|e \cup s|} = \frac{|e \cap s|}{|e|+|s|-|e \cap s|} \geq \delta$, 有 $|e \cap s| \geq (|e| + |s|) * \frac{\delta}{1+\delta}$。又因为 $|e \cap s|$ 是一个整数, 有 $|e \cap s| \geq \lceil (|e| + |s|) * \frac{\delta}{1+\delta} \rceil$。
(2) Cosine Similarity：因为 $\cos(e,s) = \frac{|e \cap s|}{\sqrt{|e| \cdot |s|}} \geq \delta$，所以 $|e \cap s| \geq \lceil \sqrt{|e| \cdot |s|} * \delta \rceil$。
(3) Dice Similarity：因为 $\textsc{dice}(e,s) = \frac{2|e \cap s|}{|e|+|s|} \geq \delta$，所以 $|e \cap s| \geq \lceil \frac{|e|+|s|}{2} * \delta \rceil$。
(4) 编辑距离：仅当两个字符串 $r$ 和 $s$ 共享足够多$q$-gram 时，它们才可能相似[70]。
(5) Edit Similarity：因为 $\textsc{eds}(e,s) = 1 - \frac{\textsc{ed}(e,s)}{\max(\mathrm{len}(e),\mathrm{len}(s))} \geq \delta$, $\textsc{ed}(e,s) \leq \max(\mathrm{len}(e), \mathrm{len}(s)) * (1-\delta)$。根据编辑距离的定义, 有 $|e \cap s| \geq \max(|e|,|s|) - \max(\mathrm{len}(e), \mathrm{len}(s)) * (1-\delta) * q$, 所以 $|e \cap s| \geq \lceil \max(|e|,|s|) - (\max(|e|,|s|) + q - 1) * (1-\delta) * q \rceil$。

因此，可以把各种相似函数转化为 Overlap Similarity 并给出一个统一的过滤条件：如果 $|e \cap s| < T$，可以过滤掉元组 $\langle e, s \rangle$。特别的，给定一个相似函数和一个相应的阈值，只要能够推导出两个字符串在该相似度函数及其阈值下对应的 Overlap Similarity 阈值，本章提出的方法就可以应用于该相似度函数。以上研究的五种相似度函数经常被用于信息抽取和记录链接[26,27] 问题中。

### 2.2.3 有效子字符串

有个观察是文档 $D$ 中的一些子字符串不可能与任何实体相似。例如，考虑表2.1中的文档和实体字典，假设使用编辑距离并设定阈值$\tau = 1$。子字符串 "*surauijt chadhurisigmod*" 的长度为 23，然而实体字典中所有实体的长度都在 9 到 11 之间，因此这个子字符串不可能与任何实体相似。接下来讨论如何过滤掉这样的子字符串。

给定一个实体 $e$ 以及一个子字符串 $s$，如果 $s$ 和 $e$ 相似，那么 $s$ 的标记数目 $|s|$ 一定在范围 $[\bot_e, \top_e]$ 内，也就是说 $\bot_e \leq |s| \leq \top_e$。其中 $\bot_e$ 和 $\top_e$ 分别是 $|s|$ 的上界和下界，它们在不同的相似度函数下的值为：

- Jaccard Similarity: $\bot_e = \lceil |e| * \delta \rceil$，$\top_e = \lfloor \frac{|e|}{\delta} \rfloor$。

- Cosine Similarity: $\perp_e = \lceil |e| * \delta^2 \rceil$，$\top_e = \lfloor \frac{|e|}{\delta^2} \rfloor$。
- Dice Similarity: $\perp_e = \lceil |e| * \frac{\delta}{2-\delta} \rceil$，$\top_e = \lfloor |e| * \frac{2-\delta}{\delta} \rfloor$。
- 编辑距离: $\perp_e = |e| - \tau$，$\top_e = |e| + \tau$。
- Edit Similarity: $\perp_e = \lceil (|e| + q - 1) * \delta - (q - 1) \rceil$，$\top_e = \lfloor \frac{|e|+q-1}{\delta} - (q - 1) \rfloor$。

其中 $\delta$ 是 Jaccard Similarity、Cosine Similarity、Dice Similarity 以及 Edit Similarity 下的阈值，$\tau$ 是编辑距离下的阈值。引理 2.2给出了这些界的正确性证明。

**引理** 2.2:　给定一个实体 $e$ 和任意一个子字符串 $s$，有:

- *Jaccard Similarity:* 如果 $\text{JAC}(e, s) \geq \delta$，那么 $\lceil |e| * \delta \rceil \leq |s| \leq \lfloor \frac{|e|}{\delta} \rfloor$。
- *Cosine Similarity:* 如果 $\text{COS}(e, s) \geq \delta$，那么 $\lceil |e| * \delta^2 \rceil \leq |s| \leq \lfloor \frac{|e|}{\delta^2} \rfloor$。
- *Dice Similarity:* 如果 $\text{DICE}(e, s) \geq \delta, \lceil |e| * \frac{\delta}{2-\delta} \rceil \leq |s| \leq \lfloor |e| * \frac{2-\delta}{\delta} \rfloor$。
- 编辑距离*:* 如果 $\text{ED}(e, s) \leq \tau$，那么 $|e| - \tau \leq |s| \leq |e| + \tau$。
- *Edit Similarity:* 如果 $\text{EDS}(e, s) \geq \delta$，那么 $\lceil (|e| + q - 1) * \delta - (q - 1) \rceil \leq |s| \leq \lfloor \frac{|e|+q-1}{\delta} - (q - 1) \rfloor$。

**证明** (1) Jaccard Similarity：因为 $|e| \geq |e \cap s|$，所以 $\frac{|e|}{|s|} \geq \frac{|e|}{|e|+|s|-|e \cap s|} \geq \frac{|e \cap s|}{|e|+|s|-|e \cap s|} \geq \delta$。因此有 $|s| \leq \lfloor \frac{|e|}{\delta} \rfloor$。又因为 $|s| \geq |e \cap s|$，所以 $\frac{|s|}{|e|} \geq \frac{|s|}{|e|+|s|-|e \cap s|} \geq \frac{|e \cap s|}{|e|+|s|-|e \cap s|} \geq \delta$。因此有 $|s| \geq \lceil |e| * \delta \rceil$。因此 $\lceil |e| * \delta \rceil \leq |s| \leq \lfloor \frac{|e|}{\delta} \rfloor$.
(2) Cosine Similarity：因为 $|e| \geq |e \cap s|$，所以 $\frac{|e|}{\sqrt{|e| \cdot |s|}} \geq \frac{|e \cap s|}{\sqrt{|e| \cdot |s|}} \geq \delta$。因此有 $|s| \leq \lfloor \frac{|e|}{\delta^2} \rfloor$。又因为 $|s| \geq |e \cap s|$，所以 $\frac{|s|}{\sqrt{|e| \cdot |s|}} \geq \frac{|e \cap s|}{\sqrt{|e| \cdot |s|}} \geq \delta$。因此有 $|s| \geq \lceil |e| * \delta^2 \rceil$。因此 $\lceil |e| * \delta^2 \rceil \leq |s| \leq \lfloor \frac{|e|}{\delta^2} \rfloor$。
(3) Dice Similarity：因为 $|e| \geq |e \cap s|$，所以 $\frac{2|e|}{|e|+|s|} \geq \frac{2|e \cap s|}{|e|+|s|} \geq \delta$。因此 $|s| \leq \lfloor |e| * \frac{2-\delta}{\delta} \rfloor$。又因为 $|s| \geq |e \cap s|$，所以 $\frac{2|s|}{|e|+|s|} \geq \frac{2|e \cap s|}{|e|+|s|} \geq \delta$。因此 $|s| \geq \lceil |e| * \frac{\delta}{2-\delta} \rceil$。因此 $\lceil |e| * \frac{\delta}{2-\delta} \rceil \leq |s| \leq \lfloor |e| * \frac{2-\delta}{\delta} \rfloor$。
(4) 编辑距离：因为 $||e| - |s|| \leq \tau$，所以 $|e| - \tau \leq |s| \leq |e| + \tau$。
(5) Edit Similarity：因为 $\text{EDS}(e, s) = 1 - \frac{\text{ED}(e,s)}{\max(\text{len}(e), \text{len}(s))} \geq \delta$，所以 $\text{ED}(e, s) \leq \max(\text{len}(e), \text{len}(s)) * (1 - \delta)$。如果 $|e| \leq |s|$, $|\text{len}(s) - \text{len}(e)| \leq \text{ED}(e, s) \leq \max(\text{len}(e), \text{len}(s)) * (1 - \delta) = \text{len}(s) * (1 - \delta)$。因此 $\text{len}(s) \leq \frac{\text{len}(e)}{\delta}$ 并且 $|s| \leq \lfloor \frac{|e|+q-1}{\delta} - (q - 1) \rfloor$。如果 $|e| > |s|$, $\text{len}(e) - \text{len}(s) \leq \text{ED}(e, s) \leq \max(\text{len}(e), \text{len}(s)) * (1 - \delta) = \text{len}(e) * (1 - \delta)$。$\text{len}(s) \geq \text{len}(e) * \delta$ 并且 $|s| \geq \lceil (|e| + q - 1) * \delta - (q - 1) \rceil$。因此 $\lceil (|e| + q - 1) * \delta - (q - 1) \rceil \leq |s| \leq \lfloor \frac{|e|+q-1}{\delta} - (q - 1) \rfloor$。

根据引理 2.2，给定一个实体 $e$，只有那些标记数量在 $\perp_e$ 和 $\top_e$ 之间的子字符串可能与实体 $e$ 相似，其余的子字符串都可以被过滤掉。特别的，令 $\perp_E = \min\{\perp_e | e \in E\}$，$\top_E = \max\{\top_e | e \in E\}$，显然文档 $D$ 中只有标记数在 $\perp_E$ 和 $\top_E$ 之间的子字符串

才可能和实体字典 $E$ 中的实体相似，其他的子字符串都可以直接过滤掉。根据这个观察下面提出“有效子字符串”的概念。

**定义** 2.2 (**有效子字符串**)：　给定一个实体字典 $E$ 和一个文档 $D$，对于文档中的一个子字符串 $s$ 和字典中的一个实体 $e$，如果 $\perp_e \le |s| \le \top_e$，那么 $s$ 是 $e$ 的有效子字符串；如果 $\perp_E \le |s| \le \top_E$，那么 $s$ 是字典 $E$ 的有效子字符串。

举例来说，考虑表 2.1中的实体字典和文档，假设使用 Edit Similarity 并设定阈值 $\delta = 0.8$，$q = 2$。考虑实体 $e_5$ = “surajit ch’’，有 $\perp_{e_5} = \lceil(|e_5|+q-1)*\delta-(q-1)\rceil = 7$，$\top_{e_5} = \lfloor\frac{|e_5|+q-1}{\delta}-(q-1)\rfloor = 11$，因此只有标记数量在 7 到 11 之间的有效子字符串才可能和实体 $e_5$ 相似。又因为 $\perp_E = 7, \top_E = 12$，所以只有标记数量在 7 到 12 之间的有效子字符串才可能和实体字典中的实体相似，所有其它子字符串（例如“*surauijt chadhurisigmod*’’）都可以被直接过滤掉。

一个解决近似实体抽取的简单方法枚举文档中的每一个有效子字符串，然后对于每一对有效子字符串和实体元组，它计算它们的近似度并输出所有的近似元组。然而这个简单方法需要枚举巨量的元组，非常低效，因此，本章采用一个过滤加验证的框架来解决近似实体抽取问题。在过滤步骤中产生所有候选元组，候选元组由文档 $D$ 中一个有效子字符串和实体字典 $E$ 中一个实体组成，它们的 Overlap Similarity 不小于阈值 $T$；在验证步骤中通过计算候选元组的真实的相似度来验证他们是否是真正的近似元组。本文主要关注过滤步骤。

## 2.3　基于堆的过滤算法

过滤加验证框架依赖于高效的索引结构和快速的过滤算法，本节首先介绍一个索引结构（第 2.3.1节），然后提出两个基于堆的过滤算法：一个基于多个堆（第 2.3.2节），另一个基于单个堆（第2.3.3节）。

### 2.3.1　倒排索引

只有当一个有效子字符串和一个实体共享足够多的标记时，它们才有可能相似，可以利用倒排索引结构来快速的清点它们的共同标记数目。下面为整个字典中的所有实体建立一个倒排索引。倒排索引中一个条目对应一个标记，每条条目附有一个倒排列表，倒排列表中保存着包含该标记的实体的 ID，倒排列表中的 ID 是按照升序排列的。例如，图 2.2中给出了基于表 2.1中所有实体的 2-gram 的倒排索引。

| | | | | | |
|---|---|---|---|---|---|
| ka→1→4 | k_→1 | ***ra→2→5*** | ud→3 | en→4 | aj→5 |
| ***au→1→3*** | ***_c→1→5*** | ab→2 | dh→3 | nk→4 | ji→5 |
| us→1 | ***ch→1→2→3→5*** | ba→2 | hu→3 | at→4 | it→5 |
| sh→1→4 | ha→2→3 | ar→2 | ***ur→3→5*** | te→4 | ***t_→5*** |
| hi→1 | ak→2 | rt→2 | ri→3 | es→4 | |
| ik→1 | kr→2 | ti→2 | ve→4 | ***su→5*** | |

图 2.2　基于表 2.1中所有实体的 2-gram 的倒排列表

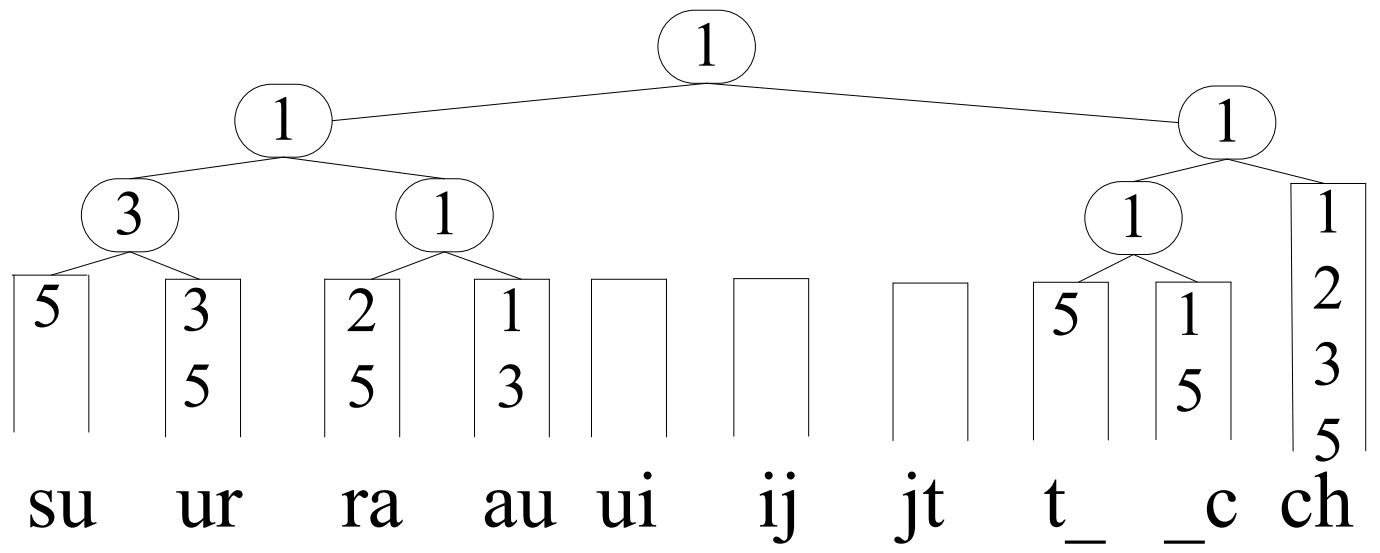

图 2.3　基于有效子字符串 “surauijt ch’’ 的小顶堆

对于文档 $D$ 中的每一个有效子字符串，首先把它切分成标记集合，对每一个标记去倒排索引中检索它相对应的倒排列表。然后对于倒排列表中的每一个实体，清点它在该文档对应的所有倒排列表中的出现次数，也就是包含该实体的倒排列表的个数。显而易见，实体在倒排列表中出现的次数恰好是 $|e \cap s|$[①]。对每个实体 $e$ 来说，它的出现次数如果不小于 $T$（即 $|e \cap s| \geq T$），那么 $\langle s, e\rangle$ 就是一对候选元组。

例如，考虑一个有效子字符串 “*surauijt ch*’’，首先产生它的标记集合 {su, ur, ra, au, ui, ij, jt, t_, _c, ch}。然后检索它们的倒排列表（图 2.2中斜体加粗的部分）。假设使用编辑距离并设定阈值 $\tau = 2$，对于实体 $e_5$，$T = \max(|e_5|, |s|) - \tau * q = 6$。因为 $e_5$ 的出现次数为 6，所以 ⟨“*surauijt ch*”, $e_5$=“surajit ch”⟩ 是一个候选元组。

简单起见，给定实体 $e$ 和有效子字符串 $s$，把 $e$ 在 $s$ 中所有标记的倒排列表中的出现次数简称为 $e$ 在 $s$（或者 $s$ 的倒排列表）中的出现次数。为了快速清点出现次数，下节提出基于堆的过滤算法。

### 2.3.2　基于多堆的方法

过滤加验证框架中的过滤步骤主要关注如何产生由文档中有效子字符串和字典中实体组成的候选元组。这些候选元组之间的 Overlap Similarity 不小于阈值 $T$。可以利用倒排索引来清点实体在有效子字符串的倒排列表中出现的次数，并输出出现次数不小于阈值 $T$ 的实体与对应的有效子字符串组成的候选元组。为了加快清点实体出现次数并避免枚举所有元组，下面提出一个基于多堆（multi-heap）的

① 因为在实体或者子字符串中可能存在重复的标记，所以把 $e$ 和 $s$ 当作 multiset。如果把它们当做 set，本章提出的方法仍然奏效。

表 2.2　基于多堆的算法的复杂度

(a) 空间复杂度

| | |
|---|---|
| 最大的堆 | $O(\top_E)$ |

(b) 时间复杂度

| | |
|---|---|
| 堆的构建 | $O\left(\sum_{l=\bot_E}^{\top_E}(\lvert D\rvert - l + 1) * l\right)$ |
| 堆的调整 | $O\left(\sum_{l=\bot_E}^{\top_E} log(l) * l * N\right)$ |

方法。

首先枚举文档 $D$ 的所有有效子字符串（标记数目在 $\bot_E$ 与 $\top_E$ 之间的子字符串）。然后对于每一个有效子字符串，产生它的标记集合并且在它的标记的非空倒排列表上面建立一个小顶堆。最开始用每个倒排列表的第一个实体 ID 来建立这个小顶堆。对于该堆的堆顶实体，清点它在这个小顶堆中的出现次数。如果它的出现次数不小于 $T$，那么由这个有效子字符串和这个实体组成的元组即是一个候选元组。接下来，弹出位于堆顶的实体。接下来把弹出实体所在的倒排列表中的下一个实体 ID 添加到小顶堆中并且调整小顶堆。然后再次清点新的堆顶实体 ID 的出现次数。这样循环多次就可以找到所有的候选元组了。

例如，考虑一个有效子字符串 "*surauijt ch*''。首先生成它的标记集合并基于每个倒排列表的第一个实体建立起一个小顶堆（图2.3）。接下来反复调整这个小顶堆，可以得到实体 ID 的升序序列 $\{1, 1, 1, 2, 2, 3, 3, 3, 5, 5, 5, 5, 5, 5\}$。之后清点每个实体的出现次数。例如，$e_1, e_2, e_3,$ 和 $e_5$ 的出现次数分别为 3，2，3 和 6。假设使用编辑距离并设置 $\tau = 2$。对于实体 $e_5$ 来说, $T = \max(|e_5|, |s|) - \tau * q = 6$。所以该子字符串和实体 $e_5$ 组成了一组候选元组。最后验证候选元组并得到最终结果集。

**复杂度分析**. 对于一个有 $l$ 个标记的有效子字符串，它相应的堆中包含最多 $l$ 个不为空的倒排列表，所以初始化堆的时间复杂度为 $O(l)$。又因为可以一次建立一个堆，所以整个算法的空间复杂度就是最大的小顶堆所占用的空间，也就是 $O(\top_E)$ (表2.2(a))。

为一个拥有 $l$ 个标记的有效子字符串构建堆的时间复杂度为 $O(l)$。由于文档 $D$ 中一共有 $|D| - l + 1$ 个包含 $l$ 个标记的有效子字符串，所以为这些有效子字符串构建堆的时间复杂度为 $O((|D| - l + 1) * l)$。因此为文档中所有有效子字符串构建堆的时间复杂度为 $O(\sum_{l=\bot_E}^{\top_E}(|D| - l + 1) * l)$ (表2.2(b))。此外，对于每个实体，需要调整包含该实体 ID 的堆。考虑包含 $l$ 个倒排列表的堆，调整一次这种堆的时间复杂度为 $O(log(l))$。又因为一个实体被 $l$ 个这样的堆所包含（图2.4），所以对于每个实体调整堆的时间复杂度是 $O(\sum_{l=\bot_E}^{\top_E} log(l) * l)$。假设文档 $D$ 的倒排列表中一共有 $N$ 个实体，那么基于多堆的方法的总时间复杂度为 $O(\sum_{l=\bot_E}^{\top_E} log(l) * l * N)$(表 2.2(b))。

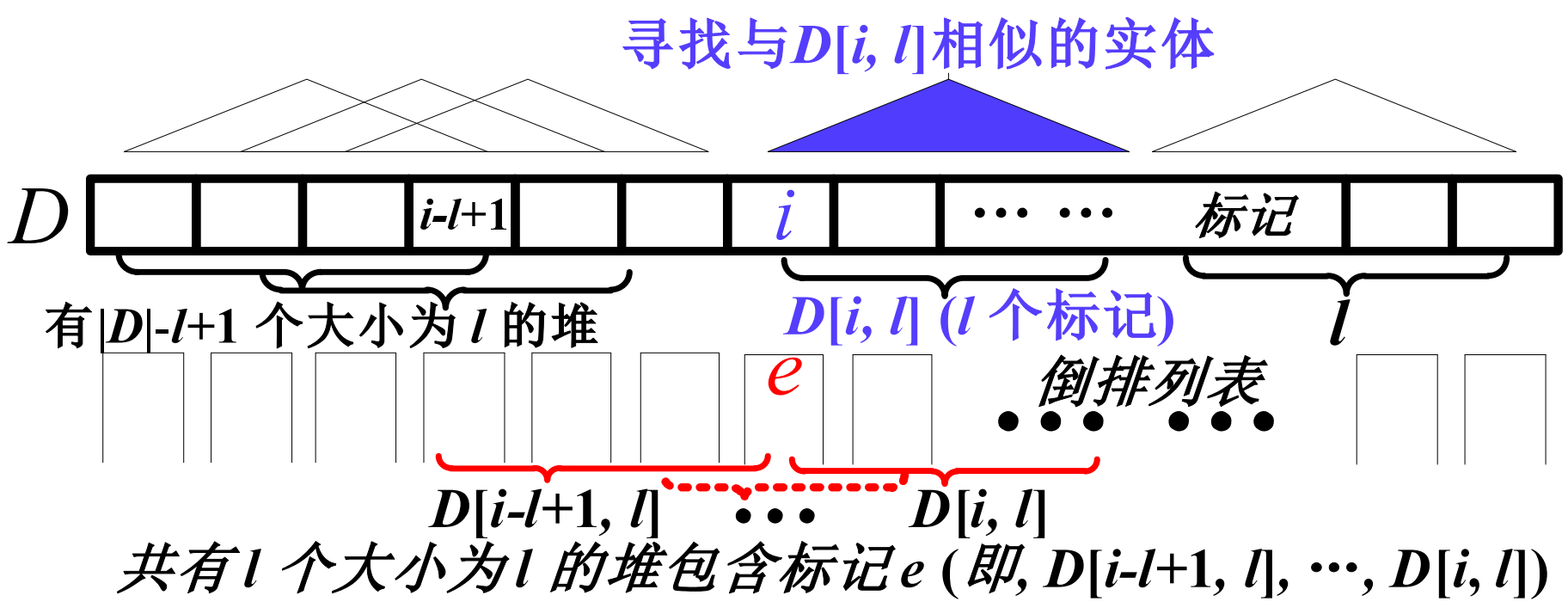

图 2.4　多堆的示意图

基于多堆的方法需要多次访问同一个倒排列表，做出大量的堆调整操作，为了解决这个问题，下面在 2.3.3节中提出一种新的方法，它对每个倒排列表只访问一次。

### 2.3.3　基于单堆的方法

本节提出一种基于单堆 (single-heap) 的方法，它首先对文档 $D$ 切分以获得它的标记列表，然后检索其中每个标记对应的倒排列表。用 $token[i]$ 来表示文档 $D$ 的第 $i$ 个标记，用 $\mathcal{IL}[i]$ 来表示第 $i$ 个标记的倒排列表。该方法在文档 $D$ 的所有非空倒排列表上建立起单个小顶堆 $H$，然后只用这一个小顶堆来寻找候选元组。

为了方便表达，下文用一个二维数组 $V[1\cdots|D|][\perp_E\cdots\top_E]$ 来记录一个实体在各个有效子字符串的倒排列表中的出现次数。正式地，用 $D[i,l]$ 来表示文档 $D$ 中从第 $i$ 个标记开始，标记总数为 $l$ 的有效子字符串。给定一个实体 $e$，用 $V[i][l]$ 来清点 $e$ 在 $D[i,l]$ 的倒排列表中的出现次数，也就是 $V[i][l] = |e\cap D[i,l]|$。首先，对于每个 $1\le i\le|D|$，$\perp_E\le l\le\top_E$，初始化 $V[i][l]$ 为 0。

对于堆顶的实体 $e$，假设它来自第 $i$ 个倒排列表，按照如下规则把数组 $V$ 中的相关状态（包含 $e$ 的有效子字符串）的值递增 1。不失一般性的，考虑有 $l$ 个标记的有效子字符串。显然，只有 $D[i-l+1,l],\ldots,D[i,l]$ 包含倒排索引中的第 $i$ 个倒排列表（图 2.5），因此 $V[i-l+1][l],\ldots,V[i][l]$ 是相关状态。同样的，对于 $\perp_E\le l\le\top_E, V[i-l+1][l],\ldots,V[i][l]$ 是相关状态。把相关状态的值递增 1，如果 $V[i][l]\ge T, \langle D[i,l],e\rangle$ 就是一个候选元组。然后，弹出位于堆顶的实体，把 $\mathcal{IL}[i]$ 中的下一个实体压入堆中，并调整堆获得下一个实体，之后清点新的堆顶实体在倒排索引中的出现次数。不断的重复执行以上步骤，最终可以找到所有的候选元组。

实际上，对于实体 $e$，只有标记数目在 $\perp_e$ 和 $\top_e$ 之间的有效子字符串才有可

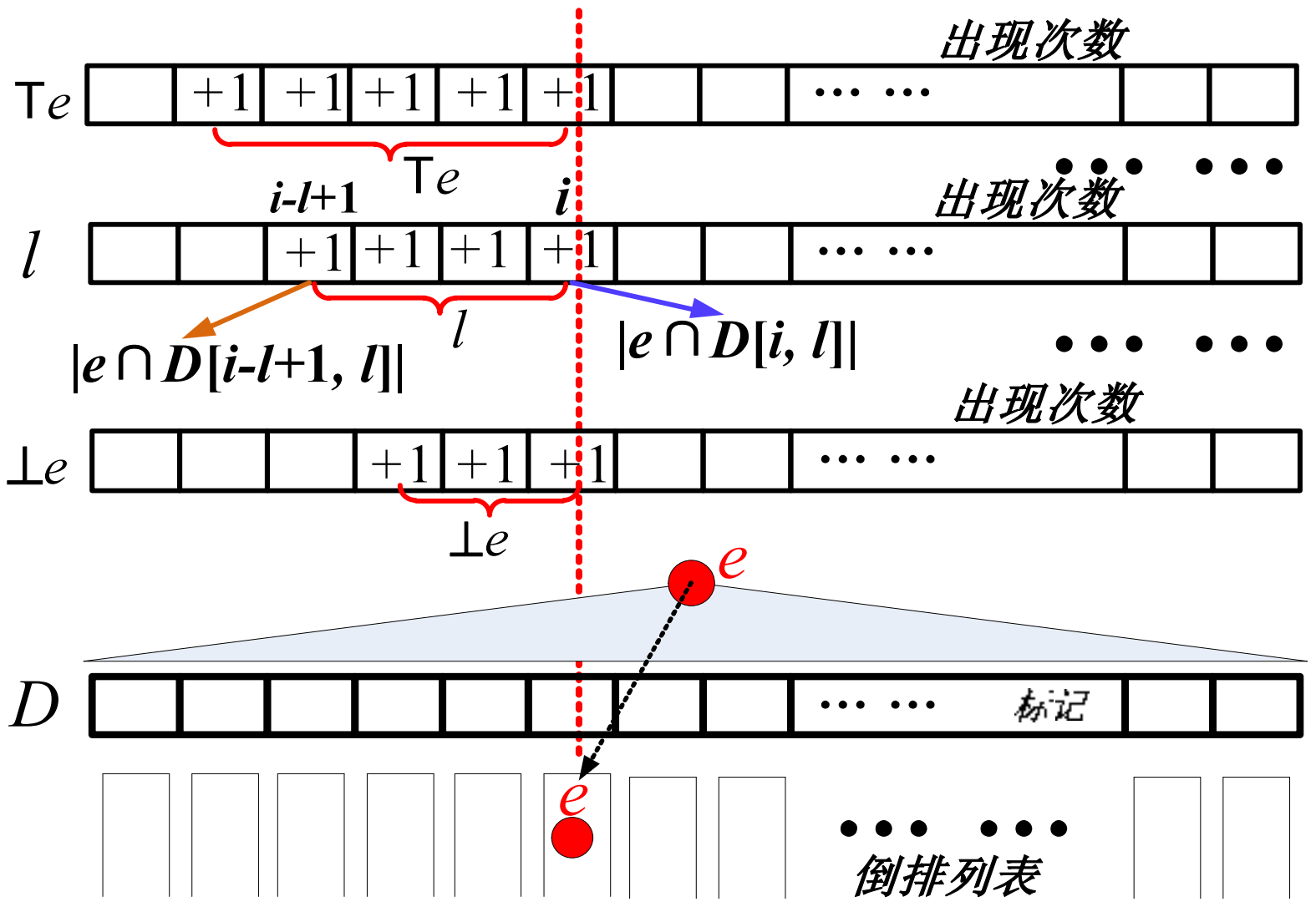

图 2.5 单堆的示意图

能和 $e$ 相似[①]。因此，只需要维护 $V[1\cdots|D|][\perp_e\cdots\top_e]$。

接下来给出一个基于单堆的方法的例子。考虑一个文档 “*venkaee shga kamunshi*”，在该文档上构造如图2.6所示的单堆，假设使用编辑距离并设置 $\tau=2$。那么有 $\perp_E=6$，$\top_E=12$。对于从文档第一个标记中选择出的实体 $e_4$，只需要为其在每个有效子字符串 $D[1,l]$ 中递增 1 次出现次数，其中 $\perp_E\le l\le\top_E$。也就是说递增 1 次 $e_4$ 在 $D[1,6],\ldots,D[1,12]$ 中的出现次数。对于从文档第二个标记中选择出的下一个实体 $e_4$，在每个有效子字符串 $D[1,l],D[2,l]$ 中递增 $e_4$ 的出现次数 1 次，其中 $\perp_E\le l\le\top_E$。如此这样下去，可以清点出所有实体的出现次数。例如，实体 $e_4$ (“venkatesh”) 在 $D[1,9]$ 中的出现次数是 5。由于它的出现次数不少于 $T=\max(|e_4|,|D[1,9]|)-\tau*q=9-2*2=5$，所以元组 $D[1,9]$(“venkaee sh”) 和 $e_4$(“venkatesh”) 是一对候选元组。实际上，由于 $\perp_{e_4}=6,\top_{e_4}=10$，所以只需要考虑 $V[1\cdots20][6\cdots10]$ 中的状态。

**复杂度分析**: 基于单个堆的方法的空间复杂度为 $O(|D|)$(表 2.3(a)) 为了清点一个实体的出现次数，其实并不需要维护二维数组，可以使用另一种方法。具体来说，首先不断从堆中弹出相同的实体 ID 直到堆顶的实体 ID 与弹出的实体 ID 不同（用 $|D|$ 的空间来存储弹出的实体 ID）。假设该实体为 $e$，递增 $e$ 在所有 $V[1\cdots|D|-l+1][l]$ 中的出现次数 1 次，其中 $\perp_e\le l\le\top_e$。通过这种方法，只需要维护一个一维数组。因此，清点实体在子字符串中的出现次数的空间复杂度为 $O(\max\{|D|-\perp_e+1|e\in E\})=O(|D|-\perp_E+1)$(表 2.3(a))。

建立堆的时间复杂度为 $O(|D|)$(表2.3(b))。为了清点每个实体的出现次数，需

① 注意，可以构建一个包含实体及其标记数量信息的散列映射，通过这个散列映射可以获得一个实体 $e$ 的标记数量 $|e|$，因此获得标记数量的时间复杂度是 $O(1)$。

表 2.3　基于单堆的算法复杂度

(a) 空间复杂度

| | |
|---|---|
| 存储单堆 | $O(\|D\|)$ |
| 清点出现次数 | $O(\|D\| - \bot_E + 1\|)$ |

(b) 时间复杂度

| | |
|---|---|
| 堆的构建 | $O(\|D\|)$ |
| 堆的调整 | $O(log(\|D\|) * N)$ |
| 清点出现次数 | $O(N * \max\{\sum_{l=\bot_e}^{\top_e} l \| e \in E\})$ |

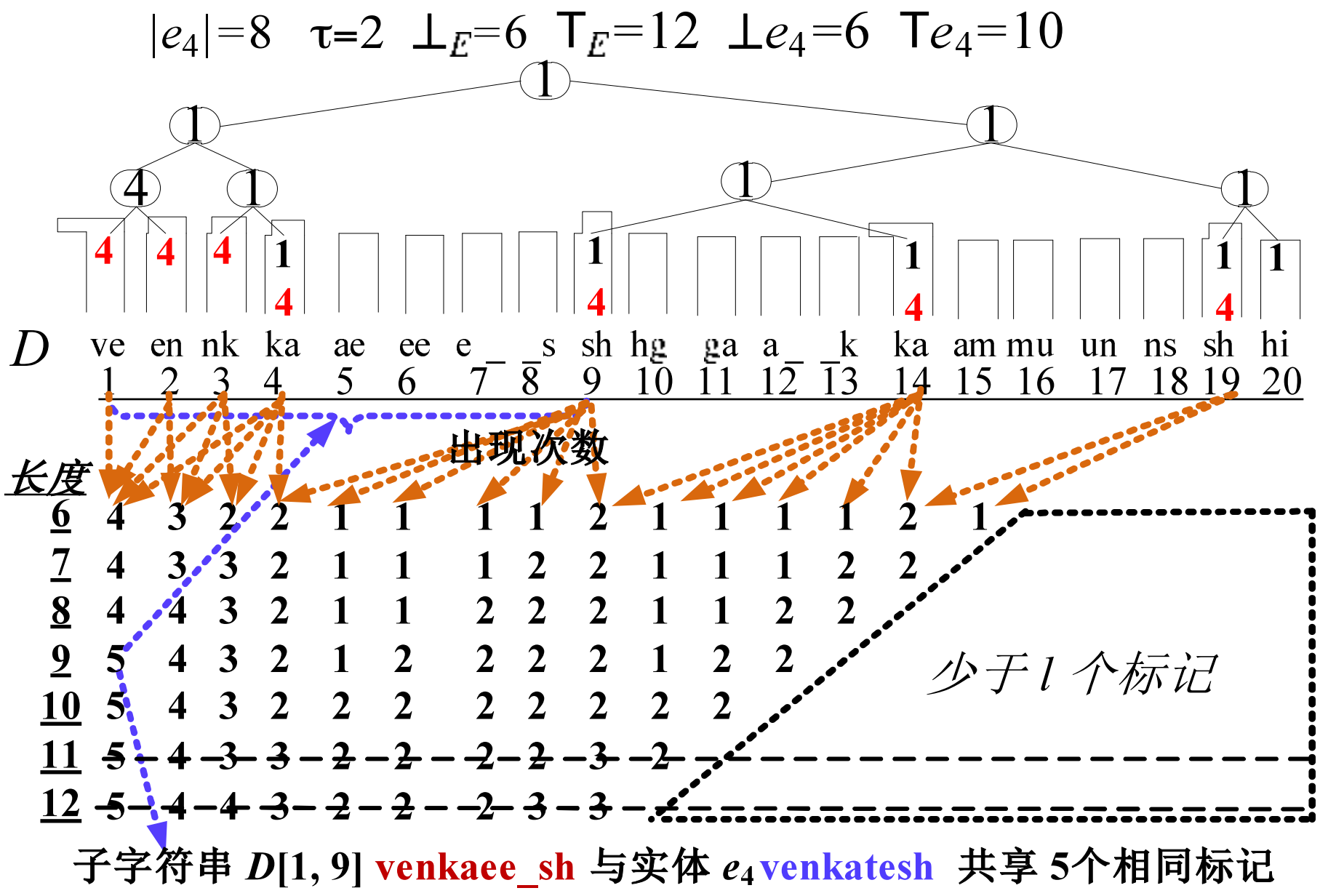

图 2.6　以文档 “venkaee shga kamunshi” 为输入的单堆算法的实例

要调整堆。调整堆的总时间复杂度为 $O(log(|D|) * N)$，其中 $N$ 是所有倒排列表中实体的总数。此外，对于每个实体，需要不断递增其出现次数。对于实体 $e$，共有 $\sum_{l=\bot_e}^{\top_e} l$ 个元素需要递增出现次数一次（图2.5），对于任何实体来说，该数目最大不超过 $\max\{\sum_{l=\bot_e}^{\top_e} l | e \in E\}$。因此，总的时间复杂度为 $O(N * \max\{\sum_{l=\bot_e}^{\top_e} l | e \in E\})$ (表 2.3(b)).

值得注意的是，基于单堆的方法只需要扫描每个倒排列表一次，它共享了不同有效子字符串之间重叠部分的计算。所以它的时间复杂度远远低于基于多堆的方法的时间复杂度，因此能够达到更高的性能（详见第 2.6节）。

## 2.4 改进基于单堆的算法

基于单堆的算法仍然需要计算每个实体在文档中的出现次数。如果实体的数量非常多，那么这个方法的代价依然很高昂。为了解决这个问题，本节改进基于单堆的算法，并提出有效的剪枝技术来跳过很多不相关的实体。

### 2.4.1 剪枝技术

本节通过估计 $|e \cap s|$ 的下界提出几种剪枝技术。

**Lazy-Count 剪枝**： 对于每个在堆顶的实体，并不立即清点其在各个有效子字符串中的出现次数，而是采用一种更加懒惰的方式：首先清点该实体在整个堆中的总出现次数，假如总出现次数足够少，可以直接剪枝掉该实体。

正式地，给定一个实体 $e$，使用一个有序的位置列表 $P_e = [p_1, \cdots, p_m]$ 来记录实体在堆中的出现情况（按升序顺序）。位置列表 $P_e$ 中的每个元素就是倒排列表中包含实体 $e$ 的标记在文档中所在的位置。通过堆结构可以很容易地得到位置列表。然后清点实体 $e$ 在堆中的总出现次数，即 $P_e$ 中元素的个数（$|P_e|$）。如果该总出现次数小于一个阈值，记作 $T_l$，那么可以直接剪枝掉该实体；否则，清点其在各个有效子字符串中的出现次数，注意其中有效子字符串的标记数目应在 $\perp_e$ 和 $\top_e$ 之间（第2.3.3节）。例如，在图 2.6中，$P_{e_1} = [4, 9, 14, 19, 20]$，$|P_{e_1}| = 5$。

接下来讨论如何计算这个阈值 $T_l$。回想 Overlap Similarity 的阈值 $T$(第2.2.2节)，$T$ 取决于 $|e|$ 和 $|s|$。为了推导只取决于 $|e|$ 的 $T$ 的下界，要用 $\perp_e$ 来替换 $|s|$(因为与 $e$ 相似的 $s$ 的大小不小于 $\perp_e$)。在不同的相似度函数下，这个新的下界 $T_l$ 的值可以通过如下公式计算：

- Jaccard Similarity：$T_l = \lceil |e| * \delta \rceil$。
- Cosine Similarity：$T_l = \lceil |e| * \delta^2 \rceil$。
- Dice Similarity：$T_l = \lceil |e| * \frac{\delta}{2-\delta} \rceil$。
- 编辑距离：$T_l = |e| - \tau * q$。
- Edit Similarity：$T_l = \lceil |e| - ((|e| + q - 1) * \frac{(1-\delta)}{\delta} * q) \rceil$。

显然 $T_l \leq T$。如果 $|P_e| < T_l \leq T$，那么 $e$ 不可能与任何子字符串相似，因此可以剪枝掉实体 $e$。举例来说，在图 2.6中，考虑实体 $e_1$ 并假设 $\tau = 1$，那么 $|e_1| = 9$，$T_l = |e_1| - \tau * q = 9 - 2 = 7$。由于 $|P_{e_1}| = 5 < T_l$，可以剪枝掉实体 $e_1$。定理 2.3给出了其正确性证明。

**引理** 2.3： 给定单堆上的一个实体 $e$，如果其在堆中的出现次数 ($|P_e|$) 比 $T_l$ 小，那么 $e$ 不可能与任何有效子字符串相似。

**证明** 由于 $D$ 和 $e$ 所有的共同标记都在堆中的倒排列表里，所以 $e$ 与 $D$ 的任何子字符串共享的相同标记数目都小于 $T_l$。因此它在任何有效子字符串中的出现次数都一定小于 $T_l$。根据 $T_l$ 的定义，$e$ 与任何有效子字符串都不可能相似。

**Bucket-Count 剪枝**：一个观察是可以把位置列表 $P_e$ 切分为几个不相交的桶，然后在每个桶中使用 Lazy-Count 剪枝。因为每个桶包含 $P_e$ 的一个子列表，所以桶的大小远小于 $|P_e|$，每个桶有更高的几率能够被剪枝掉。接下来讨论如何把 $P_e$ 切分为不相交的桶。

考虑一个实体 $e$。如果一个有效子字符串 $s$ 与 $e$ 相似，那么 $s$ 最多有 $\top_e$ 个标记并且至少与 $e$ 共享 $T_l$ 个标记。换句话说，如果 $s$ 与 $e$ 相似，那么它们的不匹配的标记数目必须不大于 $\top_e - T_l$ 个。可以利用这个性质来进行有效的剪枝。

正式的，给定 $P_e$ 中两个相邻的元素 $p_i$ 和 $p_{i+1}$，任何包含这两个元素（TOKEN$[p_i]$ 和 TOKEN$[p_{i+1}]$）的子字符串将至少包含 $p_{i+1} - p_i - 1$ 个不匹配的标记。如果 $p_{i+1} - p_i - 1 > \top_e - T_l$，那么任何包含这两个标记的子字符串一定不可能与 $e$ 相似，这种情况下，不需要清点 $e$ 在任何子字符串中的出现次数。

利用这个观察，可以把 $P_e$ 中的元素划分到不相交的桶中。如果一个桶中的元素数量小于 $T_l$，那么可以剪枝掉桶中的所有元素（使用 lazy-count 剪枝）；否则，用桶中的元素来清点实体 $e$ 在标记数量在 $\perp_e$ 到 $\top_e$ 之间的有效子字符串中的出现次数（第 2.3.3节）。

下一步介绍如何进行桶的切分。最开始，创建桶 $b_1$ 并把位置列表中的第一个元素 $p_1$ 放到该桶中。接下来考虑下一个元素 $p_2$。如果 $p_2 - p_1 - 1 > \top_e - T_l$，那么创建一个新桶 $b_2$ 并把 $p_2$ 放入 $b_2$ 桶中；否则，把 $p_2$ 放入桶 $b_1$ 中。如此循环往复，可以把所有的元素划分到不同的桶中。

下面为各个相似函数给出一个更严格的界。例如，考虑编辑距离，可以使用该条件 $p_{i+1} - p_i - 1 > \tau * q$ 来进行剪枝。这是因为 $p_i$ 和 $p_{i+1}$ 之间存在至少 $\tau * q + 1$ 个不匹配的标记（$q$-gram ），另一方面，至少需要 $\tau + 1$ 个单字符的编辑操作来形成这 $\tau * q + 1$ 个不匹配的标记（$q$-gram)，因此任何同时包含这两个元素的子字符串都不可能与实体 $e$ 相似。

例如，在图 2.6中，假设使用编辑距离并设置 $\tau = 1$。考虑 $P_{e_4} = [1, 2, 3, 4, 9, 14, 19]$，那么有 $T_l = |e_4| - \tau * q = 8 - 1 * 2 = 6$。明显，$e_4$ 可以通过 lazy-count 剪枝，因为 $|P_{e_4}| \geq T_l$。下一步检查它是否可以通过 bucket-count 剪枝。首先把 $P_{e_4}$ 中的元素划分到不同的桶中：首先创建桶 $b_1$，并把元素 $p_1$ 放入桶 $b_1$ 中。接下来对于 $p_2 = 2$，因为 $p_2 - p_1 - 1 \leq \tau * q = 2$，所以把 $p_2$ 放到桶 $b_1$ 中。同样的，$p_3 = 3$ 和 $p_4 = 4$ 也被添加到桶 $b_1$ 中。对于 $p_5 = 9$，因为 $p_5 - p_4 - 1 > \tau * q$，所以

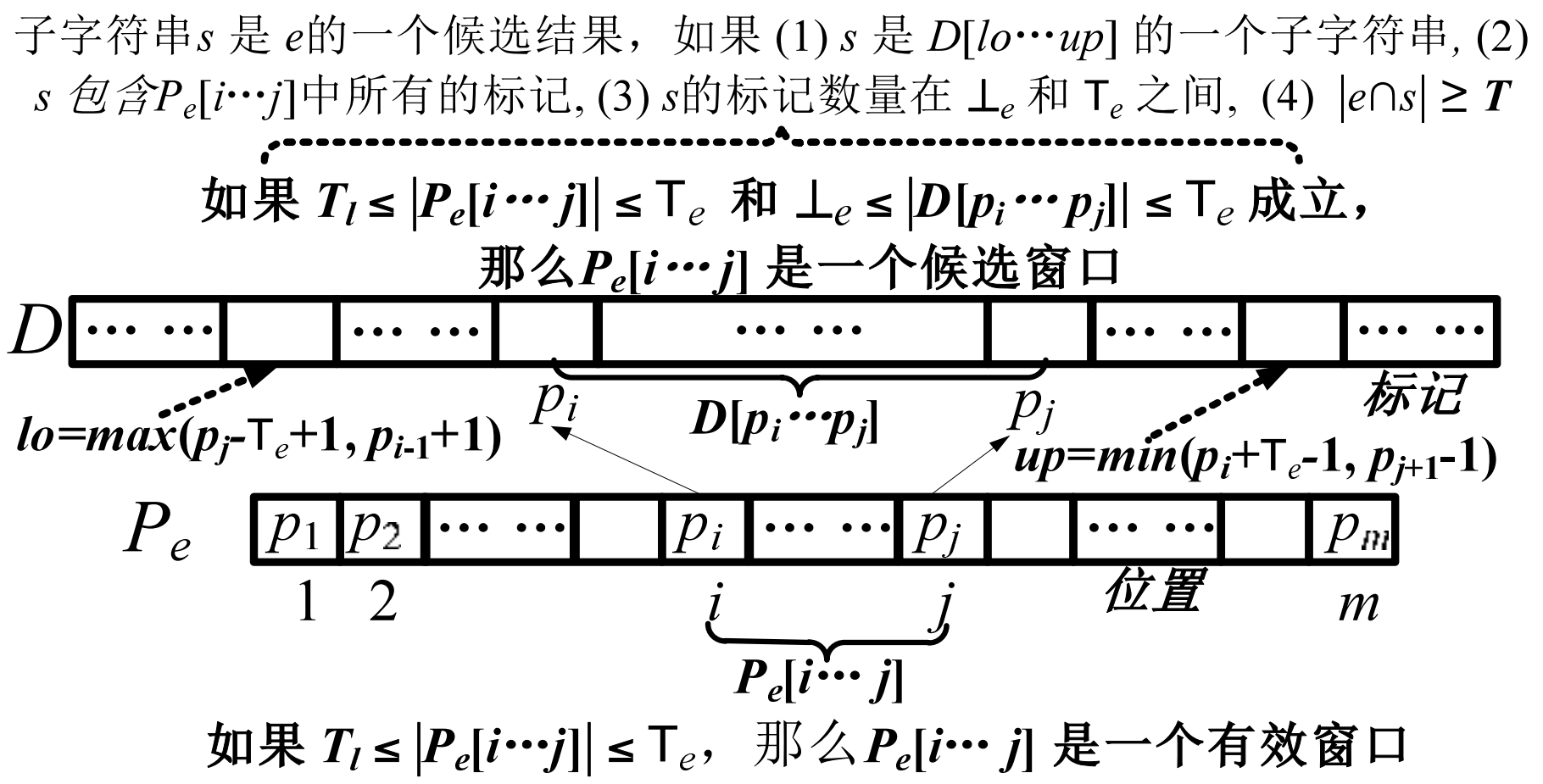

图 2.7　候选窗口和有效窗口

创建一个新的桶 $b_2$ 并把 $p_5$ 加入到桶 $b_2$ 中。不断重复以上步骤，最后可以得到四个桶 $b_1 = [1, 2, 3, 4]$, $b_2 = [9]$, $b_3 = [14]$ 以及 $b_4 = [19]$。由于这些桶的大小都小于 $T_l$，所以可以直接剪枝掉所有这些桶。因此不需要清点实体 $e_4$ 在任何有效子字符串中的出现次数。

此外还可以推广这个思想：给定两个元素 $p_i$ 和 $p_j(i < j)$，如果 $p_j - p_i - (j - i) > \top_e - T_l$，任何包含这两个标记（TOKEN$[p_i]$ 和 TOKEN$[p_j]$）的子字符串都不可能与实体 $e$ 相似。接下来将介绍如何利用该性质来做进一步的剪枝。

**Batch-Count 剪枝**：受 bucket-count 剪枝的启发，其实并不需要枚举位置列表 $P_e$ 中的每个元素来清点 $e$ 在每个有效子字符串中的出现次数。相反，只需要检查 $P_e$ 的子列表，看这些子列表是否能够产生候选元组。如果可能产生候选元组，查找该子列表中的候选元组；否则，剪枝掉该子列表。

正式的，如果一个有效子字符串 $s$ 和实体 $e$ 是相似的，它们必须共享足够多的相同标记 ($|e \cap s| \geq T_l$)。换句话说，也就是只需要检查元素数目不小于 $T_l$ 的子列表。考虑一个包含 $|P_e[i \cdots j]| = j - i + 1 \geq T_l$ 个元素的子列表 $P_e[i \cdots j]$。用 $D[p_i \cdots p_j]$ 来表示恰好只包含标记 $token[p_i], token[p_{i+1}], \cdots, token[p_j]$（图 2.7）的子字符串。如果 $|D[p_i \cdots p_j]| = p_j - p_i + 1 > \top_e$, 那么任何包含 $D[p_i \cdots p_j]$ 中所有标记的有效子字符串都有大于 $\top_e$ 个标记，因此可以剪枝掉 $P_e[i \cdots j]$。相反的，如果 $\bot_e \leq |D[p_i \cdots p_j]| \leq \top_e$，那么 $D[p_i \cdots p_j]$ 可能和实体 $e$ 相似。这个剪枝技术要比基于不匹配标记的剪枝技术要强很多。这是因为如果 $p_j - p_i - (j - i) > \top_e - T_l$，那么就有 $p_j - p_i + 1 > \top_e$。但是如果 $p_j - p_i + 1 > \top_e$, $p_j - p_i - (j - i) > \top_e - T_l$ 则不一定成立。除此之外，因为 $|P_e[i \cdots j]| \leq |D[p_i \cdots p_j]|$, 所以 $|P_e[i \cdots j]|$ 应不大于 $\top_e$，因此 $T_l \leq |P_e[i \cdots j]| \leq \top_e$。

根据这个观察，首先产生 $P_e$ 的大小（元素个数）在 $T_l$ 到 $\top_e$ 之间的子列表，即

$T_l \le |P_e[i \cdots j]| \le \top_e$。然后对每个这样的子列表 $P_e[i \cdots j]$，如果 $|D[p_i \cdots p_j]| > \top_e$，那么就剪枝掉这个子列表；否则，如果 $\bot_e \le |D[p_i \cdots p_j]| \le \top_e$，那么就在这个子列表中寻找实体 $e$ 的候选子字符串 (对于一个子字符串 $s$，如果 $|e \cap s| \ge T$ 并且 $\bot_e \le |s| \le \top_e$，那么 $s$ 就是 $e$ 的一个候选子字符串)。对于实体 $e$ 的每一个候选子字符串 $s$，$\langle s, e\rangle$ 是一对候选元组。接下来讨论如何在 $P_e$ 上找到 $e$ 的候选子字符串。为方便表述，首先介绍两个概念。

**定义 2.3 (有效窗口和候选窗口)**： 考虑一个实体 $e$ 和它的位置列表 $P_e = [p_1 \cdots p_m]$，对于任何 $1 \le i \le j \le m$，则称子列表 $P_e[i \cdots j]$ 为 $P_e$ 的一个窗口；如果 $T_l \le |P_e[i \cdots j]| \le \top_e$，则称 $P_e[i \cdots j]$ 为一个有效窗口；如果 $P_e[i \cdots j]$ 是一个有效窗口，并且 $\bot_e \le |D[p_i \cdots p_j]| \le \top_e$，那么则称 $P_e[i \cdots j]$ 为一个候选窗口。

有效窗口限制了一个窗口的长度，候选窗口限制了一个有效子字符串的标记数目。如果一个有效子字符串是实体 $e$ 的一个候选子字符串，那么它肯定包含一个候选窗口 (图 2.7)。例如，考虑表2.1中的文档和实体，有 $P_{e_4} = [10, 17, 33, 34, 43, 58, 59, 60, 61, 66, 71, 76, 81, 86]$。假设使用编辑距离，并设置 $\tau = 2$。那么有 $|e_4| = 8$，$T_l = |e_4| - \tau * q = 4$，$\bot_{e_4} = |e_4| - \tau = 6$ 以及 $\top_{e_4} = |e_4| + \tau = 10$。$P_{e_4}[1 \cdots 4] = [10, 17, 33, 34]$，$P_{e_4}[1 \cdots 5] = [10, 17, 33, 34, 43]$ 和 $P_{e_4}[6 \cdots 9] = [58, 59, 60, 61]$ 是三个有效窗口。然而由于 $p_4 - p_1 + 1 = 34 - 10 + 1 > \top_{e_4}$，所以 $D[p_1 \cdots p_4]$ 的标记数目大于 $\top_{e_4}$，因此 $P_{e_4}[1 \cdots 4]$ 并不是候选窗口，而且任何包含 $P_{e_4}[1 \cdots 4]$ 的子字符串一定包含多于 $\top_{e_4}$ 个标记。尽管 $p_9 - p_6 + 1 \le \top_{e_4}$，$P_{e_4}[6 \cdots 9]$ 仍然不是候选窗口，这是因为 $p_9 - p_6 + 1 < \bot_{e_4}$。

值得注意的是，在使用 Jaccard Similarity, Cosine Similarity 和 Dice Similarity 的情况下，因为它们只依赖于 $|e \cap s|$，所以可以优化这个剪枝条件。给定一个有效窗口 $P_e[i \cdots j]$, 令 $s = D[p_i \cdots p_j]$。有 $|P_e[i \cdots j]| \ge |e \cap s|$[①]。以 Jaccard Similarity 为例，假如 $s$ 和 $e$ 是相似的，那么 $\frac{|e \cap s|}{|e \cup s|} \ge \delta$，有 $|D[p_i \cdots p_j]| = |s| \le |e \cup s| \le \frac{|e \cap s|}{\delta} \le \frac{\min(|e|, |P_e[i \cdots j]|)}{\delta}$，因此可以给出 $|D[p_i \cdots p_j]|$ 的一个更加严格的界，对于 Jaccard Similarity 来说，有 $\bot_e \le |D[p_i \cdots p_j]| \le \frac{\min(|e|, |P_e[i \cdots j]|)}{\delta}$；对于 Dice Similarity 来说，有 $\bot_e \le |D[p_i \cdots p_j]| \le \min(|e|, |P_e[i, j]|) * \frac{2-\delta}{\delta}$; 对于 Cosine Similarity 来说，有 $\bot_e \le |D[p_i \cdots p_j]| \le \frac{\min(|e|, |P_e[i \cdots j]|)}{\delta^2}$。

接下来介绍如何从候选窗口 $P_e[i \cdots j]$ 中找到 $e$ 的候选子字符串。包含 $D[p_i \cdots p_j]$ 中所有标记的子字符串都可能是 $e$ 的候选子字符串，可以通过如下方式找到这些子字符串。因为这些子字符串一定会包含 $token[p_i]$，它们的 ‘‘最大的开始位置’’ 是 $p_i$，‘‘最大的结束位置’’ 是 up $= p_i + \top_e - 1$。同样的，因为这些子

① 这是因为 $D[p_i \cdots p_j]$ 可能包含重复的标记，所以 $|P_e[i \cdots j]| \ge |e \cap s|$，$|P_e[i \cdots j]|$ 可能也大于 $|e|$。

字符串一定会包含 $token[p_j]$, 它们的 ''最小的开始位置'' 是 $\mathsf{lo} = p_j - \top_e + 1$，''最小的结束位置'' 是 $p_j$ 。所以只需要在子字符串 $D[p_{start} \cdots p_{end}]$ 中找候选子字符串，其中 $\mathsf{lo} \le p_{start} \le p_i, p_j \le p_{end} \le \mathsf{up}$。如果 $\perp_e \le |s| = p_{end} - p_{start} + 1 \le \top_e$ 并且 $|e \cap s| \ge T$，那么子字符串 $s = D[p_{start} \cdots p_{end}]$ 就是 $e$ 的候选子字符串（这里使用阈值 $T$ 是因为此时已经知道 $s = D[p_{start} \cdots p_{end}]$ 的值）。

然而, 这种方法可能会产生很多重复的候选子字符串。例如, 假设 $p_j - \top_e + 1 < p_{i-1} + 1$，那么 $D[p_{i-1}, \top_e] = D[p_{i-1} \cdots (p_{i-1} + \top_e - 1)]$ 可能是由列表 $P_e[i \cdots j]$ 产生的候选子字符串。在这种情况下，因为 $\perp_e \le p_j - p_i + 1 \le p_j - p_{i-1} + 1 \le \top_e$，$T_l \le |P_e[i \cdots j]| \le |P_e[i-1 \cdots j]| = p_j - p_{i-1} + 1 \le \top_e$, 所以 $P_e[(i-1) \cdots j]$ 也是一个候选窗口。因此列表 $P_e[(i-1) \cdots j]$ 也会产生候选子字符串 $D[p_{i-1}, \top_e]$。对于列表 $P_e[i \cdots j]$, 为了避免和 $P_e[i-1 \cdots j]$ 以及 $P_e[i \cdots j+1]$ 产生重复的候选子字符串，这里不扩展 $P_e[i \cdots j]$ 到小于 $p_{i-1} + 1$ 或者大于 $p_{j+1} - 1$ 的位置，并且这里将设置 $\mathsf{lo} = \max(p_j - \top_e + 1, p_{i-1} + 1), \mathsf{up} = \min(p_i + \top_e - 1, p_{j+1} - 1)$。通过这种方式，该方法将不会产生重复的候选子字符串。

总结来说，为了找到实体 $e$ 的候选子字符串，首先产生实体的位置列表；然后产生相应的有效窗口和候选窗口；接下来从候选窗口中找到它的候选子字符串；最后，每个被找到的候选子字符串与实体 $e$ 组成一个候选元组。

### 2.4.2　快速地找到候选窗口

给定一个实体 $e$，因为它有非常多的有效窗口 ($\sum_{l=T_l}^{\top_e} |P_e| - l + 1$ 个), 为了找到候选窗口而枚举所有的有效窗口将会非常耗时。为了提高性能，本节提出一个高效的方法来寻找候选窗口。

**基于伸展和平移的方法**: 为了表述简单, 首先介绍一个概念 "可能的候选窗口''。对于一个有效窗口 $P_e[i \cdots j]$，如果 $p_j - p_i + 1 \le \top_e$，则称其为一个可能的候选窗口。根据这个概念, 介绍两种操作：伸展（span）和平移（shift）。给定一个当前的有效窗口 $P_e[i \cdots j]$，可以利用这两种操作按照如下方式来产生新的有效窗口（图 2.8 ）。

- 伸展：如果 $p_j - p_i + 1 \le \top_e$，对于 $k \ge j$, $P_e[i \cdots k]$ 有可能是一个可能的候选窗口。伸展当前的有效窗口 $P_e[i \cdots j]$ 来产生所有以 $i$ 为开始位置的可能的候选窗口：$P_e[i \cdots (j+1)], \ldots, P_e[i \cdots x]$, 其中 $x$ 满足 $p_x - p_i + 1 \le \top_e$ 和 $p_{x+1} - p_i + 1 > \top_e$。对于 $j \le k \le x$, 如果 $p_k - p_i + 1 \ge \perp_e$，那么 $P_e[i \cdots k]$ 就是一个候选窗口。相反地，如果 $p_j - p_i + 1 > \top_e$，对于 $k \ge j$，因为 $p_k - p_i + 1 \ge p_j - p_i + 1 > \top_e, P_e[i \cdots k]$ 不可能是一个候选窗口，因此不需要伸展 $P_e[i \cdots j]$。

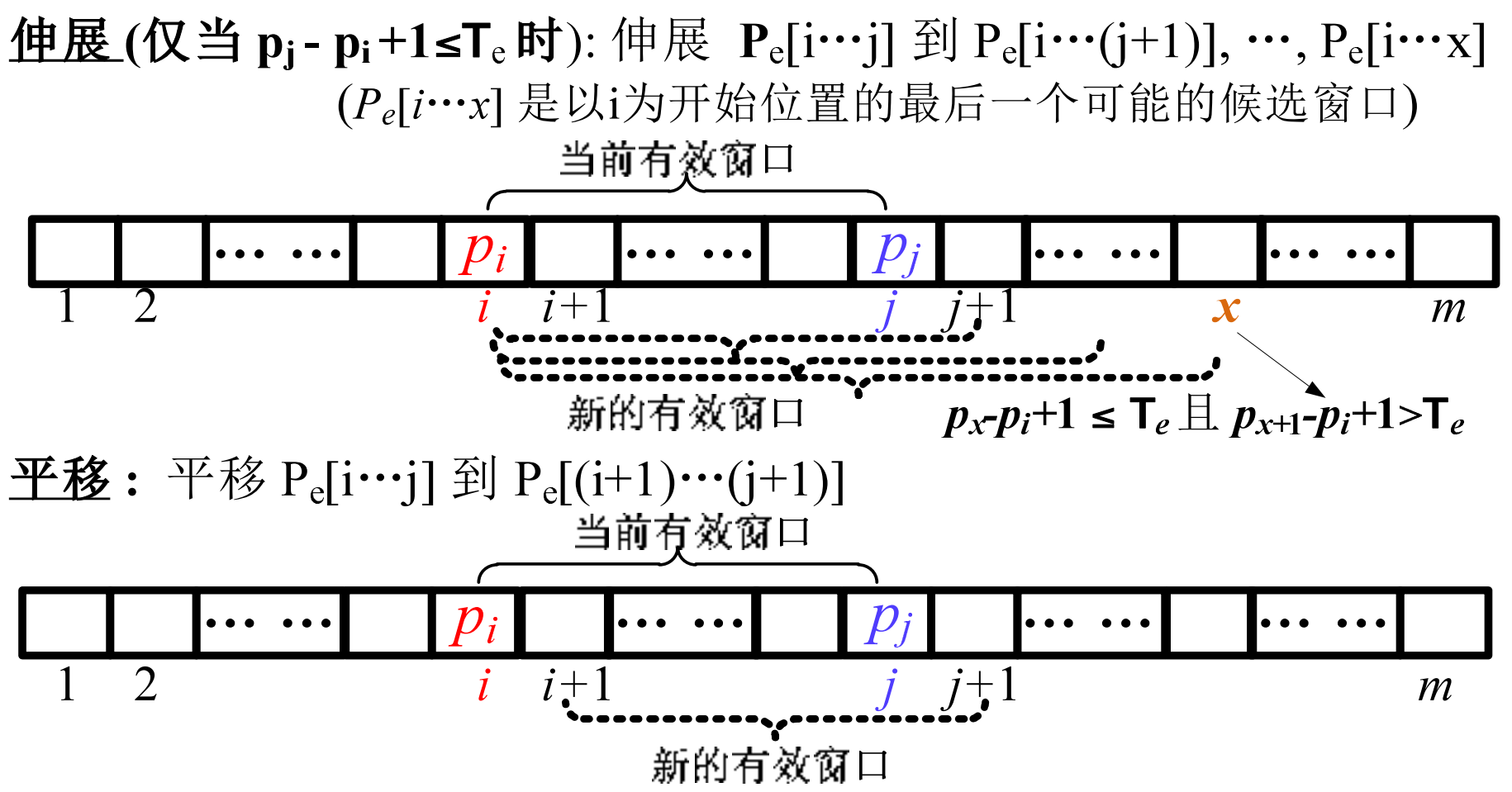

图 2.8　伸展操作和平移操作

- 平移：平移当前的有效窗口到一个新的有效窗口 $P_e[(i+1)\cdots(j+1)]$。

下面利用这两种操作按照如下方式来找到候选窗口。最开始取得第一个有效窗口 $P_e[1\cdots T_l]$，然后在 $P_e[1\cdots T_l]$ 上做伸展操作和平移操作。对于由伸展操作产生的新的有效窗口，检查它们是否是候选窗口；对于由平移操作生成的有效窗口，对其做伸展操作和平移操作。如此反复，可以找到 $P_e[1\cdots T_l]$ 中所有的候选窗口。接下来用一个例子来展示这个方法是如何工作的。对于 $e_4$(“venkatesh”), 有 $P_{e_4}=[10,17,33,34,43,58,59,60,61,66,71,76,81,86]$。假设编辑距离阈值 $\tau=2$，有 $|e_4|=8, T_l=|e_4|-\tau*q=4, \bot_{e_4}=|e_4|-\tau=6, \top_{e_4}=|e_4|+\tau=10$。其中第一个有效窗口是 $P_{e_4}[1\cdots4]=[10,17,33,34]$。因为 $p_4-p_1+1=34-10+1>\top_{e_4}$，所以不需要对其做伸展操作。接下来对其做一个平移操作来获取下一个窗口 $P_{e_4}[2\cdots5]$。因为 $p_5-p_2+1=43-17+1>\top_{e_4}$，所以再做一个平移操作，当平移到有效窗口 $P_{e_4}[6\cdots9]$ 时，因为 $p_9-p_6+1=61-58+1<\bot_{e_4}\le\top_{e_4}$，所以 $P_{e_4}[6\cdots9]$ 不是一个候选窗口，并且接下来需要做一个伸展操作。因为 $p_{10}-p_6+1=9\le\top_{e_4}$，$p_{11}-p_6+1=14>\top_{e_4}$，所以有 $x=10$，因此得到一个有效窗口 $P_{e_4}[6\cdots10]$，并且它还是个候选窗口。接下来平移到 $P_{e_4}[7\cdots10]$。这样反复，可以找到所有的候选窗口：$P_{e_4}[6\cdots10]$ 和 $P_{e_4}[7\cdots10]$ (图 2.9)。

给定一个有效窗口 $P_e[i\cdots j]$，如果 $p_j-p_i+1>\top_e$，平移操作可以剪枝掉所有开始于 $i$ 的有效窗口，例如 $P_e[i\cdots k]$，其中 $j<k\le i+\top_e-1$。然而，这种方法仍然需要扫描大量的有效窗口，为了进一步提高性能，下面提出一种基于二分查找的方法，它可以更快的跳过更多的有效窗口。

**基于二分伸展和二分平移的方法**：首先给出其基本思想。给定一个有效窗口 $P_e[i\cdots j]$，如果 $p_j-p_i+1>\top_e$，那么不平移到 $P_e[(i+1)\cdots(j+1)]$，相反，这

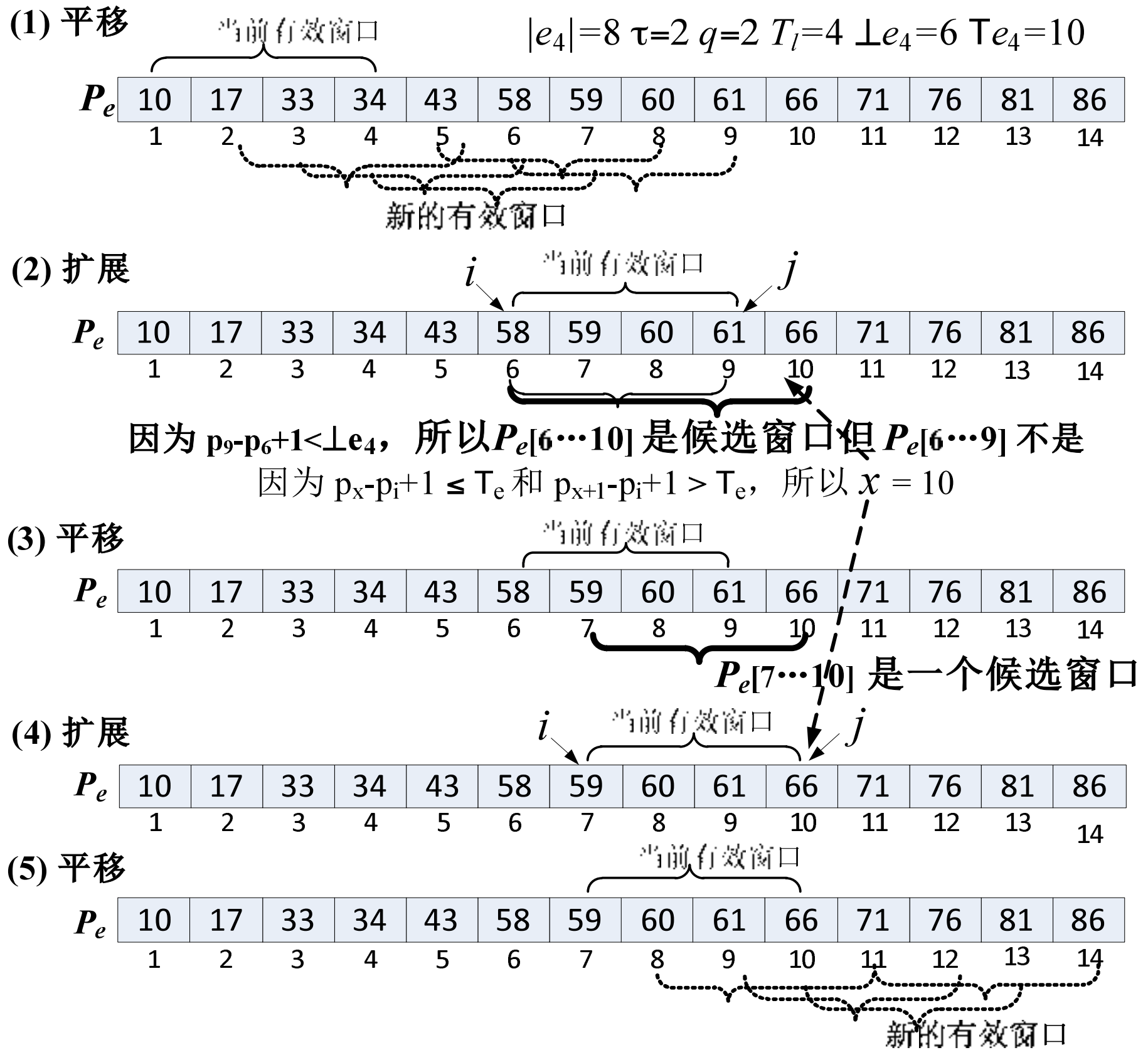

图 2.9　伸展操作和平移操作的一个实例

里希望直接平移到 $i$ 后的第一个可能的候选窗口，记作 $P_e[mid\cdots(mid+j-i)]$，其中 $mid$ 满足 $p_{mid+j-i}-p_{mid}+1\leq\top_e$ 并且对于任何 $i\leq mid'<mid$, 满足 $p_{mid'+j-i}-p_{mid'}+1>\top_e$。类似地，如果 $p_j-p_i+1\leq\top_e$，那么不反复伸展它到 $P_e[i\cdots(j+1)],P_e[i\cdots(j+2)],\ldots,P_e[i\cdots x]$。相反，这里希望直接伸展到从 $i$ 开始的最后一个可能的候选窗口，记为 $P_e[i\cdots x]$，其中 $x$ 满足 $p_x-p_i+1\leq\top_e$，并且对于任何 $x'>x$，满足 $p_{x'}-p_i+1>\top_e$。

如果伸展操作函数 $F(x)=p_x-p_i+1$ 和平移操作函数 $F'(mid)=p_{mid+j-i}-p_{mid}+1$ 是单调的，那么可以使用二分查找的方法来快速寻找 $x$ 和 $mid$。

对于伸展操作，显然 $F(x)=p_x-p_i+1$ 是单调的，因为 $F(x+1)-F(x)=p_{x+1}-p_x>0$。下一步给出查找范围的下限和上限。首先，显然 $x\geq j$；此外，因为 $p_i+j\leq p_{i+j}$，有 $p_x\leq p_i+\top_e-1\leq p_{i+\top_e-1}$ 以及 $x\leq i+\top_e-1$。因此可以通过在 $j$ 和 $i+\top_e-1$ 之间做二分查找来快速找到 $x$。

然而 $F'(mid)=p_{mid+j-i}-p_{mid}+1$ 并不单调。但是，观察到 $F''(mid-1)-F''(mid)=p_{mid}-p_{mid-1}-1\geq 0$，所以 $F''(mid)=(p_j+(mid-i))-p_{mid}+1$ 是单调的。更重要的是，对于任何 $i\leq mid\leq j$, 因为 $(p_j+(mid-i))\leq p_{mid+j-i}$，所以 $F''(mid)<F'(mid)$。因此如果 $F''(mid-1)>\top_e$,那么有 $F'(mid-1)>\top_e$，这种情

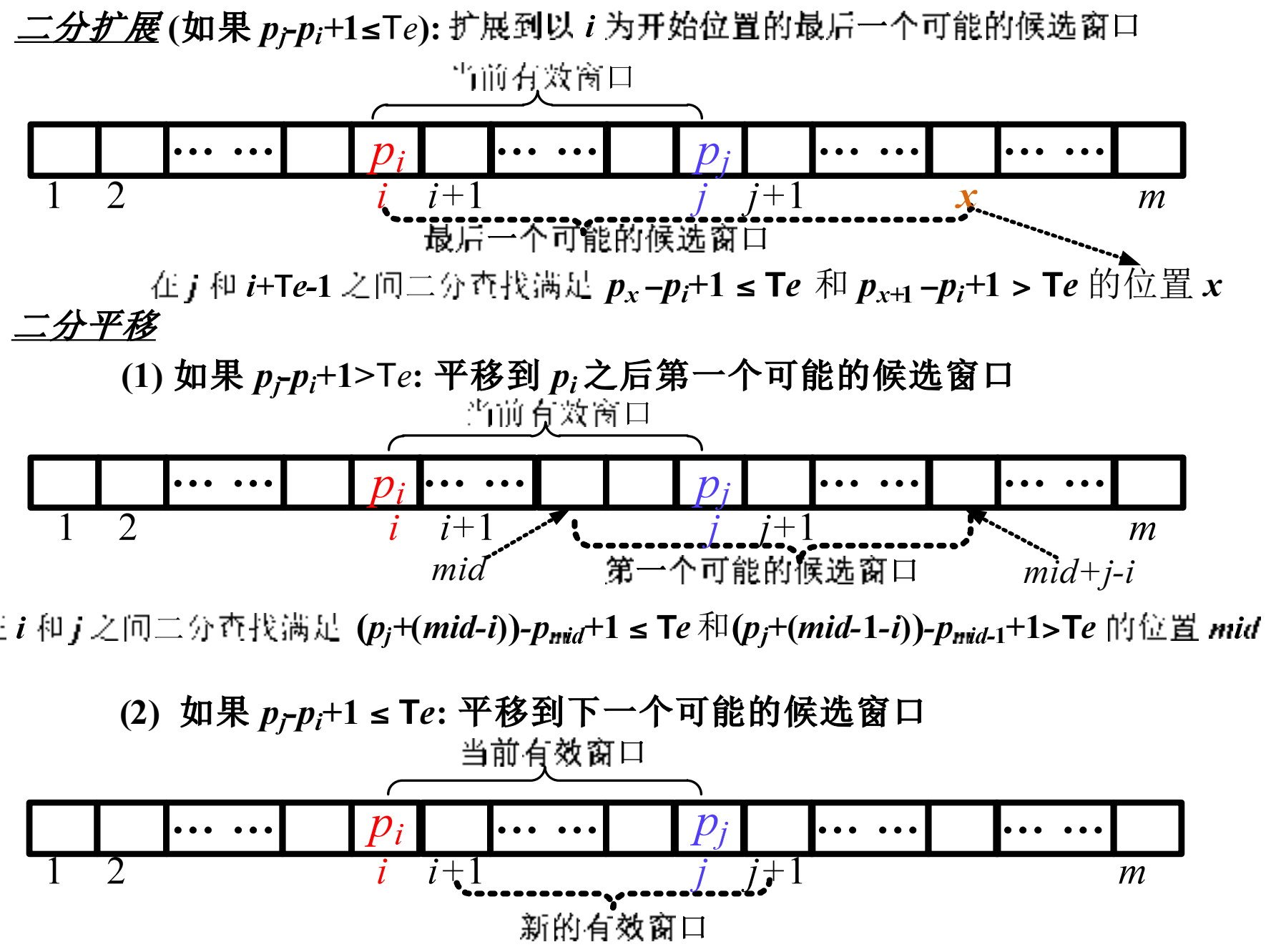

图 2.10　二分伸展操作和二分平移操作

况下 $P_e[(mid-1)\cdots(mid-1+j-i)]$ 不可能是候选窗口。但是如果 $F''(mid) \leq \top_e$，那么 $P_e[(mid)\cdots(mid+j-i)]$ 可能是个候选窗口。通过这种方式，可以在 $i$ 和 $j$ 之间通过二分查找满足条件 $F''(mid-1) > \top_e$ 和 $F''(mid) \leq \top_e$ 的 $mid$ 的位置。如果 $F'(mid) \leq \top_e$，那么已经找到最后一个可能的候选窗口；否则，继续在 $mid+1$ 和 $mid+1+j-i$ 之间二分查找 $mid'$，如此反复，可以找到最后一个可能的候选窗口。

因此，给定一个有效窗口 $P_e[i\cdots j]$，可以使用二分伸展操作和二分平移操作来查找候选窗口（图 2.10）。

- 二分伸展：如果 $p_j - p_i + 1 \leq \top_e$，那么首先在 $j$ 和 $i+\top_e-1$ 之间二分查找满足 $p_x - p_i + 1 \leq \top_e$ 和 $p_{x+1} - p_i + 1 > \top_e$ 的 $x$, 然后直接扩展到 $P_e[i\cdots x]$。
- 二分平移：如果 $p_j - p_i + 1 > \top_e$, 那么在 $i$ 和 $j$ 之间二分搜索满足 $(p_j + (mid - i)) - p_{mid} + 1 \leq \top_e$ 和 $(p_j + (mid-1-i)) - p_{mid-1} + 1 > \top_e$ 的 $mid$。如果 $p_{mid+j-i} - p_{mid} + 1 > \top_e$，那么在 $mid+1$ 和 $mid+1+j-i$ 之间反复地做二分平移操作。相反，如果 $p_j - p_i + 1 \leq \top_e$，那么平移到一个新的有效窗口 $P_e[(i+1)\cdots(j+1)]$。

给定一个有效窗口 $P_e[i\cdots j]$，引理 2.4证明，二分平移操作可以跳过很多不必要的有效窗口（非候选窗口），比如 $P_e[(i+1)\cdots(j+1)]$，…，$P_e[(mid-1)\cdots(mid-1+j-i)]$。例如，考虑图2.11中的位置列表。假设 $\tau = 2$，那么有 $T_l = 4, |e_4| = 8, \top_{e_4} = 10$。对于第一个有效窗口 $P_{e_4}[1\cdots4]$，平移操作会将它依次平移到 $P_{e_4}[2\cdots5], P_{e_4}[3\cdots6], \cdots, P_{e_4}[6\cdots9]$ 并检查它们是否是候选窗口。而二分平

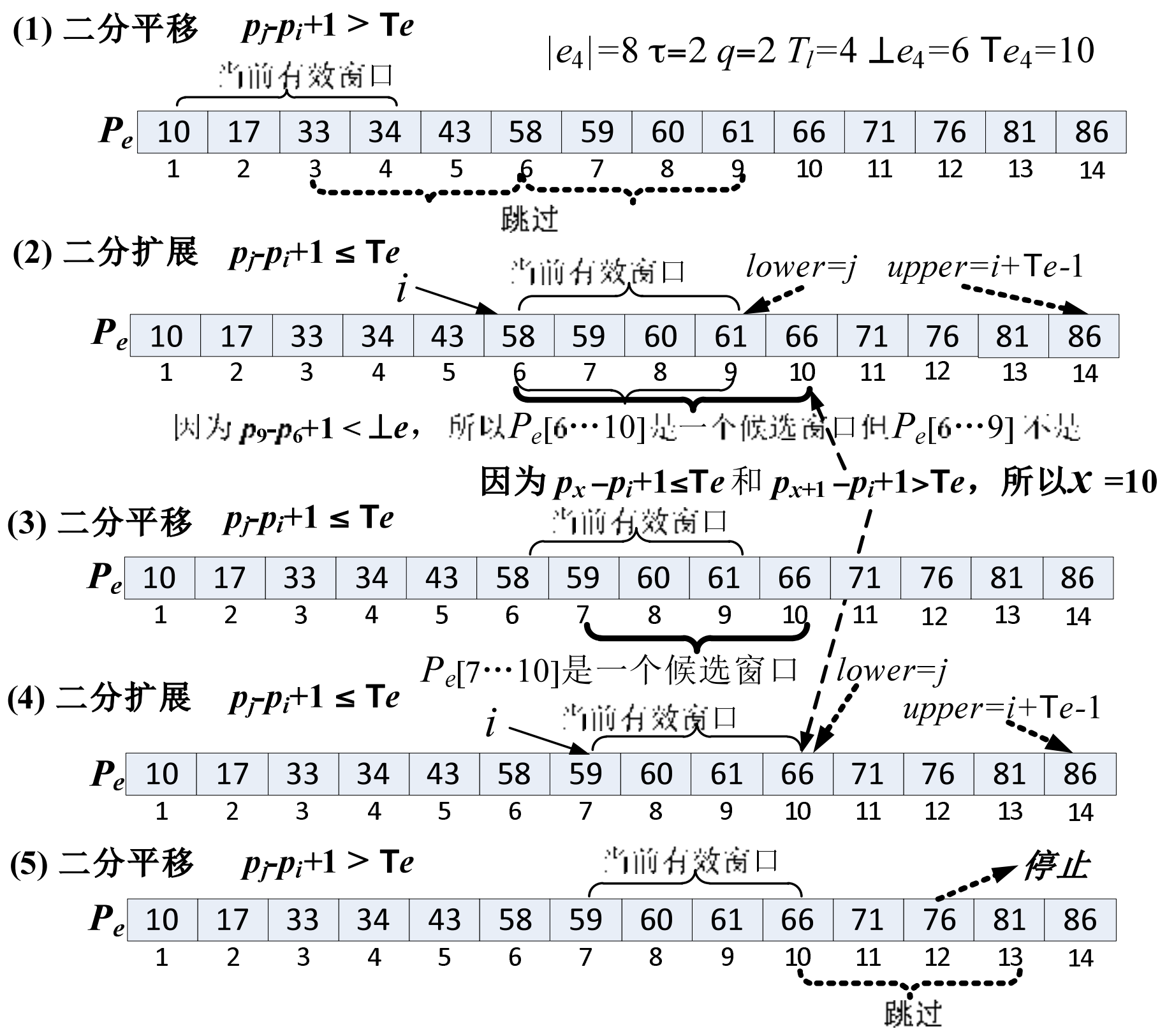

图 2.11　二分伸展操作和二分平移操作的一个实例

移操作可以直接将它平移到 $P_{e_4}[3\cdots6]$，然后再平移到 $P_{e_4}[6\cdots9]$，因此它可以跳过很多有效窗口。

**引理** 2.4：　给定满足 $p_j-p_i+1>\top_e$ 的有效窗口 $P_e[i\cdots j]$，如果 $(p_j+(mid-i))-p_{mid}+1\le\top_e$ 并且 $(p_j+(mid-1)-i)-p_{mid-1}+1>\top_e$，那么 $P_e[(i+1)\cdots(j+1)]$, $\ldots$, $P_e[(mid-1)\cdots((mid-1)+j-i)]$ 都不是候选窗口。

**证明** 用反证法证明该引理。假设存在一个有效窗口 $P_e[k\cdots(j+k-i)]$ 也是一个候选窗口，其中 $i\le k\le mid-1$，那么 $p_{j+k-i}-p_k+1\le\top_e$。因为 $p_j+k-i\le p_{j+k-i}$，所以 $p_j+k-i-p_k+1\le\top_e$，因此 $p_j+k-i+(mid-1-k)-(p_k+(mid-1-k))+1\le\top_e$，所以 $p_j+(mid-1)-i-(p_{mid-1})+1\le\top_e$。这与 $p_j+(mid-1)-i-p_{mid-1}+1>\top_e$ 互相矛盾。因此 $P_e[(i+1)\cdots(j+1)],\ldots,P_e[(mid-1)\cdots(j+mid-1-i)]$ 都不是候选窗口。

二分伸展操作可以直接伸展到最后一个可能的候选窗口 $P_e[i\cdots x]$，它有两个优点：首先，在很多应用中，用户希望找到最相似的元组（共享公共标记越多越好），而二分伸展操作可以快速的找到这样的子字符串；第二，不需要在 $P_e[i\cdots(j+1)],\ldots,P_e[i\cdots x]$ 中一个一个的寻找 $e$ 的候选子字符串，相反，由于在 $lo=p_j-\top_e+1$

**Algorithm 2.1**: 查找候选窗口

**Input**: $e$: 一个实体；$P_e$:$e$ 的位置列表；$T_l$: 阈值；$\top_e$: 标记数目的上界。

1 **begin**<br>
2 　$i = 1$;<br>
3 　**while** $i \le |P_e| - T_l + 1$ **do**<br>
4 　　$j = i + T_l - 1$;<br>
5 　　**if** $p_j - p_i + 1 \le \top_e$ **then**<br>
6 　　　BinarySpan$(i, j)$;<br>
7 　　　$i = i + 1$;<br>
8 　　**else** $i$ = BinaryShift$(i, j)$;<br>
9 **end**

**Procedure** BinarySpan $(i, j)$

**Input**: $i$: 起始位置；$j$: 结束位置。

1 **begin**<br>
2 　$lower = j; upper = i + \top_e - 1$ ;<br>
3 　**while** $lower \le upper$ **do**<br>
4 　　$mid = \lceil (upper + lower)/2 \rceil$;<br>
5 　　**if** $p_{mid} - p_i + 1 > \top_e$ **then** $upper = mid - 1$;<br>
6 　　**else** $lower = mid + 1$;<br>
7 　$mid = upper$ ;<br>
8 　从中 $D[i \cdots mid]$ 寻找候选窗口;<br>
9 **end**

和 $up = p_{i+x-j} + \top_e - 1$ 之间可能有很多候选子字符串，可以成批地寻找它们。具体说来，可以按照候选子字符串的标记数目对它们分组，同一组中的实体拥有相同数量的标记。考虑有 $g$ 个标记的组，假设 $T_g$ 是用 $|e|$ 和 $g$ 计算出的阈值。如果 $|P_e[i \cdots x]| < T_g$，那么可以直接剪枝掉该组中的所有候选子字符串。

可以使用这两个二分查找操作来替换平移操作和伸展操作以跳过一些有效窗口。算法2.1利用这两种操作查找候选窗口。它首先初始化第一个有效窗口(第2到4行)。然后，它用这两个二分操作来扩充当前的有效窗口直至最后一个有效窗口。如果当前有效窗口的标记的数目不大于 $\top_e$，它调用子程序 BinarySpan (第6行) 并做一个二分平移操作（第7行）；否则，它调用子程序 BinaryShift (第8行)。BinarySpan 通过二分查找来找到最后一个开始于 $p_i$ 的候选窗口（第3到6行），然后检索候选窗口 (第8行)。BinaryShift 通过二分查找来找到位于 $p_i$ 之后的第一个可能的候选窗口。如此反复，它可以找到所有的候选窗口。图2.11给出了一个例子来解释这个算法。

**Procedure** BinaryShift ($i, j$)

**Input**: $i$: 起始位置； $j$: 结束位置。
**Output**: 新的起始位置。

```
begin
    lower = i; upper = j;
    while lower ≤ upper do
        mid = ⌈(lower + upper)/2⌉;
        if (p_j + (mid − i)) − p_mid + 1 > ⊤_e then
            lower = mid + 1;
        else upper = mid − 1;
    i = lower; j = i + T_l − 1;
    if p_j − p_i + 1 > ⊤_e then return BinaryShift (i, j);
    else return i;
end
```

**时间复杂度**：不同于基本的基于单堆的算法，改进的基于单堆的算法不需要使用数组来清点出现次数，它从每个位置列表 $P_e$ 中寻找候选窗口，对于任意位置列表 $P_e$，最坏情况下，它的大小在 $\bot_E$ 和 $\top_E$ 之间的子列表都是候选窗口。因此，寻找候选窗口的时间复杂度是 $|P_e| * \top_E$。这个方法为每个位置列表都需要寻找候选窗口，由于位置列表长度之和和倒排列表长度之和是相同的，因此利用改进的方法寻找候选窗口的时间复杂度是 $O(N * \top_E)$。

## 2.5 Faerie 算法

本节给出一个基于单堆的算法来快速寻找结果，并称之为 Faerie。

首先为给定字典 $E$ 中所有的实体构建一个倒排索引，然后对文档 $D$，获取它的标记和对应的倒排列表。接下来，根据 $D$ 的标记的倒排列表构建一个单堆。通过堆从小到大的遍历实体，对每个堆中的实体 $e$，获取它的位置列表 $P_e$。如果 $|P_e| < T_l$，根据 lazy-count 剪枝技术可以剪枝掉 $e$；否则，使用上文中的两种二分操作来寻找候选窗口。然后，根据候选窗口生成候选元组。最后验证候选元组并确定最终结果，算法给出了 Faerie 的伪代码。

Faerie 算法首先为预定义的实体构建一个倒排索引（第 2 行），然后切分文档，获取对应的倒排列表（第 3 行）并构建小顶堆（第 4 行）。Faerie 用 $\langle e_i, p_i \rangle$ 来表示堆顶的元素，其中 $e_i$ 是当前 ID 最小的实体，$p_i$ 是 $e_i$ 所在的倒排列表的位置。Faerie 用位置列表 $P_e$ 来保存文档中所有包含 $e$ 的倒排列表的位置（第5行）。然后，对每个堆顶元素 $\langle e_i, p_i \rangle$，如果 $e_i = e$，那么 Faerie 把 $p_i$ 添加到 $P_e$ 中（第7到8行），其中

**Algorithm 2.4**: Faerie 算法

**Input**: 实体字典 $E = \{e_1, e_2, \ldots, e_n\}$，一个文档 $D$，一个近似函数与阈值。
**Output**: $\{\langle s, e\rangle |$ $s$ 和 $e$ 在给定近似函数与阈值下相似，$s$ 是 $D$ 的子字符串，$e \in E\}$。

1 **begin**
2 　把 $E$ 中的实体切分为标记并创建倒排索引;
3 　把文档 $D$ 切分为标记集合，获取 $D$ 中标记所对应的倒排列表;
4 　根据 $D$ 的倒排列表构建小顶堆 $H$;
5 　为堆 $H$ 顶的元素 $e$ 初始化位置列表 $P_e = \phi$;
6 　**while** $(\langle e_i, p_i\rangle = H.top)! = \phi$ **do**
7 　　**if** $e_i == e$ **then**
8 　　　$P_e \cup = \{p_i\}$;
9 　　**else**
10 　　　为实体 $e$ 推导相应的阈值 $T_l$;
11 　　　**if** $|P_e| \geq T_l$ **then**
12 　　　　用算法2.1寻找候选窗口;
13 　　　　从候选窗口中获取候选元组;
14 　　　$e = e_i$; $P_e = \{p_i\}$;
15 　　调整堆 $H$;
16 　验证候选元组;
17 **end**

$e$ 是最后一个从堆中弹出的实体；否则，Faerie 按照如下方式处理 $e$ 的位置列表。根据给定近似函数和阈值，Faerie 为实体 $e$ 计算阈值 $T_l$。如果 $|P_e| \geq T_l$，那么这个实体可能存在候选元组。Faerie 根据算法2.1生成候选窗口（第12行），并根据候选窗口寻找候选元组（第13行）。接下来 Faerie 调整堆来为下一个实体生成候选元组（第15行）。最后，Faerie 验证候选元组并得到最后结果（第16行）。

接下来给出一个例子来贯穿 Faerie。考虑表2.1中的实体字典和文档。首先构建一个单堆（图2.6）。然后调整堆来为每个实体产生位置列表。如图2.11所示，考虑 $e_4$（“venkatesh’’）的位置列表。假设 $\tau = 2$，那么 $|e_4| = 8$，$\perp_E = 6$，$\top_E = 12$，$\perp_{e_4} = 6$，$\top_{e_4} = 10$，$T_l = 4$。然后使用二分平移和二分伸展操作来获取候选窗口（$P_e[6 \cdots 10]$ 和 $P_e[7 \cdots 10]$），接下来根据候选窗口来生成候选元组（例如 $\langle D[58, 9]$=“*venkaeesh*’’，$e_4$=“venkatesh”$\rangle$）。最后验证候选元组并得到最终结果。

### 2.5.1　正确性和完备性

本节给出定理2.1证明 Faerie 算法的正确性和完备性。

表 2.4　实验数据集的详细信息

| 数据集 | 大小 | 平均字符数量 | 平均单词数量 | 备注 |
|---|---|---|---|---|
| DBLP Dict | 100,000 | 21.1 | 2.77 | 作者名字 |
| DBLP Docs | 10,000 | 123.3 | 16.99 | 文章元信息 |
| PubMed Dict | 100,000 | 52.96 | 6.98 | 文章标题 |
| PubMed Docs | 10,000 | 235.8 | 33.6 | 文章元信息 |
| WebPage Dict | 100,000 | 66.89 | 8.5 | 文章标题 |
| WebPage Docs | 1,000 | 8949 | 1268 | 网页 |

**定理** 2.1 (**正确性和完备性**)：　Faerie 能够从文档中正确地、完备地抽取出所有近似子字符串。

**证明**　首先证明 Faerie 算法能够完备地找到所有结果，即给定一个子字符串 $s$ 和一个实体字符串 $e$ 组成的近似元组 $\langle s, e\rangle$，Faerie 算法能够找到这个元组作为一个结果。因为 $s$ 和 $e$ 近似，根据引理2.3，$|e \cap s| \geq T_l$。对于位置列表 $P_e$，有 $|P_e| \geq |e \cap s| \geq T_l$。因此，位置列表 $P_e$ 能够通过 lazy-count 剪枝，Faerie 算法将会从该位置列表中寻找候选窗口。因为 $s$ 与 $e$ 相似，根据引理2.2，有 $\perp_e \leq |s| \leq \top_e$。不失一般性的，假设 $P_e[i]$ 和 $P_e[j]$ 分别是 $s$ 与 $e$ 的第一个和最后一个共同标记所在的位置。Faerie 算法一定能找到 $P_e[i \dots j]$ 作为一个候选窗口，因为 $T_l \leq |P_e[i \dots j]| \leq |s| \leq \top_e$。在验证步骤中，Faerie 算法一定能找到 $\langle s, e\rangle$ 作为一个结果。因此 Single 算法满足完备性。

接下来证明 Faerie 算法的正确性，即 Faerie 算法返回的所有子字符串和实体元组都一定是近似元组。因为 Faerie 算法包含一个验证步骤，这个步骤只返回那些近似元组，因此 Faerie 算法也满足正确性。

## 2.6　实验

本节报告实验结果，实验的目的是衡量算法的性能。

**实验设置**：实验把 Faerie 同现有的最好的方法 NGPP[26]（编辑距离近似函数下最好的方法）和 ISH[27]（Jaccard Similarity 和 Edit Similarity 近似函数下最好的方法）作比较。实验中从 ‘‘近似连接’’ 项目网站中下载了 NGPP 的可执行程序并自行实现了 ISH 算法。实验报告这两个方法的最佳性能。所有的算法都是用 C++ 语言实现，并使用 GCC 4.2.4 加上-O3 编译优化选项进行编译。所有的实验都是在一个拥有 Intel Core 2 Quad X5450 3.0GHz 处理器和 4GB 内存的 Ubuntu 机器上进行。

**实验数据集**：实验中运用了三个真实的数据集：DBLP[①]，PubMed[②]和 ACM WebPage[③]。DBLP 是一个计算机科学出版物的数据集，实验中随机选取了 100,000 个作者名作为字典实体，10,000 条论文记录作为文档。PubMed 是一个生物医药出版物的数据集，实验中随机选取了 100,000 篇文章标题作为字典实体，10,000 条出版记录作为文档。WebPage 是一个计算机科学刊物网页数据集，实验中随机选取了 100,000 个论文标题作为字典实体，1,000 个网页作为文档（每个网页包含数千个标记）。表2.4给出了数据集的详细情况。实验中在实体和文档中都不区分不同的属性，字典中的每个实体就是一个完整的字符串。

### 2.6.1　比较基于多堆的方法和基于单堆的方法

本节把基于多堆的方法同基于单堆的方法（没有加上第 2.4节中的剪枝技术）进行比较。实验中采用不同的相似函数，在所有三个数据集上测试了它们在这两种方法下的性能。图 2.12展示了实验结果。从图中可以看到基于单堆的方法比基于多堆的方法快 1 到 2 数量级，甚至有时要好 3 个数量级。比如，在编辑距离近似函数下，当阈值 $\tau = 3$ 时，基于多堆的方法在 DBLP 数据集上花费了超过 10,000 秒，然而基于单堆的方法只花费了 180 秒。在 EDS 近似函数下，当阈值 $\delta = 0.9$ 时，基于多堆的方法在 PubMed 数据集上花费了超过 14,000 秒，然而基于单堆的方法仅仅花费了 600 秒。基于单堆的方法要好于基于多堆的方法有如下两个原因。首先，基于多堆的方法需要多次扫描了文档中的倒排列表，而基于单堆的方法仅仅扫描了它们一次。第二，以多堆为基础的方法构造的堆比基于单堆的方法构建的堆更加宽广（堆调整代价与堆的宽度正相关），并且实施了更多的堆调整操作。因为基于单堆的方法要远优于基于多堆的方法，所以在以下的实验中，只关注基于单堆的方法。

### 2.6.2　剪枝技术的有效性

本节检测剪枝技术的有效性。实验首先评估不同的剪枝技术（lazy-count 剪枝，bucket-count 剪枝和 binary span and shift 剪枝。由于 batch-count 剪枝是 binary span and shift 剪枝的一个特例，因此实验中只展示 binary span and shift 的结果）下候选元组的数量。在本文中，候选元组的数目是指用于清点出现次数的数组中非零元素的个数，因为这些元素是需要被验证的。图2.13展示了试验结果。实验中在 DBLP 数据集上测试了编辑距离函数，在 WebPage 数据集上测试了 Jaccard Similarity 函

① http://www.informatik.uni-trier.de/Simley/db

② http://www.ncbi.nlm.nih.gov/pubmed

③ http://portal.acm.org

数，在 PubMed 数据集上测试了 Edit Similarity 函数。值得注意的是，图中 y 轴的刻度是指数增长的。比如有一亿候选元组，图中对应的 y 轴的读数就是 8（因为 $10^8 = 100,000,000$）。在实验中，本章提出的方法在 DBLP 数据集上在 $\tau = 0, 1, 2, 3, 4$ 时分别采用了参数 $q = 16, 8, 5, 4, 3$；在 PubMed 数据集上在 $\delta = 1, 0.95, 0.9, 0.85, 0.8$ 时分别采用了参数 $q = 26, 11, 7, 5, 4$。

可以观察到本章提出的剪枝技术可以剪枝掉大量的候选元组。比如，在 DBLP 数据集下，当 $\tau = 3$ 时，没用任何剪枝技术的方法有 110 亿候选元组；lazy-count 剪枝将其数量减少到 8 亿 6 千万；bucket-count 剪枝将其数量进一步减少到 6 亿；binary span and shift 剪枝则仅仅只有 2 亿候选元组。在 WebPage 数据集下，当 $\delta = 0.9$ 时，binary span and shift 剪枝把候选元组的数量从 100 亿减少至 35。在 PubMed 数据集下，当 $\delta = 0.85$ 时，binary span and shift 剪枝把候选元组的数量从 1800 亿减少至一亿两千万。这其中的主要原因是剪枝技术计算了实体和子字符串共享标记数目的上界。如果这个上界比共享标记数目阈值还要少的话，那么可以直接剪枝掉这个子字符串；如果对于任意一个子字符串，这个实体的上界都小于阈值，那么可以直接剪枝掉该实体。实验结果反映了剪枝技术的有效性。

接下来评估剪枝技术对性能的提升。图2.14展示了试验结果，从图中可以观察到剪枝技术可以提升性能。比如，在 DBLP 数据集下，当 $\tau = 3$ 时，没有运用任何剪枝技术的方法耗时是 180 秒，lazy-count 剪枝将时间减少到了 43 秒；binary span and shift 剪枝将时间进一步减少至 25 秒。在 PubMed 数据集下，当 $\delta = 0.9$ 时，剪枝技术可以将时间从 600 秒减少至 8 秒。这表明剪枝技术可以提高性能。余下的实验关注基于单堆的方法，并使用 binary span and shift 剪枝。

### 2.6.3　与现有的最好的方法的比较

本节将 Faerie 同现有的最好的方法 NGPP[26] (仅支持编辑距离函数) 和 ISH[27] (仅支持 Edit Similarity 和 Jaccard Similarity 函数) 进行比较。实验中调整 NGPP 和 ISH 的参数 (比如 NGPP 的前缀长度）以报告它们最佳的性能。图 2.15展示了该实验结果。可以看到 Faerie 取得了最佳的性能。特别是 Faerie 在 Edit Similarity 和 Jaccard Similarity 函数下要优于 ISH 1 至 2 个数量级。比如，在 PubMed 数据集下，当 Edit Similarity 阈值 $\delta = 0.9$ 时，ISH 的耗时是 1000 秒，而 Faerie 将耗时减少至 8 秒。这是因为 Faerie 在标识重叠的地方共享了计算。另外，本章提出的剪枝技术可以减少大量不必要的有效子字符串，从而减少了候选元组的数值。尽管 NGPP 在较小编辑距离阈值时取得了高性能，但是它在对较大的编辑距离阈值时效率很低。这是因为它需要枚举实体的删集，阈值越大，实体需要枚举的删集就越多。在 Jaccard

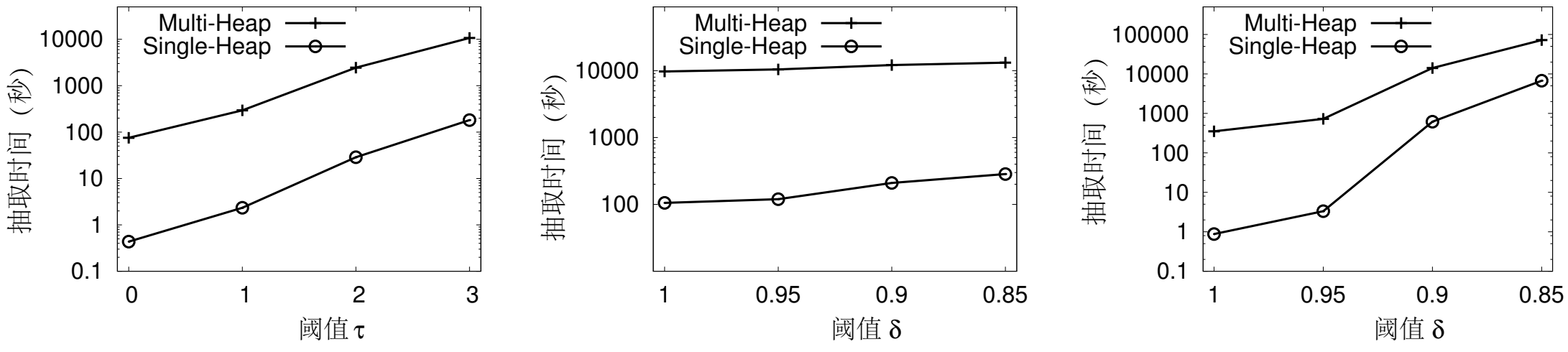

(a) ED (DBLP)　(b) JAC (WebPage)　(c) EDS (PubMed)

图 2.12　基于多堆的方法与基于单堆的方法的性能比较

Similarity 函数下，由于每个实体很小数量的标识（平均数量是 8），不同的阈值下的 $T_l$ 和 $\top_e$ 几乎相同（当 $\delta$=1 时，$T_l = 8$；当 $\delta$=0.8 时，$T_l = 10$），因此 Faerie 在不同的 Jaccard Similarity 阈值下性能变动非常小。

此外实验还比较了不同算法的索引大小。注意，因为 NGPP 使用编辑距离阈值 $\tau$ 来产生删集，所以它在不同的阈值下，有不同的索引大小。当编辑距离阈值越大时，因为实体有越多的删集，所以索引的大小就越大。在 DBLP 数据集下，当 $\tau = 3$ 时，NGPP 占用了约 43MB 大小的索引空间，而 Faerie 的索引大小仅仅只有 7MB（$q = 4$ 时）。这个结果与[26] 中的结果是一致的：基于$q$-gram 的方法比基于 neighborhood 的方法 (NGPP) 的索引大小更小。在 WebPage 数据集下，当 Jaccard Similarity 的阈值 $\delta = 0.9$ 时，ISH 的索引占用了 18MB 的空间（取参数 $k = 3$），而 Faerie 的索引仅占用了 4MB。

### 2.6.4　字典大小的可扩展性

本节评估了 Faerie 在各种相似度函数下的可扩展性。实验中逐渐增加字典中的实体数量，从表 2.4中的文档集合中抽取实体，图 2.16展示了五个近似函数下的实验结果。可以观察到随着字典大小的增加，Faerie 的扩展性很好。比如在 DBLP 数据集下，当 $\tau = 3$ 时，Faerie 在 20,000 个实体下的耗时是 6 秒，在 100,000 个实体下的耗时是 25 秒。在 WebPage 数据集下，因为每个实体的标记数目更少，Faerie 在不同的阈值下的运行时间变动变动很小。在 PubMed 数据集下，使用$q$-gram 做标记，评估 Edit Similarity, Dice Similarity 和 Cosine Similarity 近似函数下算法的可扩展性。对于 Edit Similarity，当 $\delta = 0.85$ 时，Faerie 在 20,000 个实体下的耗时是 9 秒，在 100,000 个实体下的耗时是 48 秒。

此外实验还评估了索引大小随着实体字典大小增加时的变化，表2.5展示了实验结果。可以看到，随着实体数量增加，Faerie 的索引大小很小，扩展性也很好。

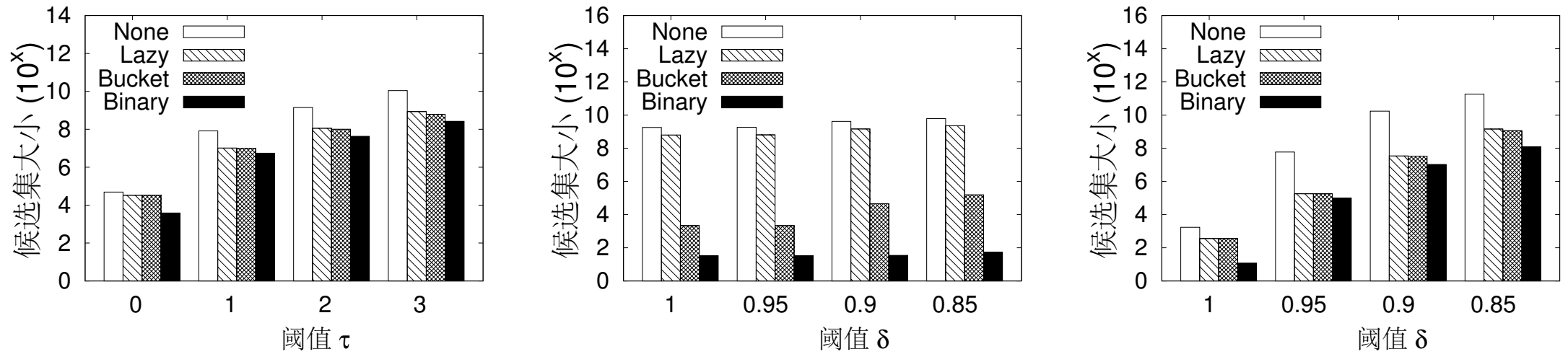

(a) ED (DBLP)　(b) JAC (WebPage)　(c) EDS (PubMed)

图 2.13　不同剪枝技术下的候选元组数目

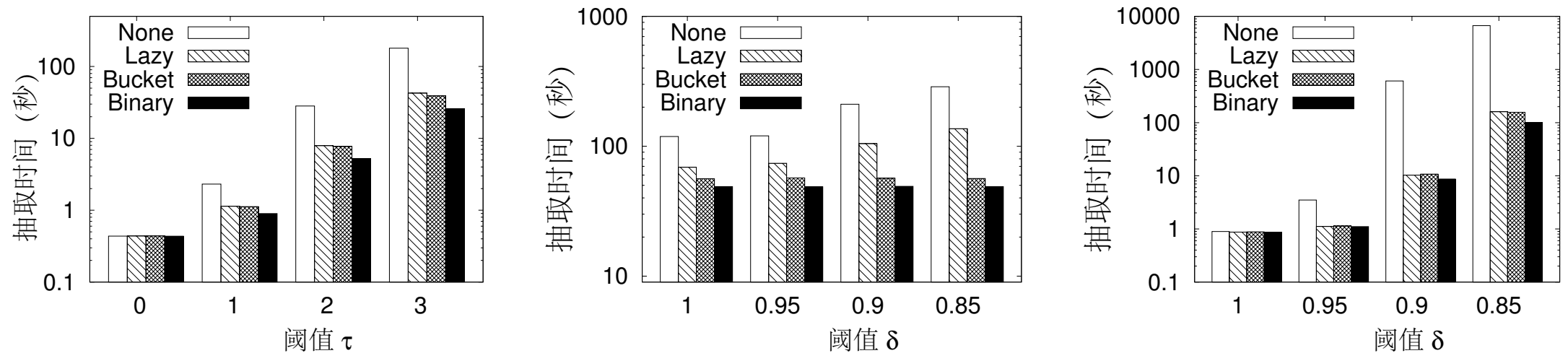

(a) ED (DBLP)　(b) JAC (WebPage)　(c) EDS (PubMed)

图 2.14　不同剪枝技术下的算法性能

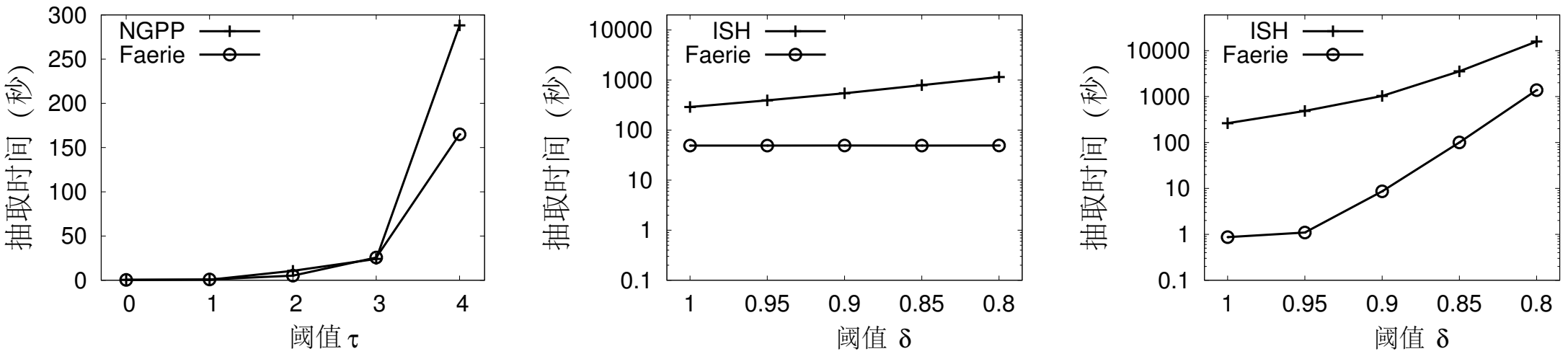

(a) ED (DBLP)　(b) JAC (WebPage)　(c) EDS (PubMed)

图 2.15　与现有的最好的方法的比较

## 2.7　相关工作

关于近似实体抽取的研究[25–27,71–73]。Wang 等人[26] 提出了一种基于 Neighborhood-Generation 的方法来解决近似函数编辑距离约束下的近似实体抽取问题。他们首先将字符串按照规则分割成多个片段，以确保如果两个字符串相似，那么一定存在两个片段的编辑距离不大于 1。然后他们通过删除这些片段中的一个字符来产生 Neighborhoods，这样的话，当且仅当两个片段有相同的 Neighborhoods 时，这两个片段的编辑距离才不大于 1。但是这个方法不能支持基于字词的近似函数。Chakrabarti 等人[27] 提出了一种基于 Inverted Signature 的哈希表来解决近似实体抽取问题。他们首先选择权重最大的一些标记作为签名，然后将实体字典编排成 0-1 矩阵，接下来他们为文档也建立一个矩阵，并利用该矩阵去找到候选元组。通过运用更加严格的临界值，Lu 等人[72] 提出利用 Signature-based Inverted Lists 来提高 Inverted Signature-based Hash-table 的性能。然而这些方法并不能支持编辑距

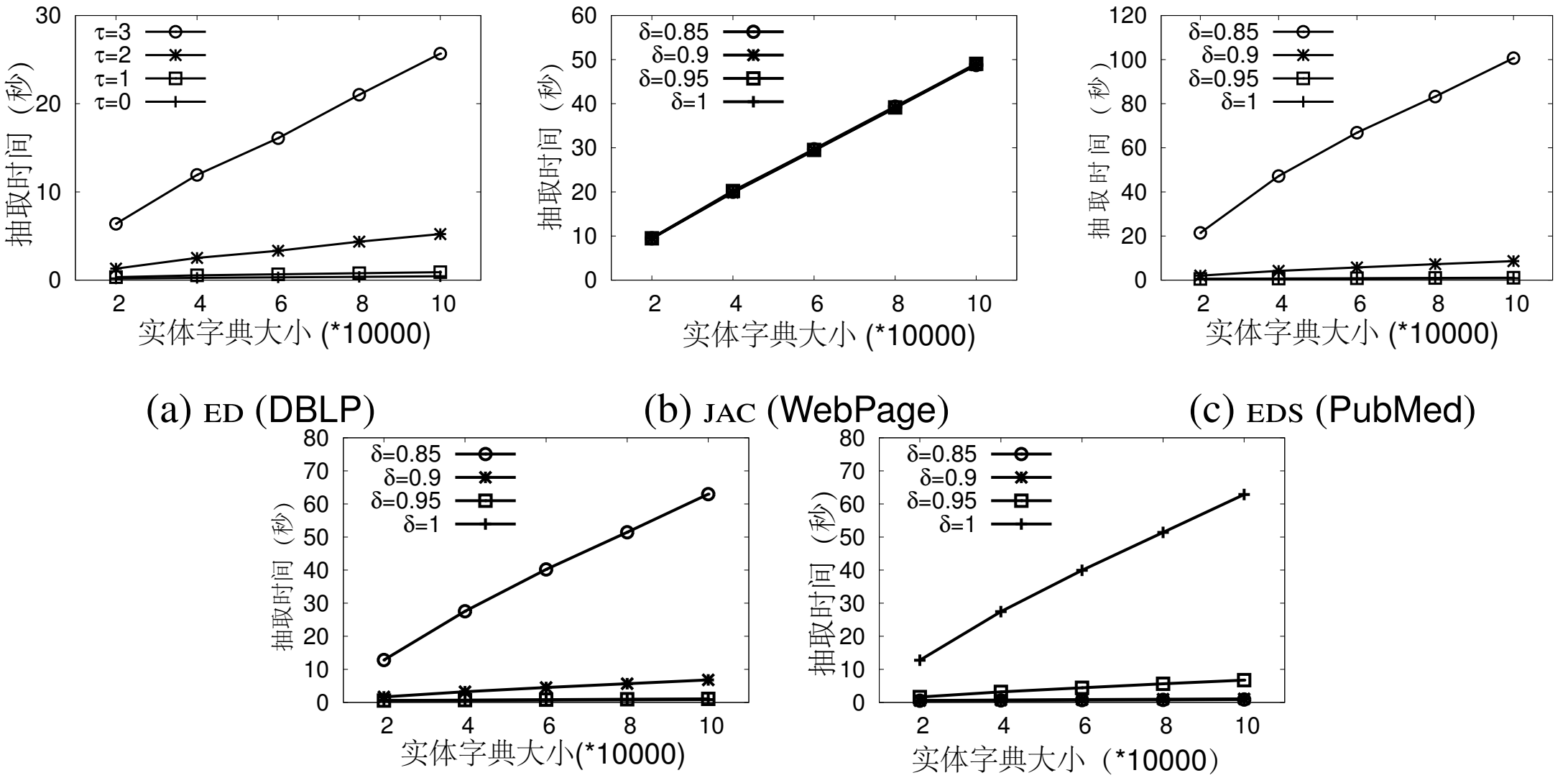

(a) ED (DBLP)　(b) JAC (WebPage)　(c) EDS (PubMed)

(d) DICE (PubMed)　(e) COS (PubMed)

图 2.16　不同数据集、不同近似函数下算法可扩展性实验

表 2.5　索引的可扩展性

($a$) DBLP ($q = 5$)

| 实体字典大小 | 20k | 40k | 60k | 80k | 100k |
|---|---|---|---|---|---|
| 索引大小 (MB) | 1.6 | 3.22 | 4.9 | 6.5 | 8.2 |
| 堆和数组大小 (KB) | 4.5 | 4.5 | 4.5 | 4.5 | 4.5 |

($b$) WebPage

| 实体字典大小 | 20k | 40k | 60k | 80k | 100k |
|---|---|---|---|---|---|
| 索引大小 (MB) | 0.8 | 1.63 | 2.45 | 3.3 | 4.2 |
| 堆和数组大小 (KB) | 38 | 38 | 38 | 38 | 38 |

($c$) PubMed ($q = 7$)

| 实体字典大小 | 20k | 40k | 60k | 80k | 100k |
|---|---|---|---|---|---|
| 索引大小 (MB) | 4.5 | 9.2 | 14.1 | 18.3 | 22.8 |
| 堆和数组大小 (KB) | 7.2 | 7.2 | 7.2 | 7.2 | 7.2 |

离差异函数。另外，Agrawal 等人[73] 运用 Inverted Lists 来进行“一对一”的实体抽取。Chandel 等人[25] 研究了 Batch Top-K 的基于实体字典的实体识别。Chaudhuri 等人[71] 提出通过挖掘大量的文件集合来扩大参考实体字典。

还有很多工作着重于研究字符串近似搜索问题[27,28,35,42,43,45,46,74–77] 和字符串近似连接问题[28–30,36,40,70,78]。尽管可以扩展这些方法来解决近似抽取问题，但是他们需要枚举出文档中的所有的子字符串，并且还不能利用子字符串的重叠部分来进行共享计算。这些问题导致它们的效率极低。目前已有的工作（NGPP[26] 和 ISH[27]）已经表明，对于近似实体抽取问题，基于抽取的方法要优于基于近似连接的方法。

因此本文只和现有的最好的算法 NGPP[26] 以及 ISH[27] 比较就足够了。除此之外，关于估计近似查询字符串的选择性问题也已经有了很多相关研究[79–81]。

## 2.8　本章小结

本章研究了近似实体抽取问题。论文提出了一个统一的框架来支持不同的相似函数并设计了基于堆的过滤算法来从文档中高效的抽取与实体近似的子字符串。此外，论文研究了一个基于单堆的算法，通过在文档中的标记所对应的倒排列表上构建一个堆并扫描这些倒排列表一次，它能够共享不同子字符串重叠部分的计算。论文还提出了几个快速有效的剪枝技术来剪枝掉大量的不必要候选元组。最后论文设计了基于二分查找的技术来提高性能。论文在真实数据集上进行实验，实验结果表明论文提出的方法取得了高性能，并且比现有最好的方法效率更好。

# 第 3 章　基于序列相似性的近似连接方法

## 3.1　引言

近似连接是指在两个数据集中找到所有近似的数据对。这个问题在现实中有很多应用，如数据集成、数据清洗、协同过滤和实体解析等。由于现实世界中很多数据都能够用序列来表示，例如 DNA，蛋白质，人名，字符串等，本章研究基于序列相似性的近似连接问题，即找到两个序列数据集中所有相似的序列对。一个字符串也是一个序列，之后不加区分的使用序列和字符串。

近似连接问题的暴力算法需要枚举 $O(n^2)$ 个字符串元组，在数据集较大时，这是不可接受的。为了避免枚举所有的字符串元组，现有的方法都采用过滤加验证的框架。在过滤阶段，现有的方法对每个字符串生成一些特征 (signature) 并保证仅当两个字符串至少共享一个相同的特征时，它们才可能相似。在验证阶段，现有的方法验证在过滤阶段得到的候选元组。

然而现有解决该问题的方法要么只对短的字符串效果比较好，要么只对长的字符串效果比较好。这是因为现有的方法在产生的特征没有考虑字符串自身的长度。本章提出一个基于划分的算法来解决字符串近似连接问题，并命名为 PassJoin。它把字符串平均的划分为片段作为特征，另外也选取一些子字符串作为特征。由于在划分字符串和选取子字符串时考虑了字符串本身的长度，这个方法在短字符串数据集和长字符串数据集下的表现都非常好。在 EDBT 大数据融合竞赛中，该方法以绝对优势取得了冠军①，并比亚军快 10 倍[82,83]。

之后按照如下次序组织本章。首先第 3.2节给出问题的形式化定义并介绍相关工作，然后第 3.3节提出了基于划分的算法框架来解决字符串近似连接问题。接下来第 3.4节提出了有效的子字符串选取算法，之后第 3.5节研究快速验证候选元组的技术。第3.6节讨论如何扩展 PassJoin 以支持 Edit Similarity 、异连接和 MapReduce，第3.7节报告实验结果。最后第 3.8节对本章进行总结。

## 3.2　预备知识

### 3.2.1　问题定义

给定两个字符串集合，字符串近似连接找到其中所有的近似字符串元组。本章研究基于序列相似性的近似连接，并使用编辑距离和 Edit Similarity 来衡量两个

① http://www2.informatik.hu-berlin.de/~wandelt/searchjoincompetition2013/Results.html

字符串的相似性。下文首先基于编辑距离介绍相关算法和技术，然后在第3.6.1节扩展这些算法和技术来支持 Edit Similarity。在本章，如果两个字符串的编辑距离不大于一个给定的阈值 $\tau$，那么就称它们是相似的。接下来正式给出基于编辑距离的字符串近似连接问题定义。

**定义** 3.1 (**字符串近似连接**)：　给定两个字符串集合 $\mathcal{R}$ 和 $\mathcal{S}$ 以及一个编辑距离阈值 $\tau$，字符串近似连接找到 $\mathcal{R} \times \mathcal{S}$ 中所有满足 $\text{ED}(r, s) \le \tau$ 的近似字符串元组 $\langle r, s\rangle$。

不失一般性的，下文首先关注自连接的情况 (即 $\mathcal{R} = \mathcal{S}$ 的情况)，之后在第3.6节扩展支持异连接 (即 $\mathcal{R} \neq \mathcal{S}$ 的情况)。例如，考虑表 3.1(a) 中的字符串，假设编辑距离阈值是 $\tau = 3$，那么 ⟨“kaushik chakrab”, “caushik chakrabar”⟩ 就是一对近似的元组，因为它们的编辑距离不大于 $\tau$。

### 3.2.2　相关工作

关于近似连接有很多相关工作[28–30,36,40,70,78,84]。与本章提出的方法最相关的包括 TrieJoin[31], All-Pairs-Ed[29], ED-Join[30] 和 PartEnum[28]。All-Pairs-Ed 是一个基于 $q$-gram 的方法。它首先为每个字符串生成 $q$-gram，然后根据预先定义的顺序选择前 $q\tau + 1$ 个 $q$-gram 作为前缀。之后它剪枝掉所有在前缀中没有相同 $q$-gram 的字符串元组。最后验证剩下的候选元组并得到结果。ED-Join 通过基于位置的和基于内容的失配过滤 (mis-match) 技术改进了 All-Pairs-Ed，它可以减少前缀中 $q$-gram 的数量。研究表明 ED-Join 的性能超过了 All-Pairs-Ed[29]。TrieJoin 使用一个 Trie[85] 结构、通过前缀过滤技术来解决近似连接问题。PartEnum 提出了一个高效的框架，称之为 Part-Enum，来解决汉明距离下的近似连接问题。研究表明 All-Pairs-Ed 和 PartEnum 都比 ED-Join 和 TrieJoin[31] 的性能差。因此本章实验只与 ED-Join 和 TrieJoin 进行了对比。

Gravano 等人[70] 提出了基于 $q$-gram 的方法来利用 SQL 语句在数据库管理系统中解决近似连接问题。Sarawagi 等人[78] 提出了基于倒排索引的算法来解决近似连接问题。Chaudhuri 等人[36] 提出了高效近似连接的原子操作。Xiao 等人[35] 提出了 PPJoin+ 算法，它通过基于位置的前缀过滤和后缀过滤来改进 AllPair[29] 算法。Xiao 等人[40] 研究了 top-$k$ 近似连接，它不需要给定阈值，可以直接找到集合中最相似的 $k$ 个元组。

除此之外，Jacox 等人[86] 研究了度量空间上的近似连接问题。因为这些方法都没有 ED-Join 和 TrieJoin[31] 效率高，所以本章实验不与它们比较。最近，Wang 等人[87] 设计了一个新的近似函数来同时容忍字符错误和字词错误，他们还设计了高效的算法来支持该相似函数限制下的近似连接。

表 3.1　一个字符串集合

(a) 字符串集合

| 字符串 $s$ |
| --- |
| avataresha |
| caushik chakrabar |
| kaushik chakrab |
| kaushuk chadhui |
| kausic chakduri |
| vankatesh |

(b) 按照长度排序 (升序)

| ID | 字符串 $s$ | 长度 |
| --- | --- | --- |
| $s_1$ | vankatesh | 9 |
| $s_2$ | avataresha | 10 |
| $s_3$ | kaushik chakrab | 15 |
| $s_4$ | kaushuk chadhui | 15 |
| $s_5$ | kausic chakduri | 15 |
| $s_6$ | caushik chakrabar | 17 |

(c) 按照长度排序 (降序)

| ID | 字符串 $s$ | 长度 |
| --- | --- | --- |
| $s_6$ | caushik chakrabar | 17 |
| $s_5$ | kausic chakduri | 15 |
| $s_4$ | kaushuk chadhui | 15 |
| $s_3$ | kaushik chakrab | 15 |
| $s_2$ | avataresha | 10 |
| $s_1$ | vankatesh | 9 |

其他相关工作还有字符串近似检索[41–44,46,69,76]，给定一个查询字符串和一个数据字符串集合，字符串近似检索找到所有与查询串相似的数据字符串。Navarro[88] 研究了近似子字符串匹配问题，它给定一个查询字符串和一个文档字符串，找到文档字符串中所有与查询串近似的子字符串。这两个问题与字符串近似连接是不同的，字符串近似连接给定两个字符串集合，找到其中所有近似的字符串元组。关于近似查询的选择性估计[79–81] 以及近似抽取[26,27,60,73] 也有一些研究。

## 3.3　基于划分的框架

本节首先介绍一个将字符串划分为多个不相交的片段的划分方法 (第 3.3.1节)，然后提出一个基于划分的算法框架 (第 3.3.2节)。

### 3.3.1　划分方法

给定一个字符串 $s$，把它划分为 $\tau+1$ 个不相交的片段，每个片段的长度不小于 1①。比如，考虑字符串 $s_1$=“vankatesh’’，假设 $\tau=3$，可以将 $s_1$ 划分为 $\tau+1=4$ 个片段，例如 {“va”,“nk”,“at”, “esh”}。

考虑两个字符串 $r$ 和 $s$，如果 $s$ 没有任何子字符串和 $r$ 的片段匹配，根据鸽巢原理，$s$ 不可能与 $r$ 相似。引理3.1给出了其证明。换句话说，如果 $s$ 与 $r$ 近似，那么 $s$ 一定包含一个与 $r$ 的一个片段完全匹配的子字符串。例如，考虑表 3.2中的字符串，假设 $\tau=3$，$s_1$ =“vankatesh’’ 的 4 个片段为：{“va”, “nk”, “at”, “esh”}。由于 $s_3, s_4, s_5, s_6$ 的任何子字符串与 $s_1$ 的任何片段都不匹配，所以它们不可能与 $s_1$ 相似。

**引理** 3.1：　给定编辑距离阈值 $\tau$，包含 $\tau+1$ 个片段的字符串 $r$ 和另一个字符串 $s$，如果 $s$ 与 $r$ 相似，那么 $s$ 一定包含一个子字符串与 $r$ 的一个片段完全匹配。

① 　字符串 $s$ 的长度 ($|s|$) 必须大于 $\tau$，即 $|s| \geq \tau+1$，否则容错没有意义。

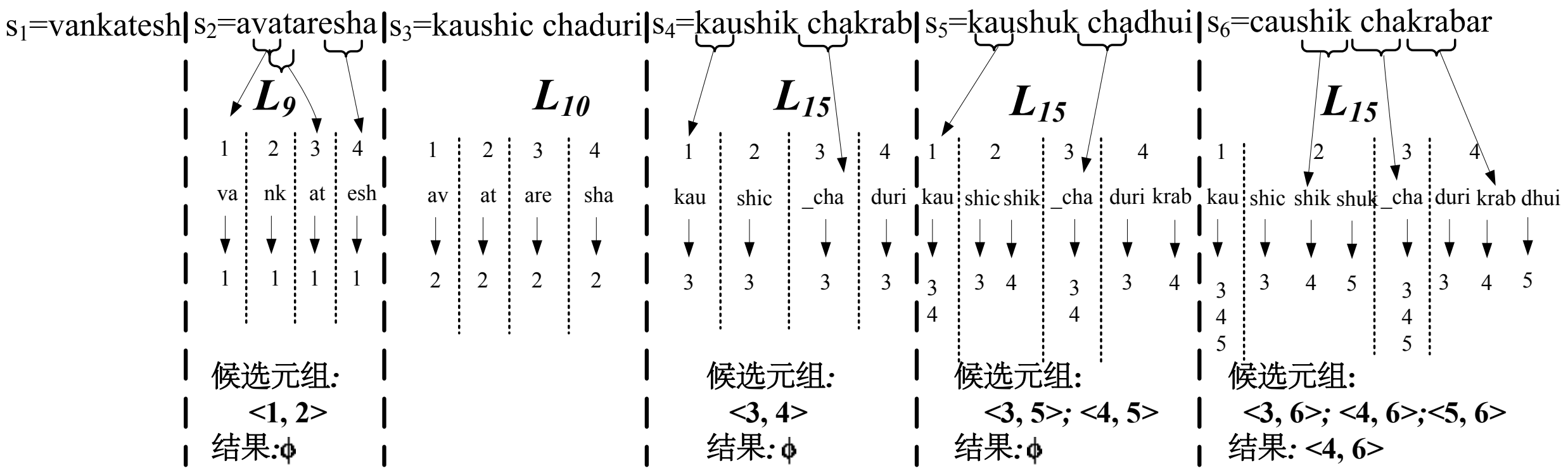

图 3.1　基于划分的方法的例子

**证明** 使用反证法证明。假设 $s$ 不包含任何子字符串与 $r$ 的任何片段相匹配，换句话说，$r$ 的任何片段都与 $s$ 的子字符串不相同，那么，对于任何从 $r$ 到 $s$ 的转换 $\mathcal{T}$，$r$ 中每个片段上都存在至少一个编辑操作。也就是说任何转换 $\mathcal{T}$ 中都存在至少 $\tau+1$ 个编辑操作。这与 $s$ 和 $r$ 近似互相矛盾，因此 $s$ 一定包含一个子字符串与 $r$ 的一个片段完全匹配。 □

给定一个字符串，有很多种策略把它切分为 $\tau+1$ 个片段。一个好的切分策略能够减少候选元组的数量，因此能够提高算法的性能。直观的来说，$r$ 的一个片段越短，这个片段就会有更高的概率出现在其他的字符串中，那么就会有更多的字符串与 $r$ 形成候选元组，因此剪枝效果更差。根据这个观察，在切分中不希望保留较短的片段。换句话说，每个片段应该有差不多相同的长度。因此，下面提出一个平均划分的策略。考虑一个长度为 $|s|$ 的字符串 $s$，在平均划分策略里，每个片段的长度都是 $\lfloor\frac{|s|}{\tau+1}\rfloor$ 或者 $\lceil\frac{|s|}{\tau+1}\rceil$。因此两个片段最大的长度差是 1。令 $k=|s|-\lfloor\frac{|s|}{\tau+1}\rfloor*(\tau+1)$，在平均划分策略里，最后 $k$ 个片段的长度是 $\lceil\frac{|s|}{\tau+1}\rceil$，前 $\tau+1-k$ 个片段的长度是 $\lfloor\frac{|s|}{\tau+1}\rfloor$。例如，考虑 $s_1$=“vankatesh” 并假设 $\tau=3$，那么有 $k=1$，$s_1$ 的 4 个片段是：{“va”,“nk”,“at”, “esh”}。

虽然可以设计其他的划分策略，但是选择一个好的划分策略是很耗时的过程。注意选择划分策略的时间也要包含在近似连接的运行时间中。论文中使用平均划分策略，论文将选取好的划分策略作为一个未来工作。本章之后提出的技术也能够应用到其它任何划分策略中。

### 3.3.2　基于划分的框架

如果一个字符串 $s$ 没有任何子字符串与另一个字符串 $r$ 的任何片段匹配，那么可以剪枝掉这个元组 $\langle s,r\rangle$。利用这个观察可以剪枝掉大量的不相似的字符串元

组。为了达到这个目的，下面提出一个基于划分的算法框架来解决字符串近似连接问题，并命名为 PassJoin。图 3.2 展示了该算法框架。

为了方便表述，首先介绍一些标记。令 $\mathcal{S}_l$ 表示所有长度为 $l$ 的字符串集合，$\mathcal{S}_l^i$ 表示 $\mathcal{S}_l$ 中所有字符串的第 $i$ 份片段的集合。$\mathcal{L}_l^i$ 表示基于片段集合 $\mathcal{S}_l^i$ 中所有片段的倒排索引。给定第 $i$ 份的一个片段 $w$，$\mathcal{L}_l^i(w)$ 表示片段 $w$ 所对应的倒排列表，即第 $i$ 份片段为 $w$ 的字符串集合，PassJoin 利用这些倒排索引按如下的方法来解决字符串近似连接问题。

PassJoin 首先按照字符串的长度升序进行排序。对于相同长度的字符串，它按照字母表顺序对它们排序。然后 PassJoin 按顺序依次访问每个字符串。考虑当前访问的字符串 $s$，PassJoin 利用相关的倒排索引从已经访问过的字符串中找出与 $s$ 近似的字符串。为了避免检测相同的字符串元组两次，PassJoin 只为已经访问过的字符串创建倒排索引。根据长度过滤[70]，PassJoin 只需要检查在倒排索引 $\mathcal{L}_l^i$ 中的字符串是否与 $s$ 相似，其中 $|s|-\tau \le l \le |s|$，$1 \le i \le \tau+1$）。不失一般性的，考虑倒排索引 $\mathcal{L}_l^i$。PassJoin 按照如下方式找到 $\mathcal{L}_l^i$ 中与 $s$ 相似的字符串。

- **子字符串选取**：假如 $s$ 与 $\mathcal{L}_l^i$ 中的一个字符串相似，那么 $s$ 必须包含一个子字符串与 $\mathcal{L}_l^i$ 中的一个片段匹配。一个直观的方法枚举 $s$ 的所有子字符串，对于每一个子字符串，该方法检查它是否出现在 $\mathcal{L}_l^i$ 中。实际上，并不需要考虑 $s$ 的所有子字符串，其实只需要选取它的一部分子字符串来检测（用 $\mathcal{W}(s,\mathcal{L}_l^i)$ 表示这些子字符串）并使用这些被选取的子字符串来寻找候选元组。第 3.4节将讨论如何生成 $\mathcal{W}(s,\mathcal{L}_l^i)$。对于每个被选择的子字符串 $w \in \mathcal{W}(s,\mathcal{L}_l^i)$，需要检查它是否出现在 $\mathcal{L}_l^i$ 中。如果出现了，对于每个 $r \in \mathcal{L}_l^i(w)$，$\langle r,s\rangle$ 都是一对候选元组。

- **验证**：为了验证一对候选元组 $\langle r,s\rangle$ 是否是一对相似元组，一个简单的方法计算它们的真实编辑距离。然而这个方法相当耗时，因此第 3.5节提出了几个高效的技术来进行快速验证。

在找到与 $s$ 相似的字符串后，需要把 $s$ 划分为 $\tau+1$ 个不相交的片段并把这些片段插入到倒排索引 $\mathcal{L}_{|s|}^i(1 \le i \le \tau+1)$ 中。然后访问 $s$ 之后的字符串。如此反复，可以找到所有的近似元组。注意，对于所有 $k < |s|-\tau$，可以移除倒排索引 $\mathcal{L}_k^i$。因此该算法同时最多维护 $(\tau+1)^2$ 个倒排索引 $\mathcal{L}_l^i$，其中 $|s|-\tau \le l \le |s|$，$1 \le i \le \tau+1$。

为了连接两个字符串集合 $\mathcal{R}$ 和 $\mathcal{S}$，首先分别对这两个集合中的字符串进行排序。然后为其中一个集合的所有片段建立索引，例如 $\mathcal{S}$。接下来，按顺序访问 $\mathcal{R}$ 中的每个字符串，对 $\mathcal{R}$ 中每个长度为 $|r|$ 的字符串 $r$，使用 $\mathcal{S}$ 中长度在 $[|r|-\tau,|r|+\tau]$ 范围内的字符串对应的倒排索引来寻找近似元组。找到 $r$ 的近似元组之后，可以

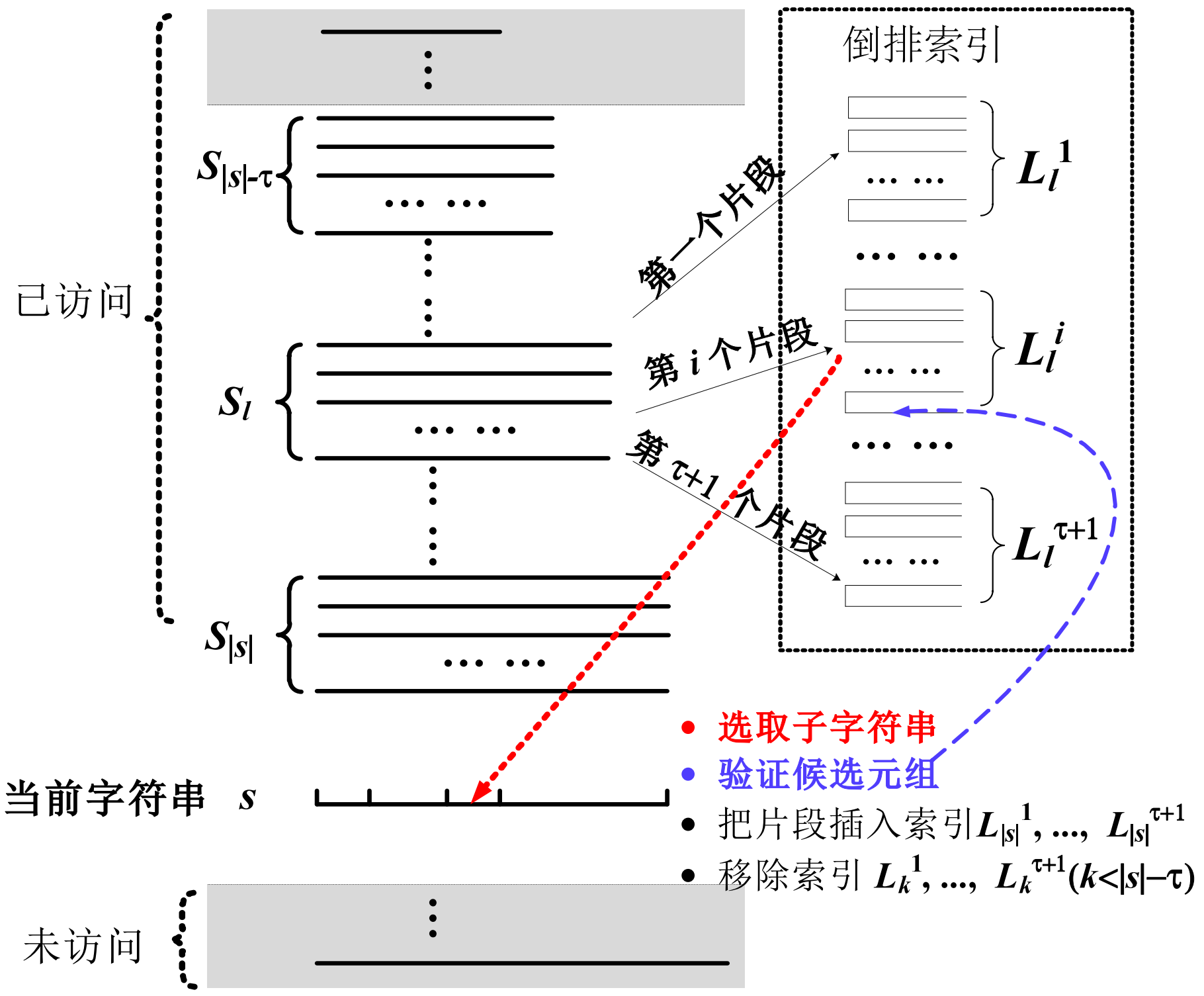

图 3.2　基于划分的算法框架

移除长度小于 $|r| - \tau$ 的字符串对应的倒排索引。本章主要关注索引可以被全部加载进内存的情况。

例如，考虑表 3.1 中的字符串并假设 $\tau = 3$，可以按照如下方式寻找相似元组（如图 3.1 所示）。对于第一个字符串 $s_1$ = “vankatesh”，把它切分为 $\tau + 1$ 个片段并把这些片段插入到长度为 9 的字符串的倒排索引中，即 $\mathcal{L}_9^1$，$\mathcal{L}_9^2$，$\mathcal{L}_9^3$ 和 $\mathcal{L}_9^4$。接下来，对于 $s_2$ = “avataresha”，枚举它的所有子字符串并检查每个子字符串是否出现在 $\mathcal{L}_{|s_2|-\tau}^i$，$\cdots$，$\mathcal{L}_{|s_2|}^i$ 中，其中 $1 \le i \le \tau + 1$。这里在 $\mathcal{L}_9^1$ 中找到了 “va”，在 $\mathcal{L}_9^3$ 中找到了 “at” 并且在 $\mathcal{L}_9^4$ 中找到了 “esh”。对于片段 “va”，因为 $\mathcal{L}_9^1(\text{va}) = \{s_1\}$，所以元组 $\langle s_2, s_1\rangle$ 是一对候选元组。然后验证这个元组，因为它们的编辑距离大于 $\tau$，所以它们不是一对相似元组。接下来划分 $s_2$ 成 4 个片段并把它们分别插入 $\mathcal{L}_{|s_2|}^1$，$\mathcal{L}_{|s_2|}^2$，$\mathcal{L}_{|s_2|}^3$ 和 $\mathcal{L}_{|s_2|}^4$ 中。重复以上步骤可以找到所有的相似元组。

图 4.1 给出了 PassJoin 的伪代码。PassJoin 首先以字符串长度为第一顺序，字母序为第二顺序对字符串进行排序（第 2 行）。然后 PassJoin 按顺序依次访问每个字符串（第 3 行）。对于每个倒排索引 $\mathcal{L}_l^i$，其中 $|s| - \tau \le l \le |s|$，$1 \le i \le \tau + 1$，PassJoin 从 $s$ 中选取子字符串（第 4 行）并检查每个选择的子字符串 $w$ 是否出现在 $\mathcal{L}_l^i$ 中（第 5 行）。如果出现了，对于倒排列表 $\mathcal{L}_l^i(w)$ 中的任何字符串 $r$，字符串元组 $\langle r, s\rangle$ 都是一个候选元组。然后 PassJoin 验证这个元组（第 7 行）。最后，PassJoin 把

**Algorithm 3.1**: PassJoin ($\mathcal{S}, \tau$)

**Input**: $\mathcal{S}$: 一个字符串集合； $\tau$: 一个给定的编辑距离阈值
**Output**: $\mathcal{A} = \{(s \in \mathcal{S}, r \in \mathcal{S}) \mid \text{ED}\,(s, r) \le \tau\}$

1 **begin**
2 　按照字符串长度为第一顺序，字母序为第二顺序对 $\mathcal{S}$ 中的字符串排序;
3 　**for** $s \in \mathcal{S}$ **do**
4 　　**for** $\mathcal{L}_l^i$ $(|s| - \tau \le l \le |s|, 1 \le i \le \tau + 1)$ **do**
5 　　　$\mathcal{W}(s, \mathcal{L}_l^i)$ = SubstringSelection($s$, $\mathcal{L}_l^i$);
6 　　　**for** $w \in \mathcal{W}(s, \mathcal{L}_l^i)$ **do**
7 　　　　**if** $w$ 在 $\mathcal{L}_l^i$ 中 **then** Verification($s, \mathcal{L}_l^i(w), \tau$);
8 　　划分 $s$ 并把它的片段插入 $\mathcal{L}_{|s|}^i$ 中;
9 **end**

**Procedure** SubstringSelection ($s$, $\mathcal{L}_l^i$)

**Input**: $s$: 一个字符串; $\mathcal{L}_l^i$: 倒排索引
**Output**: $\mathcal{W}(s, \mathcal{L}_l^i)$: 选择的子字符串

**begin**
　$\mathcal{W}(s, \mathcal{L}_l^i) = \{w \mid w$ 是 $s$ 的一个子字符串$\}$;
**end**

**Procedure** Verification ($s$, $\mathcal{L}_l^i(w)$, $\tau$)

**Input**: $s$: 一个字符串; $\mathcal{L}_l^i(w)$: 倒排列表; $\tau$: 编辑距离阈值
**Output**: $\mathcal{A} = \{(s \in \mathcal{S}, r \in \mathcal{S}) \mid \text{ED}\,(s, r) \le \tau\}$

**begin**
　**for** $r \in \mathcal{L}_l^i(w)$ **do**
　　**if** $ED(s, r) \le \tau$ **then** $\mathcal{A} \leftarrow \langle s, r \rangle$;
**end**

图 3.3　PassJoin 算法

$s$ 划分为 $\tau+1$ 个片段并把这些片段插入到倒排索引 $\mathcal{L}_{|s|}^i(1 \le i \le \tau+1)$ 中（第 8 行）。这里子程序 SubstringSelection 选择所有的子字符串。子程序 Verification 采用动态规划算法计算两个字符串真实的编辑距离以验证候选元组。为了提高算法性能，3.4节提出了有效的技术来优化子字符串的选取，第3.5节提出提高验证效率。

**复杂度分析**：首先分析空间复杂度。PassJoin 的索引结构包含片段以及片段的倒排列表，首先给出片段的空间复杂度。对于 $\mathcal{S}_l$ 中的每个字符串，PassJoin 产生 $\tau + 1$ 个片段，因此片段的数量最多有 $(\tau + 1) \times |\mathcal{S}_l|$ 个，其中 $|\mathcal{S}_l|$ 是 $\mathcal{S}_l$ 中字符串的数量。

因为能用一个整数来编码一个片段，所以存储片段的空间复杂度是

$$O\Big(\max_{l_{min} \le j \le l_{max}} \sum_{l=j-\tau}^{j} (\tau + 1) \times |\mathcal{S}_l|\Big)$$

。其中 $l_{min}$ 和 $l_{max}$ 分别代表最短的字符串长度以及最长的字符串长度。

接下来给出倒排列表的空间复杂度。对于 $\mathcal{S}_l$ 中的每个字符串，因为它的第 $i$ 个片段对应 $\mathcal{L}_l^i$ 中的一个元素，所以有 $|\mathcal{S}_l| = |\mathcal{L}_l^i|$，因此倒排列表的空间复杂度（即倒排列表的长度之和）是

$$O\Big(\max_{l_{min} \le j \le l_{max}} \sum_{l=j-\tau}^{j} \sum_{i=1}^{\tau+1} |\mathcal{L}_l^i| = \max_{l_{min} \le j \le l_{max}} \sum_{l=j-\tau}^{j} (\tau + 1) \times |\mathcal{S}_l|\Big)$$

。

最后给出算法的时间复杂度。PassJoin 首先对字符串按照长度进行分组，然后对每个组中的字符串进行排序。因此排序复杂度是 $O(\sum_{l_{min} \le l \le l_{max}} |\mathcal{S}_l| log(|\mathcal{S}_l|))$。对于每个字符串 $s$，需要为每一个 $|s| - \tau \le l \le |s|$ 和 $1 \le i \le \tau + 1$ 选择一个子字符串集合 $\mathcal{W}(s, \mathcal{L}_l^i)$，选择复杂度是 $O\Big(\sum_{s \in \mathcal{S}} \sum_{l=|s|-\tau}^{|s|} \sum_{i=1}^{\tau+1} \mathcal{X}(s, \mathcal{L}_l^i)\Big)$，其中 $\mathcal{X}(s, \mathcal{L}_l^i)$ 是 $\mathcal{W}(s, \mathcal{L}_l^i)$ 的选取时间，等于 $O(\tau)$ (第 3.4节)。因此选择复杂度是 $O(\tau^3|\mathcal{S}|)$。对于每个子字符串 $w \in \mathcal{W}(s, \mathcal{L}_l^i)$，需要验证 $\mathcal{L}_l^i(w)$ 中的字符串是否与 $s$ 相似。验证复杂度是 $O\Big(\sum_{s \in \mathcal{S}} \sum_{l=|s|-\tau}^{|s|} \sum_{i=1}^{\tau+1} \sum_{w \in \mathcal{W}(s, \mathcal{L}_l^i)} \sum_{r \in \mathcal{L}_l^i(w)} \mathcal{V}(s, r)\Big)$, 其中 $\mathcal{V}(s, r)$ 是验证元组 $\langle s, r \rangle$ 的复杂度，等于 $O(\tau * \min(|s|, |r|))$(第 3.5节)。本章接下来的内容研究如何缩小 $\mathcal{W}(s, \mathcal{L}_l^i)$ 的大小以及如何降低验证代价 $\mathcal{V}(s, r)$。

## 3.4 优化子字符串选取

对于 $\mathcal{S}$ 中的任何子字符串 $s$ 和一个长度 $l$ ($|s| - \tau \le l \le |s|$)，PassJoin 需要从 $s$ 中选取一个子字符串集合 $\mathcal{W}(s, l) = \cup_{i=1}^{\tau+1} \mathcal{W}(s, \mathcal{L}_l^i)$ 并利用 $\mathcal{W}(s, l)$ 中的子字符串来寻找与 $s$ 相似的候选字符串。PassJoin 需要保证用 $\mathcal{W}(s, l)$ 寻找候选元组的完备性。也就是说，任何近似元组必须作为一对候选元组被找到。接下来给出完备性的正式定义。

**定义** 3.2 (**完备性**)：　一个子字符串选取方法选取的子字符串集合 $\mathcal{W}(s, l)$ 满足完备性，如果对任何字符串 $s$ 和在 $|s| - \tau$ 与 $|s|$ 之间的一个长度 $l$，任意长度为 $l$、与 $s$ 相似并在 $s$ 之前被访问的字符串 $r$ 一定包含一个片段 $r_m$ 与 $s$ 的一个子字符串 $s_m$ 完全匹配，其中 $r_m$ 是 $r$ 的第 $i(1 \le i \le \tau + 1)$ 份片段，$s_m \in \mathcal{W}(s, \mathcal{L}_l^i)$。

一个简单的子字符串选取方法直接把 $s$ 的所有子字符串都加入到 $\mathcal{W}(s,l)$ 中，因为 $s$ 有 $|s|-i+1$ 个长度为 $i$ 的子字符串，所以 $s$ 的子字符串总数为 $\sum_{i=1}^{|s|}(|s|-i+1)=\frac{|s|*(|s|+1)}{2}$。对于长的字符串，它们的子字符串数量非常多，所以枚举所有的子字符串的代价将非常高昂。

直观的，$\mathcal{W}(s,l)$ 的大小越小，算法的性能就越高。因此希望选取到大小更小的子字符串集合。本节提出多个方法来选取子字符串集合 $\mathcal{W}(s,l)$。因为 $\mathcal{W}(s,l)=\cup_{i=1}^{\tau+1}\mathcal{W}(s,\mathcal{L}_l^i)$ 并且 PassJoin 使用倒排索引 $\mathcal{L}_l^i$ 来快速过滤，因此下文讨论如何针对倒排索引 $\mathcal{L}_l^i$ 选取子字符串集合 $\mathcal{W}(s,\mathcal{L}_l^i)$。

**基于长度的子字符串选取方法**（Length-based）：因为在 $\mathcal{L}_l^i$ 中的片段拥有相同的长度（用 $l_i$ 表示），所以基于长度的子字符串选取方法选择 $s$ 中所有长度为 $l_i$ 的子字符串。用 $\mathcal{W}_\ell(s,\mathcal{L}_l^i)$ 表示该方法选择的子字符串集合。令 $\mathcal{W}_\ell(s,l)=\cup_{i=1}^{\tau+1}\mathcal{W}_\ell(s,\mathcal{L}_l^i)$，基于长度的方法满足完备性，这是因为它选择了所有长度为 $l_i$ 的子字符串。$\mathcal{W}_\ell(s,\mathcal{L}_l^i)$ 的大小是 $|\mathcal{W}_\ell(s,\mathcal{L}_l^i)|=|s|-l_i+1$，该方法选取的子字符串总数是 $|\mathcal{W}_\ell(s,l)|=(\tau+1)(|s|+1)-l$。

**基于位移的子字符串选取方法**（Shift-based）：然而，基于长度的方法并不考虑子字符串所在的位置。为了利用位置信息，Wang 等人[26] 提出了一个基于位移的方法来解决实体抽取问题，可以按照如下方式扩展他们的方法以支持近似连接问题。因为在 $\mathcal{L}_l^i$ 中的片段拥有相同的长度，所以它们也有相同的起始位置，用 $p_i$ 表示。其中 $p_1=1$，$p_i=p_1+\sum_{k=1}^{i-1}l_k\ (i>1)$。基于位移的方法选取 $s$ 的所有长度为 $l_i$ 并且起始位置在 $[p_i-\tau,p_i+\tau]$ 中的片段，用 $\mathcal{W}_f(s,\mathcal{L}_l^i)$ 来表示其选择的片段集合。令 $\mathcal{W}_f(s,l)=\cup_{i=1}^{\tau+1}\mathcal{W}_f(s,\mathcal{L}_l^i)$，$\mathcal{W}_f(s,\mathcal{L}_l^i)$ 的大小为 $|\mathcal{W}_f(s,\mathcal{L}_l^i)|=2\tau+1$，该方法选取的片段总数是 $|\mathcal{W}_f(s,l)|=(\tau+1)(2\tau+1)$。

这个方法背后的基本思想如下。假设 $s$ 的一个起始位置小于 $p_i-\tau$ 或者大于 $p_i+\tau$ 的子字符串 $s_m$ 匹配了 $\mathcal{L}_l^i$ 中的一个片段，那么对于 $\mathcal{L}_l^i(s_m)$ 中的任意一个字符串 $r$，可以把 $s(r)$ 切分为三个部分：匹配的部分 $s_m(r_m)$、在匹配之前左边的部分 $s_l(r_l)$、在匹配之后右边的部分 $s_r(r_r)$。因为 $r_m$ 的起始位置是 $p_i$，$s_m$ 的起始位置是小于 $p_i-\tau$ 的或者大于 $p_i+\tau$ 的，所以 $s_l$ 和 $r_l$ 的长度之差一定大于 $\tau$。假如在转换这两个字符串时匹配对齐 $s_m$ 和 $r_m$（即首先把 $r_l$ 转化为 $s_l$，然后匹配 $r_m$ 和 $s_m$，最后把 $r_r$ 转化为 $s_r$），那么这个转换一定包含多于 $\tau$ 个编辑操作，所以可以不考虑这样的转换，即不考虑把这样的子字符串 $s_m$ 与片段 $r_m$ 匹配。也就是说可以不选取这样的子字符串 $s_m$，因此基于位移的方法满足完备性。

然而基于位移的方法仍然选取了大量不必要的子字符串。例如，考虑两个子字符串 $s_1=$“vankatesh” 和 $s_2=$“avataresha”，假设 $\tau=3$，那么“vankatesh” 被划

分为 4 个片段 {va, nk, at, esh}。$s_2$ = “avataresha’’ 包含一个子字符串 “at’’，它和 “vankatesh’’ 中的第三个片段相匹配，基于位移的方法将会选取它作为一个子字符串。然而，实际上是可以剪枝掉它的，原因如下。假设按匹配的部分把两个字符串切分成三个部分，例如 “vankatesh’’ 被切分为 {“vank” , “at”, “esh”}，“avataresha’’ 被切分为 {“av”, “at”, “aresha”}。显然左边部分 (“vank’’ 和 “av”) 最小的编辑距离 (即长度之差）是 2，右边部分 (“esh’’ 和 “aresha”) 最小的编辑距离（即长度之差）是 3。因此，假如按照匹配的部分 “at’’ 转换这两个字符串的话，它们不会相似。因此可以剪枝掉子字符串 “at’’，即不选择这个子字符串。

### 3.4.1　位置敏感的子字符串选取方法

本节提出一个位置敏感 (position-aware) 的子字符串选取方法。注意到 $\mathcal{L}_l^i$ 中所有的片段拥有相同的长度 $l_i$ 以及相同的起始位置 $p_i$。不失一般性的，考虑 $\mathcal{L}_l^i$ 中的一个片段 $r_m$，倒排列表 $\mathcal{L}_l^i(r_m)$ 中所有的字符串拥有相同的长度 $l$ (其中 $l \le |s|$)，考虑一个包含片段 $r_m$ 的字符串 $r$。假设 $s$ 有一个子字符串 $s_m$ 匹配 $r_m$，接下来给出 $s_m$ 可能的起始位置。仍然切分 $s(r)$ 为三个部分，匹配的部分 $s_m(r_m)$, 左边的部分 $s_l(r_l)$，以及右边的部分 $s_r(r_r)$。假如在 $r$ 和 $s$ 的转化中匹配 $r_m = s_m$，也就是说首先用 $d_l = \text{ED}(r_l, s_l)$ 个编辑操作把 $r_l$ 转化为 $s_l$，然后匹配 $r_m$ 和 $s_m$，最后用 $d_r = \text{ED}(r_r, s_r)$ 个编辑操作把 $r_r$ 转化为 $s_r$，这个转化中总的编辑操作数目是 $d_l + d_r$。如果在这个转换下 $s$ 和 $r$ 相似，那么 $d_l + d_r \le \tau$。根据这个观察，如图 3.4 所示，可以计算出 $s_m$ 的最小的起始位置（$p_{min}$）以及最大的起始位置（$p_{max}$）。

**最小的起始位置**: 用 $p$ 来表示 $s_m$ 的起始位置，假设它不大于 $p_i$。令 $\triangle = |s| - |r|$，$\triangle_l = p_i - p$。如图 3.4(a) 所示，有 $d_l = \text{ED}(r_l, s_l) \ge \triangle_l$ 以及 $d_r = \text{ED}(r_r, s_r) \ge \triangle_l + \triangle$。假如在该转换下 $s$ 与 $r$（或 $\mathcal{L}_l^i(r_m)$ 中任意一个字符串）相似，那么有

$$\triangle_l + (\triangle_l + \triangle) \le d_l + d_r \le \tau$$

，即 $\triangle_l \le \lfloor\frac{\tau-\triangle}{2}\rfloor$，$p = p_i - \triangle_l \ge p_i - \lfloor\frac{\tau-\triangle}{2}\rfloor$。因此 $p_{min} \ge p_i - \lfloor\frac{\tau-\triangle}{2}\rfloor$。因为 $p_{min} \ge 1$，所以 $p_{min} = \max(1, p_i - \lfloor\frac{\tau-\triangle}{2}\rfloor)$.

**最大的起始位置**: 假设 $s_m$ 的起始位置 $p$ 大于 $p_i$，令 $\triangle = |s| - |r|$，$\triangle_r = p - p_i$。如图 3.4(b) 所示，有 $d_l = \text{ED}(r_l, s_l) \ge \triangle_r$ 以及 $d_r = \text{ED}(r_r, s_r) \ge |\triangle_r - \triangle|$。假如 $\triangle_r \le \triangle$，$d_r \ge \triangle - \triangle_r$。因此 $\triangle = \triangle_r + (\triangle - \triangle_r) \le d_l + d_r \le \tau$，在这种情况下，$\triangle_r$ 的最大值是 $\triangle$，否则的话 $\triangle_r > \triangle$，$d_r \ge \triangle_r - \triangle$。假如 $s$ 与 $r$（或者 $\mathcal{L}_l^i(r_m)$ 中任意一个字符串）相似，有

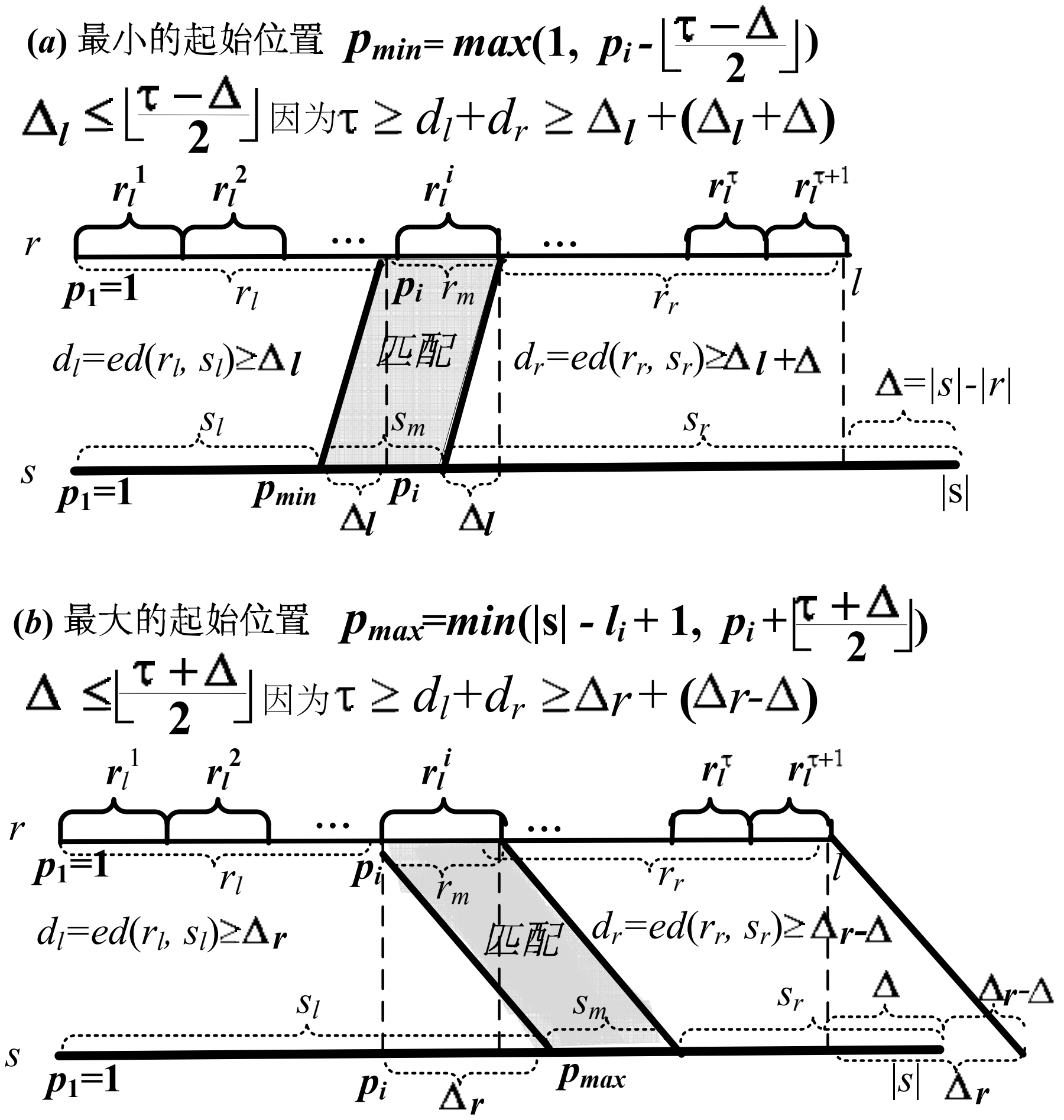

图 3.4　位置敏感的子字符串选取方法

$$\triangle_r + (\triangle_r - \triangle) \le d_l + d_r \le \tau$$

, 也就是说 $\triangle_r \le \lfloor\frac{\tau+\triangle}{2}\rfloor$, 以及 $p = p_i + \triangle_r \le p_i + \lfloor\frac{\tau+\triangle}{2}\rfloor$。因此 $p_{max} \le p_i + \lfloor\frac{\tau+\triangle}{2}\rfloor$。因为片段长度为 $l_i$, 根据这个界, 有 $p_{max} \le |s| - l_i + 1$。因此 $p_{max} = \min(|s| - l_i + 1, p_i + \lfloor\frac{\tau+\triangle}{2}\rfloor)$。

例如, 考虑字符串 $r$="vankatesh''。假设 $\tau = 3$, "vankatesh'' 被划分为 4 个片段 {va, nk, at, esh}。对于字符串 $s$="avataresha'', 有 $\triangle = |s| - |r| = 1$, $\triangle_l \le \lfloor\frac{\tau-\triangle}{2}\rfloor = 1$ 以及 $\triangle_r \le \lfloor\frac{\tau+\triangle}{2}\rfloor = 2$。对于第一个片段 "va'', 有 $p_1 = 1$。因此 $p_{min} = \max(1 p_1 - \lfloor\frac{\tau-\triangle}{2}\rfloor) = 1$, $p_{max} = 1 + \lfloor\frac{\tau+\triangle}{2}\rfloor = 3$。所以只需要为第一个片段枚举以下子字符串 "av'', "va'', "at''。类似的, 需要为第二个片段枚举子字符串 "va'', "at'', "ta'', "ar'', 为第三个片段枚举子字符串 "ta'', "ar'', "re'', "es'' 以及为第四个片段枚举子字符串 "res'', "esh'', "sha''。可以看到, 位置敏感的方法可以在基于位移的方法的基础上减少很多不必要的子字符串 (从 28 个子字符串减少到 14 个)。

针对倒排索引 $\mathcal{L}_l^i$, 位置敏感的子字符串选取方法选择长度为 $l_i$、起始位置在

$[p_{min}, p_{max}]$ 中的子字符串，把它选择的子字符串集合记作 $\mathcal{W}_p(s, \mathcal{L}_l^i)$。令 $\mathcal{W}_p(s,l) = \cup_{i=1}^{\tau+1}\mathcal{W}_p(s, \mathcal{L}_l^i)$，$\mathcal{W}_p(s, \mathcal{L}_l^i)$ 的大小为 $|\mathcal{W}_p(s, \mathcal{L}_l^i)| = \tau + 1$，该方法选取的子字符串总数为 $|\mathcal{W}_p(s,l)| = (\tau+1)^2$。定理 3.1 证明了位置敏感的选取方法满足完备性。

**定理** 3.1：　位置敏感的子字符串选取方法满足完备性。

**证明** 对于任何字符串 $s$，考虑任意长度为 $l(|s| - \tau \le l \le |s|)$、与 $s$ 相似并在 $s$ 之前访问过的字符串 $r$。考虑任何从 $s$ 到 $r$ 的包含 $|\mathcal{T}| \le \tau$ 个编辑操作的转换 $\mathcal{T}$，根据引理 3.1，该转换中，$s$ 一定有一个子字符串 $s_m$ 和 $r$ 的一个片段 $r_m$ 相匹配。把 $r$ $(s)$ 切分为三个部分：左边在匹配片段之前的部分 $r_l$ $(s_l)$，匹配的片段 $r_m$ $(s_m)$ 以及右边在匹配片段之后的部分 $r_r$ $(s_r)$。假设 $r_m$ 是 $r$ 的第 $i$ 个片段，因此 $r \in \mathcal{L}_l^i(r_m)$。接下来证明 $s_m \in \mathcal{W}_p(s, \mathcal{L}_l^i) \subseteq \mathcal{W}_p(s,l)$。

首先，因为 $s_m = r_m$，$|s_m| = |r_m| = l_i$，假设 $s_m$ 在 $s$ 中的起始位置为 $p$，所以只需要证明 $p \in [p_{min}, p_{max}]$。又因为 $[p_{min}, p_{max}] = [1, |s| - l_i + 1] \cap [p_i - \lfloor\frac{\tau-\triangle}{2}\rfloor, p_i + \lfloor\frac{\tau+\triangle}{2}\rfloor]$，所以只需要证明 $p \in [1, |s| - l_i + 1]$ 并且 $p \in [p_i - \lfloor\frac{\tau-\triangle}{2}\rfloor, p_i + \lfloor\frac{\tau+\triangle}{2}\rfloor]$.

首先证明 $p \in [1, |s| - l_i + 1]$。显然，对于任何子字符串，其最小起始位置是 1。因为 $s_m$ 的长度是 $l_i$，其最大的起始位置是 $|s| - l_i + 1$。因此 $p$ 一定在 $[1, |s| - l_i + 1]$ 内。

然后证明 $p \in [p_i - \lfloor\frac{\tau-\triangle}{2}\rfloor, p_i + \lfloor\frac{\tau+\triangle}{2}\rfloor]$，用反证法证明。假设 $p \notin [p_i - \lfloor\frac{\tau-\triangle}{2}\rfloor, p_i + \lfloor\frac{\tau+\triangle}{2}\rfloor]$，因为 $\mathcal{T}$ 把 $s_l$ 转化为 $r_l$，匹配 $s_m$ 和 $r_m$ 并把 $s_r$ 转化为 $r_r$。有 $\tau \ge |\mathcal{T}| \ge \text{ED}(s_l, r_l) + \text{ED}(s_m, r_m) + \text{ED}(s_r, r_r) \ge |p_i - p| + 0 + \Big|(|r| - p_i) - (|s| - p)\Big|$。

如果 $p < p_i - \lfloor\frac{\tau-\triangle}{2}\rfloor$，那么有

$$
\begin{aligned}
|\mathcal{T}| &\ge |p_i - p| + |(|r| - p_i) - (|s| - p)| \ge (\lfloor\frac{\tau-\triangle}{2}\rfloor + 1) + (\triangle + \lfloor\frac{\tau-\triangle}{2}\rfloor + 1) \\
&\ge (2\lfloor\frac{\tau-\triangle}{2}\rfloor + 1) + (\triangle + 1) \ge \tau - \triangle + (\triangle + 1) \ge \tau + 1 > \tau
\end{aligned}
$$

如果 $p > p_i + \lfloor\frac{\tau+\triangle}{2}\rfloor$，那么有

$$
\begin{aligned}
|\mathcal{T}| &\ge |p_i - p| + |(|r| - p_i) - (|s| - p)| \ge (\lfloor\frac{\tau+\triangle}{2}\rfloor + 1) + (\lfloor\frac{\tau+\triangle}{2}\rfloor + 1 - \triangle) \\
&\ge (2\lfloor\frac{\tau+\triangle}{2}\rfloor + 1) + (1 - \triangle) \ge \tau + \triangle + (1 - \triangle) \ge \tau + 1 > \tau
\end{aligned}
$$

在全部两个情况下，都有 $|\mathcal{T}| > \tau$，这与 $|\mathcal{T}| \le \tau$ 互相矛盾，因此 $p \in [p_i - \lfloor\frac{\tau-\triangle}{2}\rfloor, p_i + \lfloor\frac{\tau+\triangle}{2}\rfloor]$。

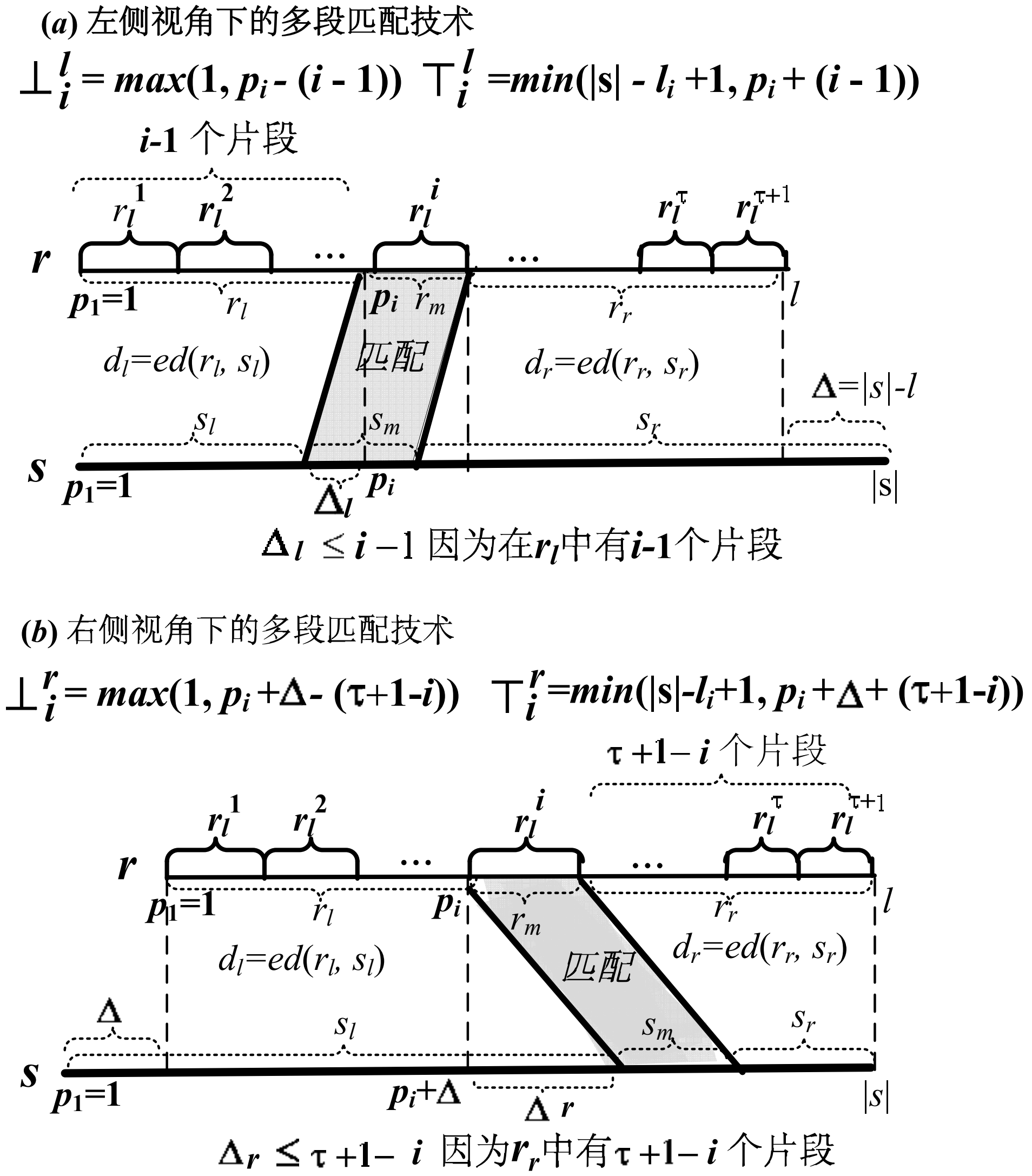

图 3.5　多段匹配敏感的子字符串选取

综上，$p \in [p_{min}, p_{max}]$。因此对于任何字符串 $s$，任何长度为 $l(|s| - \tau \le l \le |s|)$，与 $s$ 相似并在 $s$ 之前访问过的字符串 $r$ 一定包含一个片段 $r_m$ 与 $s$ 的一个子字符串 $s_m$ 匹配，其中 $r_m$ 是第 $i$ 份片段，$s_m \in \mathcal{W}_p(s, \mathcal{L}_l^i)$，定理得证。 □

### 3.4.2　多段匹配敏感的子字符串选取

观察到 $s$ 可能有多个子字符串与字符串 $r$ 的片段匹配，在这种情况下，可以扔掉其中一些子字符串。例如考虑 $r$ = “vankatesh” 以及它的 4 个片段 {va，nk，at，esh}。对于字符串 $s$=“avataresha”，它包含 3 个子字符串 va，at，esh 与 $r$ 的片段相匹配，实际上可以扔掉其中的一些子字符串。利用这个观察，本节提出一个多段匹配敏感 (Multi-match-aware) 的子字符串选取方法。

对于字符串 $r$，假设字符串 $s$ 有一个子字符串 $s_m$ 与 $r$ 的一个片段 $r_m$ 相匹配。如果知道 $s$ 在 $s_m$ 之后一定还有一个子字符串与 $r_m$ 之后的另一个片段相匹配，那么可以扔掉子字符串 $s_m$。例如，$s$=“avataresha” 有一个子字符串 “va” 与 $r$ = “vankatesh” 中的一个片段相匹配，考虑这三个部分 $r_m = s_m =$ “va”，$r_l = \phi$ 和 $s_l$= “a” 以及 $r_r$=“nkatesh” 和 $s_r$=“taresha”。因为 $d_l \geq 1$，假如在这个转换下 $s$ 和 $r$ 相似，那么 $d_r \leq \tau - d_l \leq \tau - 1 = 2$。因为 $r_r$ 中还有 3 个片段，因此跟据鸽巢原理，$s_r$ 一定有一个子字符串与 $r_r$ 中的一个片段相匹配。因此可以扔掉子字符串 “va” 并使用下一个匹配的子字符串来寻找相似元组。接下来泛化这个想法。

假设 $s$ 有一个起始位置为 $p$ 的子字符串与 $\mathcal{L}_l^i$ 中的一个片段 $r_m$ 相匹配。如图 3.5 所示，仍然考虑这两个字符串的三个部分：$s_l, s_m, s_r$ 以及 $r_l, r_m, r_r$。令 $\triangle_l = |p_i - p|$，有 $d_l = \text{ED}(r_l, s_l) \geq \triangle_l$。因为 $s_l$ 有 $i-1$ 个片段，如果在把 $r_l$ 变换为 $s_l$ 的过程中，每个片段只有一个编辑错误，那么有 $\triangle_l \leq i-1$。如果 $\triangle_l \geq i$, 有 $d_l = \text{ED}(r_l, s_l) \geq \triangle_l \geq i$, $d_r = \text{ED}(r_r, s_r) \leq \tau - d_l \leq \tau - i$（如果 $s$ 与 $r$ 在该转换下近似）。因为 $r_r$ 包含 $\tau + 1 - i$ 个片段, 根据鸽巢原理 $s_r$ 一定包含一个子字符串与 $r_r$ 中的一个片段相匹配。可以用与定理 3.1类似的证明方法证明这个结论。这样的话，因为对于 $\mathcal{L}_l^i(r_m)$ 中的任何字符串 $r, s$ 一定有一个子字符串与右边部分 $r_r$ 中的一个片段相匹配，所以可以扔掉子字符串 $s_m$，用下一个与片段相匹配的子字符串来找到与 $s$ 近似的字符串。总结来说，如果 $\triangle_l = |p - p_i| \leq i-1$，那么为 $\mathcal{L}_l^i$ 保留起始位置为 $p$ 的子字符串，也就是说针对 $\mathcal{L}_l^i$ 选取的子字符串的最小的起始位置是 $\perp_i^l = \max(1, p_i - (i-1))$，最大的起始位置是 $\top_i^l = \min(|s| - l_i + 1, p_i + (i-1))$。

例如，考虑有 4 个片段 {va, nk, at, esh} 的字符串 $r$=“vankatesh” 和字符串 $s$=“avataresha”。对于第一个片段，有 $\perp_i^l$=1-0=1 以及 $\top_i^l$=1+0=1。所以为第一个片段选取的子字符串只有 “av”。对于第二个片段，有 $\perp_i^l$=3-1=2 以及 $\top_i^l$=3+1=4，所以为第二个片段选取的子字符串有 “va”，“at” 和 “ta”。类似的，对于第三个片段有 $\perp_i^l$=5-2=3 和 $\top_i^l$=5+2=7，对于第四个片段有 $\perp_i^l$=7-3=4 和 $\top_i^l$=7+3=10。

以上观察是从左侧视角得出的。类似的，从右侧视角利用类似的思想可以推导出另一个起始位置的范围。因为右边部分 $r_r$ 有 $\tau + 1 - i$ 个片段, 所以在 $r_r$ 中最多有 $\tau + 1 - i$ 个编辑错误。如果以右侧视角把 $r$ 变换为 $s$，如图 3.5(b) 所示，$r$ 上的位置 $p_i$ 应该和 $s$ 上的位置 $p_i + \triangle$ 相对齐。假如 $s$ 上的位置 $p$ 和 $r$ 上的位置 $p_i$ 相匹配，令 $\triangle_r = |p - (p_i + \triangle)|$，有 $d_r = \text{ED}(s_r, r_r) \geq \triangle_r$。因为在右边部分 $r_r$ 中有 $\tau + 1 - i$ 个片段，所以有 $\triangle_r \leq \tau + 1 - i$。因此针对 $\mathcal{L}_l^i$ 选取的子字符串最小的起始位置是 $\perp_i^r = \max(1, p_i + \triangle - (\tau + 1 - i))$, 最大起始位置是 $\top_i^r = \min(|s| - l_i + 1, p_i + \triangle + (\tau + 1 - i))$。

仍然考虑以上的例子，假设 $\tau = 3$，$\triangle = 1$。对于第四个片段，有 $\perp_i^r = 7 + 1 -$

$(3+1-4)=8$ 和 $\top_i^r=7+1+(3+1-4)=8$。所以为第四个片段选取的子字符串只有 “sha’’。类似的，对于第三个片段，有 $\bot_i^r=5$ 和 $\top_i^r=7$。所以，为第三个片段选取的子字符串有 “ar”, “re’’ 和 “es’’。

更有趣的是，可以同时使用这两个技术。也就是说，对于 $\mathcal{L}_l^i$，只需要选取长度为 $l_i$，起始位置在 $\bot_i=\max(\bot_i^l,\bot_i^r)$ 到 $\top_i=\min(\top_i^l,\top_i^r)$ 之间的子字符串，用 $\mathcal{W}_m(s,\mathcal{L}_l^i)$ 表示该子字符串集合。令 $\mathcal{W}_m(s,l)=\cup_{i=1}^{\tau+1}\mathcal{W}_m(s,\mathcal{L}_l^i)$，如定理 3.2所示，被选取的子字符串数量是 $|\mathcal{W}_m(s,l)|=\lfloor\frac{\tau^2-\Delta^2}{2}\rfloor+\tau+1$。

**引理** 3.2： $|\mathcal{W}_m(s,l)|=\lfloor\frac{\tau^2-\Delta^2}{2}\rfloor+\tau+1$。

**证明** 因为 $\mathcal{W}_m(s,l)=\cup_{i=1}^{\tau+1}\mathcal{W}_m(s,\mathcal{L}_l^i)$，所以 $|\mathcal{W}_m(s,l)|=\sum_{i=1}^{\tau+1}|\mathcal{W}_m(s,\mathcal{L}_l^i)|=\sum_{i=1}^{\tau+1}(\top_i-\bot_i+1)=\sum_{i=1}^{\tau+1}\Big(\min(|s|-l_i+1,p_i+(i-1),p_i+\Delta+(\tau+1-i))-\max(1,p_i-(i-1),p_i+\Delta-(\tau+1-i))+1\Big)$。

因为 $(p_{i+1}-(i+1))-(p_i-i)=p_{i+1}-p_i-1\geq 0$, 所以 $p_i-i$ 是个单调递增的函数。因此对于任何 $i\in[1,\tau+1]$, 有 $p_i-(i-1)\geq p_1-(1-1)=1$ 和 $p_i+\Delta+(\tau+1-i)=p_i-i+\Delta+\tau+1\leq p_{\tau+1}-(\tau+1)+\Delta+\tau+1=p_{\tau+1}+\Delta=p_{\tau+1}+|s|-|r|=p_{\tau+1}+|s|-(p_{\tau+1}+l_{\tau+1})=|s|-l_{\tau+1}\leq|s|-l_i<|s|-l_i+1$, 因此

$$|\mathcal{W}_m(s,l)|=\sum_{i=1}^{\tau+1}(\top_i-\bot_i+1)=$$
$$\sum_{i=1}^{\tau+1}\Big(\min(p_i+(i-1),p_i+\Delta+(\tau+1-i))-\max(p_i-(i-1),p_i+\Delta-(\tau+1-i))+1\Big)$$

考虑 $\bot_i=\max(p_i-(i-1),p_i+\Delta-(\tau+1-i))$。如果 $p_i-(i-1)\geq p_i+\Delta-(\tau+1-i)$, 有 $\bot_i=p_i-(i-1)$。在这种情况下 $i\leq\lfloor\frac{\tau-\Delta}{2}\rfloor+1$。相反, 如果 $p_i-(i-1)<p_i+\Delta-(\tau+1-i)$, 有 $\bot_i=p_i+\Delta-(\tau+1-i)$，这种情况下 $i>\lfloor\frac{\tau-\Delta}{2}\rfloor+1$。

类似的，对于 $\top_i=\min(p_i+(i-1),p_i+\Delta+(\tau+1-i))$, 如果 $p_i+(i-1)\leq p_i+\Delta+(\tau+1-i)$，有 $\top_i=p_i+(i-1)$，在这种情况下 $i\leq\lfloor\frac{\tau+\Delta}{2}\rfloor+1$。相反的，如果 $p_i+(i-1)>p_i+\Delta+(\tau+1-i)$，有 $\top_i=p_i+\Delta+(\tau+1-i)$，这种情况下 $i>\lfloor\frac{\tau+\Delta}{2}\rfloor+1$。

因此，为了计算 $\bot_i$-$\top_i$+1，可以把 $i\in[1,\tau+1]$ 分为 $i\leq\lfloor\frac{\tau-\Delta}{2}\rfloor+1$，$\lfloor\frac{\tau-\Delta}{2}\rfloor+2\leq i\leq\lfloor\frac{\tau+\Delta}{2}\rfloor+1$ 和 $\lfloor\frac{\tau+\Delta}{2}\rfloor+2\leq i\leq\tau+1$ 三段。

$$|\mathcal{W}_m(s,l)|=\sum_{i=1}^{\tau+1}(\top_i-\bot_i+1)=\sum_{i=1}^{\lfloor\frac{\tau-\Delta}{2}\rfloor+1}\Big((p_i+(i-1)-(p_i-(i-1))+1\Big)+$$
$$\sum_{i=\lfloor\frac{\tau-\Delta}{2}\rfloor+2}^{\lfloor\frac{\tau+\Delta}{2}\rfloor+1}\Big((p_i+(i-1))-(p_i+\Delta-(\tau+1-i))+1\Big)+$$
$$\sum_{i=\lfloor\frac{\tau+\Delta}{2}\rfloor+2}^{\tau+1}\Big((p_i+\Delta+(\tau+1-i))-(p_i+\Delta-(\tau+1-i))+1\Big)$$
$$=\sum_{i=1}^{\lfloor\frac{\tau-\Delta}{2}\rfloor+1}(2i-1)+\sum_{i=\lfloor\frac{\tau-\Delta}{2}\rfloor+2}^{\lfloor\frac{\tau+\Delta}{2}\rfloor+1}(\tau-\Delta+1)+\sum_{i=\lfloor\frac{\tau+\Delta}{2}\rfloor+2}^{\tau+1}(2\tau-2i+3)$$
$$=\tau^2+\Delta\tau-\Delta^2+\Delta+1+\lfloor\frac{\tau+\Delta}{2}\rfloor^2+\lfloor\frac{\tau-\Delta}{2}\rfloor^2+2\lfloor\frac{\tau-\Delta}{2}\rfloor-2\tau\lfloor\frac{\tau+\Delta}{2}\rfloor$$

如果 $\tau+\vartriangle$ 是偶数，那么 $\tau-\vartriangle$ 也一定是偶数。因此 $\lfloor\frac{\tau-\vartriangle}{2}\rfloor=\frac{\tau-\vartriangle}{2}$，$\lfloor\frac{\tau+\vartriangle}{2}\rfloor=\frac{\tau+\vartriangle}{2}$，

$$|\mathcal{W}_m(s,l)|=\frac{\tau^2-\vartriangle^2}{2}+\tau+1=\lfloor\frac{\tau^2-\vartriangle^2}{2}\rfloor+\tau+1$$

如果 $\tau+\vartriangle$ 是奇数，那么 $\tau-\vartriangle$ 也一定是奇数。因此 $\lfloor\frac{\tau-\vartriangle}{2}\rfloor=\frac{\tau-\vartriangle-1}{2}$，$\lfloor\frac{\tau+\vartriangle}{2}\rfloor=\frac{\tau+\vartriangle-1}{2}$，

$$|\mathcal{W}_m(s,l)|=\frac{\tau^2-\vartriangle^2+1}{2}+\tau=\lfloor\frac{\tau^2-\vartriangle^2}{2}\rfloor+\tau+1$$

综上，引理得证。 □

除此之外，定理 3.2 和定理 3.2 分别证明了左侧视角以及同时使用左侧视角和右侧视角的多段匹配敏感的选取方法满足完备性。

**引理** 3.3：　左侧视角的多段匹配敏感的选取方法满足完备性。

**证明** 对于任何字符串 $s$，考虑长度为 $l$（其中 $|s|-\tau\le l\le|s|$）、与 $s$ 相似并在 $s$ 之前被访问的字符串 $r$ 以及任何包含 $|\mathcal{T}|\le\tau$ 个编辑操作的从 $r$ 到 $s$ 的转换 $\mathcal{T}$，根据引理3.1，在转换 $\mathcal{T}$ 中，$s$ 一定包含一个子字符串 $s_m$ 与 $r$ 的片段 $r_m$ 完全匹配，假设在转换 $\mathcal{T}$ 中，$r_m$ 和 $s_m$ 是 $r$ 和 $s$ 最后一对匹配的片段和子字符串。不失一般性的，假设 $s_m$ 的起始位置是 $p$，$r_m$ 是第 $i$ 份片段，因此 $r\in\mathcal{L}_l^i(r_m)$。根据定理3.1，有 $|s_m|=l_i$ 和 $p\in[1,|s|-l_i+1]$。

因为 $[\perp_i^l,\top_i^l]=[1,|s|-l_i+1]\cap[p_i-(i-1),p_i+(i-1)]$，所以只需要证明 $p\in[p_i-(i-1),p_i+(i-1)]$。

接下来用反正法证明这点。假设 $p\notin[p_i-(i-1),p_i+(i-1)]$，那么有 $\text{ED}(s_l,r_l)\ge|p-p_i|\ge i$。因为 $\mathcal{T}$ 把 $s_l$ 转化为 $r_l$，匹配 $s_m$ 和 $r_m$，并把 $s_r$ 转化为 $r_r$，所以有 $\tau\ge|\mathcal{T}|\ge\text{ED}(s_l,r_l)+\text{ED}(s_r,r_r)\ge i+\text{ED}(s_r,r_r)$，因此 $\text{ED}(s_r,r_r)\le\tau-i$。另一方面，因为在 $r_r$ 中有 $\tau+1-i$ 个片段，根据引理3.1，$r_r$ 中一定存在一个片段与 $s_r$ 的一个子字符串匹配。这与 $r_m$ 是 $r$ 中最后一个存在匹配的片段的假设互相矛盾，因此 $p\in[p_i-(i-1),p_i+(i-1)]$。

所以，给定一个字符串 $s$，对任何长度为 $l$（其中 $|s|-\tau\le l\le|s|$）、与 $s$ 相似并在 $s$ 之前被访问的字符串 $r$，一定存在一个片段 $r_m$ 与 $s$ 的一个子字符串 $s_m$ 完全匹配，其中 $r_m$ 是第 $i$ 份片段并且 $s_m\in\mathcal{W}_r(s,\mathcal{L}_l^i)$。定理得证。 □

类似的可以证明使用右侧视角的多段匹配敏感的选取方法也满足完备性。接下来证明同时使用左侧视角和右侧视角的多段匹配敏感的选取方法也满足完备

性。针对每个倒排索引 $\mathcal{L}_l^i$，该方法选取一个子字符串集合 $\mathcal{W}_m(s,\mathcal{L}_l^i)$，它包含 $s$ 中所有起始位置在 $[\perp_i,\top_i]$ 内，长度为 $l_i$ 的子字符串，其中 $\perp_i = \max(\perp_i^l,\perp_i^r)$，$\top_i = \min(\top_i^l,\top_i^r)$，$\perp_i^l = \max\big(1,p_i-(i-1)\big)$，$\top_i^l = \min\big(|s|-l_i+1,p_i+(i-1)\big)$，$\perp_i^r = \max\big(1,p_i+\triangle-(\tau+1-i)\big)$，$\top_i^r = \min\big(|s|-l_i+1,p_i+\triangle+(\tau+1-i)\big)$。$\mathcal{W}_m(s,l)$ 是它们的并集，即 $\mathcal{W}_m(s,l)=\cup_{i=1}^{\tau+1}\mathcal{W}_m(s,\mathcal{L}_l^i)$。接下来证明定理3.2。

**定理** 3.2:　多段匹配敏感的子字符串选取方法满足完备性。

**证明** 对于任何字符串 $s$，考虑长度为 $l$（其中 $|s|-\tau\le l\le|s|$）、与 $s$ 相似并在 $s$ 之前被访问的字符串 $r$ 以及任何包含 $|\mathcal{T}|\le\tau$ 个编辑操作的从 $r$ 到 $s$ 的转换 $\mathcal{T}$。对于 $s$ 的任何与 $r$ 的片段 $r_m$ 相匹配的子字符串 $s_m$，根据定理3.1，$s_m$ 的起始位置 $p$ 一定在 $[1,|s|-l_i+1]$ 内并且 $|s_m|=l_i$（假设 $r_m$ 是 $r$ 的第 $i$ 个片段）。接下来只需要证明 $\mathcal{T}$ 中一定存在一个 $i$，使得 $r$ 的第 $i$ 个片段与 $s$ 的一个子字符串 $s_m$ 匹配并且 $s_m$ 的起始位置 $p\in[\max\big(p_i-(i-1),p_i+\triangle-(\tau+1-i)\big),\min\big(p_i+(i-1),p_i+\triangle+(\tau+1-i)\big)]$，即要证 $p\in[p_i-(i-1),p_i+(i-1)]$ 和 $p\in[p_i+\triangle-(\tau+1-i),p_i+\triangle+(\tau+1-i)]$ 成立。

根据引理3.1，$\mathcal{T}$ 中一定存在 $s$ 的一个子字符串与 $r$ 的一个片段匹配，假设在 $\mathcal{T}$ 中，$r_m$ 和 $s_m$ 是 $r$ 和 $s$ 第一对匹配的片段和子字符串。不失一般性的，假设 $s_m$ 在 $s$ 中的起始位置是 $p$，$r_m$ 是 $r$ 的第 $k$ 份片段，即 $r\in\mathcal{L}_l^k(r_m)$。可以把 $s$ $(r)$ 切分为三个部分匹配之前的部分 $s_l$ $(r_l)$，匹配的部分 $s_m$ $(r_m)$ 以及匹配之后的部分 $s_r$ $(r_r)$。根据引理3.3 (右侧视角), 有 $p\in[p_k+\triangle-(\tau+1-k),p_k+\triangle+(\tau+1-k)]$。如果 $p\in[p_k-(k-1),p_k+(k-1)]$，可以令 $i=k$，定理得证。否则，假设 $p\notin[p_k-(k-1),p_k+(k-1)]$，有 $\textsc{ed}(s_l,r_l)\ge|p-p_k|\ge k$。因为 $\mathcal{T}$ 把 $s_l$ 转换为 $r_l$，匹配 $s_m$ 和 $r_m$，然后转换 $s_r$ 为 $r_r$，所以 $\tau\ge|\mathcal{T}|\ge\textsc{ed}(s_l,r_l)+\textsc{ed}(s_r,r_r)\ge k+\textsc{ed}(s_r,r_r)$，因此 $\textsc{ed}(s_r,r_r)\le\tau-k$。另一方面，因为 $r_r$ 中存在 $\tau+1-k$ 个片段，根据引理3.1，在转换 $\mathcal{T}$ 中，$r_r$ 一定包含一个片段与 $s_r$ 中的一个子字符串完全匹配。

假设在 $\mathcal{T}$ 中 $r'_m$ 和 $s'_m$ 是 $r_r$ 和 $s_r$ 中第一对匹配的片段和子字符串。不失一般性的，假设 $s'_m$ 在 $s$ 中的起始位置是 $p'$ 并且 $r'_m$ 是 $r$ 的第 $j$ 份片段 $(j>k)$，即 $r\in\mathcal{L}_l^j(r'_m)$。把 $s_r$ $(r_r)$ 切分为三部分：匹配之前的部分 $s'_l$ $(r'_l)$，匹配的部分 $s'_m$ $(r'_m)$ 以及匹配之后的部分 $s'_r$ $(r'_r)$。接下来用反证法证明 $p'\in[p_j+\triangle-(\tau+1-j),p_j+\triangle+(\tau+1-j)]$。假设 $p'\notin[p_j+\triangle-(\tau+1-j),p_j+\triangle+(\tau+1-j)]$，那么有 $\textsc{ed}(r'_r,s'_r)\ge|(|s|-p')-(|r|-p_j)|=|p_j+(|s|-l)-p'|=|(p_j+\triangle)-p'|\ge\tau+1-j+1$。因为 $\mathcal{T}$ 把 $s_l$ 转化为 $r_l$，匹配 $s_m$ 和 $r_m$，把 $s'_l$ 转化为 $r'_l$，匹配 $s'_m$ 和 $r'_m$，把 $s'_r$ 转化为 $r'_r$，所以有 $\tau\ge|\mathcal{T}|\ge\textsc{ed}(r_l,s_l)+\textsc{ed}(r'_l,s'_l)+\textsc{ed}(r'_r,s'_r)\ge k+\textsc{ed}(r'_l,s'_l)+\tau+1-j+1$，因此 $\textsc{ed}(r'_l,s'_l)\le\tau-k-(\tau+1-j+1)=j-k-2$。另一方面，因为 $r'_l$ 中存在 $j-k-1$

个片段，根据引理3.1，在 $\mathcal{T}$ 中 $r'_l$ 一定包含一个片段与 $s'_l$ 的一个子字符串匹配。这与 $r'_m$ 和 $s'_m$ 是 $r_r$ 和 $s_r$ 中第一份匹配的片段和子字符串的假设互相矛盾，因此 $p' \in [p_j + \triangle - (\tau + 1 - j), p_j + \triangle + (\tau + 1 - j)]$。假如 $p' \in [p_j - (j-1), p_j + (j-1)]$，可以令 $i = j$，引理得证。否则，有 $p' \in [p_j + \triangle - (\tau + 1 - j), p_j + \triangle + (\tau + 1 - j)]$ 以及 $p' \notin [p_j - (j-1), p_j + (j-1)]$。然后不断重复以上证明，直到该定理得证，或者抵达 $r$ 中最后一份与 $s$ 中子字符串（用 $s''_m$ 表示）匹配的片段（用 $r''_m$ 表示）。在后一种情况，有 $r''_m$ 是 $r$ 的第 $i$ 份片段，$s''_m$ 的起始位置为 $p''$。根据以上证明，有 $p'' \in [p_i + \triangle - (\tau + 1 - i), p_i + \triangle + (\tau + 1 - i)]$。根据引理3.3的证明，有 $p'' \in [p_i - (i-1), p_i + (i-1)]$。因此 $p'' \in [p_i - (i-1), p_i + (i-1)] \cap [p_i + \triangle - (\tau + 1 - i), p_i + \triangle + (\tau + 1 - i)] = [\bot_i, \top_i]$。

综上，对任何字符串 $s$，任何长度为 $l$（其中 $|s| - \tau \le l \le |s|$）、与 $s$ 相似并在 $s$ 之前被访问的字符串 $r$ 一定包含一个片段 $r_m$ 与 $s$ 的一个子字符串 $s_m$ 完全匹配，其中 $r_m$ 是第 $i$ 份片段，$s_m \in \mathcal{W}_m(s, \mathcal{L}^i_l)$。定理得证。 □

仍然考虑以上的例子，对于第一个片段，有 $\bot_i = 1 - 0 = 1$ 和 $\top_i = 1 + 0 = 1$，所以为第一个片段选取子字符串 “av’’。对于第二个片段，有 $\bot_i = 3 - 1 = 2$ 和 $\top_i = 3 + 1 = 4$。因此为第二个片段选取子字符串 “va’’，“at’’ 和 “ta’’。对于第三个片段，有 $\bot_i = 5 + 1 - (3 + 1 - 3) = 5$ 和 $\top_i = 5 + 1 + (3 + 1 - 3) = 7$。因此为第三个片段选取子字符串 “ar’’，“re’’ 和 “es’’。对于第四个片段，有 $\bot_i = 7 + 1 - (3 + 1 - 4) = 8$ 和 $\top_i = 7 + 1 + (3 + 1 - 4) = 8$。因此为第四个片段选取子字符串 “sha’’。多段匹配敏感的子字符串选取方法只选取了 8 个子字符串。

### 3.4.3 子字符串选取方法的比较

本节比较不同方法选取的子字符串集合。令 $\mathcal{W}_\ell(s, l)$，$\mathcal{W}_f(s, l)$，$\mathcal{W}_p(s, l)$，$\mathcal{W}_m(s, l)$ 分别代表使用基于长度的选取方法、基于位移的选取方法、位置敏感的选取方法以及多段匹配敏感的选取方法所选取的子字符串集合。根据对各个子字符串集合大小的分析，有 $|\mathcal{W}_m(s, l)| \le |\mathcal{W}_p(s, l)| \le |\mathcal{W}_f(s, l)| \le |\mathcal{W}_\ell(s, l)|$。接下来给出引理 3.4来进一步证明 $\mathcal{W}_m(s, l) \subseteq \mathcal{W}_p(s, l) \subseteq \mathcal{W}_f(s, l) \subseteq \mathcal{W}_\ell(s, l)$。

**引理** 3.4：对于任何字符串 $s$ 和长度 $l$，有 $\mathcal{W}_m(s, l) \subseteq \mathcal{W}_p(s, l) \subseteq \mathcal{W}_f(s, l) \subseteq \mathcal{W}_\ell(s, l)$。

**证明** 如果 $\tau = 0$，这四个方法都只选取子字符串 $s$，因此 $\mathcal{W}_\ell(s, l) = \mathcal{W}_f(s, l) = \mathcal{W}_p(s, l) = \mathcal{W}_m(s, l) = \{s\}$。接下来考虑 $\tau > 0$ 的情况。

针对倒排索引 $\mathcal{L}^i_l$，首先所有方法选取的子字符串长度是相同的，即 $l_i$。接下来考虑它们的起始位置。

$(i)$ **首先证明** $\mathcal{W}_f(s,l) \subseteq \mathcal{W}_\ell(s,l)$。对 $\mathcal{W}_\ell(s,l)$，其起始位置在 $[1,|s|-l_i+1]$ 中。对 $\mathcal{W}_f(s,l)$，其起始位置在 $[\max(1,p_i-\tau),\min(|s|-l_i+1,p_i+\tau)]$ 中。为了证明 $\mathcal{W}_f(s,l) \subseteq \mathcal{W}_\ell(s,l)$，只需要证明

$$[1,|s|-l_i+1] \supseteq [\max(1,p_i-\tau),\min(|s|-l_i+1,p_i+\tau)]$$

。显然，$\max(1,p_i-\tau) \geq 1$ 和 $\min(|s|-l_i+1,p_i+\tau) \leq |s|-l_i+1$ 成立，因此 $\mathcal{W}_f(s,l) \subseteq \mathcal{W}_\ell(s,l)$。

$(ii)$ **其次证明** $\mathcal{W}_p(s,l) \subseteq \mathcal{W}_f(s,l)$。对 $\mathcal{W}_p(s,l)$，其起始位置为：

$$[\max(1,p_i-\lfloor\frac{\tau-\triangle}{2}\rfloor),\min(|s|-l_i+1,p_i+\lfloor\frac{\tau+\triangle}{2}\rfloor)] = \\ [p_i-\lfloor\frac{\tau-\triangle}{2}\rfloor,p_i+\lfloor\frac{\tau+\triangle}{2}\rfloor] \cap [1,|s|-l_i+1]$$

对 $\mathcal{W}_f(s,l)$，因为 $[\max(1,p_i-\tau),\min(|s|-l_i+1,p_i+\tau)] = [p_i-\tau,p_i+\tau] \cap [1,|s|-l_i+1]$，所以为了证明 $\mathcal{W}_p(s,l) \subseteq \mathcal{W}_f(s,l)$，只需证明

$$[p_i-\tau,p_i+\tau] \supseteq [p_i-\lfloor\frac{\tau-\triangle}{2}\rfloor,p_i+\lfloor\frac{\tau+\triangle}{2}\rfloor]$$

。因为 $0 \leq \triangle \leq \tau$，所以 $p_i-\tau \leq p_i-\lfloor\frac{\tau-\triangle}{2}\rfloor$ 和 $p_i+\tau \geq p_i+\lfloor\frac{\tau+\triangle}{2}\rfloor$ 成立。因此 $\mathcal{W}_p(s,l) \subseteq \mathcal{W}_f(s,l)$。

$(iii)$ **接下来证明** $\mathcal{W}_m(s,l) \subseteq \mathcal{W}_p(s,l)$。对 $\mathcal{W}_m(s,l)$，其起始位置为：

$$[\bot_i,\top_i] = [\max(\bot_i^l,\bot_i^r),\min(\top_i^l.\top_i^r)] = \\ [\max(1,p_i-(i-1),p_i+\triangle-(\tau+1-i)),\min(|s|-l_i+1,p_i+(i-1),p_i+\triangle+(\tau+1-i))] = \\ [\max(p_i-(i-1),p_i+\triangle-(\tau+1-i)),\min(p_i+(i-1),p_i+\triangle+(\tau+1-i)] \cap [1,|s|-l_i+1]$$

为了证明 $\mathcal{W}_m(s,l) \subseteq \mathcal{W}_p(s,l)$，只需证明

$$[p_i-\lfloor\frac{\tau-\triangle}{2}\rfloor,p_i+\lfloor\frac{\tau+\triangle}{2}\rfloor] \supseteq [\bot_i,\top_i]$$

首先证明 $\bot_i = \max(p_i-(i-1),p_i+\triangle-(\tau+1-i)) \geq p_i-\lfloor\frac{\tau-\triangle}{2}\rfloor$。如果 $p_i-(i-1) \geq p_i+\triangle-(\tau+1-i)$，那么有 $\bot_i = p_i-(i-1)$。这种情况下有 $i \leq \lfloor\frac{\tau-\triangle}{2}\rfloor+1$。也就是说 $i-1 \leq \lfloor\frac{\tau-\triangle}{2}\rfloor$。显然 $\bot_i = p_i-(i-1) \geq p_i-\lfloor\frac{\tau-\triangle}{2}\rfloor$。相反的，如果

$p_i-(i-1)<p_i+\triangle-(\tau+1-i)$，那么有 $\bot_i=p_i+\triangle-(\tau+1-i)$。在这种情况下，$i>\lfloor\frac{\tau-\triangle}{2}\rfloor+1$。也就是说 $i-1>\lfloor\frac{\tau-\triangle}{2}\rfloor$。因为 $\bot_i=p_i+\triangle-(\tau+1-i)=p_i+(i-1)-(\tau-\triangle)$，所以 $\bot_i\geq p_i-\lfloor\frac{\tau-\triangle}{2}\rfloor$。

然后证明 $\top_i=\min(p_i+(i-1),p_i+\triangle+(\tau+1-i))\leq p_i+\lfloor\frac{\tau+\triangle}{2}\rfloor$。如果 $p_i+(i-1)\leq p_i+\triangle+(\tau+1-i)$，那么有 $\top_i=p_i+(i-1)$。这种情况下有 $i\leq\lfloor\frac{\tau+\triangle}{2}\rfloor+1$。也就是说 $i-1\leq\lfloor\frac{\tau+\triangle}{2}\rfloor$。显然 $\top_i=p_i+(i-1)\leq p_i+\lfloor\frac{\tau+\triangle}{2}\rfloor$。相反的，如果 $p_i+(i-1)>p_i+\triangle+(\tau+1-i)$，那么有 $\top_i=p_i+\triangle+(\tau+1-i)$。在这种情况下有 $i>\lfloor\frac{\tau+\triangle}{2}\rfloor+1$。也就是说 $i-1>\lfloor\frac{\tau+\triangle}{2}\rfloor$。因为 $\top_i=p_i+\triangle+(\tau+1-i)=p_i-(i-1)+\tau+\triangle$，所以 $\top_i\leq p_i+\lfloor\frac{\tau+\triangle}{2}\rfloor$。因此有 $\mathcal{W}_m(s,l)\subseteq\mathcal{W}_p(s,l)$。

因此 $\mathcal{W}_m(s,l)\subseteq\mathcal{W}_p(s,l)\subseteq\mathcal{W}_f(s,l)\subseteq\mathcal{W}_\ell(s,l)$，引理得证。 □

除此之外，定理 3.3证明了所有满足完备性的子字符串选取方法所产生的子字符串集合中，$\mathcal{W}_m(s,l)$ 的大小最小。

**定理** 3.3：多段匹配敏感的选取方法所选取的子字符串集合 $\mathcal{W}_m(s,l)$ 在所有满足完备性的子字符串选取方法所产生的子字符串集合中大小最小。

**证明** 对于 $\mathcal{W}_m(s,l)$ 中的每个子字符串，若一个子字符串选取方法满足完备性，可以证明它要么选择该子字符串，要么选择至少另外一个不同的子字符串来替代它。进一步可以证明，对于 $\mathcal{W}_m(s,l)$ 中任意两个子字符串，它们的替代子字符串都不相同。 □

### 3.4.4 子字符串选取算法

基于以上讨论，可以通过规避不必要的子字符串来提高 SubstringSelection 算法。对于 $\mathcal{L}_l^i$，使用多段匹配敏感的选取选取方法来选择子字符串，其选取时间复杂度为 $O(\tau)$。图 3.6 给出了选取算法的伪代码。

例如，考虑表 3.1中的字符串。首先建立图 3.1 所示的倒排索引。考虑有 4 个片段的字符串 $s_1$ = “vankatesh”，为它的 4 个片段 {va，nk，at，esh} 建立 4 个倒排列表。然后对于 $s_2$ = “avataresha”，使用多段匹配敏感的选取选取方法来选取它的子字符串。这里为 $s_2$ 从倒排索引中只选取 8 个子字符串并利用这 8 个子字符串从倒排索引中寻找相似的字符串。类似的，可以选择子字符串并为其它字符串寻找近似字符串元组。

**Algorithm 3.2**: SubstringSelection($s, \mathcal{L}_l^i$)

**Input**: $s$: 字符串; $\mathcal{L}_l^i$: 倒排索引
**Output**: $\mathcal{W}(s, \mathcal{L}_l^i)$: 选取的子字符串集合
**begin**
　**for** $p \in [\perp_i, \top_i]$ **do**
　　把 $s$ 的以 $p$ 为起始位置、长度为 $l_i$ 的子字符串（即 $s[p, l_i]$）添加到 $\mathcal{W}(s, \mathcal{L}_l^i)$ 中;
**end**

图 3.6　SubstringSelection 算法

## 3.5　提高验证效率

在基于划分的框架中，对于字符串 $s$ 和倒排索引 $\mathcal{L}_l^i$，它需要产生 $s$ 的一个子字符串集合 $\mathcal{W}(s, \mathcal{L}_l^i)$。对于每个子字符串 $w \in \mathcal{W}(s, \mathcal{L}_l^i)$，它需要检查 $w$ 是否出现在 $\mathcal{L}_l^i$ 中。如果 $w \in \mathcal{L}_l^i$，那么对于 $\mathcal{L}_l^i(w)$ 中每个字符串 $r$，$\langle r, s\rangle$ 都是一个候选元组，它需要验证所有的候选元组。本节提出有效的技术来进行快速验证。

### 3.5.1　长度敏感的验证方法

本节提出一个长度敏感（Length-aware）的验证方法。给定一个候选元组 $\langle r, s\rangle$，一个直观的验证方法直接利用动态规划算法计算它们的编辑距离。如果它们的编辑距离不大于 $\tau$，这个元组就是一个近似元组。动态规划算法使用一个 $|r|+1$ 行和 $|s|+1$ 列的二维表 $M$ 来计算它们的编辑距离。其中对于所有 $0 \le j \le |s|$，$M(0, j) = j$，对于所有 $1 \le i \le |r|$ 和 $0 \le j \le |s|$，

$$M(i, j) = \min\big(M(i-1, j) + 1, M(i, j-1) + 1, M(i-1, j-1) + \delta\big)$$

其中如果 $r$ 的第 $i$ 个字符和 $s$ 的第 $j$ 个字符相同，那么 $\delta = 0$，否则 $\delta = 1$。该动态规划算法的时间复杂度是 $O(|r| * |s|)$。

实际上并不需要计算候选元组之间真实的编辑距离，只需要检测它们的编辑距离是否大于 $\tau$。[31] 提出了一个基于长度剪枝的改进：只有当 $|i-j| \le \tau$ 时，它才计算 $M(i, j)$ 的值，它只需要计算图 3.7(a) 中阴影部分的格子。它的基本想法是如果 $|i-j| > \tau$，那么 $M(i, j) > \tau$，因此不需要计算这些值。这个方法把验证一个候选元组的时间复杂度 $\mathcal{V}(s, r)$ 降低到 $O\big((2 * \tau + 1) * \min(|r|, |s|)\big)$。接下来通过考虑 $s$ 和 $r$ 的长度差来进一步改进性能。

首先使用一个例子来展示基本想法。考虑字符串 $r$ = “kaushuk chadhui” 和字符

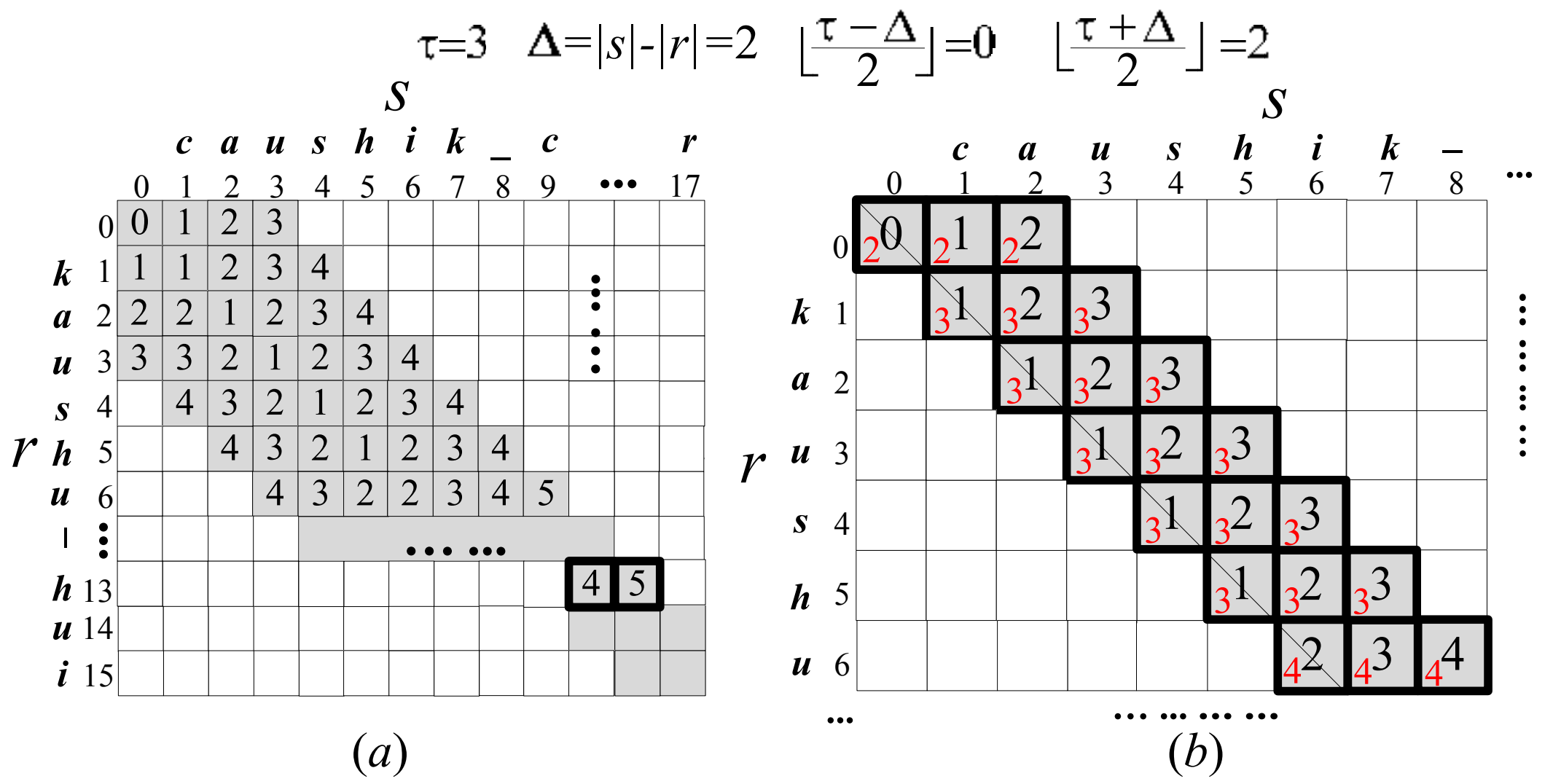

图 3.7　验证算法的例子

串 $s$ = “caushik chakrabar”。假设 $\tau = 3$，现有的方法需要计算图3.7(a) 中阴影部分的所有格子。实际上，不需要计算 $M(2, 1)$ 的值，即 “ka” 和 “c” 的编辑距离。这是因为如果从 $r$ 到 $s$ 的一个转换首先用至少 1 个编辑操作（长度差）把 “ka” 转化为 “c”，然后用至少 3 个编辑操作（长度差）把 “ushuk chadhui” 转换为 “aushik chakrabar”，这个转换需要至少 4 个编辑操作，这大于给定的阈值 $\tau = 3$。换句话说，即使不计算 $M(2, 1)$ 的值，也知道任何包含 $M(2, 1)$ 的转换（即首先把 “ka” 转换为 “c”）需要的编辑操作都大于 $\tau$。实际上，只需要计算图 3.7(b) 中高亮部分的值。

接下来提出一种长度敏感的验证方法。不失一般性的，令 $|s| \geq |r|$ 以及 $\triangle = |s| - |r| \leq \tau$ (否则的话，它们的编辑距离一定大于 $\tau$)。

如果一个从 $r$ 到 $s$ 的转化首先使用 $d_1$ 个编辑操作转化 $r$ 的前 $i$ 个字符为 $s$ 的前 $j$ 个字符，然后使用 $d_2$ 个编辑操作转化 $r$ 的其它字符为 $s$ 的其它字符，那么称这个转化包含 $M(i, j)$。根据长度差，有 $d_1 \geq |i-j|$ 以及 $d_2 \geq |(|s|-j)-(|r|-i)| = |\triangle+(i-j)|$。如果 $d_1 + d_2 > \tau$，那么不需要计算 $M(i, j)$ 的值，因为包含 $M(i, j)$ 的转化的编辑操作数目都大于 $\tau$。为了检测 $d_1 + d_2 > \tau$ 是否成立，考虑如下情况：

(1) 如果 $i \geq j$，有 $d_1 + d_2 \geq i - j + \triangle + i - j$。如果 $i - j + \triangle + i - j > \tau$，也就是说 $j < i - \frac{\tau-\triangle}{2}$，那么不需要计算 $M(i, j)$。换句话说，只有当 $j \geq i - \frac{\tau-\triangle}{2}$ 时，才需要计算 $M(i, j)$。

(2) 如果 $i < j$，有 $d_1 = j - i$。如果 $j - i \leq \triangle$，那么 $d_1 + d_2 \geq j - i + \triangle - (j - i) = \triangle$。因为 $\triangle \leq \tau$，所以没有位置限制，需要计算 $M(i, j)$。否则，如果 $j - i > \triangle$，有 $d_1 + d_2 \geq j - i + j - i - \triangle$。如果 $j - i + j - i - \triangle > \tau$，也就是说 $j > i + \frac{\tau+\triangle}{2}$，那么不

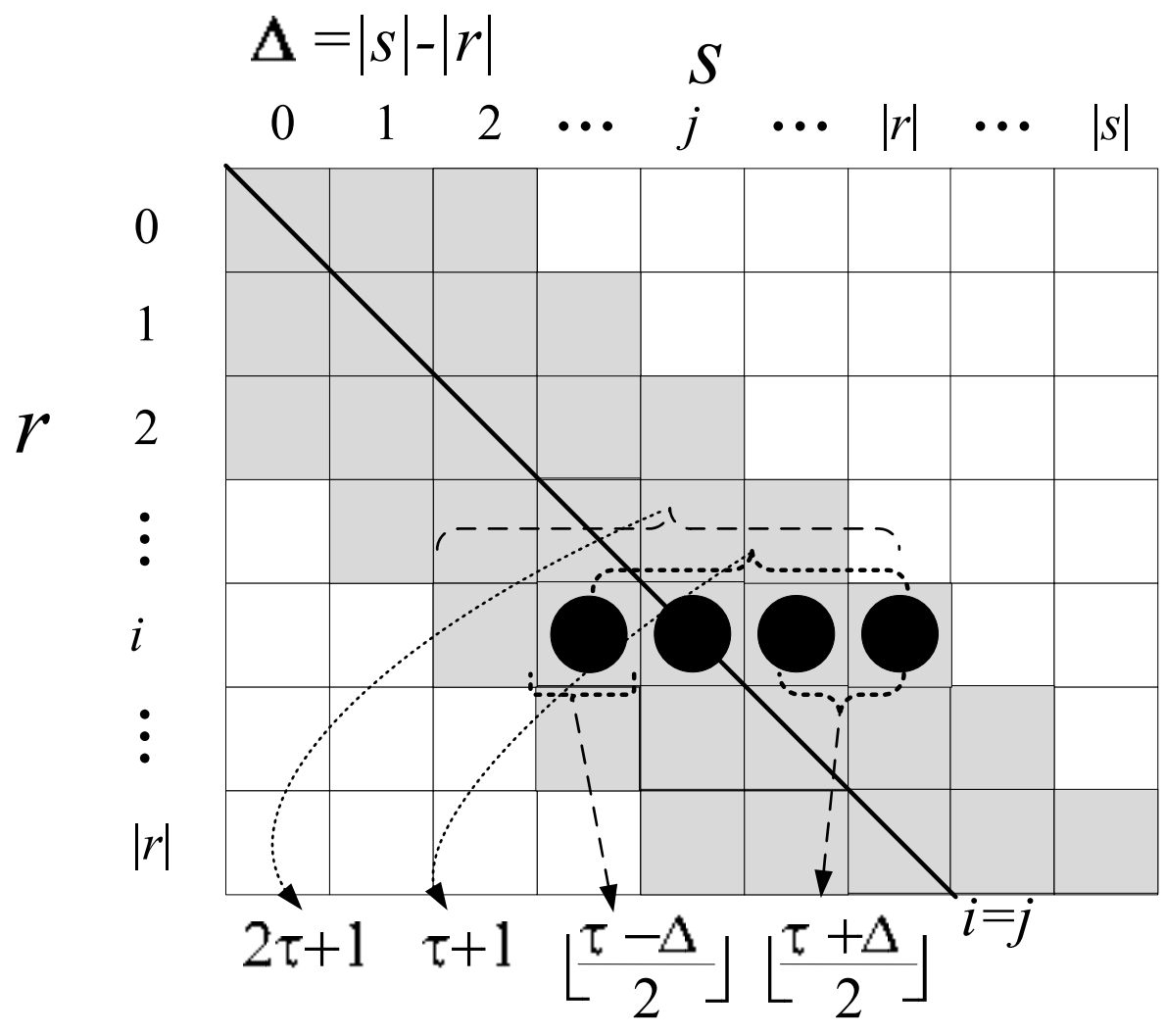

图 3.8　长度敏感验证

需要计算 $M(i,j)$。换句话说，只有当 $j \le i + \frac{\tau+\Delta}{2}$ 时，才需要计算 $M(i,j)$。

基于以上观察，对每一行 $M(i,*)$，只有当 $i - \lfloor\frac{\tau-\Delta}{2}\rfloor \le j \le i + \lfloor\frac{\tau+\Delta}{2}\rfloor$ 时，需要计算 $M(i,j)$ 。例如，在图 3.8中，只需要计算黑色圆圈中的值。因此，这个方法可以把时间复杂度 $\mathcal{V}(s,r)$ 从 $O((2\tau+1)*\min(|r|,|s|))$ 改进到 $O((\tau+1)*\min(|r|,|s|))$。

**提早终止验证**：通过提早终止验证可以进一步提高性能。考虑在行 $M(i,*)$ 中的值，一个直观的提早终止验证的方法是检查第 $i$ 行中的所有值 $M(i,*)$，如果每个值都大于 $\tau$，那么可以提早终止验证。这是因为根据动态规划算法，之后所有行中 $M(k>i,*)$ 的值都肯定大于 $\tau$，这个剪枝技术叫做*前缀剪枝*。例如，在图 3.7(a) 中，如果 $\tau=3$，在计算了 $M(13,*)$ 之后，就可以提早终止验证，因为 $M(13,*)$ 中所有的值都大于 $\tau$。但是，实际上，在计算了所有 $M(6*)$ 的值后，可以得到这两个字符串的编辑距离至少为 4（大于 $\tau=3$）的结论，因此，如图 3.7(b) 所示，可以提早终止计算而不需要计算 $M(i>6,*)$ 的值。为了达到这个目的，接下来提出一个新颖的提早终止验证的方法。

为了方便表述，首先介绍几个记号。给定一个字符串 $s$，令 $s[i]$ 代表其第 $i$ 个字符，令 $s[i:j]$ 代表其从第 $i$ 个字符开始，到第 $j$ 个字符结束的子字符串。注意 $M(i,j)$ 表示 $r[1:i]$ 与 $s[1:j]$ 之间的编辑距离。可以用 $r[i:|r|]$ 与 $s[j:|s|]$ 的长度差 $|(|s|-j)-(|r|-i)|$ 来估计它们的编辑距离的下界。使用 $E(i,j)=M(i,j)+|(|s|-j)-(|r|-i)|$ 来估计 $s$ 和 $r$ 的编辑距离，并称之为 $s$ 和 $r$ 在 $M(i,j)$ 下的*期望编辑距离* (Expected Edit Distance)。如果 $r$ 和 $s$ 在 $M(i,*)$ 中每个 $M(i,j)$ 下的期望编辑距离都大于 $\tau$，它们的编辑距离肯定大于 $\tau$，因此可以提前终止验证。为了达到这个目标，对于每个 $M(i,j)$，维护其期望编辑距离 $E(i,j)$。如果 $E(i,*)$ 中每个值都大

于 $\tau$，那么如引理3.5所述，可以提前终止验证。

**引理** 3.5：　给定字符串 $s$ 和 $r$，如果 $E(i,*)$ 中的每个值都大于 $\tau$, 那么它们的编辑距离一定大于 $\tau$.

**证明** 下面证明如果 $E(i,*)$ 中的值都大于 $\tau$，那么任何从 $r$ 到 $s$ 的转换都包含多于 $\tau$ 个编辑操作。对于任何从 $r$ 到 $s$ 的转换 $\mathcal{T}$，$\mathcal{T}$ 一定包含 $M(i,*)$ 中的一个状态。不失一般性的，假设 $\mathcal{T}$ 包含 $M(i,j)$。那么有 $d_1 = M(i,j)$，$d_2 \geq |(|s|-j)-(|r|-i)|$。因此 $|\mathcal{T}| = d_1 + d_2 \geq M(i,j) + |(|s|-j)-(|r|-i)| = E(i,j) > \tau$。因此 $\mathcal{T}$ 将包含多于 $\tau$ 个编辑操作。因此 $r$ 和 $s$ 的编辑距离大于 $\tau$。 □

例如，如图 3.7(b) 所示，图中各个格子在左下角显示了对应的期望编辑距离。当计算了 $M(6,*)$ 和 $E(6,*)$ 的值后，可以发现 $E(6,*)$ 中所有的值都大于 3，因此可以提前终止验证。通过这种方式，可以避免很多不必要的计算。注意该验证技术可以应用在任何其它算法中来验证编辑距离限制下的候选元组（例如 ED-Join 和 NGPP ）。

### 3.5.2　基于扩展的验证

考虑字符串 $s$ 中的一个被选取的子字符串 $w$，如果 $w$ 出现在倒排索引 $\mathcal{L}_l^i$ 中，对于倒排列表 $\mathcal{L}_l^i(w)$ 中每个字符串 $r$，都需要验证候选元组 $\langle s,r\rangle$。注意 $s$ 和 $r$ 共享一个相同片段 $w$，可以利用该共享片段来快速验证这个候选元组。为了达到这个目的，下面提出一种基于扩展的验证算法。

因为 $r$ 和 $s$ 共享一个相同的片段 $w$，根据该相同片段可以把它们切分为三个部分。对 $r$ 来说切分为左边部分 $r_l$，匹配部分 $r_m = w$ 和右边部分 $r_r$。类似的，可以得到字符串 $s$ 的三个部分: $s_l$，$s_m = w$，和 $s_r$。这里通过匹配子字符串 $r_m$ 和 $s_m$ 来对齐 $s$ 和 $r$，所以只需验证 $r$ 和 $s$ 在这种对齐情况下是否相似。因此首先用以上提到的方法计算 $r_l$ 和 $s_l$ 的编辑距离 $d_l = \text{ED}(r_l, s_l)$。如果 $d_l$ 大于 $\tau$，终止计算；否则，计算 $s_r$ 和 $r_r$ 的编辑距离 $d_r = \text{ED}(s_r, r_r)$。如果 $d_l + d_r$ 大于 $\tau$，扔掉这对元组；否则，把它当作一个近似元组。注意这个方法不会漏掉任何近似元组。这是因为根据定理 3.1，如果 $s$ 和 $r$ 相似，$s$ 一定有一个子字符串与 $r$ 的一个片段匹配。而且，根据动态规划算法，一定存在一个对齐 $r_m$ 和 $s_m$ 的转化，其编辑操作数是 $\text{ED}(s,r) = d_l + d_r$。因为子字符串选取方法会选择最优转换下所有可能匹配的子字符串和片段，所以这个验证方法不会丢失近似元组。另一方面，因为只输出满足 $d_l + d_r \leq \tau$ 和 $\text{ED}(s,r) \leq d_l + d_r \leq \tau$ 的元组，所以的方法不会产生任何错误的结果。

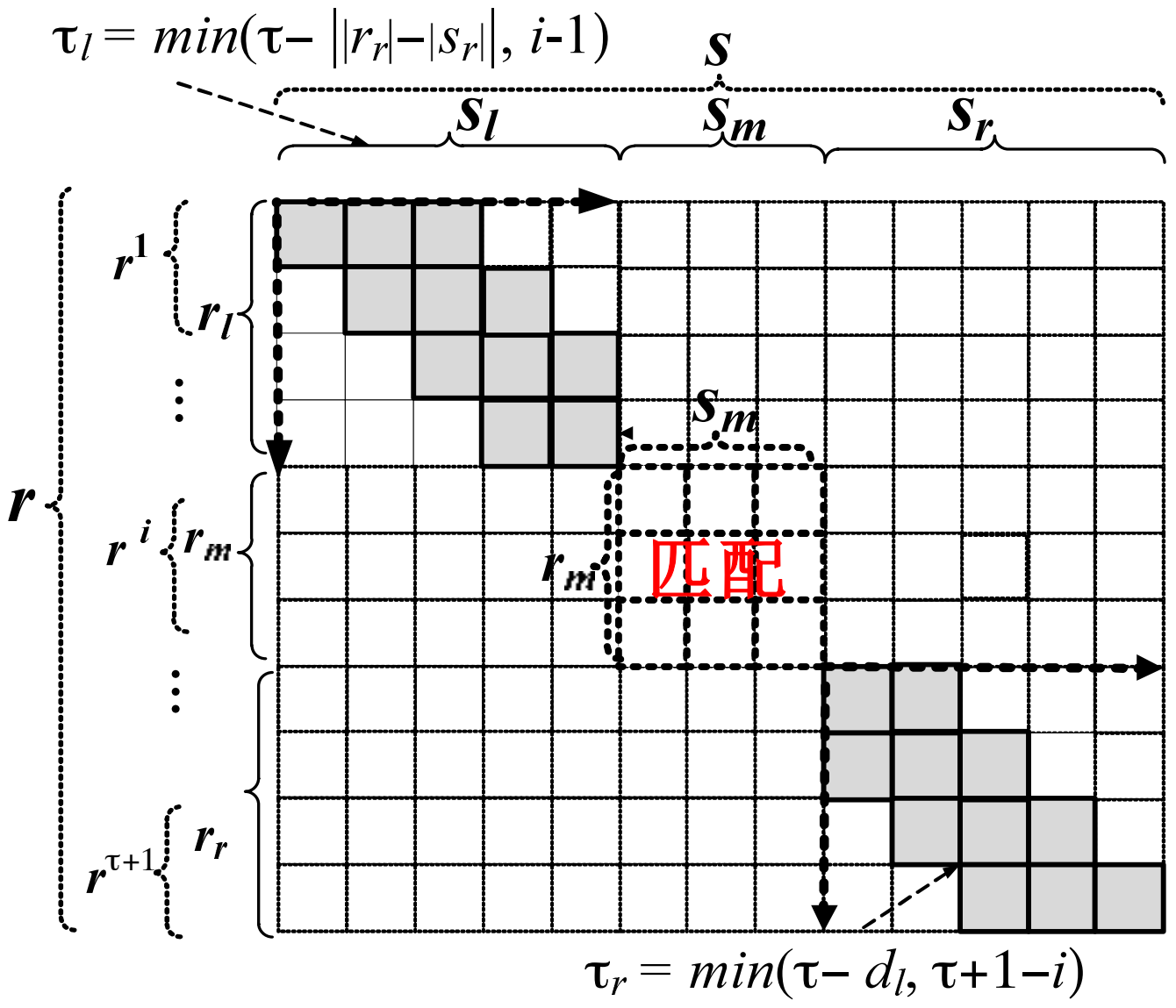

图 3.9　基于扩展的验证

**用更紧缩的界提高验证算法**：实际上，可以进一步提高验证算法。首先，对于左边部分 $r_l$ 和 $s_l$ 可以给出一个更紧的阈值 $\tau_l \le \tau$ 来做验证算法。基本想法如下：因为右边部分 $r_r$ 和 $s_r$ 之间最小的编辑距离是 $||r_r|-|s_r||$，因此，可以设置 $\tau_l = \tau-||r_r|-|s_r||$。如果左边部分 $r_l$ 和 $s_l$ 之间的编辑距离大于阈值 $\tau_l$，可以终止验证；否则，继续计算 $d_r = \text{ED}(r_r, s_r)$。类似的，对于右边部分，也可以给出一个更紧的阈值 $\tau_r \le \tau$ 来组验证算法。这是因为 $d_l$ 已经被计算出来了，所以可以设置 $\tau_r = \tau - d_l$ 为阈值来验证右边部分 $r_r$ 和 $s_r$。如果 $d_r$ 大于阈值 $\tau_r$，那么可以终止验证。

例如，如果希望验证 $s_5$ = “kaushuk chadhui” 和 $s_6$ = “caushik chakrabar’’。因为 $s_5$ 和 $s_6$ 共享一个片段 “_cha’’，所以有 $s_{5_l}$ = “kaushuk” 和 $s_{6_l}$ = “caushik’’，$s_{5_r}$ = “dhui” 和 $s_{6_r}$ = “krabar’’。假设 $\tau = 3$，因为 $\left||s_{5_r}| - |s_{6_r}|\right| = 2$，所以 $\tau_l = \tau - 2 = 1$，因此只需要验证 $s_{5_l}$ 和 $s_{6_l}$ 的编辑距离是否不大于 $\tau_l = 1$。在计算了 $M(6, *)$ 之后，可以提早终止验证，这是因为如图 3.7所示，$E(6, *)$ 中每个值都大于 1。注意，因为 $\tau_l = 1$ 和 $|s_{5_l}| - |s_{6_l}| = 0$，所以 $\perp_i = \top_i = 0$，因此只需要计算 $M(i, i)$ 的值。

接下来讨论如何为 $\tau_l$ 和 $\tau_r$ 推导一个更加紧缩的界。如果 $d_l \ge i$，根据多段匹配敏感的选取方法，可以终止验证。因此有 $\tau_l = i - 1$。结合以上过滤条件，有 $\tau_l = \min(\tau-||r_r|-|s_r||, i-1)$。因为 $||r_r|-|s_r|| = |(|r|-p_i-l_i)-(|s|-p-l_i)| = |p-p_i-\triangle| \le \tau + 1 - i$ (这是根据多段匹配敏感的选取方法得到的)，所以 $\tau - ||r_r| - |s_r|| \ge i - 1$，因此可以设置 $\tau_l = i - 1$。类似的，有 $\tau_r = \min(\tau - d_l, \tau + 1 - i)$。因为 $d_l \le \tau_l \le i - 1$，所以 $\tau - d_l \ge \tau - (i - 1)$，因此可以设置 $\tau_r = \tau + 1 - i$。

**Algorithm 3.3**: VERIFICATION($s, \mathcal{L}_l^i(w), \tau$)

**Input**: $s$: 一个字符串; $\mathcal{L}_l^i(w)$: 倒排索引; $\tau$: 编辑距离阈值

**Output**: $\mathcal{R} = \{(s \in \mathcal{S}, r \in \mathcal{S}) \mid \text{ED}(s, r) \le \tau\}$

vspace
**begin**
　$\tau_l = i - 1$;
　$\tau_r = \tau + 1 - i$;
　**for** $r \in \mathcal{L}_l^i(w)$ **do**
　　$d_l$ = VERIFYSTRINGPAIR($s_l, r_l, \tau_l$);
　　**if** $d_l \le \tau_l$ **then**
　　　$d_r$ = VERIFYSTRINGPAIR($s_r, r_r, \tau_r$);
　　　**if** $d_r \le \tau_r$ **then** $\mathcal{R} \leftarrow \langle r, s\rangle$;
**end**

**Procedure** VERIFYSTRINGPAIR ($s, r, \tau'$)

**Input**: $s$: 一个字符串; $r$: 另一个字符串; $\tau'$: 编辑距离阈值

**Output**: $d = \min(\tau' + 1, \text{ED}(s, r))$

**begin**
　以 $\tau'$ 为阈值使用长度敏感的验证技术并共享相同前缀上的计算;
　**if** 可以提前终止验证 **then** $d = \tau' + 1$;
　**else** $d = \text{ED}(s, r)$;
**end**

图 3.10　验证算法

### 3.5.3　共享计算

给定一个被选取的子字符串 $w$，在 $\mathcal{L}_l^i(w)$ 中可能有大量的字符串。当验证左边部分 $s_l$ 和 $s_r$ 时（以及右边部分 $s_r$ 和 $r_r$ 时，如果它们有共同的前缀，那么可以共享计算。接下来讨论如何共享计算。因为在 $\mathcal{L}_l^i(w)$ 中的字符串是按照字母序排好序的，所以可以按顺序访问 $\mathcal{L}_l^i(w)$ 中的字符串。假如第一个字符串是 $r_1$，它的三个部分分别是 $r_{1_l}, r_{1_m}, r_{1_r}$，那么在用动态规划算法计算 $r_{1_l}$ 与 $s_l$ 之间编辑距离时，可以存储计算中得到的二维数组。然后对于下一个字符串 $r_2$，其左边部分是 $r_{2_l}$，可以使用缓存的二维数组来验证 $r_{2_l}$ 和 $s_l$：首先计算 $r_{2_l}$ 与 $r_{1_l}$ 之间最长的公共前缀，用 $c$ 来表示其长度。然后当验证 $s_l$ 与 $r_{2_l}$ 时，由于在缓存的二维数组中已经计算出了 $s_l$ 与 $c$ 之间所有状态的值，所以可以复用它们。然后，对于在 $r_{2_l}$ 中前缀 $c$ 之后的字符，可以继续计算并把结果存在缓存的二维数组。这样可以避免很多不必要的计算。注意，这个方法只需要为当前字符串维护单一的一个二维数组，大家都复用同一个二维数组。可以将相同的想法应用在右边部分（$s_r$ 和 $r_r$）。

### 3.5.4 验证算法

根据之前提出的技术可以提高验证算法。考虑一个字符串 $s$，一个被选取的子字符串 $w$, 以及一个倒排列表 $\mathcal{L}_l^i(w)$。对于每个 $r \in \mathcal{L}_l^i(w)$, 按照如下的方式来使用基于扩展的方法验证候选元组 $\langle s, r\rangle$。首先计算 $\tau_l = i - 1$ 和 $\tau_r = \tau + 1 - i$ 的值。然后对于每个 $r \in \mathcal{L}_l^i(w)$, 使用更紧缩的界 $\tau_l$ 来计算 $r_l$ 和 $s_l$ 的编辑距离 ($d_l$)。如果 $d_l > \tau_l$，提早终止验证。否则，验证 $s_r$ 和 $r_r$ 是否在阈值 $\tau_r$ 下是近似的。当计算 $s_l$ 与 $r_l$（以及 $s_r$ 与 $r_r$）之间的编辑距离时，可以使用长度敏感的验证方法并共享相同前缀之间的计算，图3.10 展示了其伪代码。

## 3.6 进一步讨论

本节首先讨论如何支持 Edit Similarity 限制下的字符串近似连接 (第3.6.1节), 然后扩展被之前提出的技术以支持异连接 (第3.6.2节)，最后讨论如何扩展 PassJoin 在 MapReduce 框架下运行（第3.6.3节）。

### 3.6.1 支持 Edit Similarity

Edit Similarity，也叫归一化的编辑距离 (Normalized Edit Distance)，也是一个被广泛应用的来衡量两个字符串相似性的近似函数。两个字符串 $r$ 和 $s$ 的 Edit Similarity 定义为 $\textsc{eds}(r, s) = 1 - \frac{\textsc{ed}(r,s)}{max(|r|,|s|)}$。例如，$\textsc{eds}$(“kausic chakduri”, “*kaushuk chadhui*”) = $\frac{11}{17}$。给定一个 Edit Similarity 阈值 $\delta$，本节中称两个字符串相似当且仅当它们的 Edit Similarity 不小于 $\delta$。接下来给出 Edit Similarity 限制下的字符串近似连接问题正式定义。

**定义** 3.3 (Edit Similarity **限制下的字符串近似连接**)： 给定两个字符串集合 $\mathcal{R}$ 和 $\mathcal{S}$ 以及一个 Edit Similarity 阈值 $\delta$，Edit Similarity 限制下的字符串近似连接找到 $\mathcal{R} \times \mathcal{S}$ 中所有满足 $\textsc{eds}(r, s) \geq \delta$ 的近似元组 $\langle r, s\rangle$。

接下来讨论如何支持 Edit Similarity。对于两个字符串 $r$ 和 $s$，因为 $\textsc{eds}(r, s) = 1 - \frac{\textsc{ed}(r,s)}{max(|r|,|s|)}$, 所以 $\textsc{ed}(r, s) = max(|r|, |s|) \cdot (1 - \textsc{eds}(r, s))$。如果 $\textsc{eds}(r, s) \geq \delta$, 那么 $\textsc{ed}(r, s) = max(|r|, |s|) \cdot (1 - \textsc{eds}(r, s)) \leq max(|r|, |s|) \cdot (1 - \delta)$。注意在构建索引时，需要把字符串 $r$ 划分为 $\tau + 1$ 个片段。如果 $|s| > |r|$, 那么片段的数量是不确定的。为了解决这个问题，需要首先为长的字符串 ($r$) 构建索引，然后访问短的字符串 ($s$)。也就是说，基于较长字符串的片段构建索引并从较短字符串中选择子字符串。这样的话，$|s| \leq |r|$ 总是成立的。因此 $\textsc{ed}(r, s) \leq |r| \cdot (1 - \delta)$。令 $\tau = |r| \cdot (1 - \delta)$，可以用平均划分方法把 $r$ 划分为 $\tau + 1 = \lfloor |r| \cdot (1 - \delta)\rfloor + 1$ 个片段。除此之外，因为

**Algorithm 3.5**: PassJoin-NED $(\mathcal{S}, \delta)$

**Input**: $\mathcal{S}$: 一个字符串集合；$\delta$: 一个 Edit Similarity 阈值
**Output**: $\mathcal{A} = \{(s \in \mathcal{S}, r \in \mathcal{S}) \mid \text{EDS}(s, r) \geq \delta\}$

1 **begin**
2 　对 $\mathcal{S}$ 中的字符串按照长度降序排序;
3 　**for** $s \in \mathcal{S}$ **do**
4 　　**for** $|s| \leq l \leq \lfloor |s|/\delta \rfloor$ **do**
5 　　　$\tau = \lfloor (1-\delta) \cdot l \rfloor$ ;
6 　　　**for** $\mathcal{L}_l^i (1 \leq i \leq \tau + 1)$ **do**
7 　　　　$\mathcal{W}(s, \mathcal{L}_l^i) = \text{SubstringSelection}(s, \mathcal{L}_l^i)$;
8 　　　　**for** $w \in \mathcal{W}(s, \mathcal{L}_l^i)$ **do**
9 　　　　　**if** $w$ *is in* $\mathcal{L}_l^i$ **then** $\text{Verification}(s, \mathcal{L}_l^i(w), \tau)$;
10 　　把 $s$ 划分为 $\lfloor (1-\delta) \cdot |s| \rfloor + 1$ 个片段并插入到倒排索引 $\mathcal{L}_{|s|}^i$ 中;
11 **end**

图 3.11　PassJoin-NED 算法

$|r| - |s| \leq \text{ED}(r, s) \leq |r| \cdot (1-\delta)$, 所以有 $|r| \leq \lfloor \frac{|s|}{\delta} \rfloor$。因此对于字符串 $s$，只需要为长度在 $|s|$ 到 $\lfloor \frac{|s|}{\delta} \rfloor$ 之间内的字符串选择子字符串从而产生候选元组。其余子字符串选取阶段和验证阶段与原算法是一样的。

图3.11给出了支持 Edit Similarity 的字符串近似连接算法 PassJoin-NED 的伪代码。它首先对 $\mathcal{S}$ 中的字符串按照长度降序排序（第 2 行），然后依次访问每个字符串 $s$（第 3 行）。对每个可能与 $s$ 相似的字符串的长度（即 $[|s|, \lfloor \frac{|s|}{\delta} \rfloor]$）（第 4 行），算法把 Edit Similarity 阈值 $\delta$ 转化为编辑距离阈值 $\tau$ （第 5 行）。然后对每个倒排索引 $\mathcal{L}_l^i (1 \leq i \leq \tau + 1)$ （第 6 行），算法选择 $s$ 的子字符串（第 7 行）并检查被选取的子字符串 $w$ 是否存在于 $\mathcal{L}_l^i$ 中（第 8 行）。如果存在的话，那么对于任何在倒排列表 $\mathcal{L}_l^i(w)$ 中的字符串 $r$，字符串元组 $\langle r, s \rangle$ 都是一个候选元组。接下来它验证这个候选元组（第 9 行）。最后，它把 $s$ 划分为 $\lfloor (1-\delta) \cdot |s| \rfloor + 1$ 个片段并把这些片段插入倒排索引 $\mathcal{L}_{|s|}^i (1 \leq i \leq \lfloor (1-\delta) \cdot |s| \rfloor + 1)$ 中（第 10 行）。这里子程序 SubstringSelection 和 Verification 与图4.1中的子程序是一样的。

接下来给出算法的一个例子。考虑表3.1中的字符串集合并假设 Edit Similarity 阈值是 $\delta = 0.82$。如表3.1(c) 所示，首先对字符串按照降序排序。对第一个字符串 $s_6$，把它划分为 $\lfloor (1-\delta) \cdot |s_6| \rfloor + 1 = 4$ 个片段并插入到倒排索引 $\mathcal{L}_{|s_6|}$ 中。接下来，对于 $s_5$，用多段匹配敏感的选取方法针对倒排索引 $\mathcal{L}_{|s_6|}$ 从 $s_5$ 中选取子字符串，并检查是否存在任何被选取的子字符串与其相应的片段完全匹配。这里有 “chak”，因此元组 $\langle s_6, s_5 \rangle$ 是一个候选元组。然后验证这对元组，它不是一个结果。接下来，

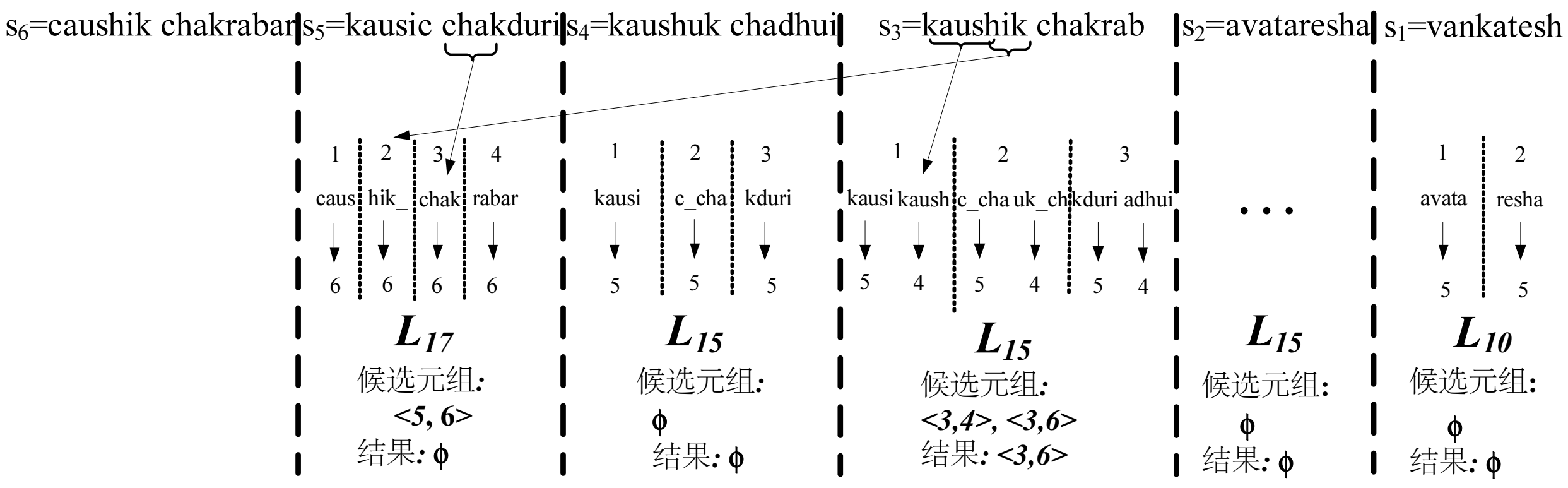

图 3.12　PassJoin-NED 算法的一个例子

---

**Algorithm 3.6**: PassJoin-RSJoin $(\mathcal{R}, \mathcal{S}, \tau)$

---

**Input**: $\mathcal{R}$: 一个字符串集合；$\mathcal{S}$: 另一个字符串集合；$\tau$: 编辑距离阈值
**Output**: $\mathcal{A} = \{(r \in \mathcal{R}, s \in \mathcal{S}) \mid \text{ED}(r, s) \le \tau\}$

**begin**
　把 $\mathcal{R}$ 和 $\mathcal{S}$ 中的字符串按长度升序排序;
　**for** $r \in \mathcal{R}$ **do**
　　划分 $r$ 并把它的片段插入到 $\mathcal{L}^i_{|r|}$ 中;
　**for** $s \in \mathcal{S}$ **do**
　　**for** $\mathcal{L}^i_l$ $(|s| - \tau \le l \le |s| + \tau, 1 \le i \le \tau + 1)$ **do**
　　　$\mathcal{W}(s, \mathcal{L}^i_l)$ = SubstringSelection$(s, \mathcal{L}^i_l)$;
　　　**for** $w \in \mathcal{W}(s, \mathcal{L}^i_l)$ **do**
　　　　**if** $w$ 在 $\mathcal{L}^i_l$ 中 **then** Verification$(s, \mathcal{L}^i_l(w), \tau)$;
**end**

---

图 3.13　PassJoin-RSJoin 算法

把 $s_5$ 划分为 $\lfloor (1-\delta) \cdot |s_5| \rfloor + 1 = 3$ 个片段。类似的，可以重复以上步骤从而找到另外两个候选元组 $\langle s_3, s_4 \rangle$ 和 $\langle s_3, s_6 \rangle$。验证它们后可以得到一个结果 $\langle s_3, s_6 \rangle$。

### 3.6.2　支持异连接

为了支持异连接，即在两个不同的集合 $\mathcal{R}$ 和 $\mathcal{S}$ 上的近似连接，需要首先对两个集合中的字符串分别排序。然后对其中一个集合中的字符串建立索引，例如 $\mathcal{R}$。接下来依次访问另一个数据集 $\mathcal{S}$ 中的所有字符串。对其中每个长度为 $|s|$ 的字符串 $s$，使用 $\mathcal{R}$ 中字符串长度在 $[|s| - \tau, |s| + \tau]$ 内的倒排索引来寻找近似元组，可以移除长度小于 $|s| - \tau$ 的字符串的倒排索引。最后验证候选字符串。注意第3.4节，对两个字符串 $r$ 和 $s$，只考虑了 $|r| \ge |s|$、把 $r$ 划分为片段并从 $s$ 中选择子字符串的

情况。实际上在 $|r| < |s|$ 时，定理 3.2 依然成立、

图 3.13展示了 PassJoin-RSJoin 的伪代码。它首先对两个集合中的字符串排序（第 2 行），然后为 $\mathcal{R}$ 中的字符串构建索引 (第 3-4 行)。接下来它依次访问 $\mathcal{S}$ 中的字符串。对每个字符串 $s$，它通过调用 SubstringSelection 算法来从 $s$ 中选择子字符串（第 7 行）并从索引中寻找候选元组。最终，它调用 Verification 算法来验证候选元组。(第 9 行)。这里子程序 SubstringSelection 与 Verification 与图4.1中的子程序是一样的。

### 3.6.3 MapReduce 下的基于划分的方法

本节提出一个 MapReduce 框架下的字符串近似连接算法 MassJoin，它是基于 PassJoin 算法的。根据基于划分的算法 PassJoin，如果两个字符串 $r$ 和 $s$ 相似，那么 $s$ 一定包含一个子字符串与 $r$ 的一个片段完全匹配。根据这个性质，可以设计一个 MapReduce 框架下的近似连接算法。它包括一个过滤步骤和一个验证步骤。下面考虑异连接的情况，即 $\mathcal{R} \neq \mathcal{S}$。图 3.14 展示了 MassJoin 的伪代码。

**过滤**： 对于任意两个字符串 $r$ 和 $s$，如果把 $r$ 分为 $\tau + 1$ 个片段并从 $s$ 中选取子字符串，那么仅当 $r$ 存在一个片段 $r_m$ 与 $s$ 的一个子字符串 $s_m$ 匹配时，它们才可能相似，其中如果 $r_m$ 是第 $i$ 份片段，那么 $s_m$ 必须属于针对倒排索引 $\mathcal{L}_l^i$ 选取的子字符串集合 $\mathcal{W}_m(s, \mathcal{L}_l^i)$（注意 $|r| = l$）。也就是说，它们不仅片段与子字符串必须完全匹配，它们对应的次序信息 $i$ 和长度信息 $|r|$ 以及 $l$ 也必须完全相同。另一方面，在 MapReduce 框架下，map 函数发送的键值对 (key-value pair) 经过 shuffle 后，拥有相同键的键值对会被归集到同一个 reduce 函数中。因此，对每个字符串，可以使用它的片段或子字符串、片段或者子字符串的次序信息和长度信息的组合作为键，这样在 reduce 阶段可以产生所有候选元组并得到最后结果。具体来说，在 map 阶段，对于每个 $r \in \mathcal{R}$，发送所有键值对 $\langle\langle r_i, i, |r|\rangle, r\rangle$，其中 $1 \le i \le \tau + 1$。对于 $s \in \mathcal{S}$，发送所有键值对 $\langle\langle s_m, i, l\rangle, s\rangle$，其中 $s_m \in \mathcal{W}_m(s, \mathcal{L}_l^i)$，$1 \le i \le \tau + 1$，$|s| - \tau \le l \le |s| + \tau$。因为两个相似的字符串一定共享一个相同的键，所以它们一定会被分配到同一个 reduce 中。注意为了减少传输代价，这里用字符串 ID 来替代字符串本身（因为 ID 比字符串本身小很多）。 reduce 函数的输入是一个键值对列表 $\langle sig, list(sid/rid)\rangle$，其中 $sig$ 是一个特征（一个片段或者一个子字符串以及对应的次序信息和长度信息），$list(sid/rid)$ 是包含该特征的字符串列表。reduce 函数首先把这个列表分为两组：$list(sid)$ 和 $list(rid)$（第 12 行）。注意，这两个子列表的笛卡尔积中的任何元组 $\langle sid, rid\rangle$ 都是一个候选元组，要存储到文件系统中作为验证步骤的输入。为了减少传输代价，

**Algorithm 3.7**: MassJoin

```
   // 过滤步骤
1  Map(⟨rid, r⟩/⟨sid, s⟩)
2      for r ∈ R do
3          for 1 ≤ i ≤ τ + 1 do
4              emit(⟨⟨r_i, i, |r|⟩, rid⟩);
5      for s ∈ S do
6          for |s| − τ ≤ l ≤ |s| + τ do
7              for 1 ≤ i ≤ τ + 1 do
8                  for S_m ∈ W_m(s, L_l^i) do
9                      emit(⟨⟨s_m, i, l⟩⟩, sid⟩);
10
11 Reduce (⟨sig, list(id)⟩)
12     把 list(id) 分离为两组 list(sid) 和 list(rid) ;
13     foreach sid ∈ list(sid) do
14         output(⟨sid, list(rid)⟩);
15
   // 验证步骤第一回合
16 Map(⟨sid, list(rid)⟩/⟨sid, s⟩)
17     emit(⟨sid, list(rid)⟩); emit(⟨sid, s⟩) ;
18
19 Reduce (⟨sid, list(list(rid)/s)⟩)
20     分离 s 和 list(list(rid)) ;
21     从 list(list(rid)) 中移除重复的 rid 得到无重复的 list(rid) ;
22     输出 ⟨s, list(rid)⟩ ;
23
   // 验证步骤第二回合
24 Map(⟨s, list(rid)⟩/⟨rid, r⟩)
25     emit(⟨rid, r⟩) ;
26     foreach rid ∈ list(rid) do
27         emit(⟨rid, s⟩) ;
28
29 Reduce (⟨rid, list(s/r)⟩)
30     从 list(s/r) 中分离出 r 和 list(s) ;
31     foreach s ∈ list(s) do
32         if ED(r, s) ≤ τ then 输出 (⟨⟨r, s⟩, ED(r, s)⟩);
33
```

图 3.14　MassJoin 算法

reduce 对每个 $sid \in list(sid)$ 生成这样的键值对 $\langle sid, list(rid)\rangle$ 并存储到 HDFS 中 (第 13 到 14 行)。map 函数和 reduce 函数的输入和输出如下。

**map:** $\langle sid/rid, string\rangle \rightarrow \langle signature, sid/rid\rangle$

**reduce:** $\langle signature, list(sid/rid)\rangle \rightarrow \langle sid, list(rid)\rangle$

**验证**：因为两个字符串可能共享多个特征，因此过滤阶段可能产生重复的候选元组。MassJoin 使用两个 MapReduce 回合来移除重复的候选元组并把字符串 ID 替换回字符串本身以验证候选元组。

在第一回合，map 函数将上一轮的输出 $\langle sid, list(rid)\rangle$ 以及数据集 $\mathcal{S}$ 作为输入，然后发送两种键值对（第 17 行）。第一种是从数据集 $\mathcal{S}$ 中生成的 $\langle sid, s\rangle$，它把 $sid$ 替换回字符串本身 $s$。第二种是从 $\langle sid, list(rid)\rangle$ 中生成的 $\langle sid, rid\rangle$。reduce 函数归集共享相同的键 $sid$ 的键值，其输入是 $\langle sid, list(list(rid)/s)\rangle$。它首先分离出字符串 $s$，然后移除 $list(rid)$ 中重复的 ID，最后输出键值对 $\langle s, list(rid)\rangle$ 存储在文件系统中 (第 20-22行)。map 函数和 reduce 函数的输入和输出如下。

**map:** $\langle sid, list(rid)\rangle \rightarrow \langle sid, list(rid)\rangle; \langle sid, s\rangle \rightarrow \langle sid, s\rangle$

**reduce:** $\langle sid, list(list(rid)/s)\rangle \rightarrow \langle s, list(rid)\rangle$

在第二回合，map 函数将另一个数据集 $\mathcal{R}$ 和上一轮的输出 $\langle s, list(rid)\rangle$ 作为输入并发射两种类型的键值对。第一种是从数据集 $\mathcal{R}$ 中产生的 $\langle rid, r\rangle$，它把 $rid$ 替换为字符串本身 $r$（第25 行）。第二种是从 $\langle s, list(rid)\rangle$ 中产生的，它对每个 $list(rid)$ 中的 $rid$ 生成一个键值对 $\langle rid, s\rangle$（第 27 行）。每个 reduce 函数将得到包含一个字符串 $r$ 与一列字符串 $list(s)$ 的列表，其中每个元组 $\langle r, s\rangle$ 都是一个候选元组。因此，它首先分离出字符串 $r$ 和列表 $list(s)$，然后验证候选元组，最后输出结果到文件系统中（第 30-32 行）。map 函数和 reduce 函数的输入和输出如下。

**map:** $\langle s, list(rid)\rangle \rightarrow \langle rid, s\rangle, \langle rid, r\rangle \rightarrow \langle rid, r\rangle;$

**reduce:**$\langle rid, list(r/s)\rangle \rightarrow \langle (r, s), \mathsf{Sim}(r, s)\rangle$

**合并键值对**：以上过滤步骤的 map 函数对字符串 $s \in \mathcal{S}$ 的每个被选取的子字符串 $s_m \in \mathcal{W}_m(s, \mathcal{L}_l^i)$（其中 $1 \le i \le \tau + 1$，$|s| - \tau \le l \le |s| + \tau$）都要发送一个键值对，根据引理 3.2，一共需要产生 $O(\tau^3)$ 个键值对，数量较多。为了减少发送的键值对数量，可以合并一些键值对。接下来讨论如何合并这些键值对。

显然，如果两个键匹配，如 $\langle r_i, i, |r|\rangle = \langle s_m, i, l\rangle$，那么其中片段 $r_i$ 和子字符串 $s_m$ 必须匹配，即 $r_i = s_m$。所以可以丢弃次序信息和长度信息，只用片段和子字符串的内容作为键。很容易发现，$s$ 针对不同倒排索引 $\mathcal{L}_l^i$ 和 $\mathcal{L}_{l'}^i$ 选取的子字符串有很多都是相同的，因此可以合并一些键值对从而减少其数量。显然，如果 $r$ 和 $s$ 相似，它们一定共享一个相同的键，所以这个方法不会漏掉结果。

然而，这个方法可能会降低过滤能力，这是因为完全匹配的片段和子字符串可能有不同的次序或者来自长度差别较大的两个字符串，这将导致产生更多的候选元组。为了达到同样的剪枝能力，可以把之前提出的过滤技术推迟到 reduce 阶段来进行剪枝。具体来说，在 reduce 阶段会检查其中的字符串 $r$ 和 $s$ 是否满足长度过滤 $|s| + \tau \le |r| \le |s| - \tau$，以及其子字符串和片段的位置是否满足多段匹配敏感技术给定的范围 $[\bot_i, \top_i]$。这样可以达到相同的过滤能力。

表 3.2　实验数据集

| 数据集 | 数据集大小 | 平均长度 | 最大长度 | 最小长度 |
|---|---|---|---|---|
| Author | 612781 | 14.826 | 46 | 6 |
| Query Log | 464189 | 44.75 | 522 | 30 |
| Author+Title | 863073 | 105.82 | 886 | 21 |

**轻量级的过滤**：最后，可以用一些轻量级的特征来进一步加强剪枝。例如，在 map 函数中可以对每个字符串将其字符分组，并将每个组中字符数量放入键值对中发送。对 reduce 函数中的每个字符串元组，通过计算它们各组中字符数量的差别可以推导出它们编辑距离的一个下界，如果该下界超过了阈值，那么就可以在 reduce 中过滤掉这对字符串。

## 3.7　实验

下面在三个真实数据集上做实验测试本章提出的技术。这三个数据集分别是 DBLP Author①, DBLP Author+Title, 以及 AOL Query Log②。DBLP Author 包含很多短字符串；DBLP Author+Title 包含很多长字符串；Query Log 是一个查询日志集合。表 3.2 给出了各个数据集的详细信息。注意 Author+Title 数据集与 ED-Join[30] 实验中使用的数据集是一样的，Author 数据集和 TrieJoin[31] 实验中使用的是一样的. 图 3.15 展示了不同数据集中字符串的长度分布。

实验将本章提出的算法与现有最优的算法（ED-Join[30] 和 TrieJoin[31]）进行了比较。因为 ED-Join 和 TrieJoin 比其它方法的性能更高，例如 PartEnum[28] 和 All-Pairs-Ed[29] ([30,31] 在实验中也证实了这一点)，所以本节只与这两种最优的方法进行比较。实验中 ED-Join ③和 TrieJoin ④的代码是从相应作者的主页下载的。

实验中所有的算法都是用 C++ 程序语言实现的并使用 GCC 4.2.4 编译器进行编译。编译中都使用了-O3 优化选项。所有的实验都是在一台 Ubuntu 服务器上进行的。这台服务器拥有一个 Intel Core 2 Quad X5450 3.00GHz 处理器和 4 GB 内存。

### 3.7.1　评估子字符串选取方法

本节评估子字符串选取技术。实验中实现了以下四种方法。(1) 基于长度的选取方法, 用 Length 来代表。它为每个片段选取与片段长度相同的所有子字符串；(2) 基于位移的选取方法, 用 Shift 来代表，它利用第3.4节中所讨论的方法，通过在

① http://www.informatik.uni-trier.de/Simley/db

② http://www.gregsadetsky.com/aol-data/

③ http://www.cse.unsw.edu.au/Simweiw/project/simjoin.html

④ http://dbgroup.cs.tsinghua.edu.cn/wangjn/

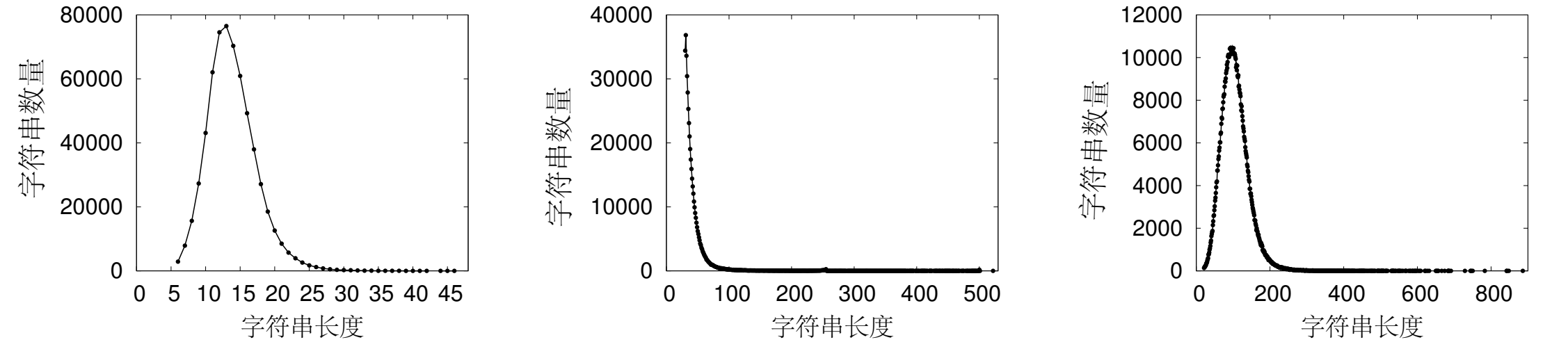

(a) Author(平均长度: 15)　(b) Query Log(平均长度: 45)　(c) Author+Title(平均长度: 105)

图 3.15　字符串长度分布

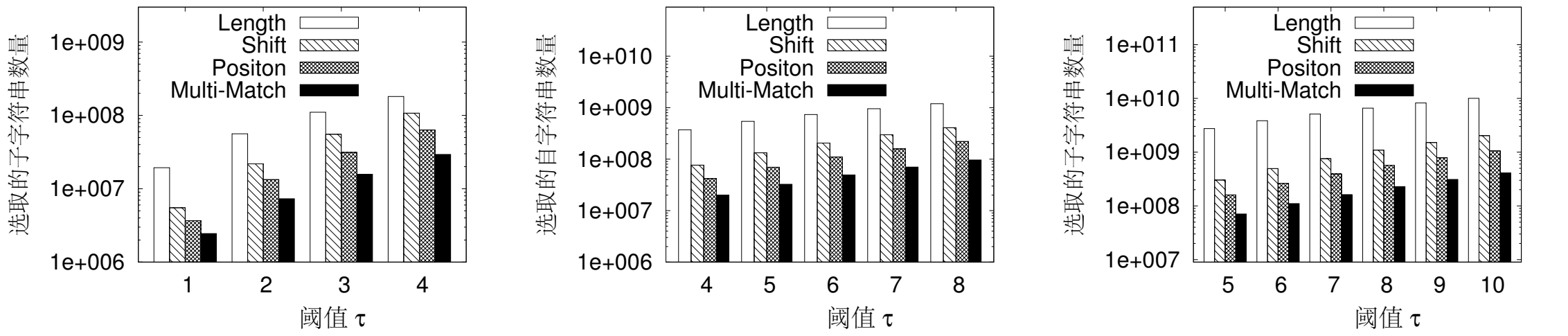

(a) Author(平均长度: 15)　(b) Query Log(平均长度: 45)　(c) Author+Title(平均长度: 105)

图 3.16　选取的子字符串数量

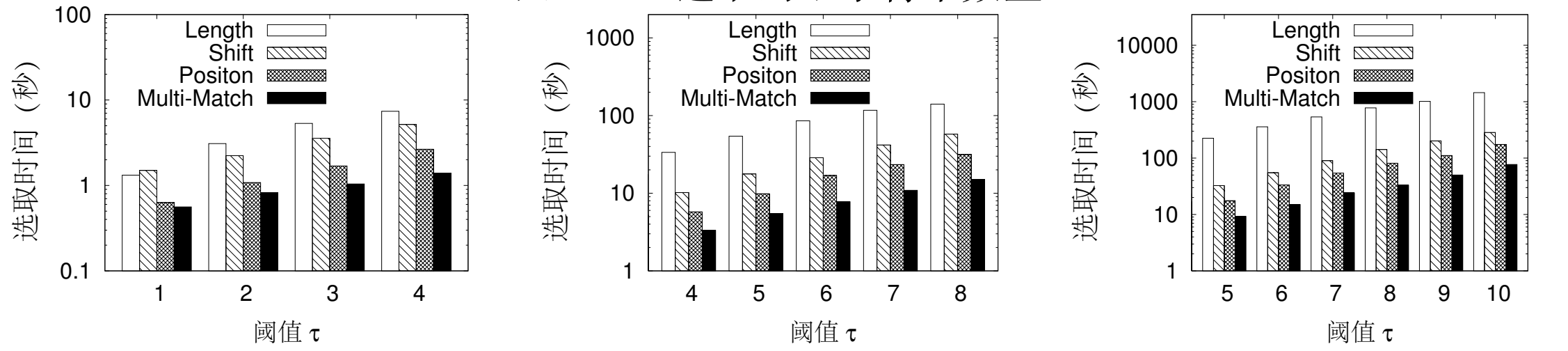

(a) Author(平均长度: 15)　(b) Query Log(平均长度: 45)　(c) Author+Title(平均长度: 105)

图 3.17　产生子字符串耗费的时间

$[-\tau, \tau]$ 的范围内挪动位置来选取子字符串。(3) 本章提出的位置敏感的选取方法，用 Position 代表. (4) 本章提出的多段匹配敏感的选取选取方法，用 Multi-match 代表。接下来首先比较它们所选取的子字符串数量。图 3.16 展示了实验结果。

可以看到 Length 选取了大量的子字符串。Position 所选取的子字符串是 Length 所选取的十分之一到四分之一，是 Shift 所选取的一半。Multi-match 进一步减少选取子字符串的数量到 Position。例如，在 Author 数据集下，当 $\tau = 1$ 时，Length 选取了约一千九百万个子字符串，Shift 选取了约五百五十万个子字符串，Position 将数目减少到三百七十万，Multi-match 进一步减少到两百四十万。根据第 3.4节的分析，对于长度为 $l$ 的字符串，基于长度的方法选取 $(\tau + 1)(|s| + 1) - l$ 个子字符串，基于位移的方法选取 $(\tau + 1)(2\tau + 1)$ 个子字符串, 位置敏感的方法选取 $(\tau + 1)^2$ 个子字符串，多段匹配敏感的方法选取 $\lfloor\frac{\tau^2-\Delta^2}{2}\rfloor + \tau + 1$ 个子字符串。如果 $|s| = l = 15$ 并且 $\tau = 1$，这四种方法所选取的子字符串数目分别是 17，6，4 和 2。显然，实验结果与理论分析是一致的。

下面比较不同的方法产生子字符串所耗费的时间，图3.17展示了这一实验结

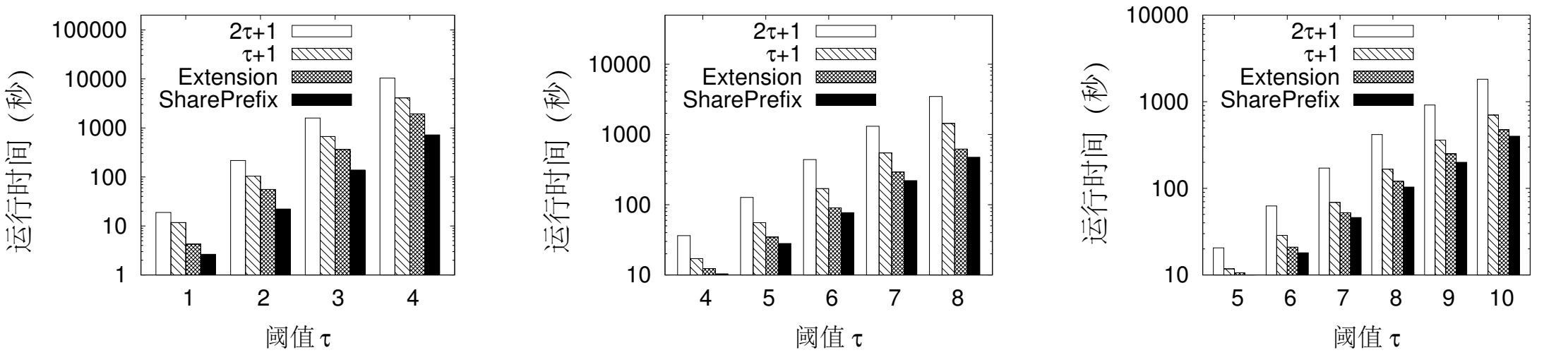

(a) Author(平均长度: 15)　(b) Query Log(平均长度: 45)　(c) Author+Title(平均长度: 105)

图 3.18　验证候选元组耗费的时间

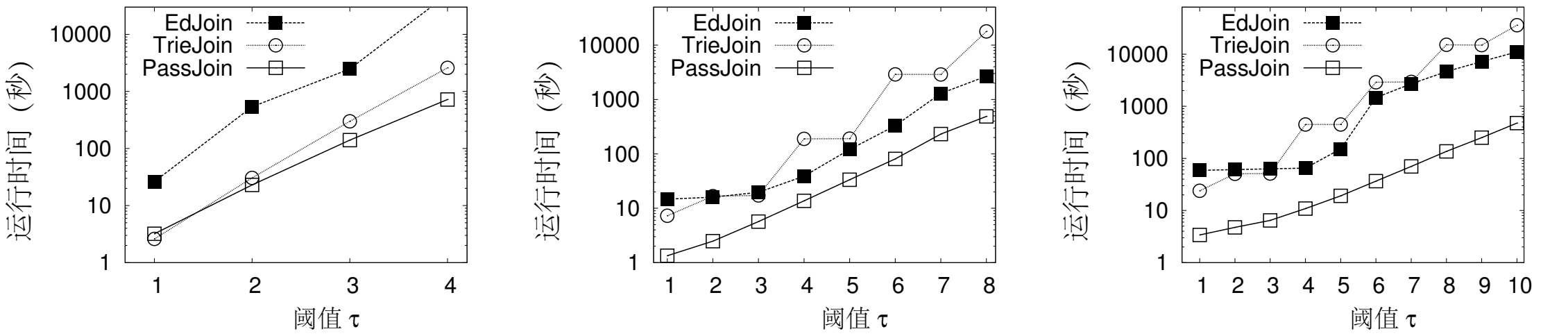

(a) Author(平均长度: 15)　(b) Query Log(平均长度: 45)　(c) Author+Title(平均长度: 105)

图 3.19　与现有最好算法的比较

果。可以看到 Multi-match 的运行时间要少于 Position，Position 反过来比 Shift 和 Length 的性能更高。这是因为选取子字符串所耗费的时间依赖于选取子字符串的数目，而 Multi-match 选取的子字符串数目是最少的。

### 3.7.2　评估验证方法

本节评估了本章提出的验证技术。实验中实现了以下四个方法。(1) 简单算法 naive，图例中用 $2\tau+1$ 表示，对于二维数组中的每一行它计算了 $2\tau+1$ 个值并利用了简单的提早终止技术（当二维数组的一行中所有的值都大于 $\tau$ 时，终止验证）。(2) 长度敏感的验证方法 length-aware，图例中用 $\tau+1$ 表示，对于二维数组中的每一行它计算了 $\tau+1$ 个值并利用了期望编辑距离技术来提早终止验证 (3) 基于扩展的验证方法 extension-based，图例中用 Extension 表示, 它使用基于扩展的框架。对于每行它也计算 $\tau+1$ 个值并使用期望编辑距离技术来提早终止验证。(4) 使用基于扩展的方法并在共同前缀中共享计算，用 SharePrefix 表示。图 3.18展示了这一实验结果。

可以看到 naive 的性能最低，因为它需要计算二维数组中很多不必要的值。length-aware 验证方法比 naive 快 2-5 倍。这是因为 length-aware 可以把验证复杂度从 $2\tau+1$ 降低到 $\tau+1$ 并使用期望编辑距离技术来提早终止验证。extension-based 取得了更好的性能并且比 length-aware 快 2-4 倍。其原因是 extension-based 可以避免共同片段上的重复计算并且它使用了一个更紧缩的阈值来验证左边部分和右边部分。SharePrefix 取得了最好的性能，因为它可以在共享前缀的字符串中避免很

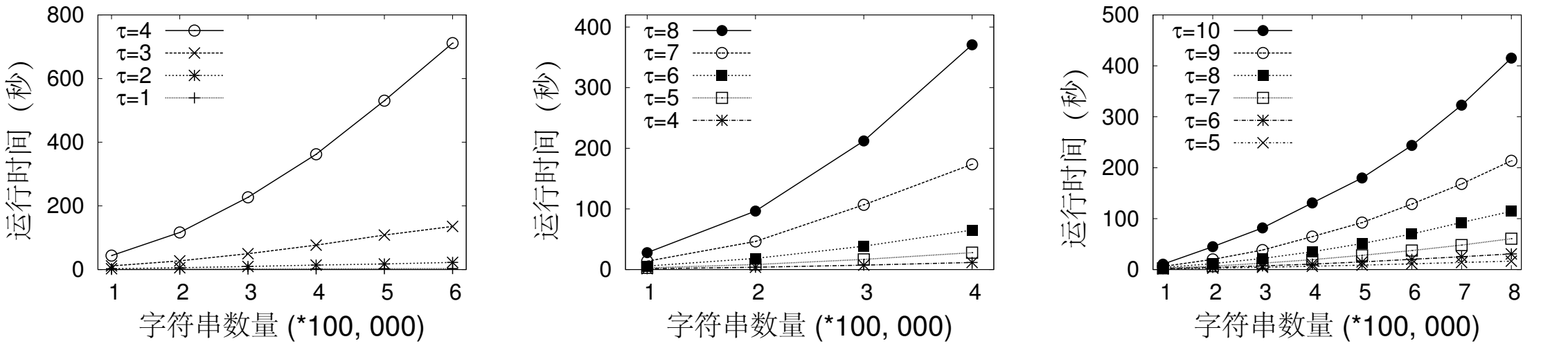

(a) Author(平均长度: 15)　(b) Query Log(平均长度: 45)　(c) Author+Title(平均长度: 105)

图 3.20　编辑距离下的可扩展性

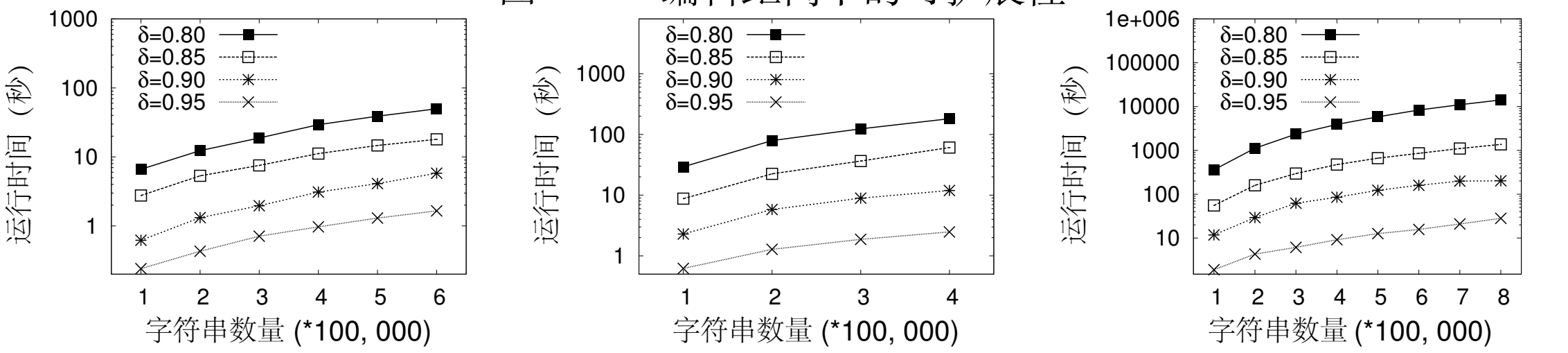

(a) DBLP Author(平均长度：15)　(b) Query Log(平均长度：45)　(c) DBLP Author+Title(平均长度：105)

图 3.21　Edit Similarity 下的可扩展性

多不必要的计算。例如，在 Author 数据集下，当 $\tau = 4$ 时，naive 耗费了约 10000 秒，length-aware 降低到约 4000 秒，extension-based 将其减少到约 2000 秒，SharePrefix 进一步改进到约 700 秒。在 QueryLog 数据集下，当 $\tau = 8$ 时，这四种验证方法所耗费的时间分别是 3500 秒，1500 秒，600 秒和 450 秒。

### 3.7.3　与现有最好方法的比较

本节比较了本章提出的算法和现有最好的算法 ED-Join[30] 和 TrieJoin[31]。因为 TrieJoin 有多种变种，实验中报告了其最好的结果。对于 ED-Join，实验调整其参数 $q$ 并报告它最好的结果。因为 TrieJoin 对于短字符串比较有效，本实验从 TrieJoin 的主页上下载了相同的数据集（即包含短字符串的 Author 数据集）并在这个数据集上与 TrieJoin 作比较。因为 ED-Join 对于长字符串比较有效，实验从 ED-Join 的主页上下载相同的数据集 (即包含长字符串的 Author+Title 数据集) 并在这个数据集上与 ED-Join 作比较。图3.19 展示了这一实验结果，其中消耗的时间包括索引时间和连接时间。

在包含短字符串的 Author 数据集上，TrieJoin 的性能超过了 ED-Join，但是 PassJoin 的性能要远远高于它们两个，特别是当 $\tau \geq 2$ 时。主要的原因如下，对于一个较大的阈值，ED-Join 一定要使用一个较小的 $q$ 来确保正确性。这样的话，因为一个较小的 $q$ 的过滤能力很低[30]，所以 ED-Join 将产生大量的候选元组。TrieJoin 使用了一个 trie 结构来利用前缀过滤寻找近似元组。如果只有一小部分字符串共享了前缀，那么 TrieJoin 的过滤能力将很低并且遍历 trie 结构的代价很高。而 PassJoin

表 3.3　索引大小（MB）

| 实验数据集 | 物理大小 | ED-Join ($q = 4$) | TrieJoin | PassJoin ($\tau = 4$) |
|---|---|---|---|---|
| Author | 8.7 | 25.34 | 16.32 | 1.92 |
| Query Log | 20 | 72.17 | 69.65 | 4.96 |
| Author+Title | 88 | 335.24 | 90.17 | 2.1 |

利用片段来过滤大量不近似的元组，片段的选择横跨整个字符串而不限制于前缀过滤。例如，当 $\tau = 4$ 时，TrieJoin 耗费了约 2500 秒. PassJoin 提高耗时到约 700 秒。ED-Join 非常慢，甚至耗费了多于 10,000 秒的时间。

在包含长字符串的 Author+Title 数据集下, PassJoin 的性能要显著优于 ED-Join 和 TrieJoin，甚至能提高 2-3 个数量级的性能。这是因为 TrieJoin 需要遍历 trie 结构，特别是对于大的阈值。ED-Join 使用一个内容失配的过滤技术来剪枝，它的过滤代价很高，而本章提出的过滤算法很高效。此外，本章提出的验证算法也比现有的方法更加快速有效。例如，当 $\tau = 8$ 时，TrieJoin 耗费了 15,000 秒，ED-Join 减少到 5000 秒，而 PassJoin 进一步改进到 130 秒。

下面比较了几个方法在不同数据集下的索引大小，其结果如表 3.3所示。可以看到，现有的方法比 PassJoin 需要更大的索引空间。例如，在 Author+Title 数据集下，ED-Join 在 $q = 4$ 时的索引大小为 335 MB，TrieJoin 占用了 90 MB 的空间，而 PassJoin 的索引大小只有 2.1 MB。这主要有两个原因。首先，对于长度为 $l$ 的每个字符串，ED-Join 会生成 $l - q + 1$ 个$q$-gram，其中 $q$ 是 gram 的长度。而 PassJoin 只需要生成 $\tau + 1$ 个片段。第二，对于长度为 $l$ 的字符串，PassJoin 只需要为长度在 $l - \tau$ 到 $l$ 范围内的字符串维护索引，而 ED-Join 需要为所有字符串维护索引，TrieJoin 需要使用一个 trie 结构来维护字符串，它需要额外的空间来存储指针和字符索引等。

### 3.7.4　可扩展性

**评估编辑距离下的可扩展性**：下面测试了 PassJoin 的可扩展性，实验调整数据集中字符串的数目并报告 PassJoin 算法的运行时间, 图 3.20展示了这一实验结果。可以看到 PassJoin 取得了接近线性的可扩展性。例如，对于 $\tau = 4$，在 Author 数据集下，当字符串数量分别为四十万，五十万和六十万时，PassJoin 的运行时间分别是 360 秒, 530 秒和 700 秒。这得益于论文提出的有效的过滤技术。

**评估编辑距离下的可扩展性**：图 3.21展示了 PassJoin-NED 算法的可扩展性试验结果。可以看到 PassJoin-NED 算法的可扩展性也非常好，随着数据集大小的增加，其运行时间几乎是呈线性增长。

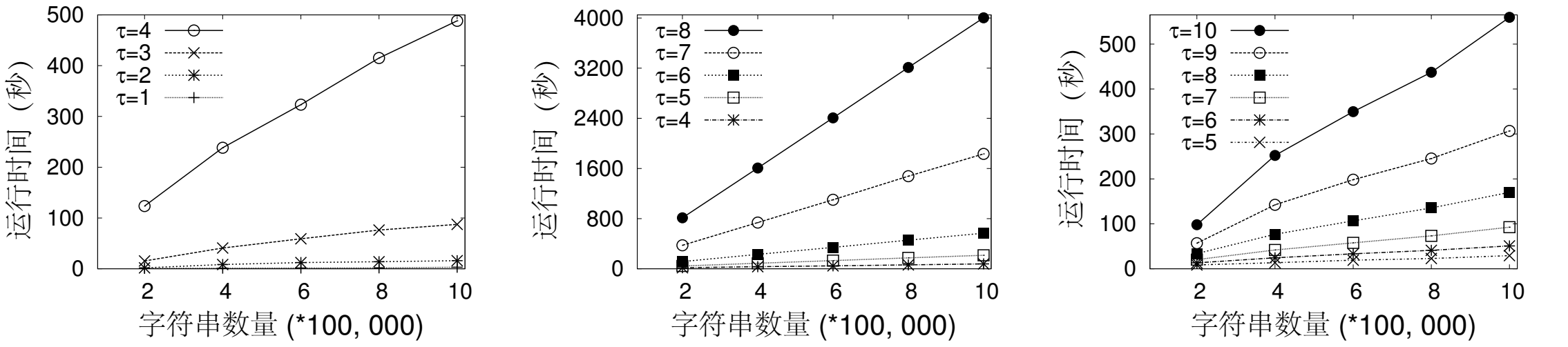

(a) CITESEERX Author(平均长度：20)　(b) Query Log 2 (平均长度：40)　(c) CITESEERX Author+Title(平均长度：107)

图 3.22　异连接的可扩展性

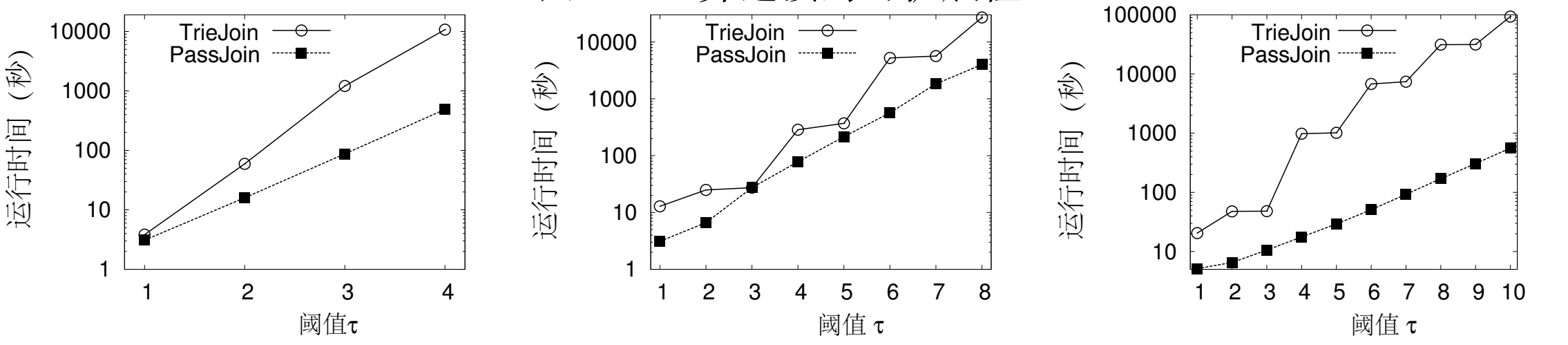

(a) CITESEERX Author(平均长度：20)　(b) Query Log 2 (平均长度：40)　(c) CITESEERX Author+Title(平均长度：107)

图 3.23　与现有最好的异连接算法的比较

**评估异连接下的可扩展性**：接下来测试异连接算法 PassJoin-RSJoin 的可扩展性。为了测试异连接算法的性能，实验中使用了另外三个数据集 CITESEERX Author ①，CITESEERX Author+Title 和 AOL Query Log 2. AOL Query Log 2 是另一个查询日志数据集，它与 AOL Query Log 数据集是不同的。实验中调整 CITESEERX Author, Query Log 2 和 CITESEERX Author+Title 数据集中字符串的数量，每次增加 200,000 条字符串，依次将它们与 DBLP Author, QueryLog 以及 DBLP Author+Title 做连接。图3.22展示了该实验结果。可以看到 PassJoin-RSJoin 在异连接的情况下可扩展性依然良好。实验还将 PassJoin-RSJoin 与 TrieJoin 做了比较。因为 ED-Join 和 Qchunk 主要关注自连接的情况，所以这里不与它们作比较。图3.23展示了该试验结果。可以看到，PassJoin-RSJoin 比 TrieJoin 的性能高。这得益与本章提出的有效的过滤技术和快速的验证技术。

### 3.7.5　评估在 MapReduce 下的性能

本节测试 MassJoin 算法在 MapReduce 框架下的性能。

**实验设置**：实验采用了两个数据集，一个是医疗文献名数据集 PubMed paper title②，另一个是 DNA 序列数据集 NCBI DNA sequence③。表3.4给出了它们的详细信息。实验程序都在一个包含 10 个计算节点的 Hadoop 系统中运行，每个节点是一台 Dell

① http://asterix.ics.uci.edu/fuzzyjoin/

② http://www.ncbi.nlm.nih.gov/pubmed

③ http://www.ncbi.nlm.nih.gov/guide/dna-rna/

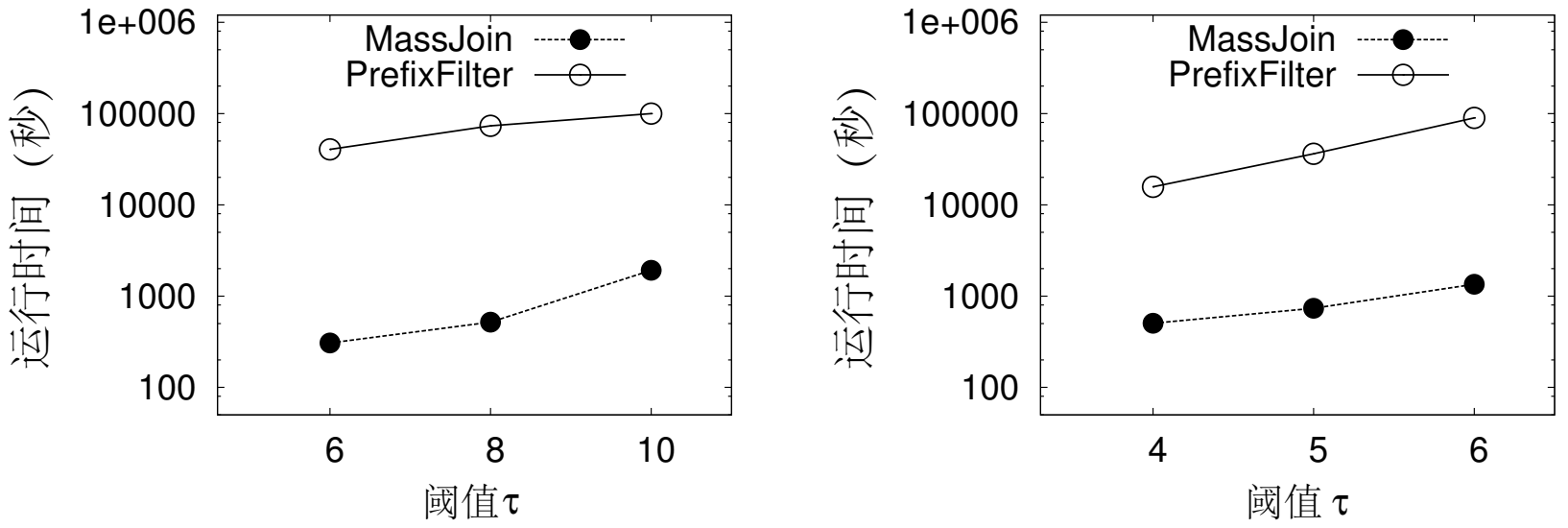

(a) PubMed Title　　(b) DNA Sequence

图 3.24　与现有方法的比较

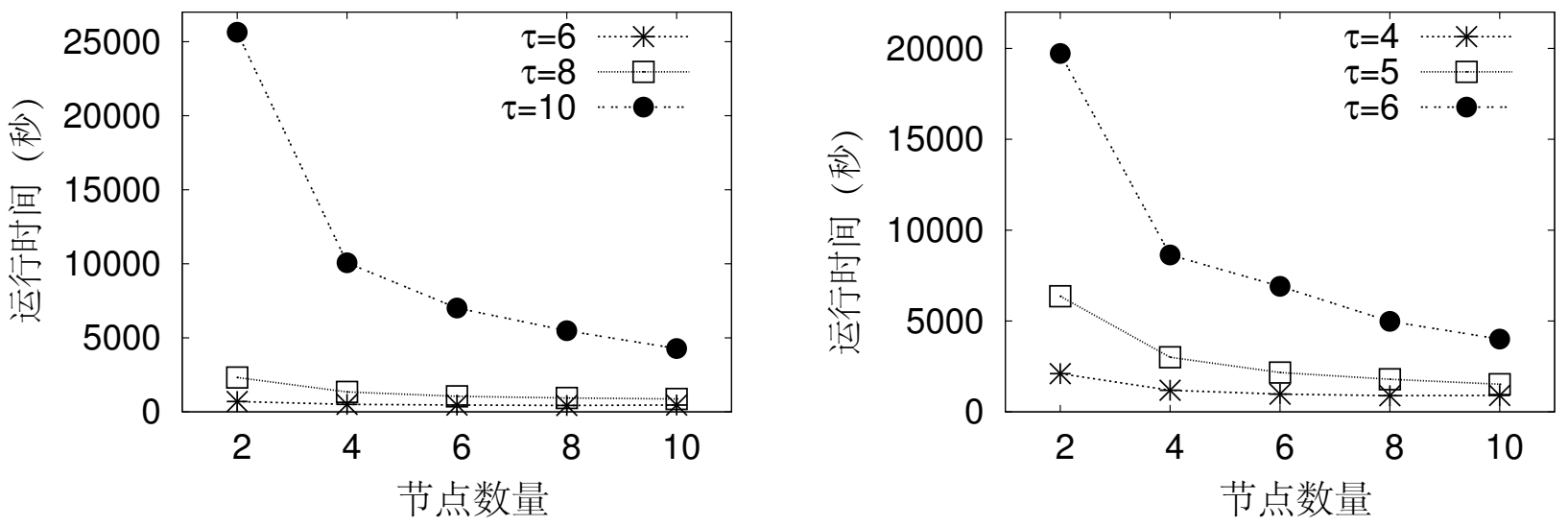

(a) PubMed Title　　(b) DNA Sequence

图 3.25　加速比：调整节点数量

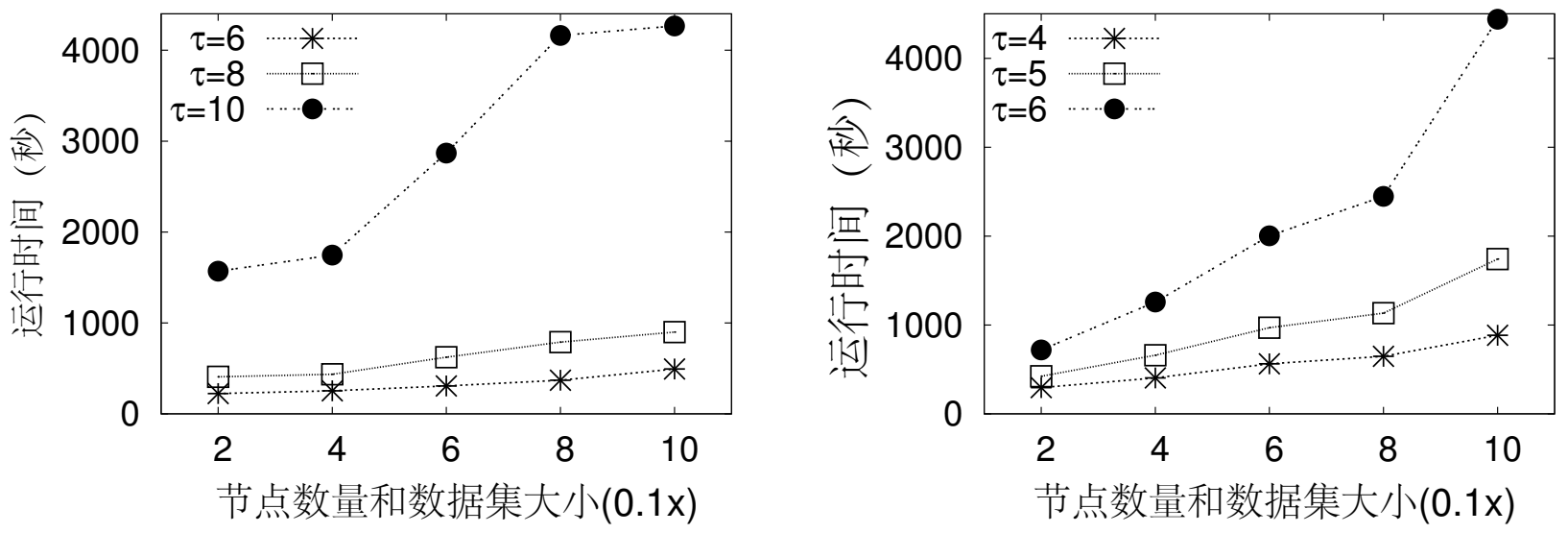

(a) PubMed Title　　(b) DNA Sequence

图 3.26　伸缩性：调整数据集大小和节点数量

表 3.4　MapReduce 实验数据集

| 数据集 | 物理大小 (MB) | 字符串数量 | 平均长度 |
|---|---|---|---|
| PubMed Title | 1494 | 10,394,374 | 144.4 |
| DNA Sequence | 2148 | 18,299,728 | 117.2 |

服务器，它包括 2 个 Intel(R) Xeon(R) E5420 2.5GHZ 处理器，16GB 内存和 1TB 硬盘。服务器运行 64-bit Ubuntu Server 10.04 操作系统，Java 1.6 和 Hadoop 1.0.4 软件。实验设置数据块大小为 16MB，每个任务分配 2GB 虚拟内存。

**与现有方法的比较**：实验首先与现有的方法做了比较，包括 `PrefixFilter`[84]，`VSMARTJoin`[89] 以及 `FuzzyJoin`[90]。对于 `PrefixFilter`，实验利用 ED-Join[30]（生成前缀）和 PassJoin 的技术（验证候选元组）扩展 `PrefixFilter` 以支持编辑距离下的近似连接。论文实现了 `VSMARTJoin` 和 `FuzzyJoin`。然而，因为它们产生大

量的键值对，导致系统报告内存溢出错误，因此本章没有报告它们的结果。因为 PrefixFilter 在本节选定的数据集下的运行时间非常长，因此这里通过随机抽样使用原始数据集 0.6 倍大小的子集做实验。图 3.24展示了实验结果，注意在数据集，PrefixFilter 在 30 个小时内完不成任务。可以看到 MassJoin 比 PrefixFilter 的性能高 1 到 2 个数量级。例如，在 PubMed paper title 数据集下，当阈值 $\tau = 6$ 时，PrefixFilter 的运行时间是 50,000 秒，而 MassJoin 只消耗了 500 秒。主要原因是因为它们的特征选择性不强，会产生大量的候选结果，因此在验证步骤会产生大量的传输代价。

**加速比**：然后评估算法的加速比，实验把节点数量从 2 个调整到 10 个并报告算法运行时间。图 3.25 展示了该实验结果。可以看到，随着集群中节点数量的增加，MassJoin 算法的性能显著的提高。这得益于论文提出的有效的过滤技术，它可以避免枚举所有的元组。

**伸缩性**：最后评估算法的伸缩性，实验同时调整数据集的大小以及集群中节点的数量，并报告运行时间。图 3.26 展示了该实验结果。值得注意的是，随着数据集大小的增加，最终结果的数量是平方次增长的，因此算法运行时间会略微增长。例如，在 PubMed title 数据集下，当阈值为 $\tau = 8$ 时，在 2 个节点的集群和 0.2 倍大小的数据集上的运行时间为 300 秒；而在 10 个节点的集群和 1 倍大小的数据集上的运行时间是 900 秒。

## 3.8 本章小结

本章研究了在编辑距离限制下的序列近似连接问题，提出了一个基于划分的方法来高效解决该问题。论文首先对字符串排序，然后按顺序访问它们。论文为已经访问过的字符串构建倒排索引，对每个字符串，论文选择它的一部分子字符串并利用选取的子字符串和倒排索引来寻找近似元组。论文提出了位置敏感的方法和多段匹配的方法来选取子字符串，论文证明了多段匹配选取方法可以最小化被选取的子字符串数量。论文基于两个字符串的长度之差一定大于它们的编辑距离的观察设计了一个高效的验证算法来验证候选元组。论文提出了基于扩展的方法和在不同的前缀上共享计算的方法来进一步提高验证算法的性能。论文讨论了如何扩展基于划分的方法以支持 Edit Similarity，异连接和 MapReduce 框架。实验表明论文提出的方法在短字符串和长字符串上都比现有的方法性能高。

# 第 4 章　基于集合相似性的近似连接方法

## 4.1　引言

现实世界中还有一大类数据可以用集合的形式来表示，例如向量、文档和图片等。本章研究基于集合相似性的近似连接方法，即给定两组集合数据集，找到其中所有相似的集合对（当这两个集合的相似度超过了一个给定的阈值时，它们才被认为是相似）。本章致力于研究三种被广泛应用的集合相似函数 Jaccard Similarity，Cosine Similarity 和 Dice Similarity。基于集合的相似性连接有很多应用，例如机器学习、推荐系统、数据挖掘等。

与序列近似连接一样，集合近似连接也是一个计算密集 (computing intensive) 的任务。现有方法当中大多采用基于前缀过滤的框架[29,35,39,40]。前缀过滤首先将每个集合中所有的元素按照全局顺序排序，然后将每个集合中最开始的几个元素作为前缀。这种方法可以保证如果两个集合的前缀不相交，那么它们一定不相似。然而，基于前缀过滤框架的剪枝能力很有限。这是因为一旦两个不相似的集合在前缀中共享一个相同的元素，这些方法就没有剪枝效果了。

为了解决这个问题，论文提出了一个新的基于划分的框架来解决集合近似连接的问题。论文设计了一种划分方法来根据全集把所有集合划分成不相交的片段（即子集）并证明仅当两个集合共享相同的片段时，它们才可能相似。论文将集合按照长度分组从而共享计算，为片段构建倒排索引来快速进行集合近似连接。为了减少访问的倒排列表的总长度从而减少候选元组的数量，论文提出混合使用片段与片段的 1-删集（1-deletion neighbourhood，移除片段中一个元素后的子集）。因为存在多种策略来分配这个混合，论文评估了不同的分配策略并设计了一个动态规划算法来选择最优的分配策略。然而为大小为 $s$ 的集合产生最优分配的时间复杂度是 $O(s^3)$，为了加快分配策略的选取，论文提出了一个近似比为 2 的贪心算法。为了进一步降低复杂度，论文设计了一种多长度分组的机制。这两个技术一起将分配选取的时间复杂度降低到 $O(s \log s)$。

接下来按照如下次序组织本章内容。第 4.2 节给出问题定义和相关工作，第 4.3 节介绍基于划分的算法框架，第 4.4 节提出基于划分的算法框架中混合使用片段与片段的 1-删集、评估不同的分配策略并设计动态规划算法来选择最优的分配，第 4.5 节给出了贪心的分配选择算法和多长度分组机制，第 4.6 节讨论如何扩展这些技术来支持 MapReduce 和 Spark，第 4.7 节和第 4.8 节分别报告实验结果和总结本章内容。

## 4.2 预备知识

### 4.2.1 问题定义

给定一个有限的全集 $\mathcal{U} = \{x_1, x_2, \ldots, x_n\}$，一个集合指的是 $\mathcal{U}$ 的一个子集。为了衡量两个集合的相似性，需要一个近似函数 Sim。本章致力于研究三种常用的近似函数 Jaccard Similarity，Cosine Similarity 和 Dice Similarity。对于任何两个集合 $\mathcal{X}$ 和 $\mathcal{Y}$，这三个近似函数分别定义如下：

$$\textsc{jac}(\mathcal{X},\mathcal{Y}) = \frac{|\mathcal{X}\cap\mathcal{Y}|}{|\mathcal{X}\cup\mathcal{Y}|} \quad \textsc{cos}(\mathcal{X},\mathcal{Y}) = \frac{|\mathcal{X}\cap\mathcal{Y}|}{\sqrt{|\mathcal{X}||\mathcal{Y}|}} \quad \textsc{dice}(\mathcal{X},\mathcal{Y}) = \frac{2|\mathcal{X}\cap\mathcal{Y}|}{|\mathcal{X}|+|\mathcal{Y}|}$$

其中 $|\mathcal{X}|$ 表示集合 $\mathcal{X}$ 的大小。根据这些定义，两个集合的相似度是在 $[0,1]$ 之间。例如，考虑表4.1中的集合，因为 $|\mathcal{X}_1 \cap \mathcal{X}_2| = 6$ 以及 $|\mathcal{X}_1 \cup \mathcal{X}_2| = 12$，所以有 $\textsc{jac}(\mathcal{X}_1,\mathcal{X}_2)= \frac{6}{12} = 0.5$。接下来，正式的定义集合近似连接问题。

**定义** 4.1 (**集合近似连接**)：　给定两组集合 $\mathcal{R}$ 和 $\mathcal{S}$，一个相似函数 Sim 和一个阈值 $\delta$，集合近似连接找出 $\mathcal{R}\times\mathcal{S}$ 中所有满足 $\text{Sim}(\mathcal{X},\mathcal{Y}) \geq \delta$ 的近似元组 $\langle\mathcal{X},\mathcal{Y}\rangle$ 。

本章首先研究自连接，即 $\mathcal{R} = \mathcal{S}$ 的情况。接下来提出的技术同样适用于异连接（第 4.6.2 节）。下面首先以 Jaccard Similarity 为例作为近似函数 Sim 来引入论文的方法，然后在第 4.6.1 节中讨论如何支持其它的近似函数。给定一个近似阈值 $\delta$，对于任意两个集合 $\mathcal{X}$ 和 $\mathcal{Y}$，如果 $\textsc{jac}(\mathcal{X},\mathcal{Y}) \geq \delta$，则称这两个集合是相似的。举个例子，考虑表 4.1 中的数据集 $\mathcal{R}$，假设阈值 $\delta = 0.73$，因为 $\textsc{jac}(\mathcal{X}_1,\mathcal{X}_5) = 0.82 \geq \delta$，集合近似连接返回近似元组 $\langle\mathcal{X}_1,\mathcal{X}_5\rangle$，所有其他元组都不相似。

Jaccard Similarity 有一个重要的特性：对于任意两个集合 $\mathcal{X}$ 和 $\mathcal{Y}$，只有当 $\delta|\mathcal{Y}| \leq |\mathcal{X}| \leq \frac{|\mathcal{Y}|}{\delta}$ 时，它们才可能近似，这通常也叫做长度过滤。这是因为 $|\mathcal{X}| \geq |\mathcal{X}\cap\mathcal{Y}|$ 和 $|\mathcal{Y}| \leq |\mathcal{X}\cup\mathcal{Y}|$，所以 $\frac{|\mathcal{X}|}{|\mathcal{Y}|} \geq \frac{|\mathcal{X}\cap\mathcal{Y}|}{|\mathcal{X}\cup\mathcal{Y}|} = \textsc{jac}(\mathcal{X},\mathcal{Y}) \geq \delta$。因此有 $|\mathcal{X}| \geq |\mathcal{Y}|\delta$。类似的，可以推断出 $|\mathcal{X}| \leq |\mathcal{Y}|/\delta$。令 Overlap Similarity 为两个集合的交集的大小，所有以上提到的三个近似函数都可以转化为 Overlap Similarity。更具体的说，因为 $\textsc{jac}(\mathcal{X},\mathcal{Y}) = \frac{|\mathcal{X}\cap\mathcal{Y}|}{|\mathcal{X}\cup\mathcal{Y}|} = \frac{|\mathcal{X}\cap\mathcal{Y}|}{|\mathcal{X}|+|\mathcal{Y}|-|\mathcal{X}\cap\mathcal{Y}|}$，所以，$\textsc{jac}(\mathcal{X},\mathcal{Y}) \geq \delta$ 当且仅当它们的 Overlap Similarity，$\text{Over}(\mathcal{X},\mathcal{Y}) = |\mathcal{X}\cup\mathcal{Y}| \geq \frac{\delta}{1+\delta}(|\mathcal{X}| + |\mathcal{Y}|)$。同样的，可以将另外两个近似函数转化为 Overlap Similarity。

### 4.2.2 基于前缀过滤的方法

为了解决集合近似连接的问题，一种暴力算法需要枚举 $O(|\mathcal{R}|^2)$ 个集合元组并计算它们的相似度，这是非常耗时的。现有的最好的方法通常采用前缀过滤（prefix

表 4.1　一组集合 $\mathcal{R}$

| *id* | | *The records* | *group* |
|---|---|---|---|
| 1 | $\mathcal{X}_1$ | $\{x_1, x_2, x_5, x_6, x_7, x_{10}, x_{11}, x_{13}, x_{14}\}$ | $\mathcal{R}_9$ |
| 2 | $\mathcal{X}_2$ | $\{x_2, x_4, x_5, x_6, x_9, x_{11}, x_{13}, x_{14}, x_{15}\}$ | |
| 3 | $\mathcal{X}_3$ | $\{x_1, x_3, x_6, x_7, x_9, x_{10}, x_{11}, x_{13}, x_{14}\}$ | |
| 4 | $\mathcal{X}_4$ | $\{x_3, x_4, x_5, x_7, x_8, x_{10}, x_{12}, x_{13}, x_{14}\}$ | |
| 5 | $\mathcal{X}_5$ | $\{x_1, x_2, x_3, x_4, x_5, x_6, x_7, x_{10}, x_{11}, x_{13}, x_{14}\}$ | $\mathcal{R}_{11}$ |

filter）框架来剪掉一些不相似的集合元组，然后验证没有被剪掉的候选元组。这种前缀过滤的方法首先按照如下方式把 Jaccard Similarity 阈值 $\delta$ 转化为 Overlap Similarity 阈值 $t$：对于任何两个集合 $\mathcal{X}$ 和 $\mathcal{Y}$，$\text{JAC}(\mathcal{X},\mathcal{Y}) \geq \delta$ 当且仅当 $\text{Over}(\mathcal{X},\mathcal{Y}) \geq \frac{\delta}{1+\delta}(|\mathcal{X}|+|\mathcal{Y}|)$。又因为根据长度过滤有 $|\mathcal{Y}| \geq |\mathcal{X}|\delta$，所以仅当 $\text{Over}(\mathcal{X},\mathcal{Y}) \geq \delta|\mathcal{X}| = t$ 时，$\text{JAC}(\mathcal{X},\mathcal{Y}) \geq \delta$。然后，它们固定一个全局顺序并对每个集合中的元素按照全局顺序排序。对于每个集合 $\mathcal{X}$，它的前缀 $\text{prefix}(\mathcal{X})$ 包含它最开始（最前面）的 $|\mathcal{X}| - \lceil t \rceil + 1$ 个元素。前缀过滤的框架可以保证仅当两个集合 $\mathcal{X}$ 和 $\mathcal{Y}$ 的前缀有相同元素时，即 $\text{prefix}(\mathcal{X}) \cap \text{prefix}(\mathcal{Y}) \neq \phi$，它们才可能相似。接下来，基于前缀过滤的方法为前缀建立一个倒排索引，在同一个倒排列表中的任意两个集合将组成一个候选元组。最后，它们验证候选元组。例如，考虑表 4.1中的数据集 $\mathcal{R}$ 并假设阈值 $\delta = 0.73$。假设全局顺序首先按照元素的频率排序，然后使用下标顺序。那么有 $\text{prefix}(\mathcal{X}_1) = \{x_1, x_2, x_5\}$，$\text{prefix}(\mathcal{X}_2) = \{x_{15}, x_9, x_2\}$，$\text{prefix}(\mathcal{X}_3) = \{x_9, x_1, x_3\}$，$\text{prefix}(\mathcal{X}_4) = \{x_8, x_{12}, x_3\}$ 以及 $\text{prefix}(\mathcal{X}_5) = \{x_1, x_2, x_3\}$。因为 $\text{prefix}(\mathcal{X}_1) \cap \text{prefix}(\mathcal{X}_2) \neq \phi$，所以 $\langle \mathcal{X}_1, \mathcal{X}_2 \rangle$ 是一对候选元组。类似的，总共可以找到有 8 个候选元组，而暴力算法有 10 个候选元组。

### 4.2.3　相关工作

**集合近似连接**：关于集合近似连接有很多已有的工作[28,29,35,39,40,64,78,84,87,91,92]。Jiang 等人[93] 对近似连接问题做了全面、详实的实验分析。Bayardo 等人[29] 提出了以上描述的前缀过滤。Xiao 等人[35] 提出了基于位置的前缀过滤（positional prefix filter）和基于位置的后缀过滤（positional suffix filter）来改进前缀过滤。此外，在转化 Overlap Similarity 阈值时，他们还推导了一个比 $t = \delta|\mathcal{X}|$ 更加紧缩的界：$t = \frac{2\delta}{1+\delta}|\mathcal{X}|$，因此缩短了前缀长度 $|\mathcal{X}| - t + 1$。Wang 等人[39] 提出在前缀中包含更多的元素来加强前缀过滤的过滤能力并为此设计了一个代价模型（cost model）。本章提出的基于划分的方法与前缀过滤框架有根本的不同，因为基于划分的方法使用元素的集合作为特征，而前缀过滤框架使用前缀中单个元素作为特征。

Sarawagi 等人[78] 设计了一个通用的算法来解决不同近似函数下的近似连接问题，包括 Jaccard Similarity 和 编辑距离。Arasu 等人[28] 提出了 PartEnum 方法来解决 DBMS 中的集合近似连接问题，该方法包括两个操作：“partition” 和 “enumeration’’。Xiao 等人[40] 扩展了前缀过滤技术来解决 top-k 集合近似连接问题。Verinica 等人[84] 和 Deng 等人[64] 提出在 MapReduce 框架下解决集合近似连接问题。Wang 等人[87] 为集合近似连接设计了一个混合近似函数。Deng 等人[65] 提出了一个基于划分的框架来解决编辑距离限制下的字符序列近似连接问题。根据[64]，直接扩展该框架来解决集合近似连接导致一个非常高昂的时间复杂度：处理一个大小为 $s$ 的集合时间复杂度是 $O(s^3)$。Li 等人[42] 为集合近似搜索研究了高效的列表合并方法。所有这些相关工作都与本章提出的方法不同：它们要么解决一个不同的问题，例如 top-k 集合近似连接和集合近似检索，要么在一个不同的环境下解决该问题，例如 MapReduce 和 DBMS。基于划分的方法可以很容易的扩展以支持 MapReduce 和 DBMS 下的集合近似连接（第 4.6.3 节）。

**集合近似连接的近似算法**：多个之前的工作[94–97] 致力于研究使用基于概率的技术来解决集合近似连接问题。位置敏感哈希（Locality Sensitive Hashing）[95] 技术，简称 LSH，是其中最流行的一个。一个 LSH scheme 是在一个哈希函数族上的分布，其中的哈希函数作用在集合上并保证两个近似的集合更可能被哈希到相同的桶中。MinHash[97] 就是 Jaccard Similarity 下的一个 LSH scheme。Satuluri 等人[94] 提出了一个贝叶斯算法 BayesLSH 来利用 LSH 技术做候选元组剪枝和相似度估计。Zhai 等人[96] 提出了一个概率算法解决极低阈值下的高维空间近似检索问题。Shrivastava[98] 提出了一个非对称的 LSH 技术来解决最大内积检索问题。所有这些方法和本章提出的方法都是不同的，因为它们找不到全部的近似元组。此外，它们还需要调整参数，这是一个冗长乏味的过程。而且当阈值较小的时候，它们的剪枝能力很低效[29]。

## 4.3 基于划分的框架

下面首先提出一种划分策略，然后讨论如何利用它来解决集合近似连接问题。

### 4.3.1 集合划分

本节提出一种基于划分的框架，它将每个集合划分为几个不相交的片段（子集）并保证两个集合近似仅当它们共享一个相同的片段。为了达到这一目标，需要确定 (1) 片段的总数和 (2) 如何把元素划分到不同的片段中。例如，考虑表 4.1 中的 4 个集合 $\mathcal{X}_1, \mathcal{X}_2$，$\mathcal{X}_3$ 和 $\mathcal{X}_4$ 并假设 $\delta = 0.73$。如图 4.1(b) 所示，把它们切分为 4

个片段，仅当两个集合共享一个片段时，它们才相似。在这里只有 3 个候选元组 $\langle\mathcal{X}_1, \mathcal{X}_3\rangle$，$\langle\mathcal{X}_1, \mathcal{X}_4\rangle$ 和 $\langle\mathcal{X}_3, \mathcal{X}_4\rangle$，而基于前缀过滤的方法将产生 4 个候选元组。

**片段份数**：对于任意两个集合 $\mathcal{X}$ 和 $\mathcal{Y}$，令 $\mathcal{X}\Delta\mathcal{Y}$ 表示它们的对称差集 (symmetric difference)，即并集与交集的差集，它们之间不同元素的个数正好是 $|\mathcal{X}\Delta\mathcal{Y}|$。如果 $\textsc{jac}(\mathcal{X},\mathcal{Y}) \geq \delta$，根据第 4.2.1 节中的讨论，有 $\mathrm{Over}(\mathcal{X},\mathcal{Y}) = |\mathcal{X}\cap\mathcal{Y}| \geq \frac{\delta}{1+\delta}(|\mathcal{X}|+|\mathcal{Y}|)$ 以及 $|\mathcal{Y}| \leq |\mathcal{X}|/\delta$。因此它们之间不同元素的数量满足：

$$
\begin{aligned}
|\mathcal{X}\Delta\mathcal{Y}| &= |\mathcal{X}\setminus\mathcal{Y}| + |\mathcal{Y}\setminus\mathcal{X}| \\
&\leq |\mathcal{X}| - \frac{\delta(|\mathcal{X}|+|\mathcal{Y}|)}{1+\delta} + |\mathcal{Y}| - \frac{\delta(|\mathcal{X}|+|\mathcal{Y}|)}{1+\delta} \\
&= \frac{1-\delta}{1+\delta}(|\mathcal{X}|+|\mathcal{Y}|) \leq \frac{1-\delta}{\delta}|\mathcal{X}|
\end{aligned}
\tag{4-1}
$$

令 $H_l = \lfloor\frac{1-\delta}{\delta}l\rfloor$，其中 $l = |\mathcal{X}|$。如引理 4.1所述，$H_l$ 是 $\mathcal{X}$ 和任何与 $\mathcal{X}$ 相似的集合之间不同元素数目的上界。

**引理** 4.1：　给定一个大小为 $l$ 的集合 $\mathcal{X}$，对于任何满足 $\textsc{jac}(\mathcal{X},\mathcal{Y}) \geq \delta$ 的集合 $\mathcal{Y}$，$|\mathcal{X}\Delta\mathcal{Y}| \leq H_l$ 成立。

**证明** 因为 $|\mathcal{X}\Delta\mathcal{Y}|$ 是一个整数，结合公式 4-1，该引理立即得证。

因为每个不同的元素最多导致一个不相交的片段不匹配，为了保证两个近似的集合共享至少一个相同的片段，对于每个大小为 $l$ 的集合 $\mathcal{X}$，需要把它划分为 $m \geq H_l + 1$ 个不相交的片段：$\mathcal{X}^1\mathcal{X}^2\cdots\mathcal{X}^m$。例如，考虑长度为 $l = 9$ 的集合 $\mathcal{X}$ 并假设 $\delta = 0.73$，把 $\mathcal{X}$ 划分为 4 个片段，那么对于任何与 $\mathcal{X}$ 近似的集合 $\mathcal{Y}$，有 $|\mathcal{X}\Delta\mathcal{Y}| \leq H_9 = 3$。

**划分策略**：如果对于任何集合，一个划分方法都能把相同的元素划分至相同次序的片段中，那么把这个划分称作一个同态划分。如引理 4.2所述，对于任何两个集合 $\mathcal{X}$ 和 $\mathcal{Y}$，同态划分有一个良好的性质：$|\mathcal{X}\Delta\mathcal{Y}| = \sum_{i=1}^{m}|\mathcal{X}^i\Delta\mathcal{Y}^i|$。

**引理** 4.2：　对于任何集合 $\mathcal{X}$ 和 $\mathcal{Y}$，如果使用同态划分把它们划分为 $m$ 个片段，那么有 $|\mathcal{X}\Delta\mathcal{Y}| = \sum_{i=1}^{m}|\mathcal{X}^i\Delta\mathcal{Y}^i|$，其中 $\mathcal{X}^i$ 和 $\mathcal{Y}^i$ 分别是 $\mathcal{X}$ 和 $\mathcal{Y}$ 的第 $i$ 个片段。

**证明** 根据同态划分的定义，(1) 首先，对于任何 $e \in \mathcal{X}^i\Delta\mathcal{Y}^i$，$e \in \mathcal{X}\Delta\mathcal{Y}$；(2) 其次，对于任何 $e' \in \mathcal{X}\Delta\mathcal{Y}$，因为当 $i \neq j$ 时，$\mathcal{X}^i$ 与任何片段 $\mathcal{X}^j$ 和 $\mathcal{Y}^j$ 都不相交，所以有且只有一个 $i \in [1,m]$ 使得 $e' \in \mathcal{X}^i\Delta\mathcal{Y}^i$。因此 $\mathcal{X}\Delta\mathcal{Y}$ 中每个元素都出现且仅出现一个 $\mathcal{X}^i\Delta\mathcal{Y}^i$ 中（$i \in [1,m]$），$|\mathcal{X}\Delta\mathcal{Y}| = \sum_{i=1}^{m}|\mathcal{X}^i\Delta\mathcal{Y}^i|$，引理得证。

因此，对于任何大小分别为 $l = |\mathcal{X}|$ 和 $s = |\mathcal{Y}|$ 的集合 $\mathcal{X}$ 和 $\mathcal{Y}$，可以把它们划分为 $m \geq H_l + 1$ 个片段。如果对每个 $i \in [1, m]$ 都有 $\mathcal{X}^i \neq \mathcal{Y}^i$，那么 $\mathcal{X}$ 和 $\mathcal{Y}$ 不可能近似，如引理 4.3所述。

**引理** 4.3：　对于任何集合 $\mathcal{X}$ 和 $\mathcal{Y}$，假设使用同态划分把它们划分为 $m \geq H_l + 1$ 个片段，如果对于每个 $i \in [1, m]$ 都有 $\mathcal{X}^i \neq \mathcal{Y}^i$，那么 $\mathcal{X}$ 和 $\mathcal{Y}$ 不可能近似。

**证明** 如果 $\mathcal{X}^i \neq \mathcal{Y}^i$，那么 $|\mathcal{X}^i \Delta \mathcal{Y}^i| \geq 1$。结合引理4.2，有 $|\mathcal{X} \Delta \mathcal{Y}| = \sum_{i=1}^{m} |\mathcal{X}^i \Delta \mathcal{Y}^i| \geq m > H_l$，因此，根据引理4.1，$\mathcal{X}$ 和 $\mathcal{Y}$ 不可能相似。引理得证。

为了得到一个同态划分，可以使用基于哈希的方法。对于每个元素 $e \in \mathcal{X}$，可以把它放入第 $((hash(e) \mod m) + 1)$ 份片段中，其中 $hash(e)$ 把一个元素 $e$ 映射为一个整数。因此不同集合中相同的元素都将被划分到相同次序的片段里。下面正式介绍划分方法。

首先从 $\mathcal{R}$ 中可以获得所有集合的全集 $\mathcal{U}$，把它划分为 $m$ 个片段 $\mathcal{U}^1$，$\mathcal{U}^2$，$\ldots$，$\mathcal{U}^m$，这些片段满足 (1) $\cup_{i=1}^{m} \mathcal{U}^i = \mathcal{U}$ 以及 (2) 对于任何 $1 \leq i \neq j \leq m$ 满足 $\mathcal{U}^i \cap \mathcal{U}^j = \phi$。对于任何集合 $\mathcal{X}$，可以利用全集来得到它的 $m$ 个片段 $\mathcal{X}^1$，$\mathcal{X}^2$，$\ldots$，$\mathcal{X}^m$，这些片段满足 (1) $\cup_{i=1}^{m} \mathcal{X}^i = \mathcal{X}$ 以及 (2) 对于任何 $1 \leq i \leq m$ 满足 $\mathcal{X}^i \subseteq \mathcal{U}^i$。用 scheme$(\mathcal{U}, m)$ 来表示这个划分方法。例如，图 4.1(a) 给出了一个 $m = 4$ 的划分方法，图 4.1(b) 展示了根据该划分方式得到的片段。对于任何集合 $\mathcal{X}$ 和 $\mathcal{Y}$，假设根据相同的划分方式 scheme$(\mathcal{U}, m)$ 来划分它们，并得到片段 $\mathcal{X}^1$，$\ldots$，$\mathcal{X}^m$ 和 $\mathcal{Y}^1$，$\ldots$，$\mathcal{Y}^m$。由于这些片段 $\mathcal{X}^i$ 和 $\mathcal{Y}^i$ 都是全集子集 $\mathcal{U}^i$ 的子集并且任何两个全集子集都是不相交的，所以如引理 4.2 所述，有 $|\mathcal{X} \Delta \mathcal{Y}| = \sum_{i=1}^{m} |\mathcal{X}^i \Delta \mathcal{Y}^i|$。

根据引理 4.1, 4.2 以及 4.3，对于任何大小为 $l$ 的集合 $\mathcal{X}$ 以及与其近似的集合 $\mathcal{Y}$，假如按照相同的划分方法 scheme$(\mathcal{U}, m)$ 把它们切分为片段，其中 $m > H_l$，那么它们一定会在同一位置共享一个相同的片段。否则的话，$|\mathcal{X} \Delta \mathcal{Y}| = \sum_{i=1}^{m} |\mathcal{X}^i \Delta \mathcal{Y}^i| \geq m > H_l$，这将导致 $\text{JAC}(\mathcal{X}, \mathcal{Y}) < \delta$。接下来根据这个结论介绍基于划分的算法。

### 4.3.2　基于划分的方法

这个方法包含以下三步。

- **划分**：直观地，一个全集子集越小，两个集合就会有更大的概率在这个位置共享一个相同的片段。有鉴于此，本章使用一个平均划分方法：对于大小为 $l$ 的集合，它平均地把全集 $\mathcal{U}$ 划分为 $m = H_l + 1$ 个大小为 $\frac{|\mathcal{U}|}{m}$ 的全集子集[①]。例如，图 4.1(a)

① 如果 $|\mathcal{U}|$ 不能被 $m$ 整除，可以把前面几个全集子集的大小设置为 $\lceil \frac{|\mathcal{U}|}{m} \rceil$ 并把剩余的全集子集大小设置为 $\lfloor \frac{|\mathcal{U}|}{m} \rfloor$。

和 4.1(b) 分别展示了通过平均划分方法划分 $\mathcal{U}$ 所得到的全集子集以及根据它所产生的片段。后文使用 scheme 来代表平均划分方法。注意下文提出的技术适用于任何划分方法。本文把划分方法的研究作为未来的工作。

- **建立索引**：为了找到共享片段的集合元组，可以使用倒排列表来索引片段，其中在同一倒排列表中的所有集合共享一个相同的片段。为了达到这个目的，首先对 $\mathcal{R}$ 中的集合按照集合大小分组和排序，令 $\mathcal{R}_l$ 包含 $\mathcal{R}$ 中所有大小为 $l$ 的集合，它们共享相同的划分方法 scheme($\mathcal{U}, m = H_l + 1$)。对于每个组 $\mathcal{R}_l$，按照如下方式建立 $m$ 个倒排索引 $\mathcal{I}_l^i$，其中 $1 \le i \le m$。对于每个集合 $\mathcal{X} \in \mathcal{R}_l$，按照 scheme($\mathcal{U}, m$) 划分它。对于每个片段 $\mathcal{X}^i$，把 $\mathcal{X}.id$ 附加到倒排列表 $\mathcal{I}_l^i[\mathcal{X}^i]$ 末尾，其中 $\mathcal{X}.id$ 是集合 $\mathcal{X}$ 的 $id$。

- **寻找近似元组**：接下来讨论如何使用索引来寻找近似元组。对于每个大小为 $s$ 的集合 $\mathcal{X} \in \mathcal{R}$，按照如下方式寻找近似元组。根据长度过滤，与其近似的集合的大小一定在 $[\delta s, s]$ 范围内[①]。对于每个可能的大小 $l \in [\delta s, s]$，按照平均划分方法 scheme($\mathcal{U}, m$) 把 $\mathcal{X}$ 划分为 $m = H_l + 1$ 个不相交的片段。然后对于每个 $1 \le i \le m$，因为倒排列表 $\mathcal{I}_l^i[\mathcal{X}^i]$ 中所有的集合都与 $\mathcal{X}$ 共享相同的片段 $\mathcal{X}^i$，所以需要访问该倒排列表并把其中所有的集合都添加到候选集 $C$ 中。最后，通过计算候选集中的集合与 $\mathcal{X}$ 的 Jaccard Similarity 来验证它们并返回所有的近似元组。

算法 4.1展示了基于划分的方法的伪代码。它的输入是一个数据集 $\mathcal{R}$ 和一个近似阈值 $\delta$, 它的输出是数据集中所有的近似元组。它首先扫描数据集 $\mathcal{R}$ 以得到 $\mathcal{R}$ 的全集 $\mathcal{U}$ (第 2 行)。然后它对所有的集合按照其大小进行分组和升序排序 (第 3 行)。接下来对于每个大小为 $s$ 的集合 $\mathcal{X} \in \mathcal{R}$ 以及每个元组 $\mathcal{R}_l$（其中 $\delta s \le l \le s$，它首先按照划分方法 scheme($\mathcal{U}, m = H_l + 1$) 把集合 $\mathcal{X}$ 划分为 $m$ 个不相交的片段。然后对每个片段 $\mathcal{X}^i$（其中 $1 \le i \le m$）它访问倒排列表 $\mathcal{I}_l^i[\mathcal{X}^i]$ 并把其中给所有的集合都加入到候选集 $C$ 中 (第 4-8 行)。之后它通过计算候选集合与 $\mathcal{X}$ 的 Jaccard Similarity 来验证它们并把所有与 $\mathcal{X}$ 近似的集合加入到结果集 $\mathcal{A}$ 中 (第 9-10 行)。最后，它按照划分方法 scheme($\mathcal{U}, m'$) 重新把 $\mathcal{X}$ 划分为 $m' = H_s + 1$ 个不相交的片段。对于每个片段 $\mathcal{X}^i$（其中 $1 \le i \le m'$）它把 $\mathcal{X}.id$ 附加到倒排列表 $\mathcal{I}_s^i[\mathcal{X}^i]$ 的末尾 (第 11 行)。最后它返回结果集 $\mathcal{A}$ (第 12 行)。

**例** 4.1：考虑表 4.1 中的数据集 $\mathcal{R}$，并假设阈值 $\delta = 0.73$。按照如下三步来寻找近似元组。

- **划分**：如图 4.1所示，对于 $\mathcal{R}_9$，因为 $m = H_9+1 = 4$，所以平均地把全集 $\mathcal{U}$ 划分为 4 个

① 这里使用自连接算法中一个常用的技巧：对于每个集合，只在大小不大于它的集合中寻找与它近似的集合

| $\mathcal{U}^1$ | $\mathcal{U}^2$ | $\mathcal{U}^3$ | $\mathcal{U}^4$ |
|---|---|---|---|
| $\{x_1, x_2, x_3, x_4\}$ | $\{x_5, x_6, x_7, x_8\}$ | $\{x_9, x_{10}, x_{11}, x_{12}\}$ | $\{x_{13}, x_{14}, x_{15}\}$ |

**(a)** $l = 9$ 时的平均划分方法 ***scheme***$(\mathcal{U}, m = 4)$

| $id$ | $X_{id}^1$ | $X_{id}^2$ | $X_{id}^3$ | $X_{id}^4$ |
|---|---|---|---|---|
| 1 | $\{x_1, x_2\}$ | $\{x_5, x_6, x_7\}$ | $\{x_{10}, x_{11}\}$ | $\{x_{13}, x_{14}\}$ |
| 2 | $\{x_2, x_4\}$ | $\{x_5, x_6\}$ | $\{x_9, x_{11}\}$ | $\{x_{13}, x_{14}, x_{15}\}$ |
| 3 | $\{x_1, x_3\}$ | $\{x_6, x_7\}$ | $\{x_9, x_{10}, x_{11}\}$ | $\{x_{13}, x_{14}\}$ |
| 4 | $\{x_3, x_4\}$ | $\{x_5, x_7, x_8\}$ | $\{x_{10}, x_{12}\}$ | $\{x_{13}, x_{14}\}$ |

**(b)** $\mathcal{R}_9$ 中的集合基于划分方法 **scheme**$(\mathcal{U}, m = 4)$ 得到的片段

| $I_9^1$ | $I_9^2$ | $I_9^3$ | $I_9^4$ |
|---|---|---|---|
| $\{x_1, x_2\} \mapsto 1$ | $\{x_5, x_6, x_7\} \mapsto 1$ | $\{x_{10}, x_{11}\} \mapsto 1$ | $\{x_{13}, x_{14}\} \to 1, 3, 4$ |
| $\{x_2, x_4\} \mapsto 2$ | $\{x_5, x_6\} \mapsto 2$ | $\{x_9, x_{11}\} \mapsto 2$ | $\{x_{13}, x_{14}, x_{15}\} \to 2$ |
| $\{x_1, x_3\} \mapsto 3$ | $\{x_6, x_7\} \mapsto 3$ | $\{x_9, x_{10}, x_{11}\} \mapsto 3$ | |
| $\{x_3, x_4\} \mapsto 4$ | $\{x_5, x_7, x_8\} \mapsto 4$ | $\{x_{10}, x_{12}\} \mapsto 4$ | |

**(c)** $\mathcal{R}_9$ 的片段倒排索引

| $\mathcal{D}_9^1$ | $\mathcal{D}_9^2$ | $\mathcal{D}_9^3$ | $\mathcal{D}_9^4$ |
|---|---|---|---|
| $\{x_1\} \to 1, 3$ | $\{x_5\} \to 2$ | $\{x_9\} \to 2$ | $\{x_{13}\} \to 1, 3, 4$ |
| $\{x_2\} \to 1, 2$ | $\{x_6\} \to 2, 3$ | $\{x_{10}\} \to 1, 4$ | $\{x_{14}\} \to 1, 3, 4$ |
| $\{x_3\} \to 3, 4$ | $\{x_7\} \to 3$ | $\{x_{11}\} \to 1, 2$ | $\{x_{13}, x_{14}\} \to 2$ |
| $\{x_4\} \to 2, 4$ | $\{x_5, x_6\} \to 1$ | $\{x_{12}\} \to 4$ | $\{x_{14}, x_{15}\} \to 2$ |
| | $\{x_5, x_7\} \to 1, 4$ | $\{x_9, x_{10}\} \to 3$ | $\{x_{13}, x_{15}\} \to 2$ |
| | $\{x_6, x_7\} \to 1$ | $\{x_9, x_{11}\} \to 3$ | |
| | $\{x_5, x_8\} \to 4$ | $\{x_{10}, x_{11}\} \to 3$ | |
| | $\{x_7, x_8\} \to 4$ | | |

**(d)** $\mathcal{R}_9$ 的 ***1***-删集倒排索引

图 4.1　$\mathcal{R}_9$ 的片段和倒排索引

不相交的全集子集：$\mathcal{U}^1 = \{x_1, x_2, x_3, x_4\}, \mathcal{U}^2 = \{x_5, x_6, x_7, x_8\}$，$\mathcal{U}^3 = \{x_9, x_{10}, x_{11}, x_{12}\}$ 和 $\mathcal{U}^4 = \{x_{13}, x_{14}, x_{15}\}$。

- **建立索引**：按照以上全集子集把 $\mathcal{R}_9$ 中的集合划分为 4 个片段，对于 $X_1 \in \mathcal{R}_9$，有 $X_1^1 = \{x_1, x_2\}$，$X_1^2 = \{x_5, x_6, x_7\}$，$X_1^3 = \{x_{10}, x_{11}\}$ 以及 $X_1^4 = \{x_{13}, x_{14}\}$。把 $X_1$ 的 id 1 附加到倒排列表 $I_9^1[X_1^1]$，$I_9^2[X_1^2]$，$I_9^3[X_1^3]$ 和 $I_9^4[X_1^4]$ 的末尾。类似地，可以划分其他集合并建立倒排索引。

- **寻找近似元组**：接下来寻找近似元组。考虑集合 $X_5$，因为 $|X_5| = 11$ 并且 $\delta|X_5| = 8.25$，所以需要探测 3 个组 $\mathcal{R}_9$，$\mathcal{R}_{10}$ 和 $\mathcal{R}_{11}$。对于 $\mathcal{R}_9$，按照 scheme$(\mathcal{U}, m)$ 划分 $X_5$ 为 $m = H_9 + 1 = 4$ 个部分：$X_5^1 = \{x_1, x_2, x_3, x_4\}$，$X_5^2 = \{x_5, x_6, x_7\}$，$X_5^3 = \{x_{10}, x_{11}\}$ 以及 $X_5^4 = \{x_{13}, x_{14}\}$。然后访问倒排列表 $I_9^1[X_5^1]$, $I_9^2[X_5^2]$，$I_9^3[X_5^3]$ 和 $I_9^4[X_5^4]$。从 $I_9^2[X_5^2]$ 中可以找到 $X_1$，从 $I_9^3[X_5^3]$ 中可以找到 $X_1$，从 $I_9^4[X_5^4]$ 中可以找到 $X_1, X_3, X_4$。因此可以得到 3 个侯选集合 $X_1$，$X_3$, 和 $X_4$。最后计算它们与 $X_5$ 的近似度，由于 $\text{JAC}(X_1, X_5) = \frac{9}{11} \geq \delta$，所以得到了一个近似元组 $\langle X_1, X_5 \rangle$。

**复杂度分析**: 首先分析时间复杂度。假设 $\mathcal{R}$ 中最大的集合大小为 $n$。获得全集并

**Algorithm 4.1**: 基于划分的框架

**Input**: $\mathcal{R}$: 数据集; $\delta$: 近似阈值;

**Output**: $\mathcal{A}$: $\{\langle \mathcal{X}, \mathcal{Y} \rangle | \textsc{jac}(\mathcal{X}, \mathcal{Y}) \geq \delta, \mathcal{X} \in \mathcal{R}, \mathcal{Y} \in \mathcal{R}\}$

**begin**
　获取 $\mathcal{R}$ 的全集 $\mathcal{U}$;
　对集合按照大小升序排序并分组;
　**foreach** 每个 $\mathcal{R}$ 中大小为 $s$ 的集合 $\mathcal{X}$ **do**
　　**foreach** $\delta s \leq l \leq s$ **do**
　　　使用平均划分方法 scheme$(\mathcal{U}, m)$ 把 $\mathcal{X}$ 划分为 $m = H_l + 1$ 个片段;
　　　**foreach** $1 \leq i \leq m$ **do**
　　　　把 $\mathcal{I}_l^i[\mathcal{X}^i]$ 中所有的集合加入到 $C$ 中;
　　**foreach** $\mathcal{Y} \in C$ **do**
　　　如果 $\textsc{jac}(\mathcal{X}, \mathcal{Y}) \geq \delta$，把 $\mathcal{Y}$ 加入到结果集 $\mathcal{A}$ 中;
　　按划分方法 scheme$(\mathcal{U}, m')$ 重新把 $\mathcal{X}$ 划分为 $m' = H_s + 1$ 个片段，并对每个 $1 \leq i \leq m'$，把 $\mathcal{X}.id$ 附加到 $\mathcal{I}_s^i[\mathcal{X}^i]$ 末尾;
　**return** $\mathcal{A}$;
**end**

对所有集合排序的时间复杂度是 $O(|\mathcal{R}|n)$ 和 $O(|\mathcal{R}| \log |\mathcal{R}|)$。对于每个集合, 基于划分的框架最多把它划分 $n - \delta * n + 1$ 次（每个可能的组一次），其时间复杂度是 $O((1-\delta)n^2)$。此外，它还需要访问倒排列表并验证候选集，其时间复杂度是 $O(|\mathcal{L}| + n|C|)$，其中 $|\mathcal{L}|$ 是被访问的倒排列表的长度总和，$|C|$ 是候选集的大小。另外，创建倒排索引的时间复杂度是 $O(|\mathcal{R}|n)$。因此整个框架总的时间复杂度是 $O(|\mathcal{R}| \log |\mathcal{R}| + |\mathcal{R}|(1-\delta)n^2 + |\mathcal{L}| + n|C|)$.

接下来分析空间复杂度。基于划分的方法需要存储倒排索引。倒排索引中条目的数量是不超过 $|\mathcal{R}|(H_n + 1)$ 的，同时倒排列表的数量又不超过倒排索引中条目的数量。此外，它还需要为每个集合存储候选集，其大小不超过 $|\mathcal{R}|$，因此总的空间复杂度是 $O(|\mathcal{R}|H_n)$.

观察到一些集合的片段的频率可能非常高，这将导致极大的候选集 $|C|$ 以及极低的算法性能。例如，在例 4.1中, $\mathcal{X}_5$ 包含一个高频的片段 $\mathcal{X}_5^4$，它将产生 3 个候选结果。接下来讨论如何减少访问倒排列表的长度 $|\mathcal{L}|$，注意它与候选集大小 $|C|$

是成正比的。

## 4.4 片段选择

为了避免使用非常频繁的片段来产生候选集，下面提出一个片段选择方法。注意如果两个片段 $\mathcal{X}^i$ 和 $\mathcal{Y}^i$ 包含两个 (或者更多) 不同的元素，那么可以跳过 $\mathcal{X}$ 的一个频繁的片段 $\mathcal{X}^j$，即不检查 $\mathcal{X}^j$ 是否与 $\mathcal{Y}^j$ 相同。这是因为即使忽略频繁的片段 $\mathcal{X}^j$，如果 $\mathcal{X}$ 和 $\mathcal{Y}$ 没有共享另外一个相同的片段，它们仍然不可能近似（因为 $|\mathcal{X} \Delta \mathcal{Y}| = |\mathcal{X}^i \Delta \mathcal{Y}^i| + |\mathcal{X}^j \Delta \mathcal{Y}^j| + \sum_{k \neq i,j} |\mathcal{X}^k \Delta \mathcal{Y}^k| \geq 2 + 0 + m - 2 = m$)。例如，在例 4.1中，因为 $\mathcal{X}_3^1$ 和 $\mathcal{X}_5^1$ 包含 2 个不同的元素，并且 $\mathcal{X}_3^2 \neq \mathcal{X}_5^2$ 以及 $\mathcal{X}_3^3 \neq \mathcal{X}_5^3$，所以 $\mathcal{X}_3$ 和 $\mathcal{X}_5$ 一定有不少于 4 个不同的元素，因此即使不去检查 $\mathcal{X}_3^4$ 是否和 $\mathcal{X}_5^4$ 相同，仍然可以直接剪枝掉这个元组。类似地，可以剪枝掉 $\mathcal{X}_4$ 和 $\mathcal{X}_5$ 组成的元组。

为了高效的探测两个片段是否包含 2 个不同的元素，下面在第4.4.1节中提出了 1-删集的概念（片段中移除一个元素后的子集)。与之前全部使用片段不同，这里能够在基于划分的框架中混合使用片段以及它们的 1-删集。因为存在多种不同的分配策略可以产生这种混合，所以下面在第4.4.2节中讨论如何评估一个分配策略并提出一个动态规划算法来选择最优的分配策略。

### 4.4.1 1-删集

给定一个非空集合 $\mathcal{Z}$, 它的 1-删集（1-deletion neighborhood）是它所有大小为 $|\mathcal{Z}| - 1$ 的子集。特别的，空集没有 1-删集。接下来给出 1-删集的正式定义。

**定义** 4.2 (**1-删集**):　给定一个非空集合 $\mathcal{Z}$，它的 1-删集是 $\mathcal{Z} \setminus \{z\}$，其中 $z \in \mathcal{Z}$。

例如，考虑例4.1中的 $\mathcal{X}_5^2 = \{x_5, x_6, x_7\}$。它的 1-删集包括 $\{x_5, x_6\}$，$\{x_6, x_7\}$ 和 $\{x_5, x_7\}$。令 $\text{del}(\mathcal{X}) = \{\mathcal{X} \setminus \{x\} | x \in \mathcal{X}\}$ 表示 $\mathcal{X}$ 的所以 1-删集组成的集合，那么有 $\text{del}(\mathcal{X}_5^2) = \{\{x_5, x_6\}, \{x_6, x_7\}, \{x_5, x_7\}\}$。

对于任意两个集合 $\mathcal{X}$ 和 $\mathcal{Y}$，如果 $\mathcal{X} \neq \mathcal{Y}$，很显然它们一定包含至少 1 个不同的元素，即 $|\mathcal{X} \Delta \mathcal{Y}| \geq 1$。此外，根据定义 4.2，可以观察到如果 $\mathcal{X} \neq \mathcal{Y}$、$\mathcal{X} \notin \text{del}(\mathcal{Y})$ 和 $\mathcal{Y} \notin \text{del}(\mathcal{Y})$ 都成立，那么它们一定包含至少 2 个不同的元素，如引理4.4所述。

**引理** 4.4:　对于任何两个集合 $\mathcal{X}$ 和 $\mathcal{Y}$，如果 $\mathcal{X} \neq \mathcal{Y}$、$\mathcal{X} \notin \text{del}(\mathcal{Y})$ 并且 $\mathcal{Y} \notin \text{del}(\mathcal{X})$，那么 $|\mathcal{X} \Delta \mathcal{Y}| \geq 2$。

**证明** 使用反正法证明该引理。假设存在两个集合 $\mathcal{X}$ 和 $\mathcal{Y}$ 满足 $\mathcal{X} \neq \mathcal{Y}$、$\mathcal{X} \notin \text{del}(\mathcal{Y})$、$\mathcal{Y} \notin \text{del}(\mathcal{X})$ 和 $|\mathcal{X} \Delta \mathcal{Y}| \leq 1$。首先，根据对称差集的定义：合集与交集的差集，有

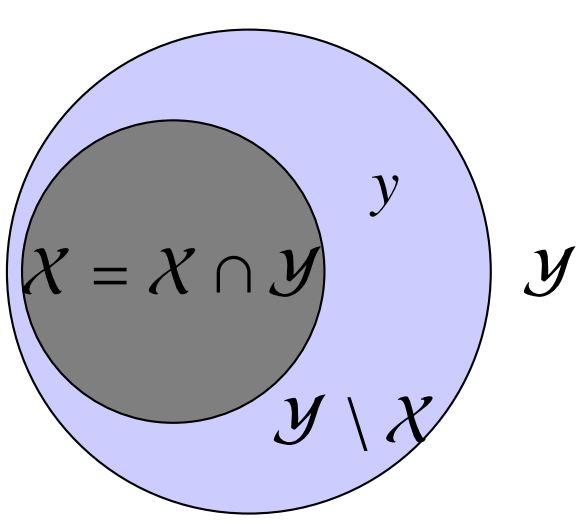

图 4.2　$\mathcal{X} \setminus \mathcal{Y} = \phi$ 和 $\mathcal{Y} \setminus \mathcal{X} = \{y\}$ 的文氏图

$|\mathcal{X} \Delta \mathcal{Y}| = |\mathcal{X} \setminus \mathcal{Y}| + |\mathcal{Y} \setminus \mathcal{X}|$。因为 $\mathcal{X} \neq \mathcal{Y}$，所以 $\mathcal{X} \setminus \mathcal{Y} \neq \phi$ 或者 $\mathcal{Y} \setminus \mathcal{X} \neq \phi$ 成立，因此 $|\mathcal{X} \Delta \mathcal{Y}| \geq 1$。又因为 $|\mathcal{X} \Delta \mathcal{Y}| \leq 1$，所以有 $|\mathcal{X} \Delta \mathcal{Y}| = |\mathcal{X} \setminus \mathcal{Y}| + |\mathcal{Y} \setminus \mathcal{X}| = 1$。因为 $|\mathcal{X} \setminus \mathcal{Y}|$ 和 $|\mathcal{Y} \setminus \mathcal{X}|$ 都是非负整数，所以要么 $\mathcal{X} \setminus \mathcal{Y} = \phi$ 并且 $\mathcal{Y} \setminus \mathcal{X} = \{y\}$（其中 $y \in \mathcal{Y}$），要么反过来。不失一般性的，假设是前一种情况 (见图4.2)。那么有 $\mathcal{Y} \setminus \{y\} = \mathcal{Y} \setminus (\mathcal{Y} \setminus \mathcal{X}) = \mathcal{X} \cap \mathcal{Y} = \mathcal{X}$。根据定义4.2，有 $\mathcal{Y} \setminus \{y\} \in \mathsf{del}(\mathcal{Y})$，这与 $\mathcal{X} \notin \mathsf{del}(\mathcal{Y})$ 互相矛盾。

例如，在例4.1中，考虑 $\mathcal{X}_5^2 = \{x_5, x_6, x_7\}$ 和 $\mathcal{X}_4^2 = \{x_5, x_7, x_8\}$。因为 $\mathcal{X}_5^2 \neq \mathcal{X}_4^2$、$\mathcal{X}_5^2 \notin \mathsf{del}(\mathcal{X}_4^2)$ 并且 $\mathcal{X}_4^2 \notin \mathsf{del}(\mathcal{X}_5^2)$，所以 $|\mathcal{X}_5^2 \Delta \mathcal{X}_4^2| \geq 2$。实际上，这里 $|\mathcal{X}_5^2 \Delta \mathcal{X}_4^2| = 2$。

**总览**：下面讨论如何利用 1-删集来解决集合近似连接问题。给定两个集合 $\mathcal{X}$ 和 $\mathcal{Y}$，其中 $l = |\mathcal{X}| \leq s = |\mathcal{Y}|$，把它们分别划分为 $m = H_l + 1$ 个片段 $\mathcal{X}^1, \mathcal{X}^2, \cdots, \mathcal{X}^m$ 和 $\mathcal{Y}^1, \mathcal{Y}^2, \cdots, \mathcal{Y}^m$。然后分 3 种情况对待 $\mathcal{X}^i$ 和 $\mathcal{Y}^i$：

**情况** 0: 跳过 $\mathcal{X}^i$ 和 $\mathcal{Y}^i$。

**情况** 1: 使用 $\mathcal{X}^i$ 和 $\mathcal{Y}^i$。如果 $\mathcal{X}^i = \mathcal{Y}^i$，那么将 $\mathcal{X}$ 和 $\mathcal{Y}$ 作为一对候选元组。

**情况** 2: 使用 $\mathcal{X}^i, \mathcal{Y}^i$ 以及它们的 1-删集。如果 $\mathcal{X}^i = \mathcal{Y}^i$，$\mathcal{X}^i \in \mathsf{del}(\mathcal{Y}^i)$ 或者 $\mathcal{Y}^i \in \mathsf{del}(\mathcal{X}^i)$，那么把 $\mathcal{X}$ 和 $\mathcal{Y}$ 作为一对候选元组。

基本框架只利用了情况 1。下面考虑如何同时使用这三种情况来取得高性能。可以使用一个 $m$ 维的向量 $(v_1, v_2, \ldots, v_m)$ 来决定每个位置使用哪种情况，其中 $v_i \in \{0, 1, 2\}$ 对应这三种情况。例如，考虑表4.1中的大小为 $s = 11$ 的集合 $\mathcal{X}_5$ 以及组 $\mathcal{R}_l$，其中 $l = 9$，并假设 $\delta = 0.73$。(2,0,2,0) 是一个 4 维向量，其表示在第 1 个片段上使用情况 2，第二个片段上使用情况 0，第三个片段上使用情况 2，以及第四个片段上使用情况 0。

接下来考虑 $\sum_{i=1}^{m} v_i$ 的值。首先假设 $\sum_{i=1}^{m} v_i \geq m = H_l + 1$。如果该向量没有匹配，即 (1) 对于 $v_i = 1$，$\mathcal{X}^i \neq \mathcal{Y}^i$; (2) 对于 $v_i = 2$，$\mathcal{X}^i \neq \mathcal{Y}^i$、$\mathcal{X}^i \notin \mathsf{del}(\mathcal{Y}^i)$ 并且 $\mathcal{Y}^i \notin \mathsf{del}(\mathcal{X}^i)$，那么 $\mathcal{X}$ 和 $\mathcal{Y}$ 不可能近似。这是因为对于情况 1, 两个片段至少有 1 个不相同的元素；并且对于情况 2, 两个片段至少有 2 个不相同的元素。因此两个集合总共至少有 $\sum_{i=1}^{m} v_i \geq m = H_l + 1$ 个不匹配的元素，即 $|\mathcal{X} \Delta \mathcal{Y}| \geq H_l + 1$。根据

引理 4.1, $\mathcal{X}$ 和 $\mathcal{Y}$ 不可能近似。

实际上，如公式 4-1所示，可以通过利用两个集合 $\mathcal{X}$ 和 $\mathcal{Y}$ 近似仅当 $|\mathcal{X}\,\Delta\,\mathcal{Y}| \le \frac{1-\delta}{1+\delta}(|\mathcal{X}|+|\mathcal{Y}|)$ 来给出一个比 $m = H_l + 1$ 更加紧缩的界。令 $H(l,s) = \lfloor \frac{1-\delta}{1+\delta}(l+s) \rfloor$，其中 $|\mathcal{X}| = l$，$|\mathcal{Y}| = s$。如果 $\sum_{i=1}^{m} v_i \ge H(l,s)+1$ 并且它们没有匹配，那么 $\mathcal{X}$ 和 $\mathcal{Y}$ 不可能近似。

接下来，考虑 $\sum_{i=1}^{m} v_i < H(l,s)+1$。这种情况下，即使存在匹配，$\mathcal{X}$ 和 $\mathcal{Y}$ 仍然不可能近似，因为 $H(l,s)$ 个不匹配的元素可以让所有被选取的片段和 1-删集都不匹配。因此在这种情况下，该方法可能漏掉结果。

因此，为了使用这个方法，需要保证 $\sum_{i=1}^{m} v_i \ge H(l,s)+1$。为了减少把两个集合当做候选元组的几率，希望 $\sum_{i=1}^{m} v_i$ 尽可能的小，因此只需要考虑满足 $\sum_{i=1}^{m} v_i = H(l,s)+1$ 的向量，这种向量被称作分配，如下面所定义的。

**定义 4.3 (分配)**：　给定两个整数 $l$ 和 $s$，一个向量 $\mathcal{V}_l^s = (v_1, v_2, \ldots, v_m)$ 是一个分配，如果它满足 (1)$m = H_l + 1$, (2)$v_i \in \{0,1,2\}$ 以及 (3) $\sum_{i=1}^{m} v_i = H(l,s)+1$。

如引理 4.5 所述，可以证明给定两个集合，对于任何分配，如果分配中不存在匹配，那么他们不可能近似。

**引理 4.5**：　给定有 $m = H_l + 1$ 个片段的两个集合 $\mathcal{X}$ 和 $\mathcal{Y}$ 以及一个分配 $\mathcal{V}_l^s$，其中 $l = |\mathcal{X}|$，$s = |\mathcal{Y}|$，如果在 $\mathcal{V}_l^s$ 中不存在匹配，那么 $\textsc{jac}(\mathcal{X},\mathcal{Y}) < \delta$。

**证明**　令 $\mathcal{V}_l^s = (v_1, v_2, \ldots, v_m)$，考虑 $v_i$ 的值 $(1 \le i \le m)$。如果 $v_i = 1$，那么 $\mathcal{X}$ 和 $\mathcal{Y}$ 的第 $i$ 份片段不匹配，第 $i$ 份片段中至少有 1 个不相同的元素；如果 $v_i = 2$，那么 $\mathcal{X}$ 和 $\mathcal{Y}$ 的第 $i$ 份 1-删集不匹配，第 $i$ 份片段中至少有 2 个不相同的元素。因此两个集合总共至少有 $\sum_{i=1}^{m} v_i \ge m = H(l,s)+1$ 个不匹配的元素，即 $|\mathcal{X}\,\Delta\,\mathcal{Y}| \ge H(l,s)+1 \ge H_l+1$。根据引理 4.1，$\mathcal{X}$ 和 $\mathcal{Y}$ 不可能近似。

**算法**：　接下来调整基于划分的框架来使用新的过滤条件。首先讨论索引结构。为了利用新的过滤条件，除了倒排索引 $\mathcal{I}_l^i$ 之外，还需要为 $\mathcal{R}_l$ 中的所有集合的第 $i$ 份片段的 1-删集建立另一个倒排索引 $\mathcal{D}_l^i$，其中 $1 \le i \le m = H_l + 1$ (见图 4.1)。对于每个集合 $\mathcal{X} \in \mathcal{R}_l$, 把它划分为 $m$ 个片段。对于每个片段 $\mathcal{X}^i$, 除了把 $\mathcal{X}.id$ 附加到倒排列表 $\mathcal{I}_l^i[\mathcal{X}^i]$ 的末尾，还需要把 $\mathcal{X}.id$ 附加到 $\mathcal{D}_l^i[\mathsf{del}(\mathcal{X}^i)]$ 中所有倒排列表的末尾，其中 $\mathcal{D}_l^i[\mathsf{del}(\mathcal{X}^i)]$ 代表所有的倒排列表 $\mathcal{D}_l^i[\mathcal{X}^i \setminus \{x\}]$，其中 $x \in \mathcal{X}^i$。同样的，$\mathcal{I}_l^i[\mathsf{del}(\mathcal{X}^i)]$ 也是。

然后对于任何大小为 $s$ 的集合 $\mathcal{X} \in \mathcal{R}$，使用两个索引 $\mathcal{I}$ 和 $\mathcal{D}$ 来寻找它的近似集合。对于每个 $l \in [\delta * s, s]$，首先用平均划分方法 $\mathsf{scheme}(\mathcal{U}, m)$ 把 $\mathcal{X}$ 划分为

**Algorithm 4.2**: 基于删集的框架

```
1 begin
      // 用以下代码替换算法 4.1 中第 7 到 8 行
2     生成一个分配 V_l^s;
3     foreach 1 ≤ i ≤ m do
4         if v_i = 1 then
5             把 I_l^i[X^i] 中所有的集合加入候选集 C 中；
6         if v_i = 2 then
7             把 I_l^i[X^i]、D_l^i[X^i] 和 I_l^i[del(X^i)] 中的所有的集合加入候选集 C 中;
      // 在算法4.1中第11行后添加以下代码
8     对于每个片段 X^i，把 X.id 附加到每个 D_s^i[del(X^i)] 的末尾;
9 end
```

$m = H_l + 1$ 个不相交的片段。之后选择一个分配 $\mathcal{V}_l^s = (v_1, v_2, \ldots, v_m)$ (下面将在第4.4.2节中讨论如何选取分配)。对于每个片段 $\mathcal{X}^i$，如果 $v_i = 1$，那么把 $\mathcal{I}_l^i[\mathcal{X}^i]$ 中所有的集合 $\mathcal{Y}$ 加入到候选集 $C$ 中，因为 $\mathcal{X}^i = \mathcal{Y}^i$。类似地，如果 $v_i = 2$，那么把 $\mathcal{I}_l^i[\mathcal{X}^i]$，$\mathcal{D}_l^i[\mathcal{X}^i]$ 和 $\mathcal{I}_l^i[\mathsf{del}(\mathcal{X}^i)]$ 中所有的集合 $\mathcal{Y}$ 都加入到候选集 $C$ 中，因为它们分别表明 $\mathcal{X}^i = \mathcal{Y}^i$，$\mathcal{X}^i \in \mathsf{del}(\mathcal{Y}^i)$ 和 $\mathcal{Y}^i \in \mathsf{del}(\mathcal{X}^i)$。根据引理4.5，它们全部都是候选元组。通过这种方法，可以调整基于划分的框架以支持混合使用片段和 1-删集，即基于删集的框架。

算法 4.2 展示了基于删集的框架的伪代码。不同于基于划分的框架那样直接访问倒排列表并把其中的条目添加到候选集 $C$ 中，它为每个大小为 $s$ 的集合 $\mathcal{X}$ 和组 $\mathcal{R}_l$ 生成一个分配 $\mathcal{V}_l^s$ (第 2 行)。然后对于该分配 $\mathcal{V}_l^s$ 中的每一个维度 $v_i$（其中 $1 \le i \le m$)，如果 $v_i = 1$, 它把倒排列表 $\mathcal{I}_l^i[\mathcal{X}^i]$ 中所有的集合加入到候选集 $C$ 中 (第 4 到 5 行)；否则的话，如果 $v_i = 2$, 它将 $\mathcal{I}_l^i[\mathcal{X}^i]$, $\mathcal{D}_l^i[\mathcal{X}^i]$ 以及 $\mathcal{I}_l^i[\mathsf{del}(\mathcal{X}^i)]$ 中的所有集合加入到候选集 $C$ 中 (第 6 到 7 行)。在建立索引阶段，除了把 $\mathcal{X}.id$ 附加到 $\mathcal{I}_l^i[\mathcal{X}^i]$ 的末尾, 它也需要将 $\mathcal{X}.id$ 附加到 $\mathcal{D}_s^i[\mathsf{del}(\mathcal{X}^i)]$ 中每个倒排列表的末尾 (第 8 行)。其余部分和基于划分的框架完全相同。

**例** 4.2:　这里使用与例 4.1中相同的设置，把 $\mathcal{R}_9$ 中的集合划分为 $m = H_9 + 1 = 4$ 个片段。如图 4.1 (c) 所示，基于它们的 1-删集建立倒排索引。接下来寻找近似元组。考虑 $\mathcal{X}_5$ 和 $\mathcal{R}_9$，假设为它们选择了分配 $\mathcal{V}_9^{11} = (2, 0, 2, 0)$。对于 $\mathcal{X}_5^1$，因为 $v_1 = 2$，

所以访问 $\mathcal{I}_9^1[\mathcal{X}_5^1]$、$\mathcal{D}_9^1[\mathcal{X}_5^1]$ 以及每个 $\mathcal{I}_9^1[\mathsf{del}(\mathcal{X}_5^1)]$，在这里没有找到候选结果。对于 $\mathcal{X}_5^2$，因为 $v_2 = 0$，所以跳过它。对于 $\mathcal{X}_5^3$，因为 $v_3 = 2$，所以访问 $\mathcal{I}_9^3[\mathcal{X}_5^3]$、$\mathcal{D}_9^3[\mathcal{X}_5^3]$ 以及每个 $\mathcal{I}_9^3[\mathsf{del}(\mathcal{X}_5^3)]$ 并得到 $\mathcal{X}_1$ 和 $\mathcal{X}_3$，然后把 $\mathcal{X}_1$ 和 $\mathcal{X}_3$ 加入到候选集 $C$ 中。对于 $\mathcal{X}_5^4$，因为 $v_4 = 0$，所以跳过它。最后，验证 $C$ 中所有的候选集合，可以找到一个近似元组 $\langle\mathcal{X}_1, \mathcal{X}_5\rangle$。

**空间复杂度**: 这里分析倒排索引的空间复杂度。对于每个集合，它最多产生 $m = H_n + 1$ 个片段和 $n$ 个 1-删集，其中 $n$ 是 $\mathcal{R}$ 中最大集合的大小。因此，在倒排列表中总共有 $O(|\mathcal{R}|n)$ 个条目。因为倒排列表的数量是不多于倒排列表中的条目数的，所以总的空间复杂度是 $O(|\mathcal{R}|n)$，相对来说这是一个比较小的复杂度。

注意基于划分的框架是基于删集的框架在所有分配都被指定为 **1** 时（即所有维度都为 1 的向量）的特例。实际上，给定一个集合和一个组，存在很多种不同的分配，一个很自然的问题是如何评估不同的分配策略以及如何选择一个最优的分配。接下来回答这个问题。

### 4.4.2　最优分配选择

下面首先讨论如何评估一个分配。如基于删集的框架所展示的，对于每个集合 $\mathcal{X}$ 和组 $\mathcal{R}_l$，需要根据分配来访问倒排列表 $\mathcal{I}_l^i[\mathcal{X}^i]$，$\mathcal{D}_l^i[\mathcal{X}^i]$ 或者 $\mathcal{I}_l^i[\mathsf{del}(\mathcal{X}^i)]$。算法访问的倒排列表长度越短，需要耗费的时间就越少，得到的候选元组也越少。因此希望最小化访问的倒排列表长度总和。为了达到这个目的，给定一个大小为 $s$ 的集合 $\mathcal{X}$ 和一个组 $\mathcal{R}_l$, 令

$$c_0^i = 0;\ \ c_1^i = |\mathcal{I}_l^i[\mathcal{X}^i]|;\ \ c_2^i = |\mathcal{I}_l^i[\mathcal{X}^i]| + |\mathcal{D}_l^i[\mathcal{X}^i]| + |\mathcal{I}_l^i[\mathsf{del}(\mathcal{X}^i)]|$$

其中 $1 \le i \le m = H_l + 1$，可以定义分配 $\mathcal{V}_l^s = (v_1, v_2, \ldots v_m)$ 的代价为 $\mathsf{cost}(\mathcal{V}_l^s) = \sum_{i=1}^m c_{v_i}^i$，也就是该分配需要访问的倒排列表的长度之和。这样的话，目标就是选择一个最小代价的分配。接下来形式化定义最优分配选择问题。

**定义** 4.4 (**最优分配策略选择**):　给定一个大小为 $s$ 的集合 $\mathcal{X}$ 和一个组 $\mathcal{R}_l$，最优分配选择问题选择一个具有最小代价的分配 $\mathcal{V}_l^s$。

例如, 图 4.4(a) 展示了集合 $\mathcal{X}_5$ 和组 $\mathcal{R}_9$ 的代价, 其最优分配是 $\mathcal{V}_9^{11} = (2, 1, 1, 0)$, 最优分配的代价是 $\mathsf{cost}(\mathcal{V}_9^{11}) = c_2^1 + c_1^2 + c_1^3 + c_0^4 = 2$。

接下来介绍一个动态规划算法来解决最优分配选择问题。令 $\mathsf{cost}(i, j)$ 为满足条件 $\sum_{k=1}^i v_k = j$（其中 $v_k \in \{0, 1, 2\}$）的 $\sum_{k=1}^i c_{v_k}^i$ 的最小值，$\mathcal{V}(i, j)$ 代表相对应的向量

| $i$ | $c_0^i$ | $c_1^i$ | $c_2^i$ |
|---|---|---|---|
| 1 | 0 | 0 | 0 |
| 2 | 0 | 1 | 3 |
| 3 | 0 | 1 | 2 |
| 4 | 0 | 3 | 4 |

**(a)** 代价

| $i$ | $\nabla_0^i = c_1^i - c_0^i$ | $\nabla_1^i = c_2^i - c_1^i$ |
|---|---|---|
| 1 | 0 | 0 |
| 2 | 1 | 2 |
| 3 | 1 | 1 |
| 4 | 3 | 1 |

**(b)** 增量代价

| cost | -1 | 0 | 1 | 2 | 3 | 4 | 联结后的向量 |
|---|---|---|---|---|---|---|---|
| 0 | $\infty$ | 0 | $\infty$ | $\infty$ | $\infty$ | $\infty$ | $\mathcal{V}(0,0) = ()$ |
| 1 | $\infty$ | 0 | 0 | 0 | $\infty$ | $\infty$ | $\mathcal{V}(1,2) = (2)$ |
| 2 | $\infty$ | 0 | 0 | 0 | 1 | 3 | $\mathcal{V}(2,2) = (2, 0)$ |
| 3 | $\infty$ | 0 | 0 | 0 | 1 | 2 | $\mathcal{V}(3,4) = (2, 0, 2)$ |
| 4 | $\infty$ | 0 | 0 | 0 | 1 | 2 | $\mathcal{V}(4,4) = (2, 0, 2, 0)$ |

**(c)** 基于 $\mathcal{X}_5$ 和 $\mathcal{R}_9$ 的 ***cost*** 表

图 4.3　最优分配策略选择的例子

$(v_1, v_2, \ldots, v_i)$。那么，最优分配选择问题就是计算一个最小代价 $\text{cost}(m, H(l,s)+1)$ 及其最优分配 $\mathcal{V}(m, H(l,s)+1)$。为了计算 $\text{cost}(i,j)$，考虑 $\mathcal{V}(i,j)$ 的最后一维 $v_i$ 的值，它可以是 0、1 或者 2，有

$$\text{cost}(i,j) = min \begin{cases} \text{cost}(i-1, j) + c_0^i \\ \text{cost}(i-1, j-1) + c_1^i \\ \text{cost}(i-1, j-2) + c_2^i \end{cases}$$

这个和

$$\text{cost}(i,j) = \min_{v \in \{0,1,2\}} \text{cost}(i-1, j-v) + c_v^i. \tag{4-2}$$

是等价的。此外，令

$$v_{\min} = \arg\min_{v \in \{0,1,2\}} \text{cost}(i-1, j-v) + c_v^i, \tag{4-3}$$

$v_{\min}$ 是 $\mathcal{V}(i,j)$ 的第 $i$ 维 (也就是最后一维)，有 $\mathcal{V}(i,j) = \mathcal{V}(i-1, j-v_{\min}) \oplus v_{\min}$，其中 $\oplus$ 是把 $v_{\min}$ 联结到 $\mathcal{V}(i-1, j-v_{\min})$ 的末尾的操作。接下来讨论初始化。对于每个 $0 \le i \le m$，初始化 $\text{cost}(i,0) = 0$，这是因为对于所有 $1 \le k \le i$，可以设置 $v_k = 0$ 从而得到 $\mathcal{V}(i,0) = \mathbf{0}^i$：一个 $i$ 维 0 向量（即所有维度的值都是 0）。对于所有 $1 \le j \le H(l,s)+1$，初始化 $\text{cost}(0,j) = \infty$；对于所有 $0 \le i \le m$，因为 $\text{cost}(i,-1)$ 没

有定义，即不存在这样的分配，所以初始化 $\text{cost}(i,-1)=\infty$。通过这样的初始化以及递推公式，可以很容易的解决最优分配选择问题。通过计算 $\text{cost}(m, H(l,s)+1)$ 的值，可以得到向量 $\mathcal{V}(m, H(l,s)+1)$，它就是最优分配。

算法 4.3 展示了最优分配选择算法 OptimalSelection 的伪代码。它的输入包括两种倒排索引 $\mathcal{I}_l$ 和 $\mathcal{D}_l$，一个大小为 $s$ 的集合 $\mathcal{X}$ 以及一个组 $\mathcal{R}_l$。它的输出是集合 $\mathcal{X}$ 和组 $\mathcal{R}_l$ 下的最优分配 $\mathcal{V}_l^s$。它首先利用 $\mathcal{X}^i, \mathcal{I}_l^i$ 和 $\mathcal{D}_l^i$ 计算所有的代价 $c_0^i, c_1^i$ 和 $c_2^i$，其中 $1 \le i \le m = H_l + 1$ (第 2 行)。然后它初始化一个代价二维表 cost 以及一个分配二位表 $\mathcal{V}$：对于所有 $0 \le i \le H_l + 1$，设定 $\text{cost}(i,0)=0$，$\text{cost}(i,-1)=\infty$ 和 $\mathcal{V}(i,0)=\mathbf{0}^i$ 以及对于所有 $1 \le j \le H(l,s)+1$，设置 $\text{cost}(0,j)=\infty$ (第 3 行)。然后对于每个 $1 \le i \le H_l+1$ 以及 $1 \le j \le H(l,s)+1$，它从 $\{0,1,2\}$ 中选择 $v_{\min}$ 以得到一个最小的代价，即它设置 $v_{\min} = \arg\min_{v\in\{0,1,2\}} \text{cost}(i-1, j-v) + c_v^i$ (第 9 行)。接下来，根据选择的 $v_{\min}$，它把 $\text{cost}(i,j)$ 设置为 $\text{cost}(i-1, j-v_{\min}) + c_{v_{\min}}^i$ 并把 $v_{\min}$ 附加到 $\mathcal{V}(i-1, j-v_{\min})$ 的末尾从而形成 $\mathcal{V}(i,j)$ (第 10-11 行)。最后，它返回 $\mathcal{V}(H_l+1, H(l,s)+1)$ 作为最优分配 $\mathcal{V}_l^s$ (第 12 行).

**例** 4.3： 考虑 $\mathcal{X}_5$ 和 $\mathcal{R}_9$，仍然使用例 4.1中的设置, 图 4.3(a) 展示了它们所有的代价。因为 $H_9+1=4$ 以及 $H(9,11)+1=4$，对于 $0 \le i \le 4$，首先初始化 $\text{cost}(i,0)=0$ 以及 $\mathcal{V}(i,0)=\mathbf{0}^i$。对于 $0 \le i \le 4$ 和 $1 \le j \le 4$，初始化 $\text{cost}(i,-1)=\text{cost}(0,j)=\infty$。注意 $\mathcal{V}(0,0)=\mathbf{0}^0=()$。然后根据公式 4-2 来填二维表 cost，其结果如图 4.3 所示。接下来联结向量。对于 $i=1$ 和 $j=2$，在 $v=0,1$ 和 2 时，$\text{cost}(i-1,j-v)+c_v^i$ 的值分别是 $\infty, \infty$ 和 0。因此 $v_{\min}=2$，$\mathcal{V}(1,2)=\mathcal{V}(1-1,2-2)\oplus 2=(2)$。类似的，可以联结其他元素。最终可以得到最优分配 $\mathcal{V}(4,4)=(2,0,2,0)$，其代价是 $\text{cost}(4,4)=2$。

**时间复杂度**: 对于每个大小为 $s$ 的集合 $\mathcal{X} \in \mathcal{R}$ 和一个组 $\mathcal{R}_l$，动态规划算法的时间复杂度是 $O((H_l+1)*(H(l,s)+1))$。因为 $l \in [\delta * s, s]$，所以对于每个集合 $\mathcal{X}$，其时间复杂度是 $O(\sum_{l=\delta*s}^{s}(H_l+1)(H(l,s)+1)) = O(s^3)$，这是很高昂的。下面在 4.5 节中讨论加速分配选择。

注意, 可以扩展删集的定义到 $q$-删集并在基于删集的框架中使用它。然而 $q$-删集的数量随着 $q$ 的增大呈指数式增长 (对于一个集合 $\mathcal{X}$ 来说，有 $\binom{|\mathcal{X}|}{q}$ 个 $q$-删集)。这不但导致存储 $q$-删集的额外代价，而且需要很多时间来产生它们。因此在本文主要关注于 1-删集并把研究基于 $q$-删集的技术作为未来工作。

## 4.5 加速分配策略选取

因为寻找最优分配策略的代价很高昂，本节讨论加速分配策略的选择。下面首先在第 4.5.1 节提出一个基于堆的、近似比为 2 的贪婪算法；然后在第 4.5.2 节

**Algorithm 4.3**: OptimalSelection($\mathcal{I}_l, \mathcal{D}_l, \mathcal{X}, s, \mathcal{R}_l$)

**Input**: $\mathcal{I}_l$ 和 $\mathcal{D}_l$: 两种倒排索引;
　　$\mathcal{X}$: 一个集合; $s$: 集合 $\mathcal{X}$ 的大小; $\mathcal{R}_l$: 一个组
**Output**: $\mathcal{X}$ 和 $\mathcal{R}_l$ 的最优分配

**begin**
　从 $\mathcal{I}_l^i, \mathcal{D}_l^i$ 和 $\mathcal{X}^i$ 中得到代价 $c_0^i, c_1^i$ 和 $c_2^i$;
　**for** $0 \leq i \leq H_l + 1$ **do**
　　$\text{cost}(i, -1) = \infty, \text{cost}(i, 0) = 0, \mathcal{V}(i, 0) = \mathbf{0}^i$;
　**for** $1 \leq j \leq H(l, s) + 1$ **do**
　　$\text{cost}(0, j) = \infty$;
　**for** $1 \leq i \leq H_l + 1$ **do**
　　**for** $1 \leq j \leq H(l, s) + 1$ **do**
　　　$v_{\min} = \arg\min_{v \in \{0,1,2\}} \text{cost}(i - 1, j - v) + c_v^i$;
　　　$\text{cost}(i, j) = \text{cost}(i - 1, j - v_{\min}) + c_{v_{\min}}^i$;
　　　$\mathcal{V}(i, j) = \mathcal{V}(i - 1, j - v_{\min}) \oplus v_{\min}$;
　**return** $\mathcal{V}(H_l + 1, H(l, s) + 1)$;
**end**

设计一个多长度集合分组机制以减少每个集合需要探测的组的数量。这两个技术一起能够将分配选择的时间复杂度从 $O(s^3)$ 降低到 $O(s \log s)$。

### 4.5.1　分配策略选择

寻找一个分配策略的目标是计算一个向量（$v_1, v_2, \cdots, v_m$）使得 $\sum_{i=1}^{m} v_i = H(l, s) + 1$ 并令 $\sum_{i=1}^{m} c_v^i$ 尽可能的小。为了达到这个目标，可以按照如下方法设计一个贪婪算法。给定一个大小为 $s$ 的集合 $\mathcal{X}$ 和一个组 $\mathcal{R}_l$，首先初始化一个 $m = H_l + 1$ 维的零向量 $\mathcal{V} = \mathbf{0}^m$。然后反复比较对 $v_i$ 增加 1 的增量代价，并选择 $i \in [1, m]$ 中增量代价最小的一个，其中 $v_i = 0$ 时的增量代价为 $c_1^i - c_0^i$，$v_i = 1$ 时的增量代价为 $c_2^i - c_1^i$①。换句话说，可以贪心地选择一个增量代价最小的维度并把选取的维度 $v_i$ 增加 1，其中 $v_i \in \{0, 1\}$ 的增量代价是 $\nabla_{v_i}^i = c_{v_i+1}^i - c_{v_i}^i$。例如，图 4.3(b) 展示了 $\mathcal{R}_9$ 中的集合的增量代价。在实施了 $H(l, s) + 1$ 次递增后，$\mathcal{V}$ 就变成了一个满足 $\sum_{i=1}^{m} v_i = H(l, s) + 1$ 和 $v_i \in \{0, 1, 2\}$（$1 \leq i \leq m$）的分配策略。然后可以返回 $\mathcal{V}$ 作为选择的分配策略 $\mathcal{V}_l^s$。这里可以利用堆来选择最小的增量代价。接下来提出一个基于堆的贪心算法。算法 4.4 展示了这个贪婪算法 GreedySelection 的伪代码。

① 注意对于 $v_i = 2$，不能继续增加它，因此不对它定义增量代价。

**Algorithm 4.4**: GreedySelection($\mathcal{I}_l, \mathcal{D}_l, \mathcal{X}, s, \mathcal{R}_l$)

```
    Input: 𝓘_l 和 𝓓_l: 两种倒排索引;
           𝓧: 一个集合; s: 𝓧 的大小; 𝓡_l: 一个组.
    Output: 𝓥_l^s : 作用在 𝓧 和 𝓡_l 上的一个分配策略。
1   begin
2       从 𝓘_l^i,𝓓_l^i 和 𝓧^i 中获取增量代价 ∇_0^i 和 ∇_1^i;
3       设置 𝓥 = (v_1, v_2, ..., v_m) 中的所有的 v_i = 0, 其中 m = H_l + 1;
4       在 ⟨v_i, ∇_{v_i}^i⟩ 上创建一个小顶堆 𝓜, 其中 1 ≤ i ≤ m;
5       foreach 1 ≤ j ≤ H(l, s) + 1 do
6           对堆 𝓜 做弹出操作, 弹出使得增量代价 ∇_{v_i}^i 最小的节点 ⟨v_i, ∇_{v_i}^i⟩;
7           对 𝓥 的第 i 维 v_i 增加 1;
8           if v_i = 1 then 将新的节点 ⟨v_i = 1, ∇_{v_i}^i⟩ 压入堆 𝓜 中
9       return 𝓥;
10  end
```

给定一个大小为 $s$ 的集合 $\mathcal{X}$ 和一个组 $\mathcal{R}_l$，首先用 $\mathcal{I}_l^i$,$\mathcal{D}_l^i$ 和 $\mathcal{X}^i$ 计算 $v_i$ 的增量代价，其中 $v_i \in \{0,1\}$，$1 \le i \le m = H_l+1$(第 2 行)。然后初始化一个 $m = H_l+1$ 维的向量 $\mathcal{V} = (v_1, v_2, \ldots, v_m)$ 并对所有 $1 \le i \le m$ 设置 $v_i = 0$(第 3 行)。接下来构建一个小顶堆 $\mathcal{M}$，$\langle v_i, \nabla_{v_i}^i \rangle$ 是 $\mathcal{M}$ 中的节点，其中 $1 \le i \le m$ (第 4 行)。之后对堆做 $H(l,s)+1$ 次弹出操作，每个弹出操作都把增量代价 $\nabla_{v_i}^i$ 最小的节点弹出 (第 5-6 行)。对于每个被弹出的节点 $\langle v_i, \nabla_{v_i}^i \rangle$，对 $\mathcal{V}$ 的第 $i$ 维 $v_i$ 增加 1。然后，如果 $v_i = 1$，那么需要更新它的增量代价。为了达到这个目的，将一个新的节点 $\langle v_i = 1, \nabla_{v_i}^i \rangle$ 压入堆 $\mathcal{M}$ 中。如果 $v_i = 2$，由于不能再继续增加 $v_i$ 的值了，因此不需要向堆 $\mathcal{M}$ 中压入新的节点 (第 7-8 行)。当一共有 $H(l,s)+1$ 个节点被弹出时，返回 $\mathcal{V}$ 作为选择的分配策略 $\mathcal{V}_l^s$ (第 9 行)。

**例** 4.4:　仍然使用与例 4.1 中相同的设置并考虑 $\mathcal{X}_5$ 和 $\mathcal{R}_9$，图 4.3 展示了它们的增量代价。首先初始化一个 4 维的零向量 $\mathcal{V} = (0,0,0,0)$ 并构建一个有 4 个节点 $\langle v_1, \nabla_0^1 = 0 \rangle$，$\langle v_2, \nabla_0^2 = 1 \rangle$，$\langle v_3, \nabla_0^3 = 1 \rangle$ 和 $\langle v_4, \nabla_0^4 = 3 \rangle$ 的小顶堆，如图 4.4 所示。注意在图中，对于所有的节点都省略了 $v_i$。接下来对堆做 $H(l,s)+1 = 4$ 次弹出操作。对于第一次弹出操作，得到节点 $\langle v_1, \nabla_0^1 \rangle$。因此对 $v_1$ 增加 1，所以 $\mathcal{V} = (1,0,0,0)$。因为 $v_1 = 1$，所以将新的节点 $\langle v_1, \nabla_1^1 = 0 \rangle$ 压入堆中。然后，对于第二次弹出操作，得到新压入的节点 $\langle v_1, \nabla_1^1 \rangle$。因此对 $v_1$ 增加 1，所以 $\mathcal{V} = (2,0,0,0)$。因为 $v_i = 2$，所

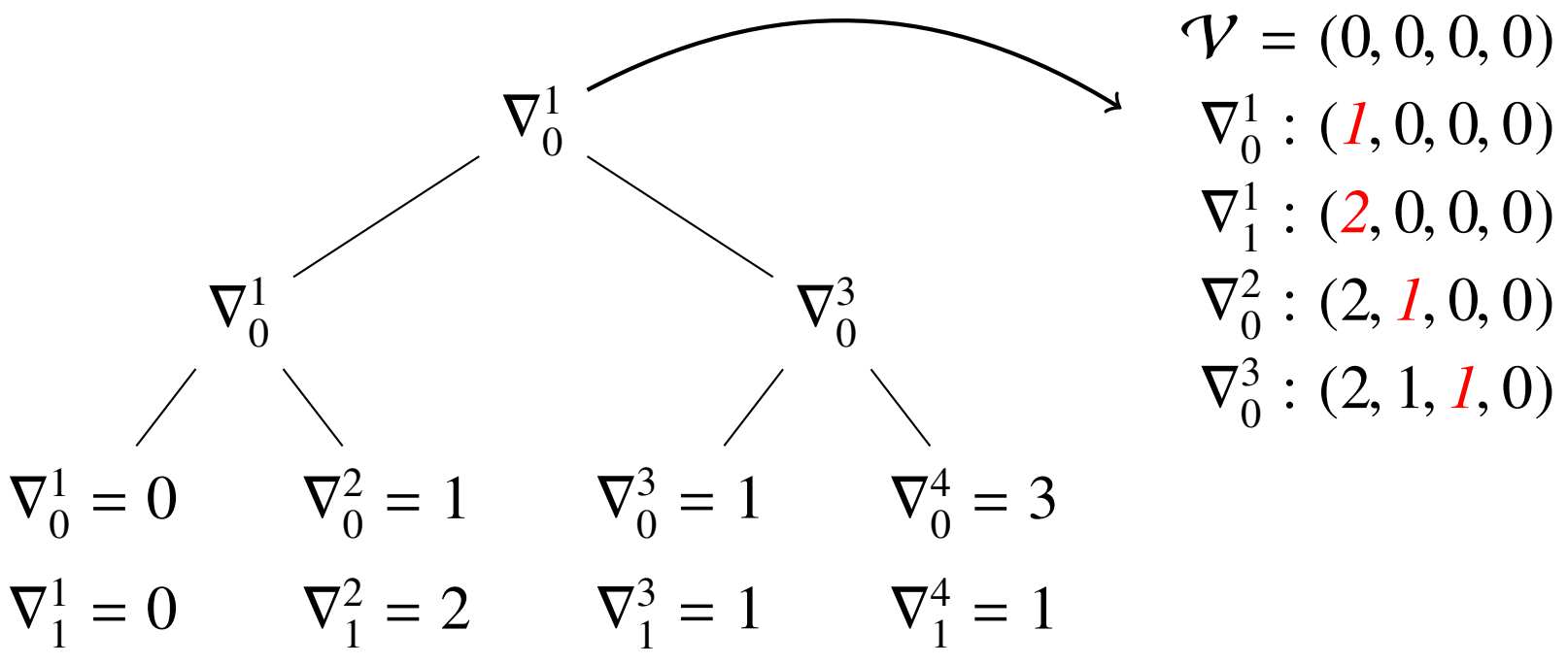

图 4.4　一个贪心选择分配策略的例子

以不压入新的节点。对于第三次弹出操作[①]，得到节点 $\langle v_2, \nabla^2_0 \rangle$。因此对 $v_2$ 增加 1，所以 $\mathcal{V} = (2, 1, 0, 0)$。因为 $v_2 = 1$，将新的节点 $\langle v_2, \nabla^2_1 \rangle$ 压入堆中。对于第四次弹出操作，得到 $\langle v_3, \nabla^3_0 \rangle$ 并对 $v_3$ 增加 1。因此有 $\mathcal{V} = (2, 1, 1, 0)$。最后返回这个向量作为选择的分配策略 $\mathcal{V}^s_l$，它的代价是 3。

引理 4.6 证明了基于堆的贪心算法选择的分配策略的代价不超过最优分配策略的代价的 2 倍。

**引理** 4.6：　基于堆得贪心算法是最优分配选择问题的近似比为 2 的算法。

**证明** 假设贪心分配策略是 $\mathcal{V}^s_l = (v_1, v_2, \cdots, v_m)$，最优分配策略是 $\mathcal{P}^s_l = (p_1, p_2, \cdots, p_m)$。首先，根据贪心分配策略 $\mathcal{V}^s_l$ 中每个维度 $v_i$ 的值把它们分成三组，其中 $1 \leq i \leq H_l + 1$。对于所有的 $v_i = 1$，可以证明 $p_i \neq 2$；对于所有 $v_i = 0$，可以证明 $p_i \neq 1$。这是因为否则的话，通过替换 $p_i$ 的值，总可以找到一个比 $\mathcal{P}^s_l$ 更优的分配策略。假设 $z_1$ 和 $z_2$ 是最优分配策略中选择的两个特征的代价，那么贪心分配策略中存在两个元素 $x$ 和 $y$ 不在最优分配策略中，因此有 $z_1 > x$ 以及 $z_2 > y$，也就是说 $2z_1 + 2z_2 > 2 * z_1 > x + y$。令 $M$ 为除了这四个特征的其他代价，有最优分配的代价 $OPT = M + z1 + z2 > M + 1/2(x + y) > 1/2(M + x + y) = 1/2 * GREEDY$，其中 $GREEDY$ 是贪心分配策略的代价。引理得证。 □

**时间复杂度**：　对于一个大小为 $s$ 的集合和一个分组 $\mathcal{R}_l$，贪心分配策略选择算法的时间复杂度是 $O((H(l, s) + 1) \log(H_l + 1)) = O(s \log s)$。然而，根据长度过滤，需要为这个集合探测 $s - \delta s + 1$ 个组并执行贪心分配策略选择算法。因此对于一个大小为 $s$ 的集合，为其选择分配策略的时间复杂度是 $O(s^2 \log s)$。这个复杂度仍然很高昂。接下来提出一个多长度分组机制来减少每个集合需要探测的组的数量。

① 注意在增量代价相同时，可以随意的选择其中一个。

### 4.5.2　多长度分组机制

如果把所有大小在 $[\frac{l_{\min}}{\delta^{k-1}}, \frac{l_{\min}}{\delta^k})$ 范围中的集合分成一个组 $\mathcal{R}_k^\delta$，其中 $_{\min}$ 是 $\mathcal{R}$ 中最小的集合大小，那么对任意集合 $\mathcal{X}$ 最多只需要探测 2 个不同的组。这是因为根据长度过滤，与 $\mathcal{X}$ 近似的集合的大小一定在 $[\delta s, s]$ 内，其中 $s = |\mathcal{X}|$。不失一般性的，假设 $\mathcal{X} \in \mathcal{R}_k^\delta$，即 $\frac{l_{\min}}{\delta^{k-1}} \le s < \frac{l_{\min}}{\delta^k}$。那么有 $\frac{l_{\min}}{\delta^{k-2}} \le \delta s < \frac{l_{\min}}{\delta^{k-1}}$，这表明所有与它近似的集合都在组 $\mathcal{R}_{k-1}^\delta$ 和组 $\mathcal{R}_k^\delta$ 中。因此对于集合 $\mathcal{X}$，只需要探测这两个组来找到所有与它近似的组。接下来形式化这个想法。

令 $l_{k+1} = \frac{l_k}{\alpha} + 1$，其中 $l_1 = l_{\min}$，$\alpha \in [\frac{1}{2}, 1]$ (之后将讨论如何设置 $\alpha$)。多长度分组机制把 $\mathcal{R}$ 中所有大小在 $[l_k, \frac{l_k}{\alpha}]$ 范围中的集合聚集为一个组 $\mathcal{R}_k^\alpha$。接下来把多长度分组机制融合进基于删集的框架中。

首先展示如何构建倒排索引 $\mathcal{I}$ 和 $\mathcal{D}$。对于组 $\mathcal{R}_k^\alpha$，平均的把全集 $\mathcal{U}$ 划分为 $m = H_{l_k} + 1$ 个不相交的片段并依照这个全集划分把组中的集合 $\mathcal{X} \in \mathcal{R}_k^\alpha$ 划分为 $m$ 个片段。对于每个片段 $\mathcal{X}^i$，把 $\mathcal{X}.id$ 附加到倒排列表 $\mathcal{I}_{l_k}^i[\mathcal{X}^i]$ 和 $\mathcal{D}_{l_k}^i[\mathsf{del}(\mathcal{X}^i)]$ 的末尾，其中 $1 \le i \le m$。然后利用这些索引来寻找近似集合元组。对于每个大小为 $s$ 的集合 $\mathcal{Y}$，探测所有大小范围 $[l_k, \frac{l_k}{\alpha}]$ 与 $[\delta s, s]$ 有交集的组 $\mathcal{R}_k^\alpha$。这表明这个组 $\mathcal{R}_k^\alpha$ 中可能包含与集合 $\mathcal{X}$ 近似的集合。对于每个这样的组 $\mathcal{R}_k^\alpha$，根据全集子集 $\mathcal{U}_{l_k}$ 把集合 $\mathcal{Y}$ 划分为 $m = H_{l_k} + 1$ 个片段。然后调用贪心分配策略选择算法或者最优分配选择算法来选择一个分配策略 $\mathcal{V} = (v_1, v_2, \ldots, v_m)$，其中 $m = H_{l_k} + 1$。注意，对于贪心分配策略选择，需要对堆做 $H(\frac{l_k}{\alpha}, s) + 1$ 次弹出操作，对于最优分配策略选择，代价二维表和分配二维表的列数是 $H(\frac{l_k}{\alpha}, s) + 1$。这是因为组 $\mathcal{R}_k^\alpha$ 包含大小最高达 $\frac{l_k}{\alpha}$ 的集合，所以选择的分配策略 $\mathcal{V}$ 需要满足 $\sum_{i=1}^m v_i = H(\frac{l_k}{\alpha}, s) + 1$。之后根据分配 $\mathcal{V}$ 可以使用与基于删集的框架一样的算法来寻找近似集合元组。

**例** 4.5：　考虑包含四种长度（$7, 8, 9, 10$）的集合的数据集，并且假设阈值为 $\delta = 0.7$。对于一个大小为 10 的集合，如果设置 $\alpha = 1$，那么它需要探测 4 个组：$\mathcal{R}_1^1, \mathcal{R}_2^1, \mathcal{R}_3^1$ 和 $\mathcal{R}_4^1$。这些组的长度范围分别是 $[7, 7], [8, 8], [9, 9]$ 和 $[10, 10]$。假如设置 $\alpha = 0.7$，那么它只需要探测 1 个组 $\mathcal{R}_1^{0.7}$，其长度范围是 $[7, 10]$。如果设置 $\alpha = 0.8$，那么需要探测两个组 $\mathcal{R}_1^{0.8}$ 和 $\mathcal{R}_2^{0.8}$，其长度范围分别为 $[7, 8]$ 和 $[9, 10]$。

下面讨论如何设置 $\alpha$。显然，$\alpha$ 不能大于 1，因为如果 $\alpha > 1$，组 $\mathcal{R}_k^\alpha$ 中集合的大小范围 $[l_k, \frac{l_k}{\alpha}]$ 是空。实际上当 $\alpha = 1$ 时，多长度分组机制就退化为之前的普通的单长渡分组机制（为每个长度建立一个分组）。而且 $\alpha$ 应该要大于 0。然而，$\alpha$ 不能无限的接近 0。这是因为当 $\alpha$ 接近 0 的时候，一个组中集合的大小范围会变得非常宽阔，从而对于这个组中大小非常大的集合，片段的数目将不够多，导致漏掉结果。

表 4.2　基于划分的框架下的参数

| Sim | $H(l,s)$ | $H_l$ | 长度过滤边界 |
| --- | --- | --- | --- |
| Jaccard | $\lfloor\frac{1-\delta}{1+\delta}(s+l)\rfloor$ | $\lfloor\frac{1-\delta}{\delta}l\rfloor$ | $[l\delta, \frac{l}{\delta}]$ |
| Cosine | $\lfloor s+l-2\delta\sqrt{sl}\rfloor$ | $\lfloor\frac{1-\delta^2}{\delta^2}l\rfloor$ | $[\delta^2 l, \frac{l}{\delta^2}]$ |
| Dice | $\lfloor(1-\delta)(s+l)\rfloor$ | $\lfloor 2\frac{1-\delta}{\delta}l\rfloor$ | $[\frac{\delta}{2-\delta}l, \frac{2-\delta}{\delta}l]$ |

正式地，考虑大小为 $s$ 的集合以及该集合探测的一个组 $\mathcal{R}_k^\alpha$。一方面 $[s\delta, s]$ 和 $[l_k, \frac{l_k}{\alpha}]$ 需要有交集，即 $s \geq l_k$ 并且 $s\delta \leq \frac{l_k}{\alpha}$。因此有 $l_k \leq s \leq \frac{l_k}{\alpha\delta}$。另一方面，分配策略 $\mathcal{V} = (v_1, v_2, \ldots, v_m)$ 满足 $\sum_{i=1}^m v_i = H(\frac{l_k}{\alpha}, s) + 1$ 和 $v_i \in \{0, 1, 2\}$，其中 $m = H_{l_k} + 1$。因为 $v_i \leq 2$，所以有 $\sum_{i=1}^m v_i \leq 2m$。因此有 $H(\frac{l_k}{\alpha}, s) + 1 \leq 2(H_{l_k} + 1)$。为了让这个不等式对于任何 $s$ 和 $l_k$ 在 $l_k \leq s \leq \frac{l_k}{\alpha\delta}$ 的限制下都成立，需要让 $\alpha \geq \frac{1}{2}$，因此 $\alpha$ 的值域是 $[\frac{1}{2}, 1]$。类似地，对于基于划分的框架，其总是选择 $\mathbf{1}^m$ 作为分配策略，可以推导出在该框架下 $\alpha$ 的值域为 $[1, 1]$，这表明多长度分组机制不能被应用到基于划分的框架中。

接下来分析在多长度分组机制下，每个集合需要探测的组的数量。考虑大小为 $s$ 的集合，需要探测所有的大小范围与 $[\delta s, s]$ 有交集的组。在这些被探测的组中，假设 $\mathcal{R}_t^\alpha$ 是其中第一个，$\mathcal{R}_{t'}^\alpha$ 是其中最后一个。因此被探测的组的数量是 $t' - t + 1$。接下来推导 $t' - t + 1$ 的上界。根据多长度分组机制，有 $l_k = \frac{l_{k-1}}{\alpha} + 1 = \frac{l_{k-2}}{\alpha^2} + \frac{1}{\alpha} + 1 = \cdots = \frac{l_{min}}{\alpha^{k-1}} + \frac{\alpha^{2-k} - \alpha}{1-\alpha}$。稍加变化，有 $k = -\log_\alpha \frac{(1-\alpha)l_k + 1}{(1-\alpha)(l_{\min} + \alpha)}$，它是随着 $l_k$ 的减少而单调递减的。另一方面因为组内集合的大小范围 $[l_k, \frac{l_k}{\alpha}]$ 与 $[\delta s, s]$ 相交，所以有 $l_k \leq s \leq \frac{l_k}{\alpha\delta}$，也就是说 $\alpha\delta s \leq l_k \leq s$。因此，被探测的组的数量 $t' - t + 1 \leq \log_\alpha \frac{(1-\alpha)\delta s + 1}{(1-\alpha)s + 1} + 1 \leq \log_\alpha \delta + 1$。

**时间复杂度**：因为每个集合需要探测的组的数量不超过 $\log_\alpha \delta + 1$，为一个大小为 $s$ 的集合选择分配策略的时间复杂度是 $O(s \log s \log_\alpha \delta)$。因为总是可以设置 $\alpha = \delta$，所以当 $\delta \geq 0.5$ 时，时间复杂度变为 $O(s \log s)$，这在实际中基本上总是可以满足的①。

## 4.6　进一步讨论

### 4.6.1　支持其他近似函数

在上文提出的技术中，只有三个参数依赖于具体的近似函数，它们是 $H(l, s)$，$H_l$ 以及长度过滤的边界。与 Jaccard Similarity 下的推导类似，也可以在 Cosine Similarity 和 Dice Similarity 下推导出这些参数，表4.2展示了这些参数。根据这些参数，可以很容易扩展上文中的技术来支持其他两个近似函数。

① 在实际中，$\delta$ 通常被设置为 0.8 到 1 之间。

### 4.6.2　支持异连接

给定两个数据集 $\mathcal{R}$ 和 $\mathcal{S}$，首先为其中一个数据集中的片段和片段的 1-删集创建倒排列表，例如 $\mathcal{R}$。然后对于另一个数据中的每个集合，例如 $\mathcal{S}$，可以使用这些索引来找到 $\mathcal{R}$ 中所有与它近似的集合。可以利用与自连接相同的方法来达到这个目：使用分配策略选择算法为这个集合选择一个分配，根据这个分配策略来探测倒排索引从而得到候选结果，并验证这些结果以得到 $\mathcal{R}$ 中所有与该集合近似的集合。也可以应用多长度分组机制来构建索引从而高效的寻找近似元组。

### 4.6.3　支持 MapReduce 和 Spark 框架

上文提出的技术可以很容易的扩展到 MapReduce 和 Spark 框架上。由于 MapReduce 算法都可以在 Spark 上运行，下面基于 MapReduce 介绍如何调整上文提到的技术，之后在 Spark 上进行实验（第 4.7 节）验证。

首先调整基于划分的框架来运行在 MapReduce 上，这只需要一个 MapReduce 回合。对于每个大小为 $l$ 的集合 $\mathcal{X}$，首先把它划分为 $H_l+1$ 个**索引**片段（用来构建倒排索引）。然后对于每个 $s \in [l*\delta, l]$，把 $\mathcal{X}$ 划分为 $H_s+1$ 个**探测**片段用来探测倒排索引从而找到与 $\mathcal{X}$ 近似的集合。给定两个集合 $\mathcal{X}$ 和 $\mathcal{Y}$，不失一般性的，假设 $l=|\mathcal{X}| \le s=|\mathcal{Y}|$。因为 $\mathcal{X}$ 和 $\mathcal{Y}$ 近似仅当它们在相同的划分方法 $\mathcal{U}_l$ 下共享一个共同片段 $\mathcal{X}^i=\mathcal{Y}^i$，其中 $\mathcal{X}^i$ 是 $\mathcal{X}$ 的一个探测片段，$\mathcal{Y}^i$ 是 $\mathcal{Y}$ 的一个索引片段，所以在 MapReduce 框架下，可以使用组合 $(\mathcal{X}^i, i, l)$ 作为键，并使用组合 $(\mathcal{X}, flag)$ 作为值，其中 $flag=0$ 表示该键值对来自一个索引片段，$flag=1$ 表示该键值对来自一个探测片段。接下来介绍如何实现 map 函数和 reduce 函数。

在 map 阶段，对于每个集合 $\mathcal{X}$，首先把它划分为 $H_l+1$ 个索引片段并为每个片段 $\mathcal{X}^i$ 发射一个键值对 $\langle(\mathcal{X}^i, i, l), (\mathcal{X}, 0)\rangle$。对于每个 $s \in [l*\delta, l]$，把 $\mathcal{X}$ 划分为 $H_s+1$ 个探测片段并为每个片段 $\mathcal{X}^i$ 发射一个键值对 $\langle(\mathcal{X}^i, i, s), (\mathcal{X}, 1)\rangle$[①]。

在 reduce 阶段，对于每个键 $(\mathcal{X}^i, i, l)$，令 $list(\mathcal{X}, flag)$ 表示这个键对应的值列表，这些值共享相同的片段 $\mathcal{X}^i$。首先根据 $flag$ 的值可以把这个列表切分为两个列表 $\mathcal{L}_0$ 和 $\mathcal{L}_1$，其中 $\mathcal{L}_0$ 包含其中所有的以 $\mathcal{X}^i$ 作为索引片段的集合，$\mathcal{L}_1$ 包含所有的以 $\mathcal{X}^i$ 为探测片段的集合。$\mathcal{L}_0 \times \mathcal{L}_1$ 中每个元组 $\langle\mathcal{X}, \mathcal{Y}\rangle$ 都是一个近似元组。因为一些集合元组共享多个相同的片段，所以在不同的 reduce 中可能存在重复的候选元组。为了避免验证重复的候选元组，对于键 $(\mathcal{X}^i, i, l)$ 中每个候选元组 $(\mathcal{X}, \mathcal{Y})$，首先检查它们是否在第 $i$ 个片段之前共享相同的片段，如果是的话（即 $\exists j < i$ 使得

① 为了保证正确性，对于不同集合中相同的元素，需要把它放在同一个片段中。为了达到这个目标，对每个元素 $e \in \mathcal{X}$，把它放在第 $((hash(e) \mod m)+1)$ 个片段中，其中 $hash(e)$ 把一个元素 $e$ 映射为一个整数。

**Algorithm 4.5**: 运行在 MapReduce 下的基于划分的框架

```
Map(𝒳)
    Generate_key_value_pair(l, 𝒳, 0);
    for s ∈ [δ ∗ l, l] do
        Generate_key_value_pair(s, 𝒳, 1)

Function Generate_key_value_pair(l, 𝒳, flag)
    令 m = H_l + 1;
    foreach e ∈ 𝒳 do
        把 e 附加到 𝒳^{i=(hash(e) mod m)+1} 的末尾
    foreach 1 ≤ i ≤ m do
        发射键值对 ⟨(𝒳^i, i, l), (𝒳, flag)⟩

Reduce(⟨(𝒳^i, i, l), list(𝒳, flag)⟩)
    根据 flag 的值把列表 list(𝒳, flag) 分为两个列表 ℒ_0 和 ℒ_1;
    foreach ⟨𝒳, 𝒴⟩ ∈ ℒ_0 × ℒ_1 do
        if 对每个 1 ≤ j < i 都有 𝒳.id ≤ 𝒴.id 和 𝒳^j ≠ 𝒴^j then
            if JAC(𝒳, 𝒴) ≥ δ then
                输出键值对 ⟨𝒳, 𝒴⟩
```

图 4.5　运行在 MapReduce 下的基于划分的框架

$\mathcal{X}^j = \mathcal{Y}^j$)，不需要验证这个候选元组。否则，验证这个元组。通过这种方式，只需要验证每个候选元组一次。

因此可以使用基于划分的框架在 MapReduce 上解决集合近似连接问题，算法 4.5 展示了其伪代码。

第 4.5节和第4.4节提出的优化技术也可以扩展以应用于 MapReduce 算法中。不同于把相同大小的集合分为同一个组，多长度分组机制把所有大小在一个范围 $[l_k, \frac{l_k}{\alpha}]$ 内的集合分为同一个组。为了在 MapReduce 算法中融合多长度分组机制，对于组 $\mathcal{R}_k$ 中每个大小为 $l$ 的集合 $\mathcal{X}$，可以在键值对中按照如下方式把 $l$ 替换为 $l_k$。首先把 $\mathcal{X}$ 划分为 $H_{l_k}+1$ 个索引片段并为每个片段 $\mathcal{X}^i$ 发射一个键值对 $\langle(\mathcal{X}^i, i, l_k), (\mathcal{X}, 0)\rangle$。然后对于每个其长度范围 $[l*\delta, l]$ 与 $[l_{k'}, \frac{l_{k'}}{\alpha}]$ 有交集的组 $\mathcal{R}_{k'}$，把 $\mathcal{X}$ 划分为 $H_{l_{k'}}+1$ 个探测片段并为每个片段 $\mathcal{X}^i$ 发射一个键值对 $\langle(\mathcal{X}^i, i, l_{k'}), (\mathcal{X}, 1)\rangle$。剩下的部分和算法 4.5 是一样的。

为了把基于 1-删集的技术融合进 MapReduce 算法中，首先采用一个 MapReduce 阶段来统计所有片段和 1-删集的频率。然后在 MapReduce 的第二回合的 map

表 4.3　实验数据集详细信息

| 数据集 $\mathcal{R}$ | $\lvert\mathcal{R}\rvert$ | $l_{\min}$ | $l_{\max}$ | $\bar{l}$ | $\lvert\mathcal{U}\rvert$ |
|---|---|---|---|---|---|
| DBLP | 873,524 | 6 | 1,538 | 94.1 | 44,798 |
| Tweets | 2,000,000 | 1 | 70 | 21.6 | 1,713,437 |
| MovieLens | 138,493 | 20 | 9,254 | 144.4 | 26,744 |

阶段，对于每个集合首先采用贪心分配策略选择算法来选择一个分配策略从而选择探测片段和探测 1-删集并为它们发射键值对。注意需要在键中加入另外一个布尔变量来区别片段和 1-删集。然后对于集合 $\mathcal{X}$ 的每个片段和 1-删集，发射一个**索引**键值对①。第二回合的其余部分和算法 4.5 是一样的。

## 4.7　实验

下面通过多个实验来评估本章提出的方法。实验的目的是评估本章提出的方法的性能并与现有最好的方法 PPJoin+[35] 和 AdaptJoin[39] 比较效率和可扩展性。

尽管还有其他集合近似连接算法，例如 `All-Pair`[29]，`Flamingo`[42] `PartEnum`[28] 和 `BayesLSH`[94]，但是之前的研究[39,93] 表明它们都没有前两个方法效率高。因此这里只报告本章提出的方法与 PPJoin+ 和 AdaptJoin 的比较结果。实验中运行的这两个方法是从其作者处取得的源代码。

实验中所有的算法都是用 C++ 实现的，源代码使用 GCC 4.8.2 编译并开始-O3 优化选项。所有的实验都是在运行着 Ubuntu 系统的服务器上进行。该服务器配置有 24 个 Intel Xeon X5670 2.93GHz 处理器和 64 GB 内存。

**实验数据集**: 在实验中使用三个真实的数据集。1) DBLP 是一个计算机科学引文信息数据集。其中每条刊物信息都切分成 3-gram，用它们的 3-gram 集合作为输入。2) Tweets 是一个推文数据集，对每条推文按照空格分词，用形成的单词集合作为输入。3) MovieLens 是一个电影评分数据集。用每个用户评过分的电影集合作为输入。表4.3展示了实验数据集的具体信息。

### 4.7.1　评估分配选择算法

本节评估不同的分配策略选择算法。实验实现了以下三种算法 (1) 基于划分的框架, Hybrid, 对于任何大小为 $s$ 的集合和组 $\mathcal{R}_l$，它总是选择分配策略 $\mathcal{V}_l^s = \mathbf{1}^m$，其中 $m = H_l + 1$。(2) 最优分配策略下的动态规划算法，用 Optimal 代表 (3) 基于堆的贪心分配策略选择算法，用 Greedy 代表。注意，对于这三个方法实验都没有使用

① 注意需要分发一个包含片段和 1-删集频率的文件到每个 map 来选择分配，如果这个文件大小大于可用内存，可以移除其中一些低频的条目，并对它们使用一个默认的频率 0。

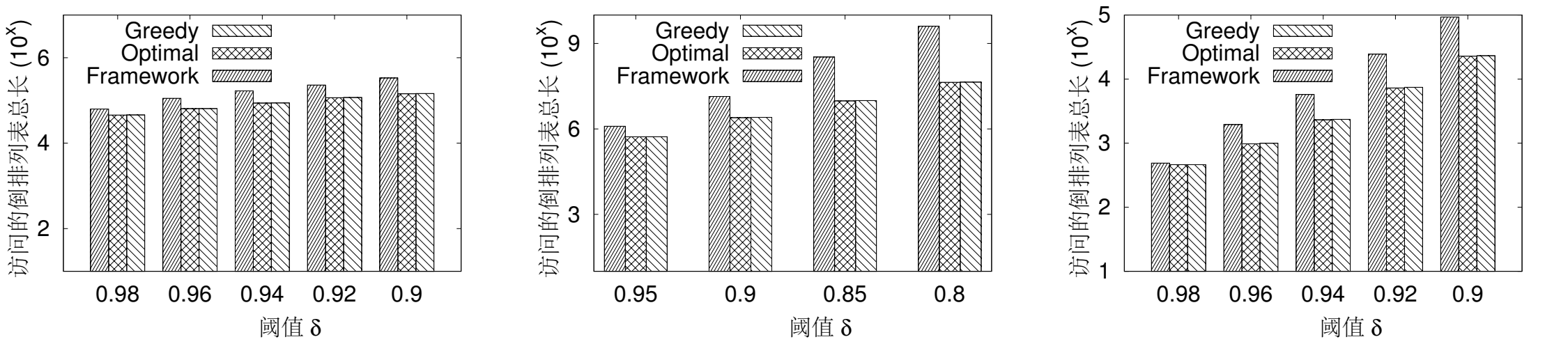

(a) DBLP (平均大小: 94)　(b) Tweets (平均大小: 22)　(c) MovieLens (平均大小: 144)

图 4.6　评估不同的分配策略选择方法：访问的倒排列表总长度

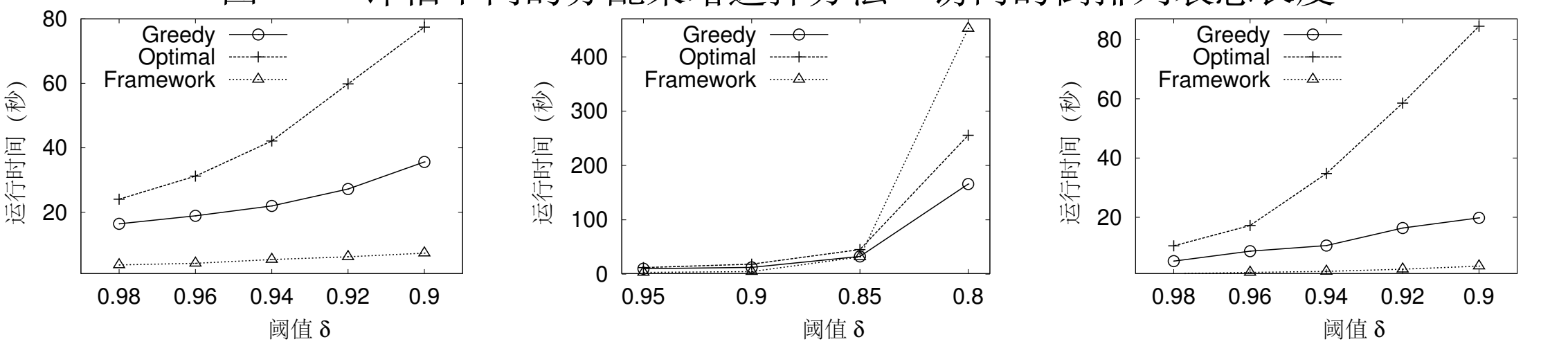

(a) DBLP (平均大小: 94)　(b) Tweets (平均大小: 22)　(c) MovieLens (平均大小: 144)

图 4.7　评估不同的分配策略选择方法：效率

多长度分组机制。根据第 4.3，4.4.2和 4.5.1节，它们为长度为 $s$ 的集合选择分配策略的时间复杂度分别是 $O(s^2)$[①]，$O(s^3)$ 和 $O(s^2 \log s)$。因为 Optimal 和 Greedy 的运行时间随着阈值的降低迅速地增加，所以本节实验只报告较大阈值下的结果从而能够在合理的时间范围内完成实验。实验首先比较了不同算法访问的倒排列表的长度总和。图4.6展示了该实验的结果，注意其中 y 轴是呈对数增长的。

可以看到 Optimal 访问的倒排列表总长最小，Greedy 访问的倒排列表总长比 Optimal 的稍大。Hybrid 访问的倒排列表总长是最大的，比 Greedy 和 Optimal 的大 10 到 100 倍，并且随着阈值的降低这个情况变得更差。例如，在 Tweets 数据集下，当 $\delta = 0.8$ 时，Hybrid, Greedy 和 Optimal 访问的倒排列表总长分别是 $4 * 10^9$, $4.4 * 10^7$ 和 $4.3 * 10^7$，这是因为 Hybrid 没有使用 1-删集，因此只能使用固定的分配策略。而 Optimal 选择了最优的分配策略使得访问的倒排列表总长最小。Greedy 是近似比为 2 的贪心算法，它访问的倒排列表总长最多是最优分配策略的 2 倍。所以该实验结果与之前的理论分析是一致的。

实验还比较了这三个方法的运行时间。图 4.7 展示了这一实验结果。可以看到 Greedy 总是比 Optimal 效率更高，这是因为 Greedy 的时间复杂度比 Optimal 的低而访问的倒排列表总长与其基本相同。当阈值大于 0.9 时，Hybrid 比 Greedy 和 Optimal 效率更高，这和分配策略选择的时间复杂度是一致的，因为当阈值较大而没有多长度分组机制时，对于 Hybrid，Optimal 和 Greedy，分配策略选择的时间复杂度分别是($O(s^2)$ , $O(s^3)$ 和 $O(s^2 \log s)$，它们主导了算法的运行时间。然而，当阈

① 集合划分的代价也包括在该复杂度内。

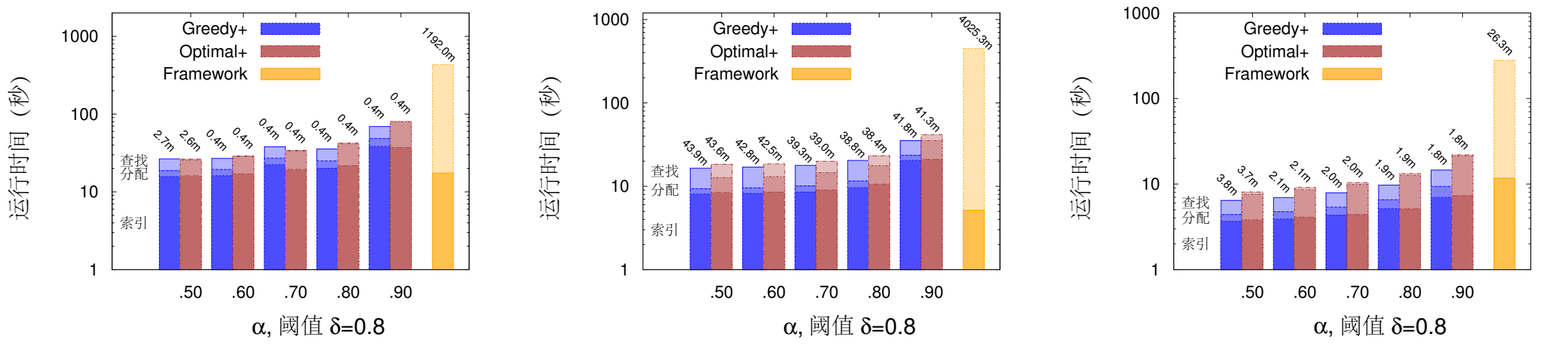

(a) DBLP (平均大小: 94)　(b) Tweet (平均大小: 22)　(c) MovieLens (平均大小: 144)

图 4.8　调整 $\alpha$ 的值来评估多长度分组机制.

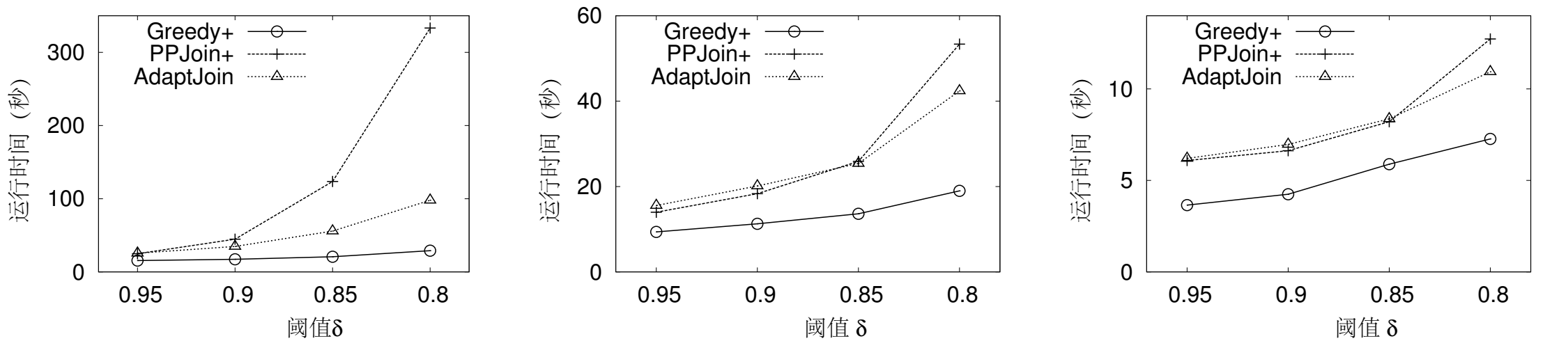

(a) DBLP (平均大小: 94)　(b) Tweet (平均大小: 22)　(c) MovieLens (平均大小: 144)

图 4.9　与现有方法的比较：效率

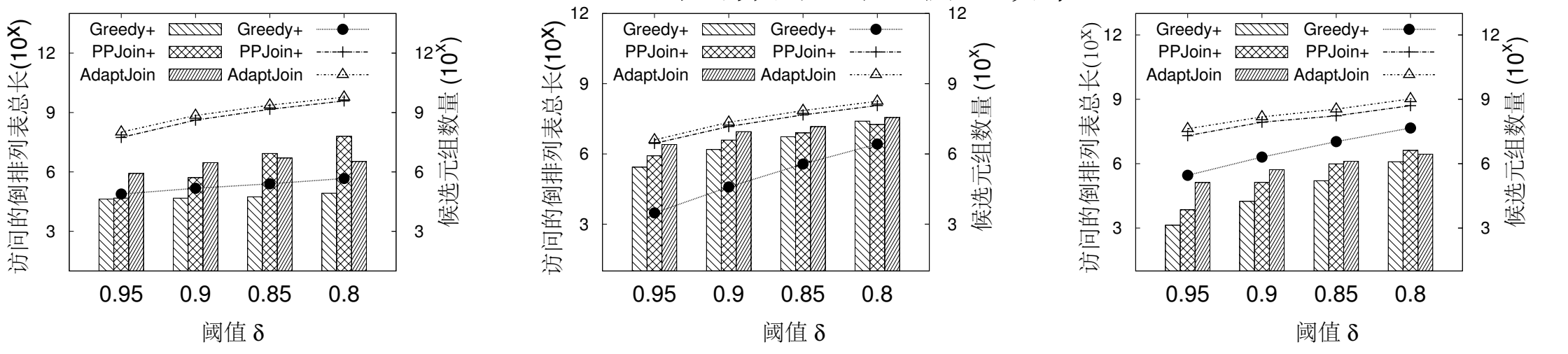

(a) DBLP (平均大小: 94)　(b) Tweet (平均大小: 22)　(c) MovieLens (平均大小: 144)

图 4.10　与现有方法的比较：访问的倒排列表总长以及候选集大小

值变得小于 0.9 时，Greedy 和 Optimal 的效率要高于 Hybrid。这是因为随着阈值的降低，Hybrid 访问的倒排列表总长以及候选元组数量爆炸式增长，从而导致验证步骤主导了运行时间。

### 4.7.2　评估多长度分组机制

本节评估多长度分组机制，实验在 0.5 到 0.9 之间调整参数 $\alpha$ 的值并报告 Greedy+ (配备了多长度分组机制的 Greedy 算法）和 Optimal+ (配备了多长度分组机制的 Optimal 算法）的运行时间以及访问的倒排列表总长。这里没有报告 $\alpha > 0.9$ 时的结果，因为在这种情况下算法运行时间相当长以至于不能在合理的时间内完成。实验还报告了 Hybrid 的结果。图4.8 展示了该实验结果，图中柱子上面的数字代表它们访问的倒排列表总长。实验中把运行时间分为三个部分：建立索引的时间，选择分配策略的时间以及寻找近似元组的时间。注意预处理时间（获取全集，对所有集合排序以及分组）仅占少于 0.5 秒。可以看到随着 $\alpha$ 的增加，Greedy+ 和

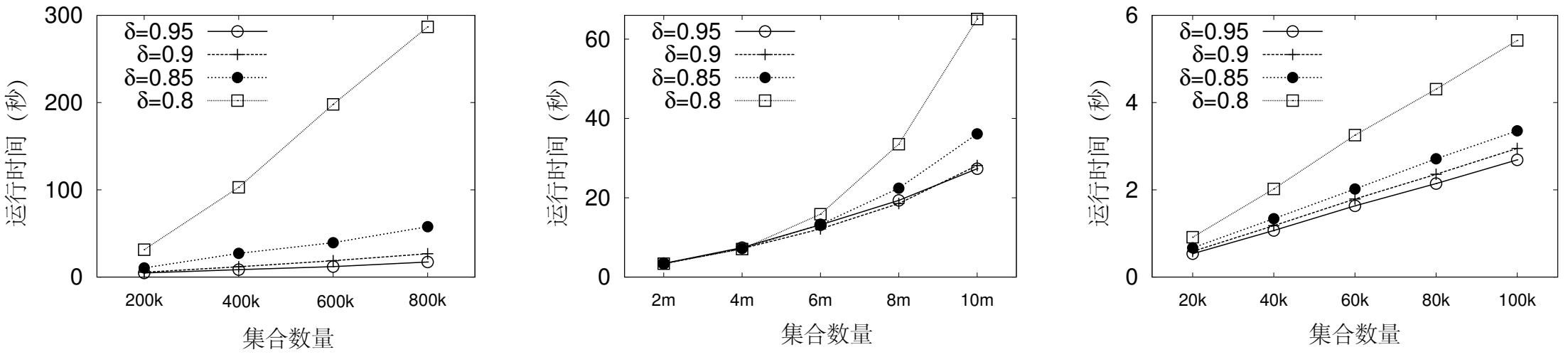

(a) Jaccard on DBLP (平均大小: 94)　(b) Jaccard on Tweet (平均大小: 22)　(c) Cosine on MovieLens (平均大小: 144)

图 4.11　不同近似函数下算法的可扩展性

Optimal+ 的运行时间都增加了。除此之外，它们都比 Hybrid 的效率高。对于同样的 $\alpha$ 值，Greedy+ 的效率比 Optimal+ 的高。例如在 MovieLens 数据集下，在 $\delta = 0.8$ 时，当 $\alpha$ 的值从 0.5 变化到 0.9 时，Greedy+ 和 Optimal+ 的运行时间分别是 6 秒，7 秒，8 秒，10 秒，14 秒，8 秒，9 秒，10 秒，14 秒，23 秒，而 Hybrid 的运行时间是 286 秒。这是因为 Greedy+ 的分配策略选择算法的时间复杂度（$O(s \log s \log_\alpha \delta)$）比 Optimal+ 的（$O(s^2 \log_\alpha \delta)$）低。另外 Greedy+ 访问的倒排列表总长最多是 Optimal+ 的两倍。此外，当 $\alpha$ 小的时候，它们的分配策略选取算法的时间复杂度都不高于 Hybrid 的（$O(s^2)$）而且它们访问的倒排列表总长比 Hybrid 所访问的要短很多，如图中所示。Greedy+ 和 Optimal+ 访问倒排列表总长随着 $\alpha$ 的增加仅仅略微的变化。这是因为它们可以为不同的集合和组选择一个好的甚至一个最优的分配策略。因为 Greedy+ 的效率最好，接下来的实验中只报告 Greedy+ 算法的结果。

### 4.7.3　与现有的方法比较

本节把本章提出的算法 Greedy+ 与现有最好的算法 PPJoin+ 以及 AdaptJoin 作比较。Greedy+ 采用基于删集的框架并采用了多长度分组机制。实验报告了运行时间、访问的倒排列表总长以及候选集大小。图 4.9 和 4.10 展示了该实验结果。可以看到，在效率实验中，本章提出的算法 Greedy+ 取得了最好的性能，它的效率是其他现有算法的 2 到 5 倍。例如，在 DBLP 数据集下，采用 Jaccard Similarity 并设置阈值 $\delta = 0.8$ 时，Greedy+ 的运行时间是 20 秒而 PPJoin+ 和 AdaptJoin 分别消耗了 330 秒和 100 秒。这是因为 Greedy+ 比 AdaptJoin 和 PPJoin+ 有更有效的剪枝技术，而且贪心分配策略选择以及多长度分组机制把分配选择的时间复杂度降低到 $O(s \log s)$。图 4.10 中的柱子展示了候选集的大小，点线展示了访问的倒排列表总长。可以看到 Greedy+ 访问的倒排列表总长最短，候选集大小也是最小。AdaptJoin 访问的倒排列表总长最长但是得到的候选集大小比 PPJoin+ 小。这是因为 PPJoin+ 和 AdaptJoin 都利用了前缀过滤框架，其中前缀中每个元素都对应一个倒排列表，而这些倒排列表是 Greedy+ 中倒排列表的超集。因此它访问的倒排列

表总长比 PPJoin+ 和 AdaptJoin 短很多。此外，通过多长度分组机制，Greedy+ 可以进一步减少访问的倒排列表总长。AdaptJoin 在前缀中囊括了一些额外的元素来加强剪枝不近似元组，因此访它问的倒排列表总长更长而得到的候选集大小更小。

实验还比较了索引大小。在 MovieLens 数据集下，当阈值 $\delta = 0.8$ 时，AdaptJoin, PPJoin+ 和 Greedy+ 的索引大小分别是 80.1 MB, 16.1 MB 和 87.2 MB。Greedy+ 的索引大小最大，因为它需要把片段和 1-删集都插入索引中。PPJoin+ 的索引大小最小，因为它只需要插入把前缀中的元素插入到索引中，而 AdaptJoin 需要把集合中每个元素都插入到索引当中。

### 4.7.4　可扩展性

实验通过调整数据集中集合的数量来评估 Greedy+ 在两种不同近似函数 Jaccard Similarity 和 Cosine Similarity 下的可扩展性。图 4.11 展示了该实验的结果。可以看到 Greedy+ 的可扩展性非常好。例如，在 Tweets 数据集下，使用 Jaccard Similarity 并设置阈值 $\delta = 0.9$ 时，实验把数据集中集合的数量从两百万调整到一千万，算法的运行时间分别是 3 秒，7 秒，12 秒，18 秒和 28 秒。这得益于本节提出的有效的过滤条件以及分配策略选择算法。

### 4.7.5　评估异连接

如第 4.6.2 节讨论的，可以扩展 Greedy+ 来支持异连接问题。本节评估异连接算法的可扩展性。对于每个数据集，实验随机的把它分为两个部分，并在这两部分上运行异连接算法。图 4.12 展示了异连接算法的运行时间和候选元组的数量。可以看到，异连接算法仍然能够取得很高的性能。这是因为异连接算法对其中一个数据集构建索引并使用该索引从另一个数据集中寻找近似元组。

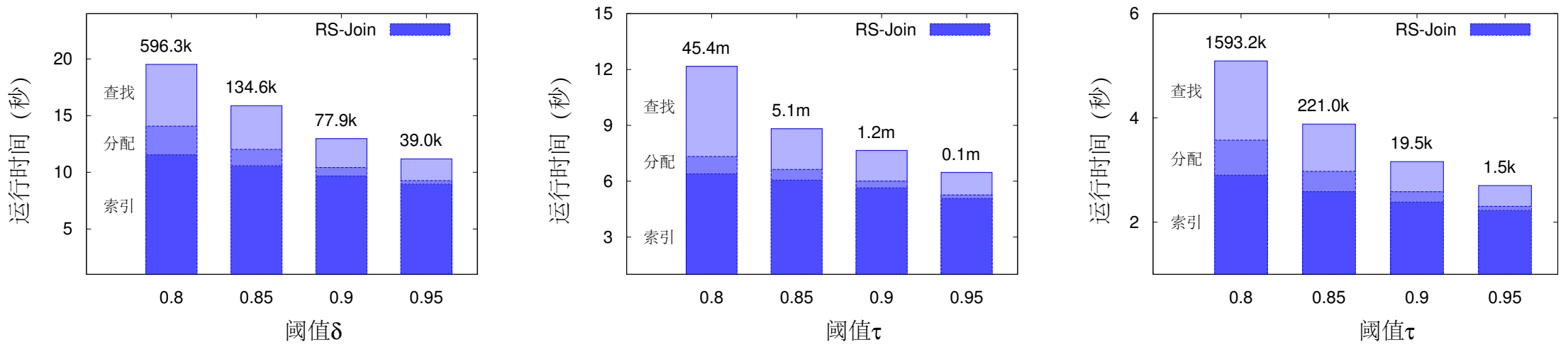

(a) RS-Join(DBLP)　(b) RS-Join(Tweet)　(c) RS-Join(MovieLens)

图 4.12　评估异连接

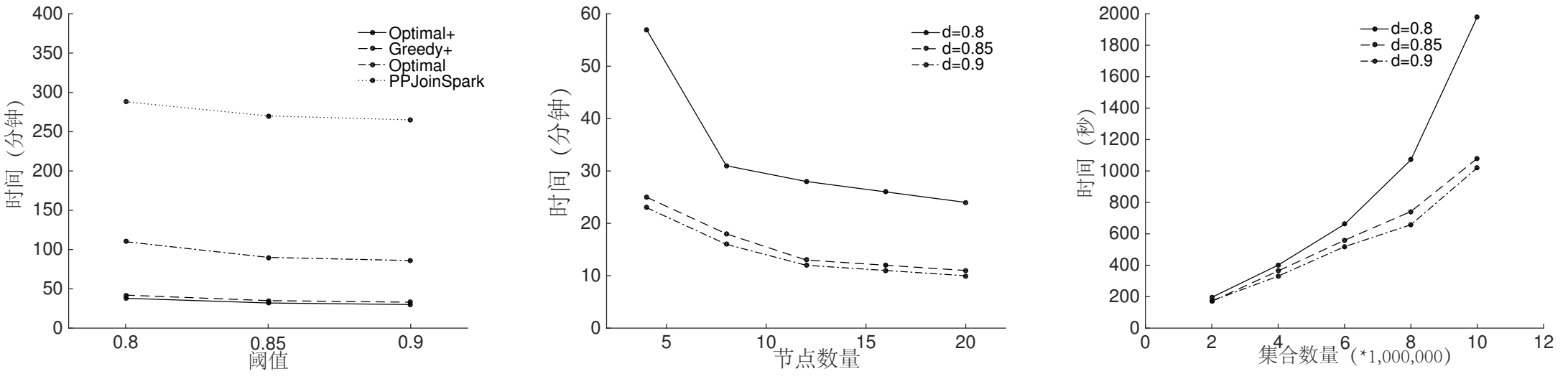

(a) 不同技术比较　(b) 加速比　(c) 可扩展性

图 4.13　Spark 下的算法性能

### 4.7.6　评估算法在 Spark 下的性能

本节评估了 MapReduce 和 Spark 框架下算法。实验中采用了一台拥有 24 个处理器的服务器，服务器运行 Ubuntu 12.04LTS 操作系统，安装有 Spark 1.6。所有的代码使用 Scala 实现。实验使用包含一千万条推文的 Tweets 数据集。实验首先衡量了论文中提出贪心分配策略选取技术和多长度分组技术在 Spark 下的性能，实验还与另外一个基于 MapReduce 的方法 `PPJoin-Spark` 进行了比较[84]，实验中从作者主页下载了 `PPJoin-Spark` 的源代码并稍加改动以在 Spark 上运行。图 4.13(a) 展示了该实验结果。可以看到，论文提出的方法效率远高于现有的方法 `PPJoin-Spark`，它比现有的方法快 5-10 倍。这是因为 `PPJoin-Spark` 是基于前缀过滤的算法，基于前缀过滤的算法的特征是集合中的单个元素，这将导致数据倾斜问题：一些 reduce 中的列表将非常长。而本文提出的方法的特征是多个元素组成的集合，它能够缓解这一问题。另外可以看到 Optimal+ 的效率比 Optimal 要高，这是因为多长度分组技术可以减少键值对的数量。Optimal+ 和 Greedy+ 的性能相差不大，这是因为在 Spark 下，选取分配策略是并行执行的，它不再主导整个算法的运行时间了。此外，实验还评估了 Optimal+ 在 Spark 下的加速比和可扩展性，图 4.13(b) 和 (c) 展示了这一实验结果，可以看到 Optimal+ 在 Spark 下也取得了不错的加速比和可扩展性。注意随着数据集中集合数量的增加，近似元组的数量是成平方增长的，所以其可扩展性和加速比很难达到线性变化。

## 4.8　本章小结

本章提出了一个基于划分的框架来解决集合近似连接问题。论文设计了一个基于全集的划分方式来把集合划分为多个片段。基于划分的方法为这些片段生成 1-删集并为这些片段和其 1-删集构建倒排索引。对每个集合，基于划分的方法访问它的一部分子集与 1-删集所对应的倒排列表来寻找近似集合。论文研究了如何衡量不同的子集与 1-删集的分配策略并设计了一个动态规划算法来选择最优的分配

策略。为了加速分配策略选择过程，论文设计了一个近似比为 2 的贪心算法来选取分配策略。论文还提出了多长度分组机制来进一步加速分配策略选择。这些技术把为一个大小为 $s$ 的集合选择分配策略的时间复杂度从 $O(s^3)$ 降低到 $O(s\log s)$。论文还讨论了如何在扩展基于划分的方法来支持异连接以及在 MapReduce 和 Spark 上运行。实验表明，论文提出的方法远远好于现有的方法。

# 第 5 章　基于关键前缀过滤的近似检索方法

## 5.1　引言

近似检索从一个给定的数据集中找到与一个给定的查询相似的所有数据。它有很多重要应用，例如查询推荐、抄袭检测和信息检索等。本章提出了一个关键前缀过滤的技术，它是基于前缀过滤的一个改进技术。由于前缀过滤技术在集合近似检索问题表现较好，所以本章主要致力于利用关键前缀过滤技术解决基于序列相似性的近似检索问题。

现有工作通常采用过滤加验证的框架[41–46]，在过滤阶段，它们利用过滤条件来剪枝掉大量与查询串不相似的数据字符串从而得到剩下的候选字符串。在验证阶段，它们验证候选字符串从而生成最后的结果。现有的方法主要致力于设计快速有效的过滤技术。其中前缀过滤[29,36] 是一个主流的过滤技术，它为每个字符串产生一个前缀并保证如果两个字符串是近似的，它们的前缀一定共享一个相同的特征。利用这个性质它可以过滤大量与查询串不相似的字符串。最近，有研究者提出基于前缀过滤的改进技术，例如基于位置的前缀过滤[30,35]，自适应前缀过滤[39] 以及非对称过滤[47]。

现有的基于前缀过滤的技术主要有两个问题。一方面它们往往选取了很多不必要的特征。值得注意的是特征的数量对剪枝能力和过滤代价有很重要的影响。一方面，减少特征的数量将会减少两个字符串之间特征相同的概率，因此剪枝能力将变强。另一方面，减少特征的数量将减少检查两个字符串是否共享相同的特征的比对代价。过滤代价也会减少。因此，减少特征的数量不但可以增加剪枝能力还能降低过滤代价。另一方面，现有的技术对连续的编辑错误效果较差。因为连续的编辑错误往往只毁掉少部分的特征，这导致现有的技术不能过滤掉由于连续编辑错误而不相似的字符串。

为了解决以上问题，论文提出了一个关键前缀过滤技术，它可以显著的减少特征的数量并仍能够找到所有的结果（第 5.3 节）。基于关键前缀过滤技术，论文提出了近似检索框架来解决字符串近似检索问题（第 5.4 节）。因为可能存在多个不同的策略来产生关键前缀，论文提出了一个动态规划算法来选择高质量的关键前缀特征，从而剪枝掉大量因为离散的编辑错误而与查询串不相似的字符串（第 5.5 节）。论文还提出了一个对齐过滤技术，它通过考虑特征之间的对齐情况来进一步剪枝掉大量因为连续的编辑错误而与查询串不相似的字符串（第 5.6 节）。

表 5.1　数据集 $\mathcal{R}$，$q = 2$，$\tau = 2$.

(a) 数据集 $\mathcal{R}$

| id | 字符串 | $\|q(r)\|$ | 包含位置信息的前缀$q$-gram 集合 pre($r$) |
|---|---|---|---|
| $r_1$ | imyouteca | 8 | $\{\langle im, 1\rangle, \langle my, 2\rangle, \langle te, 6\rangle, \langle ca, 8\rangle, \langle yo, 3\rangle\}$ |
| $r_2$ | ubuntucom | 8 | $\{\langle bu, 2\rangle, \langle un, 3\rangle, \langle nt, 4\rangle, \langle uc, 6\rangle, \langle om, 8\rangle\}$ |
| $r_3$ | utubbecou | 8 | $\{\langle bb, 4\rangle, \langle ou, 8\rangle, \langle ut, 1\rangle, \langle ub, 3\rangle, \langle co, 7\rangle\}$ |
| $r_4$ | youtbecom | 8 | $\{\langle tb, 4\rangle, \langle om, 8\rangle, \langle yo, 1\rangle, \langle ou, 2\rangle, \langle ut, 3\rangle\}$ |
| $r_5$ | yoytubeca | 8 | $\{\langle oy, 2\rangle, \langle yt, 3\rangle, \langle ca, 8\rangle, \langle yo, 1\rangle, \langle ub, 5\rangle\}$ |

(b) 全局顺序

| 〈 顺序：$q$-gram，频率 〉 |
|---|
| $\langle 1: im, 1\rangle\langle 2: my, 1\rangle\langle 3: te, 1\rangle\langle 4: bu, 1\rangle\langle 5: un, 1\rangle\langle 6: nt, 1\rangle\langle 7: uc, 1\rangle$ |
| $\langle 8: bb, 1\rangle\langle 9: tb, 1\rangle\langle 10: oy, 1\rangle\langle 11: yt, 1\rangle\langle 12: ca, 2\rangle\langle 13: om, 2\rangle\langle 14: yo, 3\rangle$ |
| $\langle 15: ou, 3\rangle\langle 16: ut, 3\rangle\langle 17: ub, 3\rangle\langle 18: co, 3\rangle\langle 19: tu, 3\rangle\langle 20: be, 3\rangle\langle 21: ec, 4\rangle$ |

(c) 查询串 $s$

| | |
|---|---|
| $s$ =yotubecom | pre($s$) = $\{\langle ot, 2\rangle, \langle om, 8\rangle, \langle yo, 1\rangle, \langle ub, 4\rangle, \langle co, 7\rangle\}$ |

## 5.2　预备知识

### 5.2.1　问题定义

字符串近似检索问题的输入是一个字符串集合，一个查询字符串，一个近似函数以及一个近似阈值，它输出数据集中所有与查询串近似的字符串。有很多近似函数被提出来来衡量两个字符串的近似度[87]，本章致力于研究编辑距离函数。两个字符串的编辑距离是把一个字符串转化为另一个字符串所需要的最少的编辑操作数目，用 ED($r, s$) 来表示。接下来给出编辑距离限制下近似检索问题的正式定义。

**定义** 5.1 (**字符串近似检索**)：　给定一个字符串集合 $\mathcal{R}$，一个查询字符串 $s$ 以及一个编辑距离阈值 $\tau$，编辑距离限制下的字符串检索从 $\mathcal{R}$ 中找到所有满足 ED($r, s$) $\leq \tau$ 的字符串 $r$。

例如考虑表5.1中包含 5 个字符串的数据集 $\mathcal{R}$ 和查询串 $s$ = “yotubecom’’。假如设定编辑距离阈值 $\tau = 2$，那么 $r_4$ = “*youtbecom*’’ 是一个结果，因为 ED(“yotubecom”, “*youtbecom*”) = $2 \leq \tau$。

### 5.2.2　基于$q$-gram 的过滤技术

$q$-gram：本节介绍基于$q$-gram 的过滤技术。一个字符串的$q$-gram 是它长度为 $q$ 的子字符串。例如，表 5.1中的查询串 $s$ = “youtbecom’’ 的 2−gram 集合是 {yo，ou，

ut，tb，be，ec，co，om}。一个位置$q$-gram（positional $q$-gram ）是附上在字符串中起始位置的$q$-gram。例如 ⟨yo, 1⟩, ⟨ou, 2⟩, · · · , ⟨om, 8⟩ 都是 $s$ 的位置$q$-gram。在上下文清楚的情况下，不加区分的使用$q$-gram 和位置$q$-gram。

**清点过滤**（Count Filter）: 因为一个编辑操作最多摧毁 $\tau$ 个 $q$-gram, 所以如果两个字符串近似，那么它们需要共享足够多的$q$-gram。利用这个性质，现有的方法提出了清点过滤技术[42]: 给定两个字符串 $r$ 和 $s$, 如果它们共享小于 $\max(|r|, |s|) - q + 1 - q\tau$ 个共同$q$-gram，那么它们不可能相似，其中 $|r|$ 是 $r$ 的长度。

**前缀过滤**（Prefix Filter）: 对于每个字符串 $s$，前缀过滤首先对$q$-gram 按照全局顺序进行排序并选取前 $q\tau + 1$ 个$q$-gram 作为前缀。给定两个字符串 $r$ 和 $s$，根据清点过滤，如果它们的前缀没有共同$q$-gram，那么它们共享的$q$-gram 数量肯定少于 $\max(|r|, |s|) - q + 1 - q\tau$。利用这个性质，现有的方法使用前缀过滤来进行剪枝[30]: 给定两个字符串 $r$ 和 $s$，如果它们的前缀没有共同$q$-gram，它们不可能相似。

**失配过滤**（Mismatch Filter）: 失配过滤[30] 选择最短的前缀使得其中所有$q$-gram 能被 $\tau + 1$ 个编辑操作毁掉。显然，失配过滤的前缀长度是介于 $\tau + 1$ 和 $q\tau + 1$ 之间的。

**长度过滤**（Length Filter）: 如果两个字符串的长度之差大于 $\tau$，它们不可能近似。

**位置过滤**（Position Filter）: 当检查两个字符串的前缀是否共享相同的特征时，只需要检查它们位置之差在 $\tau$ 之内的特征。这是因为对于任何从一个字符串到另外一个字符串的变换，如果其中两个位置差大于 $\tau$ 的特征互相对齐，那么这个转换包含的编辑操作数目一定大于 $\tau$，也就是说该转换下这两个字符串不相似。

**与现有的过滤条件的对比**: Bayardo 等[29] 提出了一个利用前缀过滤技术框架来支持近似检索和近似连接问题。Xiao 等人[35] 提出了基于位置的前缀过滤和失配过滤[30] 来减少特征的数量。尽管本章利用了同样的过滤加验证框架来支持字符串近似检索，但是本章提出的关键前缀过滤技术可以进一步大幅减少特征的数量，而且本章提出的对齐过滤技术可以快速的检测连续编辑错误。

### 5.2.3 相关工作

**字符串近似检索**: 关于字符串近似检索问题, 有很多现有工作[41–46,99,100]。它们中大部分使用了基于特征的框架: 仅当字符串与查询串共享相同的特征时, 它们才可能相似。Li 等人[42] 提出使用$q$-gram 作为特征并研究了如何快速合并$q$-gram 的倒排列表以使提高查询效率。Zhang 等人[41] 也使用$q$-gram 作为特征并利用 $B^+$-tree 的索引结构来对$q$-gram 进行索引，他们的技术支持范围查询和排序查询。Li 等人[45] 提出使用变长的$q$-gram (VGram) 作为特征来支持字符串近似检索。Chaudhuri 等人[46] 提

表 5.2　记号表

| | |
|---|---|
| q($r$) | 字符串 $r$ 的$q$-gram 集合，其中$q$-gram 按全局顺序有序 |
| pre($r$) | $r$ 的前缀$q$-gram 集合（简称前缀），大小为 $q\tau + 1$ |
| suf($r$) | $r$ 的后缀$q$-gram 集合（简称后缀），大小为 $\|\text{q}(r)\| - \|\text{pre}(r)\|$ |
| piv($r$) | $r$ 的关键前缀$q$-gram 集合（简称关键前缀），<br>即 pre($r$) 中的 $\tau + 1$ 个不相交的$q$-gram |
| last(pre($r$)) | pre($r$) 中最后一个$q$-gram （全集顺序最大的) |
| last(piv($r$)) | piv($r$) 中最后一个$q$-gram （全集顺序最大的) |
| $\widehat{\text{pre}}(r)$ | 不包括 last(piv($r$)) 的 pre($r$) |

出了一个新的近似函数并设计了高效的查询算法来支持这个函数。Hadjieleftheriou 等人[43] 解决了检索中数据更新的问题。Deng 等人[44] 研究了使用 trie 索引结构来解决 top-k 近似查询问题。

**近似连接**： 字符串近似连接从两个数据集中找到所有的近似字符串元组，关于这个问题有很多现有工作[28,29,31,35,36,39,47,65,70,93,101,102]。Jiang 等人[93] 对这些算法做了一个实验分析。AllPair[29] 利用前缀过滤来解决近似连接问题。PPJoin+[35] 提出了基于位置的前缀过滤和基于位置的后缀过滤来改进 AllPair。ED-Join[30] 提出使用基于位置的失配过滤和基于内容的失配过滤来缩短前缀长度并检测连续的编辑错误。AdaptJoin[39] 提出了一个自适应前缀过滤来动态的选择前缀。PassJoin[65] 把字符串划分为固定数目的片段作为特征并研究了如何选取最少数量的子字符串以减少候选元组的数量。Qchunk[47] 把字符串划分为多个长度相同的 chunk 作为特征并用它们探测$q$-gram 索引来生成候选元祖。Vchunk[101] 通过使用不同长度的 chunk 来改进 Qchunk。PartEnum[28] 提出了一个模型来根据特征的数量估计候选元组数量。下文扩展这个模型来评估过滤技术的过滤代价和剪枝能力，并利用它来指导设计过滤技术。TrieJoin[31] 利用 trie 结构来支持字符串近似连接。Gravano 等人[70] 在数据库管理系统中实现了近似连接操作。

**近似查询补全**: 有很多工作研究了近似查询补全问题[103–105]。近似查询补全找到数据集中存在一个前缀与查询串近似的字符串。Ji 等[103] 和 Chaudhuri 等[104] 提出使用一个 trie 索引来处理这种查询。Xiao 等[105] 提出了一个基于删集的方法来解决近似查询补全问题。Kim 等[106] 解决了 top-k 近似子字符串匹配问题。

## 5.3　关键前缀过滤

本节首先提出一个新颖的关键前缀过滤（第 5.3.1 节），然后把它与现有最好的过滤技术作比较 (第 5.3.2节 )。

### 5.3.1　关键前缀过滤

给定两个字符串 $r$ 和 $s$，首先把它们切分为$q$-gram 集合：q($r$) 和 q($s$)。然后对 q($r$) 和 q($s$) 中的$q$-gram 按照全局顺序排序，例如$q$-gram 的频率升序。不失一般性的，首先假设给定的编辑距离阈值是 $\tau$，第 5.7 节中讨论如何支持不同阈值的查询。把 q($r$) 和 q($s$) 的长度为 $q\tau+1$ 的前缀分别记做 pre($r$) 和 pre($s$)(见表 5.2 )。如果 pre($r$) $\cap$ pre($s$) $= \phi$[29]，前缀过滤能够剪枝掉元组 $\langle r, s\rangle$。

值得注意的是，前缀中的一些$q$-gram 是不必要的，可以选择前缀的一个子集来做进一步的剪枝。显然，减少$q$-gram 的数量不但能够加强剪枝能力还能减少过滤代价。为了便于表述，首先介绍几个记号，如表 5.2 所示。用 suf($r$) 表示 $r$ 的大小为 |q($r$)| − |pre($r$)| 的后缀，即 suf($r$) = q($r$) − pre($r$)。两个$q$-gram 被称作不相交的如果它们没有交叠（即它们的起始位置相距不小于 $q$）。从 pre($r$) 中选择 $\tau+1$ 个不相交的$q$-gram，并将这些$q$-gram 组成的集合记作 piv($r$)。值得注意的是，在前缀中可能存在多种这样的的$q$-gram 集合。这里用 piv($r$) 来表示其中任何一个。本章第 5.5 节将证明对于任何字符串其前缀中一定存在 $\tau+1$ 个不相交的$q$-gram 并讨论如何选取其中最好的一个。显然 piv($r$) $\subseteq$ pre($r$)。下文把 pre($r$) 中的$q$-gram 称作前缀$q$-gram，把 piv($r$) 中的$q$-gram 称作关键$q$-gram。前缀是指所有前缀$q$-gram 组成集合，即 pre($r$)，关键前缀是指关键$q$-gram，即 piv($r$)。

**例** 5.1：　考虑表 5.1 中的查询串 $s$ 以及数据字符串 $r_1$，表5.1给出了$q$-gram 的全局顺序。根据这个全局顺序，有 q($r_1$) = {im,my,te,ca,yo,ou,ut,ec} 以及 q($s$) = {ot,om,yo,ub,co,tu,be,ec}。假设 $\tau = 2$，那么它的前缀长度是 $q\tau+1 = 5$，有 pre($r_1$) = {im,my,te,ca,yo}，pre($s$) = {ot,om,yo,ub,co}，suf($r_1$) = {ou,ut,ec} 以及 suf($s$) = {tu,be,ec}。这里随机的选择 $\tau+1$ 个不相交的$q$-gram 作为关键前缀 piv($r_1$) = {im,te,ca} 以及 piv($s$) = {ot,om,ub}。

**交叉前缀过滤**（Cross Prefix Filter）：对于任何字符串 $r$ 和 $s$，如果 pre($r$) $\cap$ piv($s$) $= \phi$ 并且 piv($r$) $\cap$ pre($s$) $= \phi$，那么 $r$ 和 $s$ 不可能近似。定理5.1 给出了这个观察的正确性证明，图5.1也展示了这个观察背后的基本想法。

**引理** 5.1：　如果字符串 $r$ 和 $s$ 相似，那么 piv($r$) $\cap$ pre($s$) $\neq \phi$ 或者 pre($r$) $\cap$ piv($s$) $\neq \phi$ 成立。

**证明**　首先证明 piv($r$) $\cap$ suf($s$) $= \phi$ 或者 piv($s$) $\cap$ suf($r$) $= \phi$ 成立，用反证法证明这点。假如 piv($r$) $\cap$ suf($s$) $\neq \phi$ 并且 piv($s$) $\cap$ suf($r$) $\neq \phi$，令 $x \in$ piv($r$) $\cap$ suf($s$) 以及 $y \in$ piv($s$) $\cap$ suf($r$)。一方面有 $x \in$ piv($r$) 和 $y \in$ suf($r$)。又因为$q$-gram 按照全局顺序

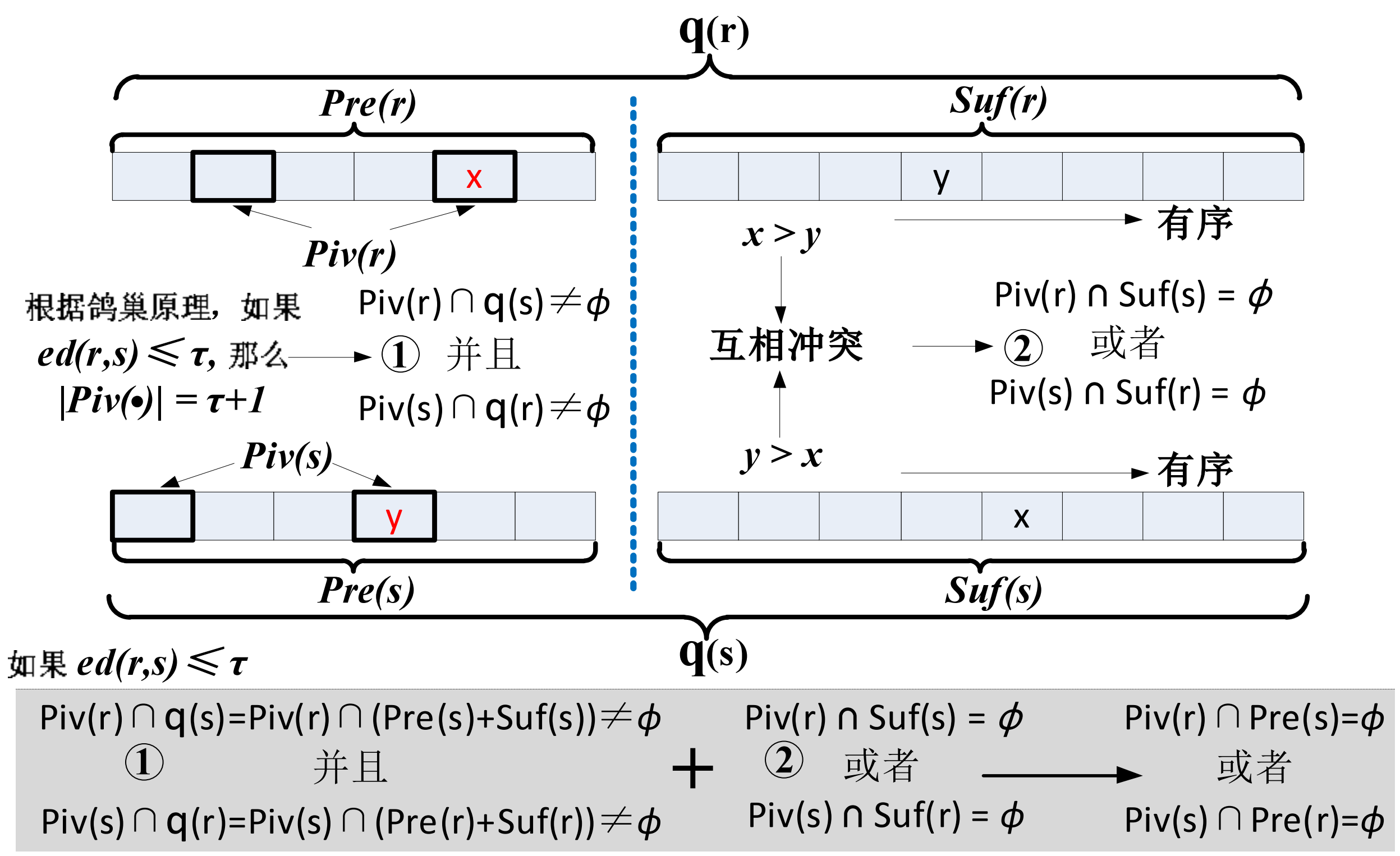

图 5.1　关键前缀过滤

排好序了，所以 $x < y$[①]。另一方面，有 $y \in \mathrm{piv}(s)$ 和 $x \in \mathrm{suf}(s)$，因此 $y < x$，这与之前的结论互相矛盾。因此 $\mathrm{piv}(r) \cap \mathrm{suf}(s) = \phi$ 或者 $\mathrm{piv}(s) \cap \mathrm{suf}(r) = \phi$ 成立。

其次，证明假如 $r$ 和 $s$ 相似，那么 $\mathrm{piv}(r) \cap \mathrm{pre}(s) + \mathrm{piv}(r) \cap \mathrm{suf}(s) \neq \phi$ 和 $\mathrm{piv}(s) \cap \mathrm{pre}(r) + \mathrm{piv}(s) \cap \mathrm{suf}(r) \neq \phi$ 成立。因为 $\mathrm{q}(s) = \mathrm{pre}(s) + \mathrm{suf}(s)$，$\mathrm{piv}(r) \cap \mathrm{q}(s) = \mathrm{piv}(r) \cap \mathrm{pre}(s) + \mathrm{piv}(r) \cap \mathrm{suf}(s)$，所以首先证明如果 $r$ 和 $s$ 相似，那么 $\mathrm{piv}(r) \cap \mathrm{q}(s) \neq \phi$，可以用反证法证明这点。假设 $\mathrm{piv}(r) \cap \mathrm{q}(s) = \phi$，因为 (1) 一个编辑操作最多改变 $\mathrm{piv}(r)$ 中一个不相交的 $q$-gram 以及 (2) $\mathrm{piv}(r)$ 包含 $\tau + 1$ 个不相交的 $q$-gram，所以把 $r$ 转化为 $s$ 需要最少 $\tau + 1$ 个编辑操作，这与 $r$ 和 $s$ 近似互相矛盾。因此 $\mathrm{piv}(r) \cap \mathrm{q}(s) \neq \phi$，即 $\mathrm{piv}(r) \cap \mathrm{pre}(s) + \mathrm{piv}(r) \cap \mathrm{suf}(s) \neq \phi$。类似地，可以证明 $\mathrm{piv}(s) \cap \mathrm{pre}(r) + \mathrm{piv}(s) \cap \mathrm{suf}(r) \neq \phi$。

根据第一个结论，有两种情况。(1) 第一种情况 $\mathrm{piv}(r) \cap \mathrm{suf}(s) = \phi$。因为 $\mathrm{piv}(r) \cap \mathrm{pre}(s) + \mathrm{piv}(r) \cap \mathrm{suf}(s) \neq \phi$，根据第二个结论有 $\mathrm{piv}(r) \cap \mathrm{pre}(s) \neq \phi$；(2) 第二种情况 $\mathrm{piv}(s) \cap \mathrm{suf}(r) = \phi$，因为 $\mathrm{piv}(s) \cap \mathrm{pre}(r) + \mathrm{piv}(s) \cap \mathrm{suf}(r) \neq \phi$，根据第二个结论有 $\mathrm{piv}(s) \cap \mathrm{pre}(r) \neq \phi$。无论哪种情况，都有 $\mathrm{piv}(r) \cap \mathrm{pre}(s) \neq \phi$ 或者 $\mathrm{pre}(r) \cap \mathrm{piv}(s) \neq \phi$ 成立，因此引理得证。　□

① 在制定全局顺序时，通过考虑 $q$-gram 的位置来避免 $x = y$ 的情况。

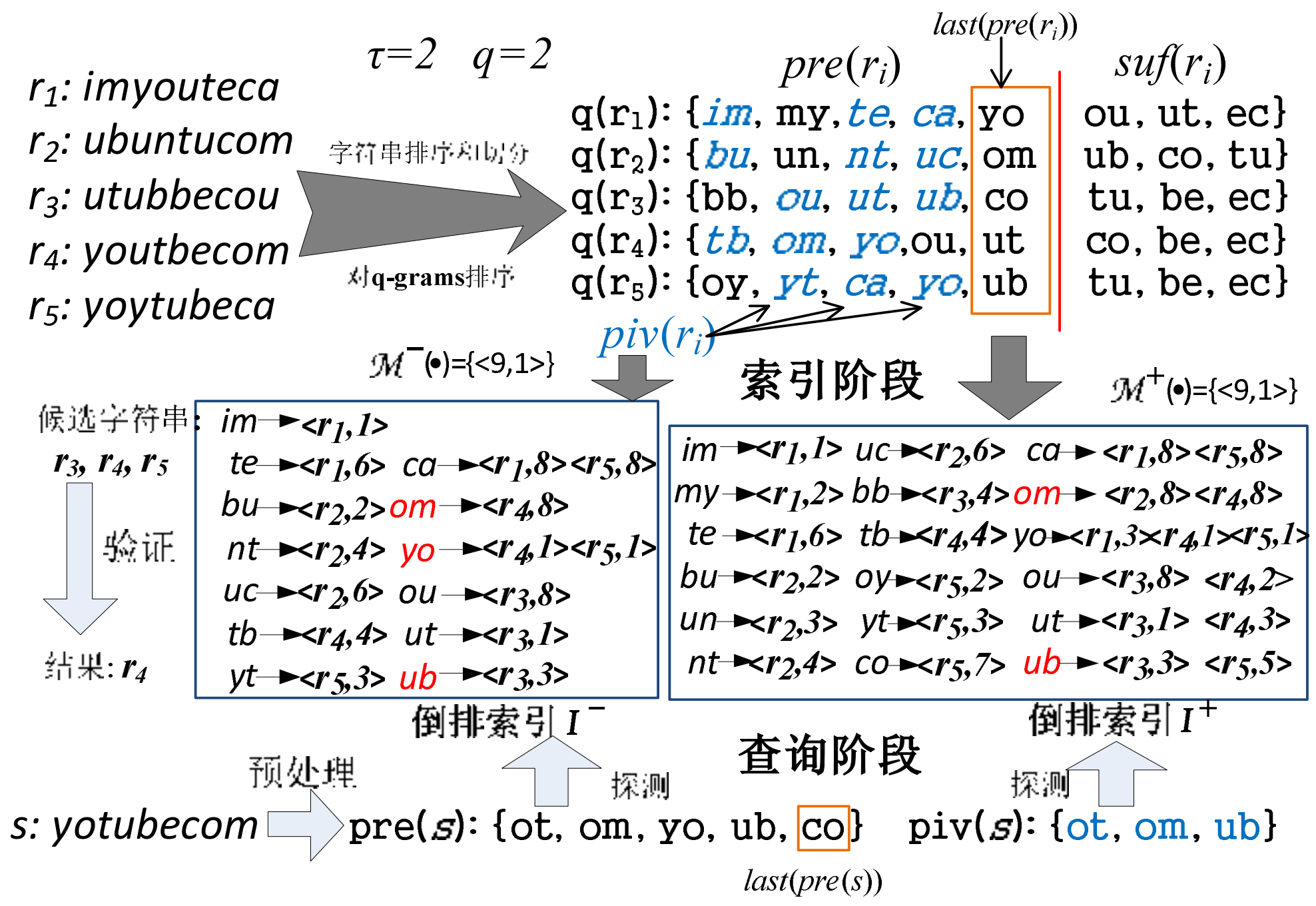

图 5.2　一个例子

考虑例5.1中的查询串 $s$ 和数据字符串 $r_1$，因为 $\text{pre}(r_1) \cap \text{pre}(s) = \{\text{yo}\} \neq \phi$，前缀过滤不能剪枝掉这个元组，然而，交叉前缀过滤能够剪枝掉这对不近似的元组，因为 $\text{pre}(r_1) \cap \text{piv}(s) = \phi$ 并且 $\text{piv}(r_1) \cap \text{pre}(s) = \phi$。

交叉前缀过滤需要检查两个交集 $\text{piv}(r) \cap \text{pre}(s)$ 和 $\text{pre}(r) \cap \text{piv}(s)$ 是否为空，因为仅当 $\text{piv}(r) \cap \text{pre}(s) = \phi$ 并且 $\text{piv}(s) \cap \text{pre}(r) = \phi$ 时，才能剪枝掉这对元组。接下来展示如何只检查其中一个交集。为了便于表述，将 $\text{pre}(r)$ 中最后一个$q$-gram 用 $\text{last}(\text{pre}(r))$ 来表示。例如，表 5.1 中 5 个字符串和查询串的最后一个$q$-gram 分别是 “yo”，“om”，“co”，“ut”，“ub” 和 “co”。根据 $\text{last}(\text{pre}(r))$ 和 $\text{last}(\text{pre}(s))$，下面提出一种新颖的关键前缀过滤并给出引理 5.2 来证明这个过滤技术的正确性。

**关键前缀过滤**：对于任何字符串 $r$ 和 $s$，有

第一种情况：当 $\text{last}(\text{pre}(r)) > \text{last}(\text{pre}(s))$ 时，如果 $\text{piv}(s) \cap \text{pre}(r) = \phi$，$r$ 和 $s$ 不可能相似；

第二种情况：当 $\text{last}(\text{pre}(r)) \leq \text{last}(\text{pre}(s))$ 时，如果 $\text{piv}(r) \cap \text{pre}(s) = \phi$，$r$ 和 $s$ 不可能相似。

**引理** 5.2：如果两个字符串 $r$ 和 $s$ 相似，那么有：

如果 $\text{last}(\text{pre}(r)) > \text{last}(\text{pre}(s))$，那么 $\text{piv}(s) \cap \text{pre}(r) \neq \phi$;

如果 $\text{last}(\text{pre}(r)) \leq \text{last}(\text{pre}(s))$，那么 $\text{piv}(r) \cap \text{pre}(s) \neq \phi$。

**证明** 考虑第一种情况，$\text{last}(\text{pre}(r)) > \text{last}(\text{pre}(s))$。首先证明 $\text{piv}(s) \cap \text{suf}(r) = \phi$。对于任何$q$-gram $g \in \text{piv}(s)$ 和 $g' \in \text{suf}(r)$，因为所有$q$-gram 都排好序了，所以有 $g \le \text{last}(\text{pre}(s)) < \text{last}(\text{pre}(r)) < g'$，即 $g \neq g'$。因此 $\text{piv}(s) \cap \text{suf}(r) = \phi$。此外，如果 $r$ 和 $s$ 相似，那么根据引理 5.1的证明，有 $\text{piv}(s) \cap \text{q}(r) = \text{piv}(s) \cap \text{pre}(r) + \text{piv}(s) \cap \text{suf}(r) \neq \phi$。因此 $\text{piv}(s) \cap \text{pre}(r) \neq \phi$。

类似地，对第二种情况，$\text{last}(\text{pre}(r)) \le \text{last}(\text{pre}(s))$，可以证明 $\text{piv}(r) \cap \text{pre}(s) \neq \phi$。引理得证。 □

**例** 5.2：　考虑表 5.1 中的查询串 $s$ 和数据字符串 $r_2$，有 $\text{pre}(r_2) = \{\text{bu,un,nt,uc,om}\}$。假设选择了 $\text{piv}(r_2) = \{\text{bu,nt,uc}\}$，因为 $\text{piv}(s) \cap \text{pre}(r_2) = \{\text{om}\} \neq \phi$，基于引理 5.1的过滤技术不能剪枝掉这对元组。然而基于引理 5.2 的关键前缀过滤可以剪枝掉这对不相似的元组，因为 $\text{last}(\text{pre}(r_2)) = \text{om} < \text{last}(\text{pre}(s)) = \text{co}$ 并且 $\text{piv}(r_2) \cap \text{pre}(s) = \phi$。

此外，可以集成失配过滤[30] 的技术来缩短前缀 $\text{pre}(r)$ 的长度。值得注意的是本章提出的关键前缀过滤在这个更短的前缀上仍然能够正常运行。论文在第 5.5 节证明即使是在这个更短的前缀中，也一定存在至少 $\tau + 1$ 个不相交的$q$-gram。因为失配过滤和本章提出的方法是正交独立的，所以不详细讨论进一步的细节。

### 5.3.2　与现有最好方法的比较

本节将关键前缀过滤技术和现有最好的基于$q$-gram 的过滤技术作比较。基于$q$-gram 的过滤的目标是尽可能多的剪枝掉不相似的字符串。它们需要分别从这两个字符串中选取一个$q$-gram 集合作为特征，用 $\text{Sig}(r)$ 和 $\text{Sig}(s)$ 来表示这两个集合。它们比较这两个$q$-gram 集合并检查这两个集合是否共享相同的特征。在设计过滤技术时，剪枝能力和过滤代价是两个非常重要的因素。

首先考虑剪枝能力。一方面，这两个特征集合大小的乘积越小，它们共享前缀的几率就越小，因此它们的剪枝能力就越强。另一方面，因为匹配的$q$-gram 的数量不能超过这两个字符串中较小的特征集合的大小 $\min(|\text{Sig}(r)|, |\text{Sig}(s)|)$，因此可以使用这两个特征集合大小的乘积和这两个特征集合中较小的大小来评估剪枝能力。接下来评估过滤代价，因为$q$-gram 集合是排好序的，如果不存在索引，可以使用 merge-join 算法来寻找匹配的$q$-gram，所以过滤代价取决于这两个字符串的特征集合大小之和：$|\text{Sig}(r)| + |\text{Sig}(s)|$。如果其中一个特征集合存在散列索引（如第 5.4 节所述，通常都会建立索引来提高性能），那么可以使用一个基于探测的方法来检测另一个特征集合中的$q$-gram 是否出现在散列索引中。在这种情况下，过滤代价取决于探测特征集合的大小。不失一般性的，使用 $\text{Sig}(r)$ 来代表探测特征集

表 5.3　与现有最好的过滤技术的比较

(a) 特征集合大小

| 过滤技术 | $\vert\mathsf{Sig}(r)\vert$ | $\vert\mathsf{Sig}(s)\vert$ |
|---|---|---|
| **前缀过滤** | $q\tau+1$ | $q\tau+1$ |
| **位置失配过滤** | $\tau+1$ to $q\tau+1$ | $\tau+1$ to $q\tau+1$ |
| Qchunk-IndexGram | $l-(\lceil\frac{l-\tau}{q}\rceil-\tau)+1$ | $\tau+1$ |
| Qchunk-IndexChunk | $\tau+1$ | $l-(\lceil\frac{l-\tau}{q}\rceil-\tau)+1$ |
| **关键前缀过滤** | $\tau+1$ to $q\tau+1$ | $\tau+1$ |

(b) 剪枝能力与过滤代价的影响因素

| 剪枝能力 | 过滤代价 |
|---|---|
| 依赖于 $\vert\mathsf{Sig}(r)\vert\times\vert\mathsf{Sig}(s)\vert$ | 有索引时依赖于 $\vert\mathsf{Sig}(r)\vert$ (探测集合的大小) |
| 以及 $\min(\vert\mathsf{Sig}(r)\vert,\vert\mathsf{Sig}(s)\vert)$ | 无索引时依赖于 $\vert\mathsf{Sig}(r)\vert+\vert\mathsf{Sig}(s)\vert$ |

合，表 5.3 比较了现有最好的基于$q$-gram 的过滤技术的剪枝能力和过滤代价，它们分别被应用在 AllPair，ED-Join，Qchunk-IndexChunk 和 Qchunk-IndexGram 算法中。

给定两个字符串 $r$ 和 $s$，前缀过滤[29] 需要从两个字符串各选取 $q\tau+1$ 个$q$-gram 作为特征。因此，两个特征集合大小的乘积，最小值以及总和分别是 $(q\tau+1)^2$，$q\tau+1$ 和 $2*(q\tau+1)$，同时探测集合的大小是 $q\tau+1$。失配过滤[30] 通过缩短前缀长度来改进前缀过滤，其前缀长度在 $[\tau+1,q\tau+1]$ 范围内，因此特征集合大小的乘积、最小值和总和分别在 $[(\tau+1)^2,q\tau+1)^2]$，$[\tau+1,q\tau+1]$ 和 $[2*(\tau+1),2*(q\tau+1)]$ 范围内，其探测集合大小在 $[\tau+1q\tau+1]$ 范围内。对于 Qchunk，它需要选择 $\tau+1$ 个 chunk 和 $l-(\lceil(l-\tau)/q)\rceil-\tau)+1$ 个$q$-gram 作为特征，其中 $l$ 是字符串长度。因此特征集合大小的乘积、最小值以及总和分别是 $(l-(\lceil(l-\tau)/q)\rceil-\tau)+1+\tau+1)*(\tau+1)$，$\tau+1$ 和 $(l-(\lceil(l-\tau)/q)\rceil-\tau)+1+\tau+1)+(\tau+1)$。在 Qchunk 中存在两种索引方式，IndexChunk 和 IndexingGram，它们的探测集合大小分别是 $l-(\lceil(l-\tau)/q)\rceil-\tau)+1$ 和 $\tau+1$。注意，Qchunk 中选择的$q$-gram 数量非常多并与字符串长度 $l$ 相关。本章提出的关键前缀过滤使用 $\mathsf{pre}(r)$ 中的前缀$q$-gram 以及 $\mathsf{piv}(r)$ 中的关键$q$-gram 作为特征。通过比较一个$q$-gram $\mathsf{last}(\mathsf{piv}(r))$，它只需要对一个前缀$q$-gram 集合与一个关键$q$-gram 集合作比较。通过集成位置失配过滤，这两个特征集合大小的乘积、最小值以及总和分别在 $[(\tau+1)^2,(q\tau+1)*(\tau+1)]$ 范围内，等于 $\tau+1$ 以及在 $[2*(\tau+1),q\tau+1+\tau+1]$ 范围内，探测集合的大小在 $[\tau+1,q\tau+1]$ 范围内。

## 5.4 基于关键前缀的过滤算法

本节提出基于关键前缀的近似检索算法 PivotalSearch，它包含离线构建索引阶段和在线处理查询阶段。

**构建索引**：给定一个数据集 $\mathcal{R}$，构建两个倒排索引：一个基于前缀中的$q$-gram，记作 $\mathcal{I}^+$；以及另一个基于关键前缀中的$q$-gram，记作 $\mathcal{I}^-$①。首先对 $\mathcal{R}$ 中的字符串按照它们的长度进行排序。对每个字符串 $r \in \mathcal{R}$，产生它的前缀 $\text{pre}(r)$，对 $\text{pre}(r)$ 中每个起始位置为 $p$ 的$q$-gram $g$，把 $\langle r, p\rangle$ 插入到倒排列表 $\mathcal{I}^+(g)$ 中。之后选取关键前缀 $\text{piv}(r)$，并对 $\text{piv}(r)$ 中每个起始位置为 $p$ 的$q$-gram $g$，把 $\langle r, p\rangle$ 插入到倒排列表 $\mathcal{I}^-(g)$ 中（第 5.5节将讨论如何选择关键前缀）。对于每个在 $\mathcal{I}^-$ ($\mathcal{I}^+$) 中的$q$-gram $g$，还需要构建散列映射 $\mathcal{M}^-$ ($\mathcal{M}^+$)：其中每个键是 $\mathcal{I}^-(g)$（$\mathcal{I}^+(g)$）中字符串的长度，值是相应字符串在列表 $\mathcal{I}^-(g)$（$\mathcal{I}^+(g)$）中的起始位置。利用这个散列映射，可以很容易的获取在倒排列表中某个长度（或者长度范围内）的所有字符串并执行长度过滤来剪枝。

**例** 5.3：考虑表5.1中的数据字符串，假设编辑距离阈值是 $\tau = 2$ 并且 gram 长度是 $q = 2$，图5.2展示了基于给定字符串的索引。下面对 $\mathcal{R}$ 中的字符串进行排序并且按顺序访问它们来构建这个索引。对于 $r_1$，把它切分为$q$-gram 集合 $\text{q}(r_1)$ 并对 $\text{q}(r_1)$ 按照$q$-gram 频率升序排序。因为 $q\tau + 1 = 5$，所以把其中前五个$q$-gram 当做前缀，即 $\text{pre}(r_1) = \{\text{im,my,te,ca,yo}\}$。随机选取 $\tau + 1 = 3$ 个不相交的$q$-gram 作为关键$q$-gram 集合 $\text{piv}(r_1) = \{\text{im,te,ca}\}$。接下来，对于这五个前缀$q$-gram，把条目 $\langle r_1, 1\rangle, \langle r_1, 2\rangle, \langle r_1, 6\rangle, \langle r_1, 8\rangle$ 和 $\langle r_1, 3\rangle$ 分别添加到 $\mathcal{I}^+(\text{im}), \mathcal{I}^+(\text{my}), \mathcal{I}^+(\text{te}), \mathcal{I}^+(\text{ca})$ 和 $\mathcal{I}^+(\text{yo})$ 的末尾。对于三个关键$q$-gram，把条目 $\langle r_1, 1\rangle$, $\langle r_1, 6\rangle$ 和 $\langle r_1, 8\rangle$ 分别添加到 $\mathcal{I}^-(\text{im}),\mathcal{I}^-(\text{te})$ 和 $\mathcal{I}^-(\text{ca})$ 的末尾。类似的，需要把 $r_2, r_3, r_4$ 和 $r_5$ 也都插入到索引中。对于每个倒排索引 $\mathcal{I}^+(g)/\mathcal{I}^-(g)$，构建一个散列映射 $\mathcal{M}^+/\mathcal{M}^-$。例如，对于倒排列表 $\mathcal{I}^+(\text{im})$，将构建一个包含条目 $\langle 9, 1\rangle$ 的散列映射 $\mathcal{M}^+$。这是因为 $\mathcal{I}^+(\text{im})$ 中所有字符串长度都为 9，而第一个长度为 9 的数据字符串在该列表中的位置是 1。

算法5.1展示了 Indexing 算法的伪代码。它的输入是一个字符串数据集 $\mathcal{R}$，一个 gram 长度 $q$ 以及一个编辑距离阈值 $\tau$，它的输出是两个倒排索引 $\mathcal{I}^+$ 和 $\mathcal{I}^-$。它首先对 $\mathcal{R}$ 中所有字符串按照长度排序，然后把所有的字符串切分为$q$-gram 并按照$q$-gram 频率对每个字符串中的$q$-gram 排序 (第2 到5行)。之后对于每个排好序的$q$-gram 集合 $\text{q}(r)$，它产生一个前缀集合 $\text{pre}(r)$，把它们插入到倒排索引 $\mathcal{I}^+$ 中并更新散列映射 $\mathcal{M}^+$ (第6 到10行)。接下来，它从 $\text{pre}(r)$ 选择 $\tau + 1$ 个关键$q$-gram 来

① 实际上只需要维护一份索引，因为关键前缀是前缀的子集。可以使用一个特殊的标志来区别索引中的一个$q$-gram 是否在关键前缀中。

**Algorithm 5.1**: PivotalSearch-Indexing ($\mathcal{R}, q, \tau$)

**Input**: $\mathcal{R}$: 一个字符串集合; $q$: Gram 长度; $\tau$: 阈值
**Output**: $\mathcal{I}^+$: 前缀索引; $\mathcal{I}^-$: 关键前缀索引

1 **begin**
2 　对 $\mathcal{R}$ 中的字符串按照长度排序;
3 　切分每个字符串 $r \in \mathcal{R}$ 为$q$-gram 集合 q($r$);
4 　设置一个全局顺序，例如 $TF/IDF$ 顺序;
5 　对 q($r$) 中的$q$-gram 按照全局顺序排序;
6 　**for** $r \in \mathcal{R}$ **do**
7 　　生成前缀 pre($r$) ;
8 　　**foreach** $q$-$gram$ $g \in pre(r)$ **do**
9 　　　把 $\langle r, g.\text{pos}\rangle$ 附加到倒排列表 $\mathcal{I}^+(g)$ 的末尾;
10 　　　**if** $\mathcal{M}^+(g)[|r|] = \phi$ **then** $\mathcal{M}^+(g)[|r|] \leftarrow |\mathcal{I}^+(g)|$
11 　　piv($r$) =PivotsSelection(pre($r$));
12 　　**foreach** $q$-$gram$ $g \in piv(r)$ **do**
13 　　　把 $\langle r, g.\text{pos}\rangle$ 附加到倒排列表 $\mathcal{I}^-(g)$ 的末尾;
14 　　　**if** $\mathcal{M}^-(g)[|r|] = \phi$ **then** $\mathcal{M}^-(g)[|r|] \leftarrow |\mathcal{I}^-(g)|$
15 　**return** $\langle \mathcal{I}^+, \mathcal{I}^- \rangle$;
16 **end**

生成 piv($r$) 并把它们插入到倒排索引 $\mathcal{I}^-$ 中，然后更新散列映射 $\mathcal{M}^-$ (第11 到14行)。最终，它返回两个索引 $\mathcal{I}^-$ 和 $\mathcal{I}^+$ (第15行)。

**检索算法**：给定一个查询串 $s$，检索算法输出 $\mathcal{R}$ 中所有与查询串近似的字符串作为结果。它首先把查询串切分为$q$-gram，用全局顺序对它们排序，产生前缀集合 pre($s$) 并选取关键前缀 piv($s$)。然后，对于 piv($s$) 中每个$q$-gram $g$，根据长度过滤，$s$ 只可能与长度在 $|s|-\tau$ 到 $|s|+\tau$ 之间的字符串近似。因此它使用散列映射 $\mathcal{M}^-$ 来检索倒排列表 $\mathcal{I}^-(g)$ 中长度为 $|s|-\tau$ 的字符串的开始位置，即 start = $\mathcal{M}^-(|s|-\tau)$，以及长度为 $|s|+\tau$ 的字符串的结束位置，即 end = $\mathcal{M}^-(|s|+\tau+1)-1$[①]。之后，检索倒排列表 $\mathcal{I}^-(g)$ 并访问长度在 [start, end] 范围内的字符串，即 $\mathcal{I}^-(g)$[start, end]。对于这个列表中每个元素 $\langle r, p\rangle$，如果 last(pre($r$)) > last(pre($s$)) 并且 $|p - g.\text{pos}| \le \tau$，那么 $r$ 就是一个候选字符串[②]。类似的，对于 pre($s$) 中的每个$q$-gram $g$，利用 $\mathcal{I}^+(g)$ 和 $\mathcal{M}^+$ 来产生候选字符串。最终，验证这些候选字符串。

**例** 5.4：考虑图 5.2 中的查询串 $s$，首先把 $s$ 切分为$q$-gram，按照全局顺序排序，有 pre($s$) = {ot,om,yo,ub,co}。随机选取关键前缀 piv($s$) = {ot,om,ub}。用 piv($s$) 中

① 如果散列映射中没有这样的长度，那么可以寻找 $\mathcal{M}^-(|s|-\tau+1)$ 做起始位置，$\mathcal{M}^-(|s|+\tau+2)$ 做结束位置。
② 为了很容易的取得 last(pre($s$)) 的值，可以具体化它。

**Algorithm 5.2**: PIVOTALSEARCH $(s, q, \tau, \mathcal{I}^+, \mathcal{I}^-)$

**Input**: $s$: 查询串; $q$: Gram 长度; $\tau$: 阈值;
$\mathcal{I}^+$: 前缀索引; $\mathcal{I}^-$: 关键前缀索引
**Output**: $\mathcal{A} = \{r \in \mathcal{R} \mid \text{ED}(r, s) \le \tau\}$

1 **begin**
2 　切分 $s$ 为$q$-gram 集合 q($s$) 并对其中的$q$-gram 按全局顺序排序;
3 　生成前缀 pre($s$);
4 　初始化候选元组集 $C = \phi$;
5 　**foreach** *q-gram* $g \in$ *pre*($s$) **do**
6 　　start =$\mathcal{M}^+(|s| - \tau)$; end =$\mathcal{M}^+(|s| + \tau + 1) - 1$;
7 　　**foreach** $i \in$ [*start*, *end*] **do**
8 　　　$\langle r, p\rangle = \mathcal{I}^+(g)[i]$;
9 　　　**if** *last*(*pre*($r$)) $>$ *last*(*pre*($s$)) & $|p - g.pos| \le \tau$ **then**
10 　　　　把 $r$ 添加到 $C$ 中
11 　piv($s$) =PivotsSelection(pre($s$));
12 　**foreach** *q-gram* $g \in$ *piv*($s$) **do**
13 　　start =$\mathcal{M}^-(|s| - \tau)$; end =$\mathcal{M}^-(|s| + \tau + 1) - 1$;
14 　　**for** $i \in$ [*start*, *end*] **do**
15 　　　$\langle r, p\rangle = \mathcal{I}^-(g)[i]$;
16 　　　**if** *last*(*pre*($r$)) $\le$ *last*(*pre*($s$)) & $|p - g.pos| \le \tau$ **then**
17 　　　　把 $r$ 添加到 $C$ 中
18 　**foreach** $r \in C$ **do**
19 　　**if** *VERIFICATION*$(s, r)$ **then** 把近似元组 $(s, r)$ 添加到 $\mathcal{A}$ 中
20 　**return** $\mathcal{A}$;
21 **end**

的关键$q$-gram 探测前缀$q$-gram 索引 $\mathcal{I}^+$。因为关键$q$-gram 'ot' 没有出现在索引中，因此只访问两个倒排列表 $\mathcal{I}^+(\text{om}) = \{\langle r_2, 8\rangle, \langle r_4, 8\rangle\}$ 和 $\mathcal{I}^+(\text{ub}) = \{\langle r_3, 3\rangle, \langle r_5, 5\rangle\}$。因为 last(pre($r_2$)), last(pre($r_3$)), last(pre($r_4$)) 和 last(pre($r_5$)) 分别是 om, co,ut 和 ub，它们都不大于 last(pre($s$)) =co，所以扔掉全部这些字符串。接下来用 pre($s$) 中的前缀$q$-gram 探测关键$q$-gram 索引 $\mathcal{I}^-$。对于位置为 8 的前缀$q$-gram om, 探测索引可以获得 $\mathcal{I}^-(\text{om}) = \{\langle r_4, 8\rangle\}$。因为 last(pre($r_4$)) =ut$\le$ last(pre($s$)) =co, 所以它能通过关键前缀过滤。因为这两个$q$-gram 的位置差别是 $8 - 8 = 0 < \tau = 2$，它能通过位置过滤。所以把 $r_4$ 加入到候选集中。按照相同的办法处理其他关键$q$-gram，最终得到了三个候选字符串 $r_3, r_4$ 和 $r_5$。验证这些候选字符串后得到一个结果 $r_4$。

算法5.2展示了检索算法的伪代码。它首先产生前缀 pre($s$) (第 2-3 行)。对每个

pre($s$) 中的前缀$q$-gram $g$, 它用散列映射 $\mathcal{M}^+$ 获取长度在 $|s|-\tau$ 到 $|s|+\tau$ 之间的字符串在倒排列表中的开始位置 start 和结束位置 end (第 6 行)。对于每个 $i \in$ [start, end], 它检索元素 $\langle r, p\rangle = \mathcal{I}^+(g)[i]$ (第 7-8 行)。如果 last(pre($r$)) >= last(pre($s$)) 并且 $|p - g.\text{pos}| \le \tau$，它将 $r$ 作为一个候选字符串 (第 9 行)。接下来它从 pre($s$) 中选取 $\tau+1$ 个关键$q$-gram 从而产生 piv($s$) (第 11 行)。对于每个关键 $q$-gram $g$, 它利用散列映射 $\mathcal{M}^-$ 获取长度在 $|s| - \tau$ 到 $|s| + \tau$ 之间的字符串在倒排列表中的起始位置 start 和结束位置 end (第 13 行)。对于每个 $i \in$ [start, end]，它检索元素 $\langle r, p\rangle = \mathcal{I}^-(g)[i]$ (第 14 到15 行)。如果 last(pre($r$)) < last(pre($s$)) 并且 $|p - g.\text{pos}| \le \tau$, 它把 $r$ 作为一个候选字符串 (第 16 行).

**复杂度**: 首先分析空间复杂度。对于每个字符串 $r \in \mathcal{R}$，最多插入 $q\tau + 1$ 个前缀$q$-gram 到倒排索引 $\mathcal{I}^+$ 中以及最多 $\tau+1$ 个关键$q$-gram 到倒排索引 $\mathcal{I}^-$ 中, 因此空间复杂度是 $O((q\tau+1+\tau+1)|\mathcal{R}|) = O(q\tau|\mathcal{R}|)$。然后分析时间复杂度。给定一个查询串 $s$, 需要产生并对$q$-gram 排序以及选取前缀。其时间复杂度是 $O(|s|+|s|\log|s|+q\tau)$。因此探测两个倒排索引的时间复杂度是 $O((q\tau+1)l_s + (\tau+1)l_p)$，其中 $l_s$ 和 $l_p$ 分别是 $\mathcal{I}^-$ 和 $\mathcal{I}^+$ 中倒排列表的平均长度。因为对 $\mathcal{R}$ 中每个字符串，需要插入 $q\tau + 1$ 个$q$-gram 到 $\mathcal{I}^+$ 中，$\tau + 1$ 个$q$-gram 到 $\mathcal{I}^-$ 中，可以估计 $l_s = \frac{\tau+1}{q\tau+1} l_p$，因此探测时间复杂度是 $O(2(\tau+1)l_p) = O(\tau l_p)$。

PivotalSearch 方法有两个挑战。第一个是如何选取高质量的关键前缀。第二个挑战是如何高效的验证候选字符串。下面分别在第5.5 节和第 5.6节应对这两个挑战。

## 5.5 关键$q$-gram 选取

本节讨论如何选取 $\tau + 1$ 个高质量的关键$q$-gram。首先证明在任何前缀$q$-gram 集合中一定存在 $\tau + 1$ 个不相交的$q$-gram (第5.5.1节)。然后，讨论如何评估不同的关键前缀（第5.5.2节)。最后设计一个动态规划算法来选取最优的关键前缀 (第5.5.3节)。

### 5.5.1 关键前缀的存在性

可以证明, 对于任何字符串 $r$, 其前缀$q$-gram 集合 pre($r$) 中一定存在至少 $\tau+1$ 个不相交的$q$-gram，如定理5.3所述。其主要原因如下。任何前缀 pre($r$) 需要满足一个条件：它需要至少 $\tau + 1$ 个编辑操作来毁掉 pre($r$) 中所有的$q$-gram。然而毁掉 pre($r$) 中所有的$q$-gram 需要在至少 $\tau + 1$ 个位置上应用编辑操作，其中任何两个位置的距离都至少是 $q$，利用这 $\tau + 1$ 个位置，可以选取 $\tau + 1$ 个不相交的$q$-gram。

**引理** 5.3: 对于任何字符串 $r$，其前缀 pre($r$) 中一定存在至少 $\tau+1$ 个不相交的$q$-gram。

**证明** 首先考虑 $|\text{pre}(r)| = q\tau + 1$。一个简单的方法选取 $\tau+1$ 个不相交的$q$-gram 首先对 $q\tau+1$ 个$q$-gram 按照它们的起始位置排序，然后根据顺序把这 $q\tau+1$ $q$-gram 分为 $\tau+1$ 个组。最后一个组包含最后一个 q-gram，对所有 $1 \le i \le \tau$, 第 $i$ 个组包含顺序早 $[1 + q\cdot(i-1), q\cdot i]$ 内的$q$-gram。任意两个组中第一个$q$-gram 的位置差别最少是 $q$，所以至少需要 $\tau+1$ 个编辑操作把所有的$q$-gram 都毁掉。因此，可以把每个组中第一个$q$-gram 作为关键$q$-gram.

接下来通过失配过滤来缩短前缀长度的情况，其中 $|\text{pre}(r)| < q\tau+1$。失配过滤需要至少 $\tau+1$ 个编辑操作来毁掉 pre($r$) 中所有的$q$-gram。用反证法证明该引理。假设 pre($r$) 包含少于 $\tau+1$ 个不相交的$q$-gram。首先对 pre($r$) 中的$q$-gram 按照其起始位置排序，然后按顺序访问这些$q$-gram。选取第一个$q$-gram 作为参照$q$-gram。跳过所有与参照$q$-gram 有交叠的$q$-gram。因为所有的$q$-gram 都是排好序的，所有包含参照$q$-gram 最后一个字符的$q$-gram。对于第一个与参照$q$-gram 没有交叠的$q$-gram，选择它为一个新的参照$q$-gram。重复以上步骤指导所有的$q$-gram 都被访问了。一方面，因为所有的参照$q$-gram 都是不相交的，参照$q$-gram 的数量比 $\tau+1$ 少。另一方面，可以在每个参照$q$-gram 的最后一个字符做一个编辑操作来毁掉 pre($r$) 中所有的$q$-gram。因此，可以用少于 $\tau+1$ 个编辑操作来毁掉 pre($r$) 中所有的$q$-gram。这与失配过滤的要求互相矛盾，因此 pre($r$) 一定包含至少 $\tau+1$ 个不相交的$q$-gram。□

### 5.5.2 评估不同的关键前缀

尽管关键前缀过滤对任何关键前缀都适用，但是一个前缀中可能存在多种不同的关键前缀，应该选取其中最好的一个。为了达到这个目的，需要评估不同关键前缀的质量。

首先讨论如何为一个查询串 $s$ 选取关键前缀。如算法5.2所示，算法对于关键前缀 piv($s$)，需要使用 piv($s$) 中的每个关键$q$-gram 来探测倒排索引 $\mathcal{I}^+$，然后扫描 piv($s$) 中$q$-gram 对应的倒排列表。扫描的倒排列表的越长，过滤代价就越高并且剪枝能力就越弱。因此应该选取倒排列表长度最短的关键前缀。为了达到的目的，可以为 pre($s$) 中的每个$q$-gram $g$ 分配一个权重 $w(g)$：$g$ 在 $\mathcal{I}^+$ 中相应的倒排列表的长度，即 $w(g) = |\mathcal{I}^+(g)|$。这样的话，目标就是从 pre($r$) 中选取 $\tau+1$ 个权重之和最小的不相交的$q$-gram 作为关键前缀。接下来形式化定义最优关键前缀选取问题。

**定义** 5.2 (**最优关键前缀选取**): 给定一个$q$-gram 前缀 pre($s$)，最优关键前缀选取问题是从 pre($s$) 中选取 $\tau+1$ 个权重之和最小的关键$q$-gram。

**例 5.5**：　考虑表5.1中的查询串 $s$，它的五个前缀$q$-gram “ot,om,yo,ub,co'' 的权重分别是 $0,2,3,2,1$。pre($s$) 有 10 个包含 $\tau+1=3$ 个$q$-gram 的子集，在它们之中，有四个子集包含三个不相交的$q$-gram。因此，一共存在四个可选的关键前缀：{yo,ub,co}, {ot,ub,co}, {yo,ub,om} 和 {ot,ub,om}，它们的权重之和分别是 $6,3,7$ 和 $4$，其中最优的关键前缀是 {ot,ub,co}。

接下来，考虑数据字符串 $r$。它的关键$q$-gram 要被插入到索引 $\mathcal{I}^-$ 中。给定一个查询，算法将用查询串的每个前缀$q$-gram 来探测索引 $\mathcal{I}^-$。直观的，如果选择数据串中低频的前缀$q$-gram 作为关键$q$-gram，那么它们比较不可能出现在查询串的前缀中，这个数据串更可能被关键前缀过滤剪枝掉。因此可以使用$q$-gram 的频率作为数据串中前缀$q$-gram 的权重，接下来的目标就是选择 $\tau+1$ 个权重之和最小的不相交的前缀$q$-gram 作为关键前缀。

**例 5.6**：　考虑表5.1中的数据串 $r_3$，它的前缀$q$-gram “bb,ou,ut,ub,co” 的权重分别是 $1,3,3,3,3$。一共存在四个包含 $\tau+1=3$ 个不相交的前缀$q$-gram 的关键前缀，分别是：{ut,ub,co}, {ut,ub,ou}, {ut,bb,co} 和 {ut,bb,ou}，其权重分别是 $9,9,7,7$。{ut,bb,co} 和 {ut,bb,ou} 都是最优关键前缀。

### 5.5.3　关键前缀选取算法

为了选取最优的关键前缀，下面设计一个动态规划算法。首先对 pre($r$) 中所有的前缀$q$-gram 按照它们的起始位置升序排序，并用 $g_k$ 来表示第 $k$ 个$q$-gram。为了方便表达，用 $\mathcal{W}(i,j)$ 来代表从前 $i$ 个$q$-gram $g_1,g_2,\cdots,g_i$ 中选取 $j$ 个不相交的$q$-gram 的最小权重，用 $\mathcal{P}(i,j)$ 来存储最小权重对应的不相交的$q$-gram 列表。这样只需要计算 $\mathcal{P}(|\mathrm{pre}(r)|,\tau+1)$ 的值就可以解决最优关键前缀选取问题。

最开始，$\mathcal{W}(i,1)$ 就是前 $i$ 个$q$-gram 的权重的最小值，即 $\mathcal{W}(i,1)=\min_{1\le k\le i} w(g_k)$，$\mathcal{P}(i,1)=\arg\min_{g_k} w(g_k)$。接下来讨论如何计算 $\mathcal{W}(i,j)$。考虑前 $i$ 个前缀$q$-gram 中第 $j$ 个关键$q$-gram 可能的位置，用 $g_k$ 来表示。因为在 $g_k$ 之前需要存在 $i-1$ 个$q$-gram，所以 $k$ 不可能比 $i$ 小。因此 $k\in[j,i]$。如果选取 $g_k$ 作为第 $j$ 个$q$-gram, 需要在 $g_k$ 之前再选取其他 $j-1$ 个$q$-gram，这些$q$-gram 不能与 $g_k$ 相交。因为这些$q$-gram 是按照位置排好序的，只需要检查第 $(j-1)$ 个$q$-gram 是否与 $g_k$ 交叠即可。令 $g_{k'}$ 表示与 $g_k$ 没有交叠的第 $(j-1)$ 个$q$-gram。因为在 $g_{k'}$ 之前需要存在 $i-2$ 个$q$-gram，$k'$ 不能比 $i-1$ 小。因此 $k'\in[i-1,k]$。值得注意的是随着 $k'$ 的增加，$\mathcal{W}(k',j-1)$ 是单调递减的。因此对每个$q$-gram $g_k$，只需要寻找与它最近且与 $g_k$ 没有交叠的$q$-gram $g'_k$。给定 $k$，可以按照如下方式高效的寻找 $k'$。首先对前缀$q$-gram 进行排序，检查它之前的$q$-gram $g_{k-1}$。如果 $g_{k-1}$ 与 $g_k$ 没有交叠，那

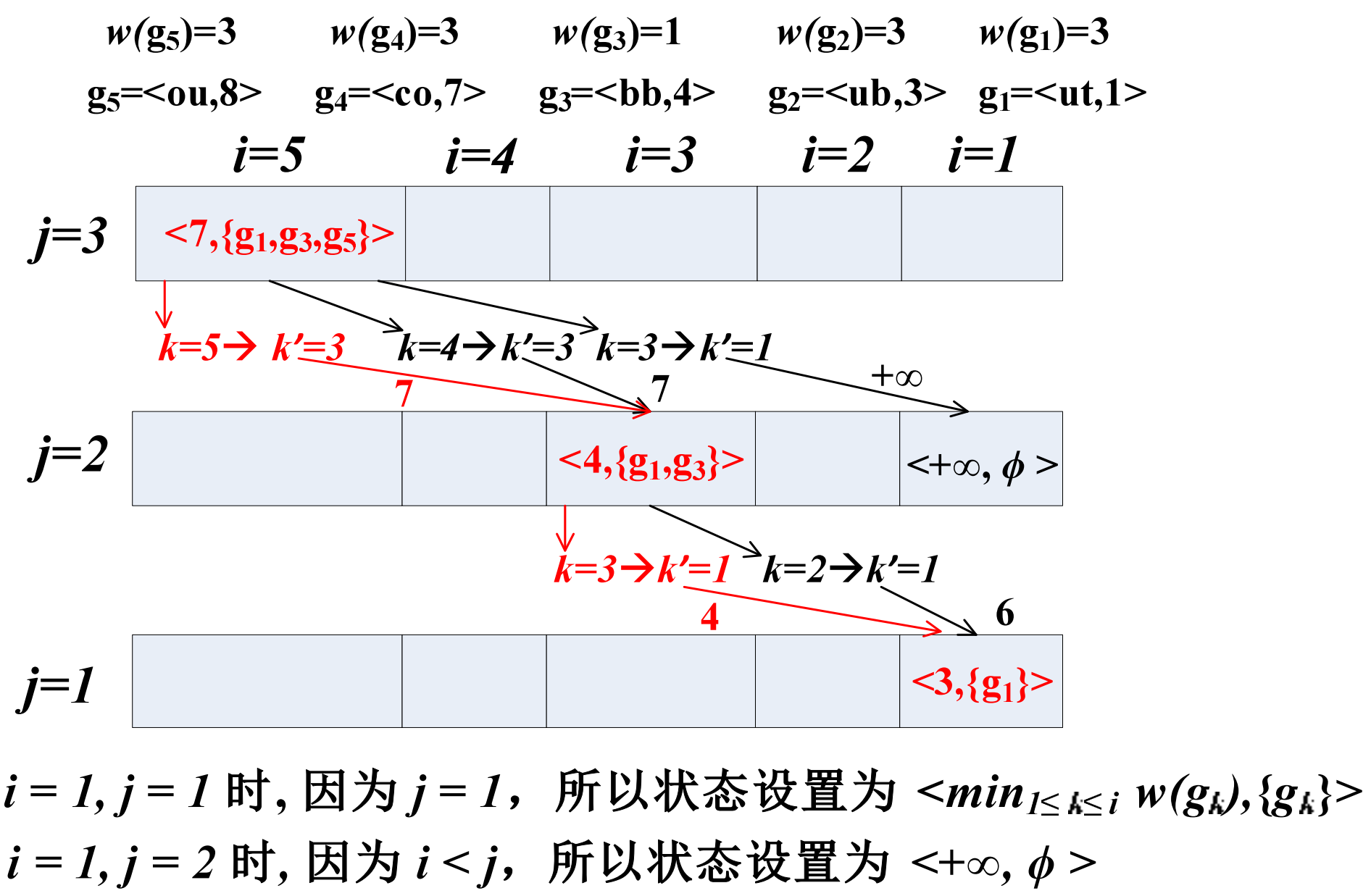

图 5.3　关键前缀选取的一个例子

么 $k' = k-1$; 否则的话，检查 $g_{k-2}$，如此反复，可以找到 $g_{k'}$。因为 $k-k'$ 最多等于 $q$，为每个 $k$ 寻找 $k'$ 的时间复杂度是 $O(q)$，对 $q\tau+1$ 个$q$-gram，总的时间复杂度是 $O(q^2\tau)$。

利用 $k$ 和 $k'$，可以推导出如下的递归公式：

$$\mathcal{W}(i,j) = \min_{i\le k\le j} \mathcal{W}(k',j-1) + w(g_k). \tag{5-1}$$

其中 $k'$ 与 $k$ 相关。根据公式5-1, 可以找到第 $i$ 个 $q$-gram $g_k$ 和第 $(i-1)$ 个$q$-gram $g_{k'}$ 以及 $\mathcal{P}(i,j) = \mathcal{P}(k',j-1)\cup\{g_k\}$。因为状态表的大小是 $O(q\tau^2)$，对每个 $j, k \in [i,j]$，动态规划算法的复杂度是 $O(q^2\tau^3)$。

基于这个递归方程，下面提出一种动态规划算法 OptimalSelection 来选择最优的关键前缀。算法5.3展示了其伪代码。OptimalSelection 算法的输入是 $i$ 和 $j$, 表示从前 $i$ 个排好序的前缀$q$-gram $g_1, g_2, \cdots, g_i$ 选取 $j$ 个关键 $q$-gram，其输出是 $\mathcal{W}(i,j)$ 和 $\mathcal{P}(i,j)$。如果 $j = 1$, 它返回前 $i$ 个$q$-gram 中最小的权重，以及相应的权重最小的$q$-gram (第 2 行)。如果 $i < j$，因为从 $i < j$ 个前缀$q$-gram 中不可能选取出 $j$ 个关键$q$-gram，该算法返回一个无限大的权重以及一个空的集合 (第 3 行)。如果之前计算过 $\mathcal{W}(i,j)$ 和 $\mathcal{P}(i,j)$ 的值，它直接返回这两个值 (第 4 行)。接下来，它选取使得 $\mathcal{W}(k',j-1)+w(g_k)$ 最小的 $k$ 和相应的 $k'$(第 5行-8 行)。最终，它返回 $\langle \mathcal{W}(k',j-1)+w(g_k), \mathcal{P}(k',j-1)\cup\{g_k\}$(第 8 行)。其时间复杂度是 $O(q^2\tau^3)$。

**Algorithm 5.3**: OptimalSelection

**Input**: $i$: 前缀长度; $j$: 需要从前缀中选取的关键$q$-gram 数量

**Output**: $\langle \mathcal{W}(i,j), \mathcal{P}(i,j) \rangle$;

**begin**

  **if** $j = 1$ **then return** $\langle min_{1 \le k \le i} w(g_k), \{g_k\} \rangle$;

  **if** $i < j$ **then return** $\langle +\infty, \phi \rangle$ ;

  **if** $\mathcal{P}(i,j) \neq \phi$ **then return** $\langle \mathcal{W}(i,j), \mathcal{P}(i,j) \rangle$;

  **for** $k = j$ 到 $i$ **do**

    找到最大的 $k' < k$ 使得$q$-gram $g_{k'}$ 与 $g_k$ 不相交;

    $\langle \mathcal{W}(k', j-1), \mathcal{P}(k', j-1) \rangle$ =OptimalSelection($k', j-1$);

    **if** $w(g_k) + \mathcal{W}(k', j-1)$ 是最小的 **then return**

    $\langle \mathcal{W}(k', j-1) + w(g_k), \mathcal{P}(k', j-1) \cup \{g_k\} \rangle$

**end**

**例** 5.7: 考虑表5.1中的查询串 $s$ 和数据字符串 $r_3$。因为 last(pre($r_3$)) = co ≤ last(pre($s$)) = co 和 piv($r_3$) ∩ pre($s$) = {ub}，所以该关键前缀不能剪枝掉 $r_3$。然而可以使用 OptimalSelection 算法来选取最优的关键$q$-gram 并过滤掉这个不相似的字符串。对于 q($r_3$)，首先对它的$q$-gram 按照它们的位置排序，有 $g_1 = \langle$ut, 1$\rangle$, $g_2 = \langle$ub, 3$\rangle$, $g_3 = \langle$bb, 4$\rangle$, $g_4 = \langle$co, 7$\rangle$ 以及 $g_5 = \langle$uu, 8$\rangle$。如图 5.3 所示，它们的权重分别是 3, 3, 1, 3 和 3。接下来，调用算法 OptimalSelection(5, 3)，它从 3, 4, 5 中选择 $k$ 使得 $\mathcal{W}(k', 2) + w(g_k)$ 最小，其中 $k'$ 分别等于 3, 3, 1。对于 $k = 3$，因为 $k' = 1 < 2$，所以 $\mathcal{W}(k', 2) = +\infty$。然后递归的调用 OptimalSelection(3, 2) 来计算 $\mathcal{W}(3, 2)$。在第一轮中，需要从 3 和 2 中选择 $k$ 的值（对应的 $k'$ 都是 1）并计算 $\mathcal{W}(1, 1)$。因为 $j = 1$，所以返回 $\mathcal{W}(1, 1) = 3$ 和 $\mathcal{P}(1, 1) = \{g_1\}$。可以发现在 $k = 3$ 时，$\mathcal{W}(1, 1) + w(g_3)$ 取得了最小值 4，所以返回 $\mathcal{W}(3, 2) = 4$ 和 $\mathcal{P}(3, 2) = \{g_1, g_3\}$。最终有 $\mathcal{W}(5, 3) = 7$ 和 $\mathcal{P}(5, 3) = \{g_1, g_3, g_5\}$。因此，有 piv($r_3$) = {bb,ou,ut}。因为 piv($r_3$) ∩ pre($s$) = $\phi$，所以最优关键前缀可以剪枝掉 $r_3$。

## 5.6　在关键$q$-gram 上的对齐过滤

关键前缀过滤对于因为离散的编辑错误而与查询串不相似的字符串很有效果，因为 $\tau$ 个离散的编辑错误可能正好毁掉 $q\tau$ 个$q$-gram。然而，它对连续的编辑错误效果不是很好，因为 $\tau$ 个连续的编辑错误可能仅仅毁掉 $\tau + q - 1$ 个$q$-gram，这比 $q\tau$ 小很多，因此关键前缀过滤可能导致很多候选字符串。例如，考虑表5.1中的字符

串 $r_5$ 和查询串 $s$。$s$ 中的的两个连续的编辑错误 'om' 仅仅毁掉 $r_5$ 中的三个$q$-gram，因此关键前缀过滤不能够剪枝掉这个与查询串不近似的字符串。为了解决这个问题，下面提出了对齐过滤来有效的检测连续的编辑错误。

### 5.6.1　转换过程中的对齐

关键前缀过滤在两种情况下会产生候选字符串：(1) $\mathsf{pre}(r) \cap \mathsf{piv}(s) \neq \phi$; 和 (2) $\mathsf{pre}(s) \cap \mathsf{piv}(r) \neq \phi$。在任何一种情况下，存在一个字符串中的关键$q$-gram 和另一个字符串中的一个前缀$q$-gram 相匹配。不失一般性的，假设一个 $s$ 的一个关键$q$-gram 和 $r$ 的一个前缀$q$-gram 匹配了。在任何包含 $\textsc{ed}(s,r)$ 个编辑操作的从 $s$ 到 $r$ 的转换中，每个 $s$ 的关键$q$-gram，比如第 $i$ 个关键$q$-gram $\mathsf{piv}_i(s)$，都将被转化为 $r$ 的一个子字符串，比如 $\mathsf{sub}_i(r)$。下面称 $\mathsf{piv}_i(s)$ 和 $\mathsf{sub}_i(r)$ 相互对齐。因为这 $\tau+1$ 个关键$q$-gram 互不相交，所以这 $\tau+1$ 个子字符串也不相交 ($\mathsf{sub}_i(r)$ 可能为空)。假设 $\mathsf{err}_i = \textsc{ed}(\mathsf{piv}_i(s), \mathsf{sub}_i(r))$，对于任何包含 $\textsc{ed}(s,r)$ 个编辑操作的从 $s$ 到 $r$ 的转换, 如引理 5.4 所述，有 $\sum_{i=1}^{\tau+1} \mathsf{err}_i \leq \textsc{ed}(s,r)$。图 5.4 展示了其中的基本想法。

**引理** 5.4:　对于任何包含 $\textsc{ed}(s,r)$ 个编辑操作的从 $s$ 到 $r$ 的转换，有 $\sum_{i=1}^{\tau+1} \mathsf{err}_i \leq \textsc{ed}(s,r)$。

**证明**　一方面，有 $\tau+1$ 个不相交的关键$q$-gram 上的编辑操作全部都在该转换的 $\textsc{ed}(s,r)$ 个编辑操作中。另一方面，每个关键$q$-gram 上的编辑操作的数目都不小于 $err_i$。因此有 $\sum_{i=1}^{\tau+1} \mathsf{err}_i \leq \textsc{ed}(s,r)$。 □

根据引理5.4，如果 $\sum_{i=1}^{\tau+1} \mathsf{err}_i > \tau$，那么 $s$ 和 $r$ 一定不相似。因为存在很多种包含 $\textsc{ed}(s,r)$ 个编辑操作的从 $s$ 到 $r$ 的转换，枚举所有这样的转换来计算 $\mathsf{err}_i$ 和检验 $\sum_{i=1}^{\tau+1} \mathsf{err}_i > \tau$ 是否成立的代价非常高昂。为了解决这个问题，观察到对于任何转换，$\mathsf{err}_i = \textsc{ed}(\mathsf{piv}_i(s), \mathsf{sub}_i(r))$ 不大于关键$q$-gram $\mathsf{piv}_i(s)$ 和 $r$ 的任何子字符串之间最小的编辑距离，用 $\mathsf{sed}(\mathsf{piv}_i(s), r)$ 来表示它并称之为 $\mathsf{piv}_i(s)$ 和 $r$ 之间的子字符串编辑距离（substring edit distance）[107]。因此有 $\mathsf{err}_i = \textsc{ed}(\mathsf{piv}_i(s), \mathsf{sub}_i(r)) \geq \mathsf{sed}(\mathsf{piv}_i(s), r)$ 和 $\sum_{i=1}^{\tau+1} \mathsf{sed}(\mathsf{piv}_i(s), r) \leq \sum_{i=1}^{\tau+1} \mathsf{err}_i$。

$g = \mathsf{piv}_i(s)$ 与 $r$ 之间的子字符串编辑距离，即 $\mathsf{sed}(g,r)$，可以用一个动态规划算法[107] 来计算。使用一个 $|g|+1 = q+1$ 行和 $|r|+1$ 列的状态二维表 $\mathcal{M}$ 来计算 $\mathsf{sed}(g,r)$，其中状态 $\mathcal{M}[i][j]$ 是 $g$ 的长度为 $i$ 的前缀与 $r$ 的结束位置为 $r[j]$ 的子字符串之间的最小的编辑距离，因此 $g$ 和 $r$ 之间的子字符串编辑距离就是

$$\mathsf{sed}(g,r) = \min_{1\leq i\leq |r|+1} \mathcal{M}[i][|g|+1]。$$

为了计算状态 $\mathcal{M}[i][j]$，首先对所有 $1 \le j \le |r|+1$ 初始化状态 $\mathcal{M}(0,j)=0$，对所有 $1 \le i \le q+1$ 初始化状态 $\mathcal{M}(i,0)=i$。然后按照如下递推公式计算状态 $\mathcal{M}[i][j]$：

$$\mathcal{M}(i,j)=\begin{cases}\mathcal{M}(i-1,j-1) & g[i]=r[j]\\ \min\begin{cases}\mathcal{M}(i-1,j)+1\\ \mathcal{M}(i,j-1)+1\\ \mathcal{M}(i-1,j-1)+1\end{cases} & g[i]\ne r[j]\end{cases}$$

计算子字符串编辑距离的时间复杂度是 $O(q|r|)$，该过滤代价很高昂。为了进一步提高性能，观察到一个关键$q$-gram $g=\mathsf{piv}_i(s)$ 不可能与 $r$ 的某些子字符串相对齐。更具体来说，根据位置过滤，如果 $s$ 和 $r$ 相似，那么 $g$ 只能与 $r$ 的位置在 $g.\mathsf{pos}-\tau$ 和 $g.\mathsf{pos}+q-1+\tau$ 之间的子字符串相对齐，记作 $r[g.\mathsf{pos}-\tau, g.\mathsf{pos}+q-1+\tau]$ (如果 $g.pos-\tau<1$，把起始位置设置为 1，如果 $g.pos+q-1+\tau>|r|$，把结束位置设置为 $|r|$)。这是因为如果 $g$ 与这个范围外的子字符串相对齐的话，那么根据长度过滤，在这个从 $s$ 到 $r$ 的转换中编辑操作的数目一定大于 $\tau$，如图5.4所示。例如，当 $g$ 和 $r[g.\mathsf{pos}-\tau]$ 之前的子字符串相对齐时，在 $\mathsf{piv}_i(s)=\mathsf{sub}_i(r)$ 之前的子字符串的长度差就比 $\tau$ 大，因此总的编辑操作数目一定大于 $\tau$。

因此，只需要计算 $g$ 和 $r[g.\mathsf{pos}-\tau, g.\mathsf{pos}+q-1+\tau]$ 之间的子字符串编辑距离[①]，所以时间复杂度是 $O(q(q+2\tau))=O(q^2+q\tau)$。此外，当 $|i-j+\tau|>\tau$ 时，在动态规划算法中可以跳过状态 $\mathcal{M}(i,j)$，即只有当 $1\le i\le |q|+1$ 和 $i\le j\le i+2\tau$ 时，才需要计算状态 $\mathcal{M}(i,j)$。这是因为如果把 $g$ 中第 $i$ 个字符 $s[g.\mathsf{pos}+i-1]$ 与 $r[g.\mathsf{pos}-\tau, g.\mathsf{pos}+q-1+\tau]$ 中第 $j$ 个字符 $r[g.\mathsf{pos}-\tau+j-1]$ 对齐，那么在这个对齐中的长度差 $|(g.\mathsf{pos}+i-1)-(g.\mathsf{pos}-\tau+j-1)|=|i-j+\tau|$ 一定不小于 $\tau$。这样计算一个子字符串编辑距离的时间复杂度可以减少到 $O(q\tau)$，并且计算 $\sum_{i=1}^{\tau+1}\mathsf{sed}(\mathsf{piv}_i(s), r[\mathsf{piv}_i(s).\mathsf{pos}-\tau, \mathsf{piv}_i(s).\mathsf{pos}+q-1+\tau])$ 的总时间复杂度是 $O(q\tau^2)$，这比直接计算 $\textsc{ed}(r,s)$ 的时间复杂度 $O(min(|r|,|s|)*\tau)$ 要小很多，因为 $min(|r|,|s|)$ 比 $q$ 大很多。

根据这个观察，下面设计一个对齐过滤（对齐过滤）。

① Li 等人[65] 提出了一些技术进一步缩小子字符串的范围，他们的技术与的技术正交独立并可以无缝融合到的方法中。

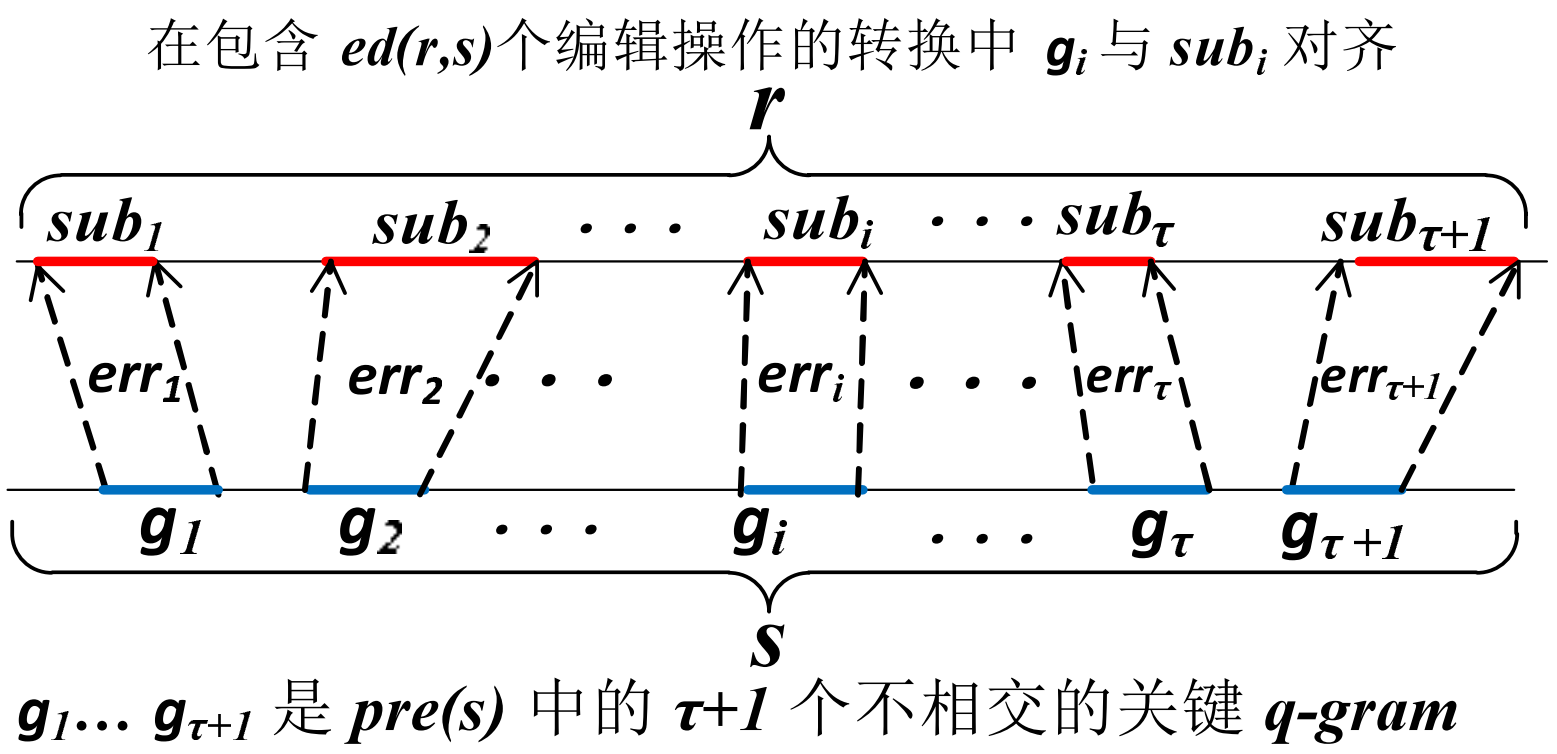

图 5.4　对齐过滤，$\sum_{i=1}^{\tau+1} \mathsf{err}_i \le \textsc{ed}(s,r)$

### 5.6.2　对齐过滤

对于任何共享一个相同关键$q$-gram 的两个字符串 $r$ 和 $s$，如果 $\sum_{i=1}^{\tau+1} \mathsf{sed}(\mathsf{piv}_i(s), r[\mathsf{piv}_i(s).\mathsf{pos} - \tau, \mathsf{piv}_i(s).\mathsf{pos} + q - 1 + \tau]) > \tau$，那么 $r$ 和 $s$ 不可能相似，如引理5.5所述。

**引理** 5.5 (**对齐过滤**)：对于任何两个字符串 $r$ 和 $s$，如果 $r$ 和 $s$ 相似，那么有

$$\sum_{i=1}^{\tau+1} \mathsf{sed}(\mathsf{piv}_i(s), r[\mathsf{piv}_i(s).\mathsf{pos} - \tau, \mathsf{piv}_i(s).\mathsf{pos} + q - 1 + \tau]) \le \tau。$$

**证明** 根据引理5.4以及长度过滤，这个引理很容易被证明. □

注意，可以通过阈值 $\tau$ 来提早终止动态规划验证，从而进一步提高子字符串编辑距离的计算。也就是说，如果对于某个 $i \in [1,q]$ 和所有 $j \in [i, i+2\tau]$ 都有 $\mathcal{M}(i,j) > \tau$，那么可以终止动态规划验证并报告子字符串编辑距离一定大于阈值 $\tau$。更进一步地，可以缩小这个阈值。具体来说，当计算第 $i$ 个关键$q$-gram 与数据字符串的子字符串编辑距离时，可以把阈值设置为 $\tau - error$，其中 $error$ 代表前 $i-1$ 个关键$q$-gram 与数据字符串的子字符串编辑距离之和。根据引理5.5，这是很直观的。可以把这个阈值作为第三个参数传给 sed 来实施提早终止验证。

### 5.6.3　验证算法

接下来把对齐过滤融合进之前的框架并提出**验证算法**（如算法5.4所示）。对于任何一对共享相同关键$q$-gram $g$ 的字符串 $r$ 和查询串 $s$，首先通过检验 $\sum_{i=1}^{\tau+1} \mathsf{sed}(\mathsf{piv}_i(s), r[\mathsf{piv}_i(s).\mathsf{pos} - \tau, \mathsf{piv}_i(s).\mathsf{pos} + q - 1 + \tau]) > \tau$ 是否成立来执行对齐过滤。如果成立，那么这一对字符串可以被对齐过滤剪枝掉，否则，需要使用动态规划算法来计算它们之间真实的编辑距离从而验证它们。

**Algorithm 5.4**: VERIFICATION

**Input**: $s$: 查询串; $r$: 数据字符串; $q$: Gram 长度;
piv($s$): $s$ 的关键前缀; $\tau$: 阈值

**begin**
  $errors = 0$;
  **for** *each q-gram* $g \in$ *piv*$(s)$ **do**
    $errors = errors + \mathsf{sed}(g, r[g.\mathsf{pos} - \tau, g.\mathsf{pos} + q - 1 + \tau], \tau - errors)$;
    **if** $errors > \tau$ **then return** *false*
  **if** $\textsc{ed}(r, s) \le \tau$ **then return** *true* **else return** *false*
**end**

例如，考虑表5.1中的字符串 $r_5$ 和查询串 $s$，因为 $\mathsf{last}(\mathsf{pre}(r_5)) \le \mathsf{last}(\mathsf{pre}(s))$，$\mathsf{piv}(r_5) \cap \mathsf{pre}(s) = \mathsf{yo}$, 所以 $r_5$ 是一个候选字符串，使用算法5.4来验证它。对于这三个关键$q$-gram，分别计算它们与查询串的子字符串编辑距离，有 $\mathsf{sed}(\mathsf{ot}, r_5[2-2, 2+2-1+2] = yoytu) = 1$ 和 $\mathsf{sed}(\mathsf{om}, r_5[8-2, 8+2-1+2] = beca) = 2$。由于总的编辑错误数目 (3) 已经大于给定的阈值 (2) 了，所以可以剪枝掉 $r_5$。

**与内容过滤比较**: ED-Join[30] 提出了一个基于内容的过滤条件来检测连续的编辑错误。它的时间复杂度是 $O(\tau|\Sigma| + l)$，其中 $\Sigma$ 是该数据集字符表的大小，$l$ 是字符串长度。显然，内容过滤的时间复杂度很高而的对齐过滤的时间复杂度 $O(q\tau^2)$ 较低，而且它与字符串的长度和字符表的大小是无关的。

## 5.7 支持动态阈值

在本节中讨论如何支持动态编辑距离阈值。与现有方法 Qchunk, AdaptJoin, Flamingo 以及 $\mathsf{B}^{ed}$tree 一样，也假设对于查询串存在一个最大阈值 $\hat{\tau}$。这是因为，如果 $|s| - q + 1 - q\tau \le 0$，基于$q$-gram 的方法将把所有的字符串都将作为候选字符串，为了让基于$q$-gram 的方法有剪枝能力，阈值不能超过 $\lfloor \frac{|s|-q}{q} \rfloor$。一个简单的方法来支持动态阈值是建立 $\hat{\tau}+1$ 个索引，即为每个 $\tau \in [0, \hat{\tau}]$ 都建立一个索引。对于一个阈值为 $\tau$ 的查询串，使用其相应的索引来处理这个查询。然而这种方法会导致很大的空间消耗。为了解决这个问题，对每个 $0 \le \tau \le \hat{\tau}$.，建立一个增量索引 $\mathcal{I}_\tau^+$ 和 $\mathcal{I}_\tau^-$。

**构建索引**：考虑一个数据集 $\mathcal{R}$ 和一个最大的编辑距离阈值 $\hat{\tau}$。首先把字符串切分为$q$-gram 并固定一个全集顺序。对于每个字符串 $r \in \mathcal{R}$，根据全局顺序对它的$q$-gram 进行排序并按顺序访问它的$q$-gram。首先选择第一个$q$-gram $g_0$ 使得至少需要一个编辑操作来毁掉所有在 $g_0$ 之前的$q$-gram（包括 $g_0$），显然 $g_0$ 一定是 $\mathsf{q}(r)$ 中第

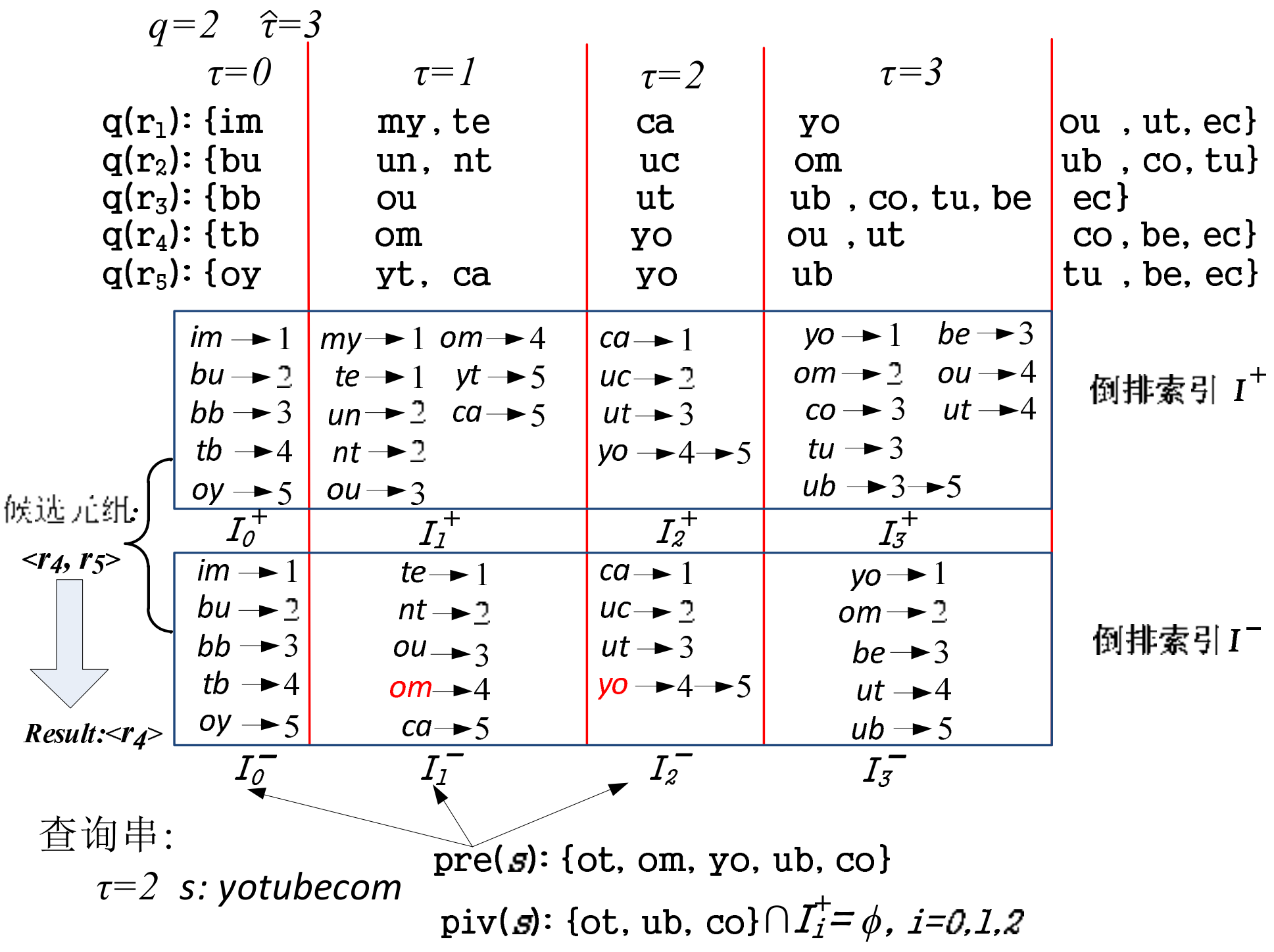

图 5.5　一个增量索引的例子

一个$q$-gram。接下来，选取第一个$q$-gram $g_1$ 使得至少需要两个编辑操作以毁掉所有在 $g_1$ 之前的$q$-gram (包括 $g_1$)。类似的，产生 $g_2, \cdots, g_{\hat{\tau}}$①。显然，$\{g_0, g_1, g_2, \cdots, g_i\}$ 是 $r$ 在阈值为 $i$ 时的一个关键前缀，在 $g_i$ 之前的$q$-gram 集合是 $r$ 在阈值为 $i$ 时的前缀。因此首先把 $g_i$ 插入到 $\mathcal{I}_i^-$，然后对所有 $1 \le i \le \hat{\tau}$，把 $g_{i-1}$ 与 $g_i$（包括 $g_i$ 但是不包括 $g_{i-1}$）之间的$q$-gram 插入到 $\mathcal{I}_i^+$ 中②。

**例 5.8**：　考虑表5.1中的数据字符串，假设最大的编辑距离阈值是 $\hat{\tau} = 3$。如图5.5所示，构建四个增量索引。以 $r_4$ 为例，q($r_4$) = {⟨tb, 4⟩, ⟨om, 8⟩, ⟨yo, 1⟩, ⟨ou, 2⟩, ⟨ut, 3⟩, ⟨co, 7⟩,⟨be, 5⟩, ⟨ec, 6⟩}。因此有 $g_0$ = tb, $g_1$ = om, $g_2$ = yo 以及 $g_3$ = ut。特别的，$g_3$ = ut 是因为毁掉 $g_3$ 之前的$q$-gram（包括 $g_3$）需要最少 4 个编辑操作并且毁掉在 $g_3$ 前一个$q$-gram (即 ou) 之前的$q$-gram 需要最少 3 个编辑操作。因此分别向 $\mathcal{I}_0^-, \mathcal{I}_1^-, \mathcal{I}_2^-, \mathcal{I}_3^-$ 中插入 {tb}, {om}, {yo}, {ut} 以及向 $\mathcal{I}_0^+, \mathcal{I}_1^+, \mathcal{I}_2^+, \mathcal{I}_3^+$ 分别插入 {tb}, {om}, {yo}, {ou, ut}。

**检索算法**：　检索算法与第5.4节中的相同，除了 (1) 为了选择最优的关键前缀，对查询串 $s$ 中每个$q$-gram $g$，它的标记权重是 $\sum_{i=0}^{\tau} |\mathcal{I}_i^+(g)|$; (2) 对于每个在前缀 pre($s$) 中的$q$-gram，使用索引 $\mathcal{I}_i^-$，对每个在关键前缀 piv($s$) 中的$q$-gram，使用索引 $\mathcal{I}_i^+$，

① ED-Join[30] 提出了一个解决方案来取得第一个$q$-gram $g_i$ 使得至少需要 $i+1$ 个编辑操来毁掉所有在 $g_i$ 之前的$q$-gram。

② 也把 $g_0$ 插入到索引 $\mathcal{I}_0^+$ 中。

表 5.4　实验数据集

| 数据集 | 数据集大小 | 平均长度 | 最长长度 | 最短长度 |
|---|---|---|---|---|
| Title | 4,000,000 | 100.6 | 386 | 54 |
| Title Query | 4,000 | 100.08 | 307 | 54 |
| DNA | 2,476,276 | 108.0 | 108 | 108 |
| DNA Query | 2,400 | 100.08 | 108 | 108 |
| URL | 1,000,000 | 28.03 | 193 | 20 |
| URL Query | 1,000 | 28.07 | 68 | 20 |

其中 $0 \le i \le \tau$。例如，考虑表5.1中的查询串 $s$，当阈值 $\tau = 2$ 时，如图5.5所示，计算 $\text{pre}(s)$ 和 $\text{piv}(s)$。$\text{pre}(s)$ 中的五个前缀$q$-gram 的权重分别是 $0, 1, 2, 0, 0$。探测增量倒排索引并得到两个候选字符串 $r_4$ 和 $r_5$。对齐过滤将剪枝掉候选字符串 $r_5$，最终将得到一个结果 $r_5$。

## 5.8　实验

本节通过实验评估本章提出的技术的性能和可扩展性。实验与四种现有最好的方法 Flamingo, AdaptJoin, Qchunk 和 $\text{B}^{ed}$tree 做了比较。实验从 AdaptJoin 和 $\text{B}^{ed}$tree 的作者处获得了这两个方法的源代码，从 “Flamingo'' 项目主页上下载了 Flamingo 的源代码[①]。实验自行实现了 Qchunk。所有的算法都是用 C++ 语言实现的，在 g++ 4.8.2 编译器上进行编译并开始-O3 编译优化选项。所有的实验都是在一台运行 Ubuntu Server 12.04 LTS 操作系统的服务器上进行的，该服务器配备一个 Intel Xeon E5-2650 2.00GHz 处理器和 16GB 内存。

**实验数据集**: 在实验中使用了三个真实数据集 PubMed[②], DNA[③]和 URL[④]。PubMed 是一个医药发表刊物数据集，使用 4,000,000 个文章题目作为数据集并随机选取了其中 4,000 条作为查询串。DNA 数据集包含 2,476,276 条 DNA 序列，随机选取了 2,400 条序列作为查询串。URL 是一个超链接数据集，它有 1,000,000 条记录，随机选取其中 1,000 条作为查询串。表 5.4 展示了实验数据集的详细信息。实验调整参数 $q$ 以取得算法最好的效率，对于所有算法，在 PubMed 数据集下，当 $\tau = 2, 4, 6, 8, 10, 12$ 时，实验分别设置 $q = 8, 8, 6, 6, 4, 4$；在 DNA 数据集下，当 $\tau = 2, 4, 6, 8, 10, 12$ 时，实验分别设置 $q = 12, 12, 12, 10, 9, 8$；在 URL 数据集下，当 $\tau = 1, 2, 3, 4, 5$ 时，实验分别设置 $q = 6, 3, 3, 2, 2$。

① http://flamingo.ics.uci.edu/
② http://www.ncbi.nlm.nih.gov/pubmed
③ http://www.ncbi.nlm.nih.gov/genome
④ http://www.sogou.com/labs/dl/t-rank.html

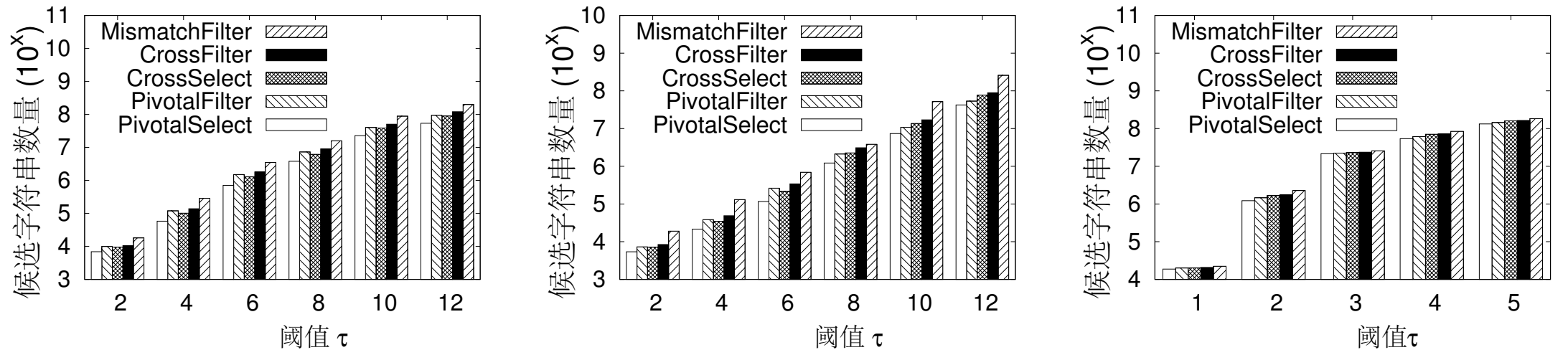

(a) Title(平均长度: 100)　(b) DNA Log(平均长度: 108)　(c) URL(平均长度: 28)

图 5.6　候选字符串数量：评估关键前缀过滤和最优关键$q$-gram 选取

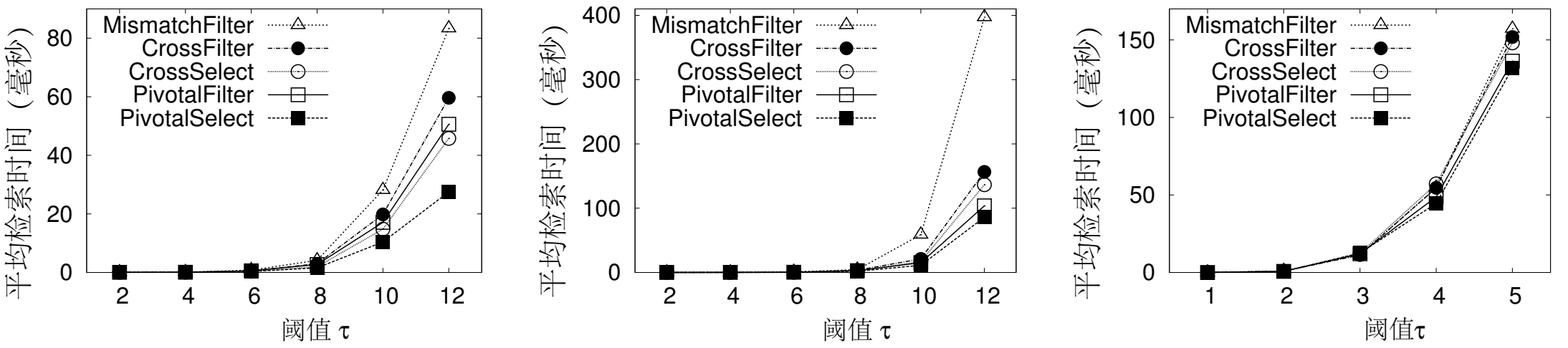

(a) Title(平均长度: 100)　(b) DNA Log(平均长度: 108)　(c) URL(平均长度: 28)

图 5.7　效率：评估关键前缀过滤和最优关键$q$-gram 选取

### 5.8.1　评估关键前缀过滤技术

本节评估过滤技术以及最优关键$q$-gram 选取技术的效率和有效性。实验实现了以下五个方法：(1) CrossFilter 只使用了交叉前缀过滤 (基于定理5.1)。(2) PivotalFilter 只使用了关键前缀过滤（基于引理5.2）。(3) CrossSelect 使用了交叉前缀过滤和最优选取算去来选取关键$q$-gram。(4) PivotalSelect 使用了关键前缀过滤和最优选取算法来选取关键$q$-gram。(5) Mismatch 是现有最好的失配过滤[30]。在验证阶段，对于所有的方法都使用动态规划算法 (配置了二维数组中每行只验证 $\tau+1$ 个格子的优化[65]) 来验证候选字符串。实验在三个数据集上测试了它们的候选字符串数量以及平均检索时间，图5.6-5.7展示了该实验结果。

从图5.6中可以看到 CrossFilter 比 Mismatch 好，CrossSelect 的候选串的数量比 CrossFilter 少。PivotalFilter 和 PivotalSelect 进一步减少了候选字符串的数量。例如在 DNA 数据集下，当编辑距离阈值 $\tau = 8$ 时, Mismatch 有九百万候选字符串，而 CrossFilter 只有三百万；CrossSelect 进一步减少到两百万，PivotalSelect 的候选字符串数量最少，只有一百万。这是因为最优关键$q$-gram 选取技术选取了最小倒排列表长度的关键$q$-gram，因此可以减少候选字符串的数量。而交叉前缀过滤和关键前缀过滤从前缀中移除了不必要的$q$-gram，因此可以减少候选字符串的数量。PivotalSelect 把这两个技术结合在一块，可以进一步减少候选字符串数量。

实验还测试这五个方法的了平均检索时间，图5.7展示了该实验结果。可以看

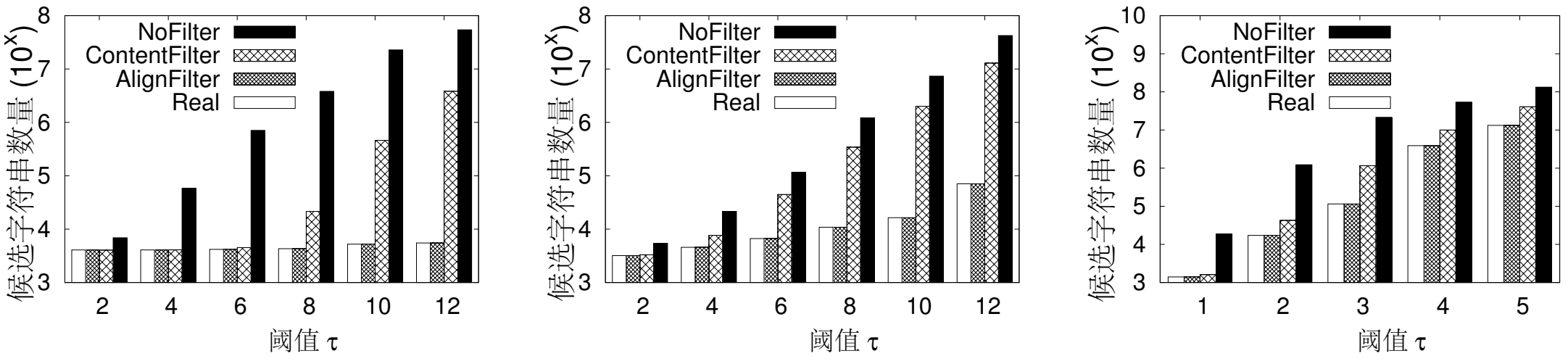

(a) Title(平均长度: 100)　(b) DNA(平均长度: 108)　(c) URL(平均长度: 28)

图 5.8　候选字符串数量：评估对齐过滤

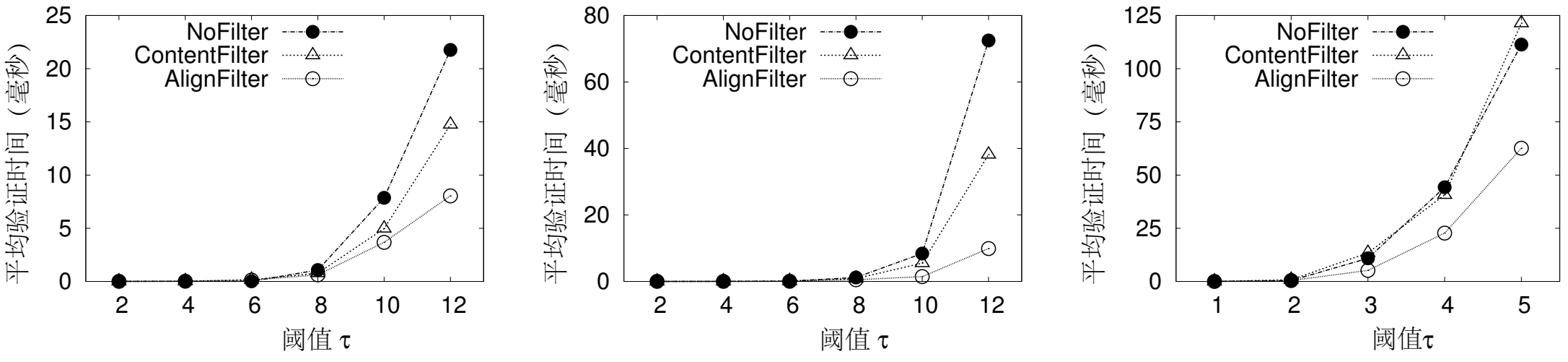

(a) Title(平均长度: 100)　(b) DNA(平均长度: 108)　(c) URL(平均长度: 28)

图 5.9　效率：评估对齐过滤

到 CrossFilter 的平均检索时间比 Mismatch 小，而 CrossSelect 和 PivotalFilter 的效率比 CrossFilter 高。PivotalSelect 取得了最高的效率。例如，在 PubMed 数据集下，当编辑距离阈值 $\tau = 12$ 时，Mismatch 和 CrossFilter 的平均检索时间分别是 82.5 毫秒和 60 毫秒，而 CrossSelect 和 PivotalFilter 的平均检索时间分别是 45 毫秒和 50 毫秒。PivotalSelect 进一步减少平均检索时间到 25 毫秒。这是因为 PivotalSelect 不但能够过滤大量的候选字符串，同时也能降低过滤代价。

### 5.8.2　评估对齐过滤

本节评估对齐过滤并和内容过滤[30] 做了比较。实验实现了以下三种算法：(1) NoFilter 利用第5.8.1节讨论的动态规划算法来验证候选字符串，(2) AlignFilter 首先使用了对齐过滤，然后使用 NoFilter 算法来验证候选字符串，(3) ContentFilter 首先使用了内容过滤，然后使用 NoFilter 算法验证候选字符串。实验使用 PivotalSelect 来产生候选字符串。比较了它们的平均验证时间和候选字符串数量，图5.8-5.9 展示了这些实验结果。

图5.8中的柱子 Real 表示真正与查询串近似的数据字符串的数量。与 NoFilter 相比, ContentFilter 减少了 1-2 个数量级的候选字符串数量。此外，AlignFilter 显著的优于 ContentFilter （2-4 个数量级）AlignFilter 的候选字符串数量与真实结果的数量非常接近。例如，在 PubMed 数据集下，当编辑距离阈值 $\tau = 12$ 时，NoFilter 有五千

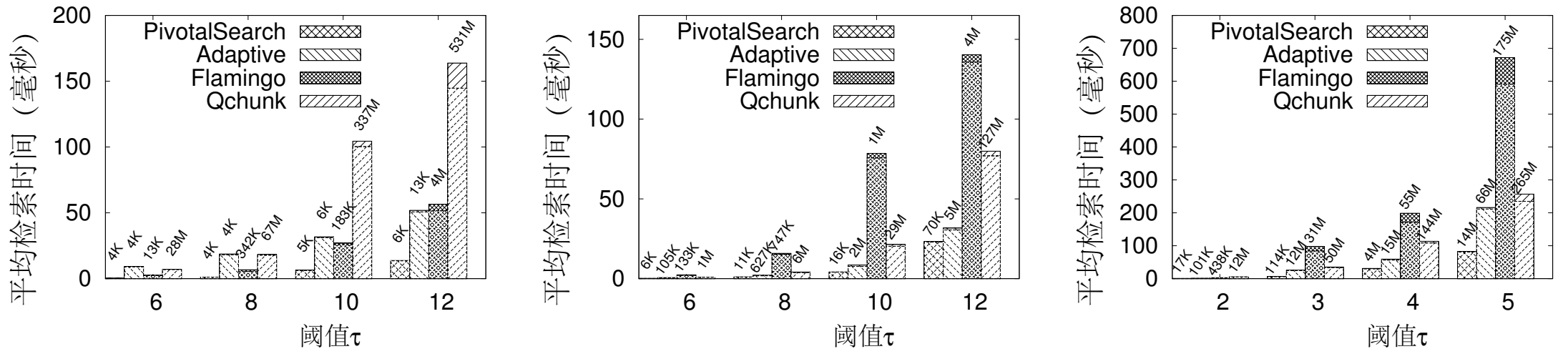

(a) Title(平均长度: 100)　(b) DNA(平均长度: 108)　(c) URL(平均长度: 28)

图 5.10　与现有最好方法的比较 (柱子上的数字代表候选字符串数量)

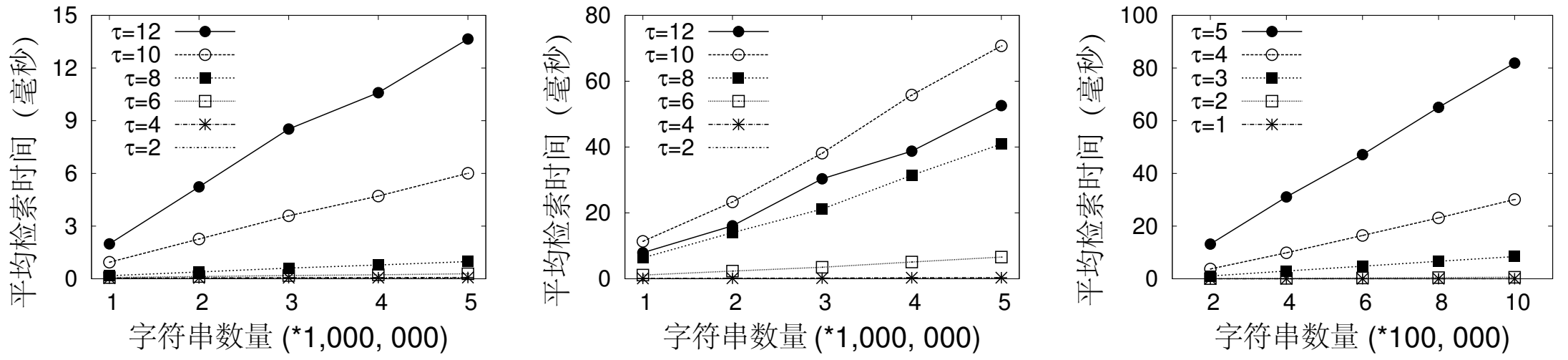

(a) Title(平均长度: 100)　(b) DNA(平均长度: 108)　(c) URL(平均长度: 28)

图 5.11　可扩展性

四百万个候选字符串，ContentFilter 有三百万个候选字符串，AlignFilter 进一步减少到 5,500 个，这和最后结果的数量 4,400 非常接近。这是因为对齐过滤能够有效的探测在低频的关键$q$-gram 上的连续编辑错误从而剪枝掉大量的不相似字符串，而 ContentFilter 只考虑了查询串和候选字符串中不同字符的数量。

实验比较了它们的平均验证时间，图5.9展示了这一实验结果。从图中可以看到 AlignFilter 比 NoFilter 的效率高 2-4 倍，比 ContentFilter 高 1-2 倍。例如，在 DNA 数据集下，当编辑距离阈值 $\tau = 12$ 时，AlignFilter 消耗了 10 毫秒，NoFilter 消耗了 72 毫秒，而 ContentFilter 消耗了 40 毫秒。这是因为对齐过滤只检查了不匹配的关键$q$-gram 上的编辑错误，这节省了很多验证代价，而 ContentFilter 需要扫描整个字符串。

### 5.8.3　与现有最好方法的比较

本节把算法 PivotalSearch （它在过滤阶段使用 PivotalSelect 算法，在验证阶段使用 AlignFilter 算法）同现有最好的算法 Qchunk, Flamingo 和 AdaptJoin 在三个数据集上做比较。实验报告平均检索时间以及候选字符串数量。图5.10展示了该实验结果，图中柱子表示平均检索时间，包括过滤时间（较矮的柱子）和验证时间（较高的柱子）。每个柱子上的数字代表候选字符串的数量。在平均检索时间上，PivotalSearch 在所有数据集上都取得了最好的结果，并且比其他现有的算法效率

高 2 到 10 倍。例如，在 URL 数据集下，当编辑距离阈值 $\tau = 5$ 时，AdaptJoin, Qchunk, Flamingo 和 PivotalSearch 的平均检索时间分别是 215 毫秒, 256 毫秒, 671 毫秒和 82 毫秒。这是因为的关键前缀过滤技术能够减少特征$q$-gram 的数量，最优关键$q$-gram 选择算法能够选取高质量的$q$-gram，对齐过滤能够检测连续的编辑错误。

本章提出的方法在过滤阶段的效率仍然是最高的。例如，在 DNA 数据集下，当编辑距离为 $\tau = 12$ 时，PivotalSearch, AdaptJoin, Flamingo 和 Qchunk 的平均过滤时间分别是 25 毫秒, 30 毫秒, 135 毫秒和 80 毫秒。这是因为 PivotalSearch 的过滤代价比现有的方法要小（见表5.3）。在验证阶段，如第5.8.2节所述，本章提出的方法也比现有的方法效率要高。

PivotalSearch 产生的候选字符串数量是最少的，其数量比其他方法要少 1 到 2 个数量级。例如，在据集下，当编辑距离阈值为时，PivotalSearch 产生了十一万的候选字符串，然而 AdaptJoin, Flamingo 和 Qchunk 分别产生了一千二百万，三千一百万和五千万的候选字符串。这是因为的关键前缀过滤技术和对齐过滤技术对离散的编辑错误和连续的编辑错误都十分的有效，这两个技术能够过滤掉绝大部分的不相似字符串。

### 5.8.4　可扩展性

本节评估 PivotalSearch 的可扩展性，实验调整数据集的大小但是使用相同的查询串。图5.11展示了该实验结果。可以看到本章提出的方法的可扩展性在这三个数据集上都很好。例如，在 PubMed 数据集下，当编辑距离阈值为 $\tau = 10$ 时，把数据集的大小从一百万调整到五百万，其平均检索时间分别是 1 毫秒，2.3 毫秒，3.5 毫秒，4.8 毫秒和 6 毫秒。这得益于有效的过滤技术，它能够剪枝掉数据集中很多与查询串不相似的字符串。

## 5.9　本章小结

本章研究了在编辑距离限制下基于序列的近似检索问题。论文提出了关键前缀过滤技术。相较现有的方法，它能够显著减少特征的数量。论文证明了关键前缀过滤比现有的过滤技术过滤代价更少，剪枝能力更强。根据关键前缀过滤，论文设计了一个过滤加验证的算法。论文提出了动态规划算法来选取最优关键前缀。为了检测连续编辑错误，论文提出了对齐过滤技术。它能够剪枝掉大量由于连续编辑错误而不相似的数据串。在真实数据集上的实验结果显示，论文提出的方法的性能显著高于现有的方法。

# 第 6 章　总结与展望

## 6.1　本文工作总结

论文研究了大数据处理中的容错技术，提出了统一的框架来支持不同类型的相似函数下的近似抽取问题；提出了基于划分的方法来解决序列近似连接和集合近似连接；提出了关键前缀过滤技术来支持序列近似检索。实验结果表明论文提出的算法和技术明显优于现有最好的方法。

论文首先研究了近似实体抽取问题。论文提出了一个统一的框架来支持不同的相似函数。论文设计了基于堆的过滤算法来从文档中高效的抽取与实体近似的子字符串。论文研究了一个基于单堆的算法，通过在文档中的标记所对应的倒排列表上构建一个堆并扫描这些倒排列表一次，它能够共享不同子字符串重叠部分的计算。论文还提出了几个剪枝技术来快速剪枝掉大量的不必要候选元组。论文设计了基于二分查找的技术来提高性能。实验结果表明论文提出的方法取得了很好的性能，并且明显优于现有最好的方法。

论文接下来研究了序列近似连接问题。论文提出了一个基于划分的方法来解决该问题。论文首先对字符串排序，然后按顺序访问它们并把它们平均切分为多个不相交的片段。论文基于已经访问过的字符串构建倒排索引。对每个字符串，论文选择它的一部分子字符串并利用选取的子字符串和倒排索引来寻找近似元组。论文提出了位置敏感的方法和多段匹配的方法来选取子字符串，论文证明了多段匹配选取方法可以最小化被选取的子字符串数量。论文基于长度之差一定大于编辑距离的观察设计了一个高效的验证算法来验证候选元组。论文提出了基于扩展的方法和在不同的前缀上共享计算的方法来进一步提高验证算法的性能。论文讨论了如何在 MapReduce 框架下执行论文的算法。实验表明论文的方法在短字符串和长字符串上都比现有的方法好。

论文之后研究了集合近似连接问题，并设计另一个基于划分的框架。论文设计了一个基于全集划分的方法把所有集合划分为多个片段。论文为这些片段生成 1-删集。然后论文为这些子集与 1-删集构建倒排索引。对每个集合，论文访问它的一部分子集与 1-删集所对应的倒排列表来寻找近似集合。论文研究了如何衡量不同的子集与 1-删集的分配策略并设计了一个动态规划算法来选择最优的分配策略。为了加速分配策略选择过程，论文设计了一个近似比为 2 的贪心算法来选取分配策略。论文还提出了多长度分组机制来进一步加速分配策略选择。这些技术一起把为一个大小为 $s$ 的集合选择分配策略的时间复杂度从 $O(s^3)$ 降低到 $O(s \log s)$。

论文还讨论了如何扩展论文的方法支持异连接以及如何在 Spark 上运行论文的算法。实验表明，论文提出的方法比现有的方法要好。

论文最后研究了在编辑距离限制下的基于序列的近似检索问题。论文提出了关键前缀过滤技术。相较现有的方法，它能够显著减少特征的数量。论文证明了关键前缀过滤比现有的过滤技术过滤代价更少，剪枝能力更强，根据关键前缀过滤，论文设计了一个过滤加验证的算法。论文提出了动态规划算法来选取最优关键前缀。为了检测连续编辑错误，论文提出了对齐过滤技术。它能够剪枝掉大量由于连续编辑错误而不相似的数据串。在真实数据集上的实验结果显示，论文提出的方法的性能显著高于现有的方法。

## 6.2　未来工作展望

论文提出了多种技术来容忍大数据处理中的错误，虽然相比现有最好的方法能够取得更高的性能，但是还存在着很多挑战性问题。在未来的工作中，可以着重探索以下问题。

首先，在 MapReduce 或者 Spark 上运行基于划分的方法来解决近似连接问题依然存在一定的数据倾斜问题，导致整个程序阻塞在某个任务上，所以探索如何优化片段选取目标函数来减轻该问题是未来重要的研究方向。

其次，虽然前缀过滤技术可以支持集合近似检索问题，然而由于前缀过滤固有的缺点，在某些特定的数据集上（例如，社交网络数据集）它们的效率十分低下。因此，研究高效、通用的集合近似检索算法也是未来研究的方向。

最后，现有的解决近似连接问题的技术的最坏时间复杂度依然是平方级的，如何降低近似连接问题的时间复杂度也是未来的一个研究方向。

# 参考文献